\documentclass[a4paper,12pt,times,numbered,print,index]{Classes/PhDThesisPSnPDF}

\input{Preamble/preamble}

\title{Yang-Baxter integrable open quantum systems}

\author{Chiara Paletta}

\university{Trinity College Dublin}

\supervisor{Dr. Marius de Leeuw}

\degreetitle{Doctor of Philosophy}

\subject{LaTeX} \keywords{{LaTeX} {PhD Thesis} {Engineering} {University of
Cambridge}}

\ifdefineAbstract
\fi

\ifdefineChapter
\fi

\begin{document}

\frontmatter

\maketitle


\begin{dedication} 

\begin{flushright}
Alle mie care nonne Chiara e Wanda\\
\vspace{1cm}

To my grandmothers
\end{flushright}
\end{dedication}


\begin{abstract}

The main result of this thesis is the use of the  {boost operator} to develop a systematic method to construct  {new integrable spin chains} with nearest-neighbour interaction and characterized by an $R$-matrix of non-difference form. This method has the advantage of being more feasible than directly solving the Yang-Baxter equation. We applied this approach to various contexts, in particular, in realm of open quantum systems, we achieved the first classification of integrable Lindbladians. These operators describe the dynamics of physical systems in contact with a Markovian environment.  Within this classification, we discovered well-known models such as the Hubbard model, as well as a novel deformation of the Hubbard model spanning three sites of the spin chain. We extensively analyzed this range 3 model and uncovered the existence of multiple Non Equilibrium Steady States (NESS). The presence of multiple NESS is connected to the existence of certain "hidden strong symmetries" in the form of quasi-local charges.  We computed the NESS analytically in the form of  a Matrix Product Operator. Additionally, we applied our method to classify models with $\alg{su}(2)\oplus \alg{su}(2)$ symmetry. In this class of models, we recovered the matrix part of the $S$-matrix of AdS$_5\times$S$^5$ derived by requiring centrally extended $\alg{su}(2|2)$ symmetry. We also obtained five models that, to the best of our knowledge, are new. Furthermore, we focus on spin $1/2$ chain. We demonstrated that all Hermitian integrable Hamiltonians can be reconducted to models of 8-Vertex type. We classified models in this class and we found two difference form models that correspond to the known XXZ and XYZ spin chain. We also found two models of non-difference form type: the one of 6-vertex type is a  known model, while the one of 8-vertex type is a newly discovered model. Furthermore, we showed that the two non difference form models satisfy the free fermion condition. This enables us to express the transfer matrix associated to some of the models in a diagonal form, simplifying the computation of the eigenvalues and the eigenvectors.

\end{abstract}


\begin{acknowledgements}      

First, I would sincerely thank my supervisor Marius de Leeuw for his continuous support during all my PhD and his invaluable advices. His ideas were always a guidance for interesting projects. I have really appreciated his support and kindness at all times, especially during the hard times of the Coronavirus, he made me feel not alone and truly part of a research group. Thanks!
\\
A big thank also to Balázs Pozsgay, for introducing me to the interesting world of open quantum system and for sharing his knowledge and expertise. Thanks for always bringing new and fascinating ideas!
\\
I am also thankful to Marco Rossi and Alessandro Torrielli, for their constant guidance in both research and life. They gave me the opportunity to approach the world of research, and I thank them for this.
\\
I will never be thankful enough to Ana Retore for taking the time to reply to all the (many) questions I had. Thanks also for the valuable comments on this thesis.
\\
I also really enjoyed working with all my other research collaborators Anton Pribytok, Paul Ryan and Eric Vernier for all the interesting discussions, it has been a real pleasure working with all of you! Thanks!
\\
I am also deeply thankful to the professors of the School of Maths, in particular: Tristan McLoughlin,  Fabrizio Nieri and Sergey Frolov, for their advices and  for the discussions. And also thanks to all the students and postdocs of the school of maths, it has been a lots of fun having fika, pizza nights and playing board games!
\\I would like to thank for discussions Berislav Buca, Benjamin Doyon, Ben Hoare, Juan Miguel Nieto Garcia, John Goold, Enej Ilievski,  Shota Komatsu, Andrea Nava, Rafael Nepomechie, Lorenzo Piroli, Vladislav Popkov and  Tomaž Prosen.
\\
I also would like to thank Sergey Frolov and Shota Komatsu for both agreeing to be part of my thesis committee and providing invaluable feedback on this work.

\end{acknowledgements}


\tableofcontents




\printnomenclature

\mainmatter


\chapter*{Introduction}
\ifpdf
    \graphicspath{{Chapter0/Figs/Raster/}{Chapter0/Figs/PDF/}{Chapter0/Figs/}}
\else
    \graphicspath{{Chapter0/Figs/Vector/}{Chapter0/Figs/}}
\fi
\addcontentsline{toc}{chapter}{Introduction}
\markboth{Introduction}{}
The study of the general behaviour of complicated non-linear systems is of interest for both physicists and mathematicians. In certain models, the presence of numerous \textit{conserved quantities} restricts their dynamics and allows for \textit{exact} solutions. While this is expected in isolated systems of non-interacting particles, it is remarkable that this phenomenon can also occur in certain \textit{interacting} theories. There are various definitions of integrability \cite{caux2011remarks}, but in this thesis, we focus on quantum models characterized by an $R$-matrix solution of the \textit{Yang-Baxter equation}. The significance of integrability lies in its ability to often provide a general mathematical approach to determine the dynamics of the system,  \cite{torrielli2016classical}.
\paragraph{Historical Remarks} Integrable systems and classical mechanics developed together as scientists aimed to find precise solutions to Newton's equations of motion. Newton's discovery of the exact solution to the Kepler's problem was remarkable, but only a few other models had exact solutions at that time. In the 19th century, Liouville categorized Hamiltonian systems as either integrable or non-integrable, where integrable referred to systems whose equations of motion could be solved using quadratures. While integrable models had been known for many years, the development of systematic methods to study them is relatively recent.
\\
\textit{Classical} integrable models can be studied via the \textit{Inverse scattering method}. This method was introduced in the 1960s by Gardner, Greene, Kruskal, and Miura, as well as Lax, Zakharov, and Fadeev. Initially, it was employed to investigate completely integrable models in classical mechanics, such as the Korteweg-de Vries equation in fluid mechanics. Over the subsequent years, the method was expanded to include certain relativistically invariant models. In the following decade, Faddeev, Korepin, Kulish, Reshetikhin, Sklyanin, Semenov Tian-Shansky, and Takhtajan began developing the \textit{quantum} version of this method.
\\
In the realm of quantum mechanics and statistical physics, various techniques have been developed by physicists such as Bethe, Onsager, Baxter, Zamolodchikov, and many others. They have played a crucial role in advancing our understanding of quantum systems and their behavior.
\\
In 1931, the young postdoc Hans Bethe made a pioneering contribution to the field of quantum integrability, \cite{Bethe:1931hc}. At that time, the general formalism for non-relativistic quantum mechanics was still being developed.  Bethe directed his attention to a specific model known as the Heisenberg spin chain\footnote{A quantum spin chain is a one-dimensional lattice, where operators can act on each site of the chain.}, nowadays considered the prototype of integrable models,  whose Hamiltonian reads
\begin{align}\label{heisspinchain}
&\mathbb{H}=\sum_l \Big( \frac{1}{4}-\vec{S}_l \cdot \vec{S}_{l+1}\Big),\,\,\,\,\,\,\,\,\,\,\,\text{with}\,\,\,\,\,\,\,\,\vec{S}_l=\frac{1}{2}\vec{\sigma}_l\,,
\end{align}
where $\vec{\sigma}_l$ are the three Pauli matrices. He was able to find the exact solution for this model by guessing the expression of the "wave function"  $\ket{\psi}$ of the system, which corresponds to the eigenstates of the spectral problem
\begin{align}
&\mathbb{H}\ket{\psi}=E\ket{\psi},
\end{align}
where $E$ are the Hamiltonian eigenvalues. Bethe's discovery of the exact solution to the spectral problem in the Heisenberg spin chain, nowadays known as the \textit{Bethe ansatz}, has had a profound impact on condensed matter theory and mathematical physics. This method has proven to be highly versatile and applicable to a wide range of diverse integrable problems.
\\
When approaching a new model, it is important to determine whether it is integrable or not. However, determining the integrability of a given model is not a straightforward task. In this thesis, we analyze this question and aim to provide a clearer understanding of the criteria and characteristics that define integrability in different systems. We focus on quantum spin chain models.
\paragraph{Classification of integrable models}Over the course of the history of quantum integrable systems, various approaches have emerged for finding solutions to the Yang-Baxter equation and, as a result, discovering integrable models. One fruitful approach in the early days was to impose specific symmetries on the solutions, \cite{Kulish:1981gi,Kulish1982}. However, finding models that do not exhibit the symmetries is a challenging task and, as time passed, different groups analyzed this problem. This lead to the discovery of a substantial number of new integrable models\footnote{More details about the available methods  will be given in chapter \ref{classificationchapter}.}. Considering the importance of integrable models, the main focus of this thesis is on the use and the \textit{development of a new method} to classify and discover integrable models in different contexts. This method makes use of the  \textbf{boost operator}, \cite{HbbBoost,Loebbert:2016cdm} to generate higher conserved charges and has the advantage of being more versatile compared to many of the other available methods\footnote{In chapter \ref{classificationchapter}, we state the pros and cons of each of the available methods and we compare them with the ones of the boost method.}.  Specifically\footnote{In chapter \ref{classificationchapter}, we list all the contexts in which the classification was performed.}, we conducted a classification of all integrable spin chains of 8-vertex type, resulting in the discovery of new solutions that encompass the $S$-matrices of the AdS$_2$ and AdS$_3$ integrable models as special cases. Although these models are very interesting, they are not the central focus of the thesis. We primarily concentrate on applying the method to the realm of \textit{open quantum systems}. This exploration leads us to the discovery of new interesting models, for instance, the \textbf{first range three deformation of the Hubbard model}.
\paragraph{One possible application of the boost method: Open quantum systems}We start by considering a total closed physical system, so that its dynamics can be represented in terms of a unitary time evolution. We are primarily concerned with studying a subsystem of the total system, which we refer to as the "system". Consequently, the total system is composed of the system of interest (the orange portion in the figure) and the surrounding environment (the blue one).

\begin{figure}[h!]
\begin{center}
	\includegraphics[height=7.cm]{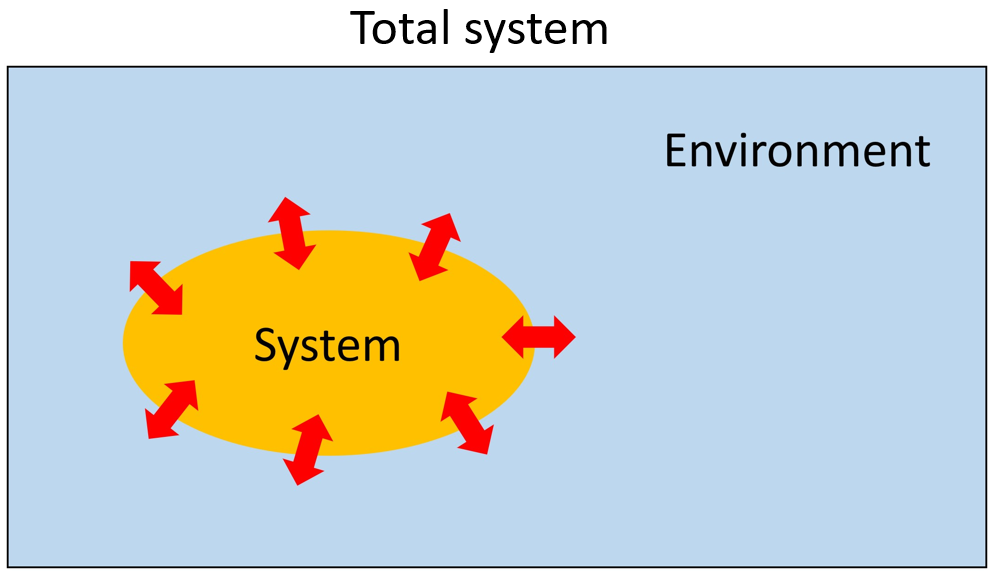}
\end{center}
	\end{figure}
In introductory physics courses, the importance of the environment is often disregarded, resulting in an idealized depiction of natural phenomena. In that case, the system is closed. However, in reality, the environment has a substantial influence, giving rise to non-trivial dynamics. The theory of open quantum systems  aims to develop a general framework to analyse the dynamical behaviour of the systems by removing the environmental degrees of freedom.\\
A complete description of the system-environment interaction is in general a very hard problem. Under some approximations, the dynamics become more tractable and it is described via the Markovian \textbf{Gorini-Kossakowski-Sudarshan-Lindblad} (GKSL) master equation, \cite{lindbladoriginal,gorinioriginal}. This equation, also known as the Lindblad equation, was independently derived by Lindblad and by Gorini, Kossakowski, Sudarshan around 1975. It takes the following form
\begin{equation}
\dot{\rho}(t)=i[\rho(t),h]+\sum_a^N \gamma_a \left[
    \ell_a\rho(t) \ell_a^\dagger-\frac{1}{2}\{\ell^\dagger_a \ell_a,\rho(t)\}\right],
\end{equation}
where $\rho$ denotes the reduced\footnote{The reduced density matrix is obtained by tracing out the degrees of freedom of the environment from the total system+environment density matrix.} density matrix of the system, $h$ is the Hamiltonian of the system, $\ell_a$ describe the effective action of the environment on the system, typically involving multiple reservoirs, and $\gamma_a$ represents the corresponding coupling constants between each reservoir and the system.
\\This equation is only valid under the \textit{Markovian approximation}\footnote{The validity of this approximation is discussed in more details in section \ref{derivlindbladeq}.}. This approximation assumes that the memory effects of the environment on the system can be neglected, meaning that the system's evolution depends solely on its present state and not on its past history. A relevant role is played by the correlation time of the environment, the measure on how quickly the property of the enviroment changes. If this time is significantly shorter than the characteristic time scale of the system's dynamics, then the approximation holds.
It is worth noting that the Markovian approximation may not be applicable in all cases, particularly when the system-environment interaction is strong or when the environment possesses long-range correlations. In such cases, non-Markovian effects may become significant, requiring more advanced theoretical frameworks to describe the dynamics of such open quantum systems. In this thesis, we will not consider this type of system. Our focus is primarily on the study of systems where the Markovian approximation is applicable.

\paragraph{Why do we study integrable open quantum systems?} Studying open quantum systems is intriguing for several reasons.
First, it is essential for addressing fundamental aspects of quantum physics, including the phenomenon of decoherence \cite{zurek2003decoherence}, which is critical for quantum technologies like quantum computers. Decoherence disrupts quantum properties such as superposition and entanglement due to interactions with the environment. Additionally, calculating non-equilibrium steady states (NESS) \cite{jakvsic2002mathematical} in quantum statistical mechanics offers insights into complex systems' behavior far from equilibrium.
\\
Moreover, due to the vast range of applications, investigating the dynamics of many-body systems is both captivating and challenging. Over the past few decades, significant progress has been made through the utilization of perturbative methods and numerical analyses. However, \textit{exact solvable} cases have gained attention in recent years. These cases allow for the study of out-of-equilibrium dynamics using exact methods, which not only contribute to the development of approximate methods but also serve to validate their precision.\\
In the literature of open quantum systems, the meaning of solvability is various. We provide a list of some of them
\begin{itemize}
\item models solvable by free fermion techniques, \cite{prosen2008third,vernier2020mixing};
\item models for which the full spectrum can be constructed, those are characterized by an evolution operator of upper triangular form, \cite{marko-lindblad-1};
\item open spin chain with the environment contribution acting on the boundary and allowing for the exact construction of the NESS, \cite{prosen-boundary-lindblad-1};
\item model for which the evolution operator is integrable in different subspaces, \cite{essler-piroli-lindblad};
\item Yang-Baxter integrable models: models that can be associated to an $R$-matrix solution of the Yang-Baxter equation, \cite{medvedyeva2016exact,ziolkowska2020yang}.
\end{itemize}
In this thesis, the focus is on the class of solvable models that are \textbf{Yang-Baxter \textit{integrable}}. The advantage of working with integrable models is that the NESS and the relaxation towards them can be computed with exact methods and the generator of the dynamics can be diagonalized involving some of the available technique, for example the Bethe Ansatz. \\
Furthermore, integrable models have played an important role in understanding the non-equilibrium dynamics in \textit{isolated} quantum systems. Isolated non-integrable many particle systems are not very interesting since they relax very quickly toward an equilibrium state. Isolated integrable system, due to the infinite number of conserved charges, equilibrate to the Generalized Gibbs Ensamble (GGE), \cite{rigol-gge} and the dynamics is described by the Generalized Hydrodynamics (GHD), \cite{doyon-GHD}.
\\
In general, when a system interacts with its environment, the property of integrability tends to be lost. However, it is remarkable that certain models exist in which integrability is preserved, allowing for the application of the Bethe ansatz technique. The availability of such integrable models is particularly valuable as they may serve to validate the accuracy of the GGE and GHD in describing the system's time evolution. However, this field of research is very new and still mostly unexplored.
\\Motivated by these reasons, very recently, different groups started to study Lindblad models that can be mapped onto known Yang-Baxter integrable interacting systems. By employing a super-operator formalism\footnote{More details can be found in section \ref{sec:ladder}.}, the Lindbladian generator of the dynamics can be written as a matrix in a higher dimensional space that corresponds to a “two-leg ladder” quantum spin chain. In one of the initial papers \cite{medvedyeva2016exact}, the authors provide a map between the well known Hubbard model\footnote{To be precise, the mapping is between the Hubbard model with an imaginary coupling constant and an open quantum system. In the latter, the system is an XX spin 1/2 chain subject to dissipation. The effective action of the environment on the system is $\ell=Z$, with $Z$ representing the third Pauli matrix. Given the importance of the result, we explicit describe it in chapter \ref{Hubbardchapter}.} and an open quantum system, while in \cite{ziolkowska2020yang} the authors did a similar job for the Hubbard model and some generalization, the Maassarani models.
\\
In this thesis, based on our initial work \cite{classificationlind} of 2020, we initiated the first \textbf{systematic classification of Yang-Baxter integrable Lindblad systems} by using the already mentioned Boost automorphism method.

\newpage

\section*{Contents}

In this thesis, we focus on quantum integrable models defined on a spin chain: a one-dimensional lattice, where each site of the chain is a Hilbert space.
\\
The main result of this thesis is the use of the \textbf{boost operator} to develop a systematic method to construct \textbf{new integrable spin chains} of non-difference form with nearest-neighbour interaction, which has the advantage of being more feasible than directly solving the Yang-Baxter equation. By applying this approach, we were able to achieve the \textbf{first systematic classification of integrable Lindbladians}. These operators describe the evolution of physical systems in contact with a Markovian environment.  Within this class of integrable Lindbladians, we discovered well-known models such as the \textbf{Hubbard model}, as well as its \textbf{first medium-range deformation}\footnote{We use the word "medium" to refer to next-to-nearest-neighbour interactions.}.  These models will be deeply analyzed in the following. We also applied the models to various contexts, for instance, to  {models with $\alg{su}(2)\oplus \alg{su}(2)$ symmetry} and with a non-difference form $R$-matrix.  Furthermore, we focus on spin $1/2$ chain with Hamiltonian of 8-vertex type. We show that all the non-difference form models we found satisfy the free fermion conditions. This enables us to express the transfer matrix associated to some of the models in a diagonal form, simplifying the computation of the eigenvalues and the eigenvectors. These new models\footnote{Although these models are very interesting, they are not the central focus of the thesis. We refer to \cite{deLeeuw:2020ahe} and \cite{de2022integrable} for additional details.} are relevant in the context of AdS/CFT, in particular they correspond to \textbf{integrable deformations} of the \textbf{AdS$_2$ and AdS$_3$ models}.
\\
The thesis is composed of nine chapters:
\begin{itemize}
\item In  chapter \ref{intro}, we introduce \textbf{quantum integrable systems} and how to characterize them. We highlight the importance of the Yang Baxter equation, which serves as the cornerstone of integrability. Having an $R$-matrix solution of the Yang-Baxter equation and the presence of an infinite set of commuting conserved charges are two ways to characterize integrable models. We show the interconnection between those two definitions. We introduce some useful definitions: Lax, monodromy and transfer matrices. We use the Heisenberg spin chain as the prototype of integrable models and we show how a direct diagonalization of the Hamiltonian becomes unfeasible as soon as the volume of the system increases. We explicitly present how to use the Bethe ansatz method, that makes use of the integrability property, to overcome the difficulties.

This chapter is mainly based on the reviews: \cite{analecture,nepomechie1998spin,levkovich2016bethe}. 

\item In chapter \ref{classificationchapter}, our focus shifts towards the classification of quantum integrable models. We begin by providing an overview of the existing methods in the literature for this purpose. We discuss the pros and cons of each approach and provide a list of the different types of models that we discovered. Then, we present the\textbf{ boost automorphism mechanism}: the novel mechanism we used to \textbf{classify integrable models} of \textbf{non-difference form}. By following \cite{HbbBoost,Loebbert:2016cdm}, we derive the boost operator. To illustrate the effectiveness of the boost automorphism mechanism, we provide a simple example that demonstrates how it can be used to classify and discover new integrable models. At the end of the chapter, we list some of the "tricks" we acquired through the experiences. These tricks and strategies have been instrumental in our classification.

This chapter is mainly based on the author's paper: \cite{classificationybandboost}.

\item In chapter \ref{integrableopenquantumsystemchapter}, we use the boost authomorphism method to \textbf{classify integrable} \textbf{open quantum systems}, which represents one of the key contributions of this thesis. We give the derivation of the Lindblad master equation by following the reviews \cite{preskill1999lecture,cappellaro2011quantum}. The dynamical evolution of the density matrix corresponding to the system is described via a Lindblad superoperator $\mathcal{L}$, which we identify as one of the conserved charges of the spin chain. We show, with an example, how to use the boost operator to provide the first systematic approach to classify integrable Lindbladians. We also give the connection between the Lindblad equation and the classical stochastic Markovian system. This connection was known in the literature, but starting from integrable Lindbladians is a new result.

This chapter serves as an introduction to understand the  results presented in the author's paper \cite{classificationlind}.

\item In chapter \ref{Lindbladclassificationmodels}, we present a collection of \textbf{integrable Lindbladians} found through the application of the boost automorphism mechanism. We classify these models into two groups: fine-tuned models and coupled models. One particular model that deserves special attention among the coupled models is model B3. We investigate the Non Equilibrium Steady States and the particle current flowing through these states. Surprisingly, this model serves as an integrable example of the \textit{pumping effect}\footnote{This effect was already observed in other systems, see for example \cite{pumping}. We provide the first case where it is observed in an integrable model.}, where a finite current persists even when the coupling constant is very small. We provide a proof of the equivalence between model B3 and the generalized Toda system associated with the non-exceptional affine Lie algebra $A_3^{(2)}$. To solve this model, we utilize the nested Algebraic Bethe ansatz, and we present the expressions for the eigenvalues and the Bethe equation. Further computational details can be found in Appendix \ref{BAB3chapter}. Lastly, we introduce model B2, which represents an integrable deformation of the Hubbard model. This model serves as bridge to the next chapter. It is worth noting that, with the exception of model B2, all the discovered models are of difference form type.

This chapter is mainly based on the author's papers: \cite{classificationybandboost} for the classification of the models and \cite{de2022bethe} for the analysis on the model B3.

\item In chapter \ref{Hubbardchapter}, we introduce the Hubbard model, a toy model used to describe the motion of electrons in the conduction band of a solid.  We present the Hubbard model in both fermionic and bosonic formulations. Additionally, we explore the relation between the Hubbard model's Hamiltonian and the Lindblad superoperator, given in \cite{medvedyeva2016exact} and we investigate all the cases where this mapping is possible. Furthermore, we introduce a new nearest-neighbour integrable elliptic model and we show how to use the bond site transformation to relate it to a range 3 integrable deformation of the Hubbard model.  We believe that this is the \textbf{first range three integrable deformation} of the \textbf{Hubbard model}. To confirm our findings, we prove the integrability of the 3-site model and highlight the unusual functional dependence of the $R$-matrix. The explicit expression of the $R$-matrix for the nearest-neighbour model is given in Appendix \ref{Rmatrixsite2Hubbard}. Details on the bond site transformation are given in Appendix \ref{bondsitematrix}.

This chapter is mainly based on the author's paper  \cite{deformhubrange3}.

\item In chapter \ref{NESSCHAPTER}, we analyze the symmetry  of the Lindblad superoperator associated with the range 3 deformation of the Hubbard model introduced in the previous chapter. We start by providing a brief introduction on the meaning of conserved quantities and symmetries in non-Hermitian models. Then, we analyze the 3-site model and we discover the presence of multiple NESS. We motivate this multiplicity with the presence of \textbf{hidden strong symmetries} in the form of quasi-local charges.  We \textbf{compute the NESS} exactly in the form of  Matrix Product Operators with fixed bond dimensions and we use this to compute the mean values of some local observables. Furthermore, we prove that the dynamics leads to the emergence of the Gibbs ensemble constructed from the hidden quasi-local charge.  Interestingly,  we did not use the fact that the 3-site model is Yang-Baxter integrable. However, the ``superintegrability'' property of the Hamiltonian \eqref{Hsigma} plays an important role. Details on the computation of the mean values are given in Appendix \ref{meanvalcompexact}.

The introduction of this chapter is based on the paper \cite{lindblad-symmetries}. The main results is based on the author's paper \cite{nessrange3}.

\item In chapter \ref{threeorfour}, we apply the boost automorphism method to classify integrable models where the local Hilbert space is of dimension 4. Since the most general Hamiltonian involves 256 free functions, it is a challenging task to solve the coupled systems of differential equations using current methods. To overcome this, we limit our focus to \textbf{models with $\alg{su}(2)\oplus \alg{su}(2)$ symmetry} and with a non-difference form $R$-matrix. We begin by presenting the procedure for constructing the ansatz for the Hamiltonian with the given symmetry and then we list the integrable models found. This ansatz is motivated by the fact that many interesting known physical models such as the Hubbard model (discussed in chapter \ref{Hubbardchapter}), exhibit this symmetry. 
We discover five models which, based on our current knowledge, are new. Furthermore, we demonstrate that the Hubbard model can be classified within our framework.

The beginning of this chapter is based on the explanation given in \cite{deLeeuw:2019vdb}. The main result is based on the author's paper \cite{classificationybandboost}.

\item In chapter \ref{6and8Vmodel}, we apply the boost automorphism method to classify integrable models where the local Hilbert space is $\mathbb{C}^2$. We restrict our ansatz to Hamiltonians of \textbf{8-vertex type or lower}.  We start with a short introduction on \textit{vertex models} and then we list the models that we found. Two of them can be reduced to a difference form type and are equivalent to the well-known XXZ and XYZ spin chains. The remaining two are of non-difference form: the 6-vertex B can be mapped to the solution A of the paper \cite{6vColored} while the 8-vertex B (to the best of our knowledge) is a \textit{new} model. To motivate the ansatz chosen, we prove that \textit{any 4x4 Hermitian integrable Hamiltonian}  can be transformed into an 8-vertex model. Models 6-vertex B and 8-vertex B are of non-difference form type and contain the AdS$_2$ and AdS$_3$ integrable models as special case. We explore this statement in Appendix \ref{adsappendixdef}.

The introduction of this chapter is based on the book \cite{eckle2019models}. The main result is based on the author's papers \cite{deLeeuw:2020ahe,classificationybandboost}.

\item In chapter \ref{ffconditionchapter}, we prove that the difference form integrable models discussed in chapter \ref{6and8Vmodel}  satisfy the \textbf{Baxter relation}, while the non difference form ones satisfy the \textbf{free fermion condition}. The latter condition is particularly significant as it enables us to express the transfer matrix associated with these models in a diagonal form, simplifying the process of computing eigenvalues and eigenvectors. We have explicitly demonstrated that the free fermion condition enables us to express the transfer matrix associated with the 6-V B model in a diagonal form, simplifying the computation of eigenvalues and eigenvectors.  This was previously hidden in the standard formalism.  Furthermore, we identify the equivalent free fermion conditions for some of the models with $\alg{su}(2)\oplus \alg{su}(2)$ symmetry discussed in chapter \ref{threeorfour}.

This chapter is mainly based on the author's paper \cite{freefermionpaper} .
\end{itemize}

The organizational structure of the thesis and the interrelation among the chapters are presented on the following page.

\begin{sidewaysfigure}[h!]
\begin{center}
	\includegraphics[height=13cm]{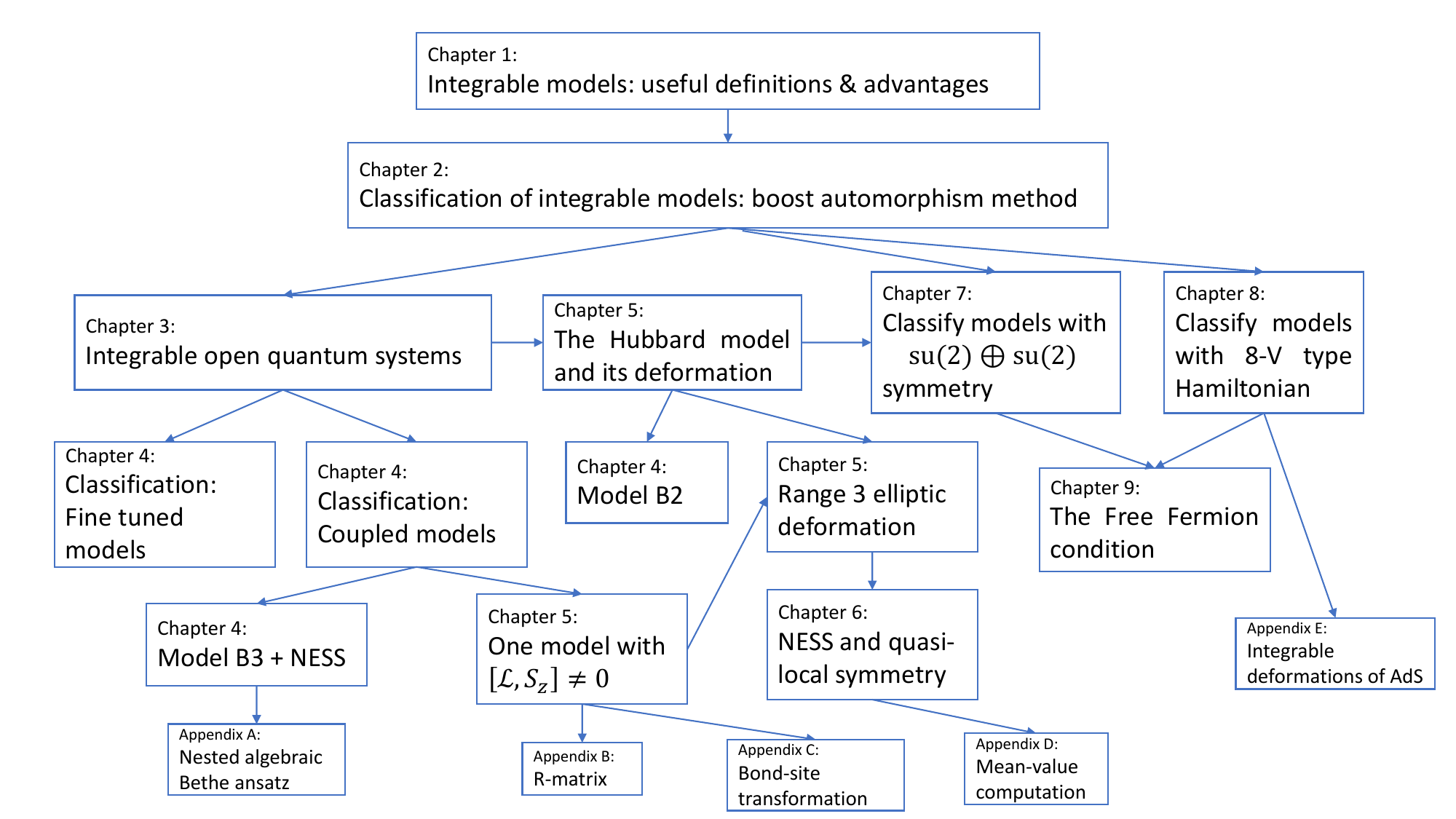}
\end{center}
	\end{sidewaysfigure}



\chapter{What is an integrable model and how can we solve it?}
\ifpdf
    \graphicspath{{Chapter1/Figs/Raster/}{Chapter1/Figs/PDF/}{Chapter1/Figs/}}
\else
    \graphicspath{{Chapter1/Figs/Vector/}{Chapter1/Figs/}}
\fi

\label{intro}

There is not a single, clear-cut definition for quantum integrability, \cite{caux2011remarks}. However, it is commonly accepted to characterize integrable models using the $R$-matrix solution of the Yang-Baxter equation and  the presence of an infinite set of commuting conserved charges. In this chapter, we  explain these two definitions in more detail and demonstrate how they are interconnected. Furthermore, we define some crucial elements like the \textit{monodromy} and the \textit{transfer matrix}, using both analytical and visual explanations to make them easier to understand.\\
Furthermore, we introduce the \textit{Heisenberg spin chain} as an example of an integrable model and we use it to highlight the importance of integrability in understanding the dynamics of a system. We solve the spectral problem of this model via the Algebraic Bethe ansatz technique and we show that this solution would not be achievable for a large system by using a brute force diagonalization.

\section{Useful Definitions}
A quantum integrable model has an infinite number of conserved charges and it is characterized by an $R$-matrix solution of the\textbf{ Yang-Baxter equation}\footnote{Sometimes the Yang Baxter equation can be found in a slightly different form recalling $R_{ij}\to \check{R}_{ij}=P_{ij} R_{ij}$ with $P$ being the permutation operator that swaps the sites $i$ and $j$.} (YBE),  \cite{baxter2016exactly}
\begin{equation}\label{eq:YBE}
R_{12}(\lambda_1,\lambda_2)R_{13}(\lambda_1,\lambda_3)R_{23}(\lambda_2,\lambda_3)=R_{23}(\lambda_2,\lambda_3)R_{13}(\lambda_1,\lambda_3)R_{12}(\lambda_1,\lambda_2).
\end{equation}
This is a matrix relation defined in  $\text{End}(V\otimes V \otimes V)$, with $V\equiv \mathbb{C}^n$ the $n$-th dimensional complex vector space. The $R$-matrix is defined in $\text{End}(V\otimes V)$ and in the indexed version $R_{ij}$, the subscripts denote which of the three spaces $R$ acts on (for example $R_{12}=R\otimes \mathbb{I}$, $\mathbb{I}$ being the identity operator in $\mathbb{C}^n$) and $\lambda_1, \lambda_2, \lambda_3$ are known as spectral parameters, each of them associated to one of the three Hilbert spaces. They can take values in $\mathbb{C}$.
\\
 The YBE is considered the crucial component of integrability (at both classical\footnote{The classical YBE takes a different expression than \eqref{eq:YBE}, we refer to  \cite{torrielli2016classical,analecture} for a detailed explanation.} and quantum levels). The primary focus of this thesis is to \textit{classify} the solutions of the quantum YBE and uncover \textbf{new} models that possess intriguing physical properties.
 \\
In some cases, the $R$-matrix can be interpreted as the \emph{scattering} matrix which serves as an operator linking the initial and final states during a scattering process.  The YBE is also called a \textit{factorization} equation: the total scattering  can be viewed as a series of two-particle interactions, with the order of scattering being irrelevant, \cite{bombardelli2016s}. When a particle interpretation is applicable, the spectral parameters are associated to the momenta of the involved particles.
\\
Pictorially, the $R$-matrix and the YBE can be represented in the following way:
\vspace{0.5cm}
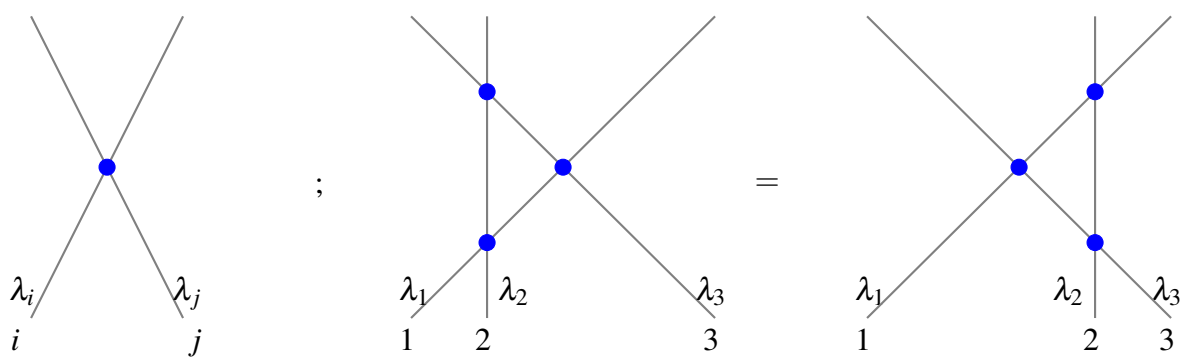
\begin{figure}[h!]
\begin{center}
\hspace*{0.em}\raisebox{-4em}{\begin{tikzpicture}
\draw[gray, thick] (-3,-2) -- (-1,2);
\draw[gray, thick] (-1,-2) -- (-3,2);
\filldraw[blue] (-2,0) circle (3pt) node[anchor=west]{};
\draw (-3,-2) node[anchor=north east] {$i$};
\draw (-0.6,-2) node[anchor=north east] {$j$};
\draw (-2.8,-1.3) node[anchor=north east] {$\lambda_i$};
\draw (-.6,-1.3) node[anchor=north east] {$\lambda_j$};
\draw (1,0) node[anchor=north east] {$;$};
\draw[gray, thick] (2,-2) -- (6,2);
\draw[gray, thick] (2,2) -- (6,-2);
\draw[gray, thick] (3,2) -- (3,-2);
\filldraw[blue] (4,0) circle (3pt) node[anchor=west]{};
\draw (2.2,-2) node[anchor=north east] {$1$};
\draw (3.2,-2) node[anchor=north east] {$2$};
\draw (6.2,-2) node[anchor=north east] {$3$};
\draw (2.4,-1.3) node[anchor=north east] {$\lambda_1$};
\draw (3.7,-1.3) node[anchor=north east] {$\lambda_2$};
\draw (6.3,-1.3) node[anchor=north east] {$\lambda_3$};
\draw (8.2,-2) node[anchor=north east] {$1$};
\draw (11.2,-2) node[anchor=north east] {$2$};
\draw (12.2,-2) node[anchor=north east] {$3$};
\draw (8.4,-1.3) node[anchor=north east] {$\lambda_1$};
\draw (11.0,-1.3) node[anchor=north east] {$\lambda_2$};
\draw (12.3,-1.3) node[anchor=north east] {$\lambda_3$};
\filldraw[blue] (3,1) circle (3pt) node[anchor=west]{};\filldraw[blue] (3,-1) circle (3pt) node[anchor=west]{};
\draw (7,0) node[anchor=north east] {$=$};
\draw[gray, thick] (8,-2) -- (12,2);
\draw[gray, thick] (12,-2) -- (8,2);
\draw[gray, thick] (11,-2) -- (11,2);
\filldraw[blue] (11,1) circle (3pt) node[anchor=west]{};\filldraw[blue] (11,-1) circle (3pt) node[anchor=west]{};\filldraw[blue] (10,0) circle (3pt) node[anchor=west]{};
\end{tikzpicture}}
\end{center}
\caption{Graphical representation of the $R$-matrix (left) and the Yang-Baxter equation (right).}
\end{figure}
\\where to each line we associated a space and a spectral parameter.
\\
From this representation of the $R$-matrix, we can construct more complicated objects like the monodromy matrix and the transfer matrix and also prove some of the identities involving them.
\\
In general, the $R$-matrix depends on two spectral parameters $(\lambda_i,\lambda_j)$ and we can distinguish two possible cases
\begin{itemize}
\item $R(\lambda_i,\lambda_j)=R(\lambda_i-\lambda_j)$ \textit{difference} form
\item $R(\lambda_i,\lambda_j)\neq R(\lambda_i-\lambda_j)$ \textit{non difference} form.
\end{itemize}
Some examples of difference form types of models include the \textit{Heisenberg} spin chain (also known as XXX), as well as the XXZ and XYZ spin chains. In section \ref{examplexxxsp}, we give a definition of these models and we use the \textit{Heisenberg} spin chain, known for its simplicity, as a prototype for integrable models. A very famous model of non-difference form type is the \textit{Hubbard} model, which describes the physics of interacting spin-1/2 fermions on  a lattice and will be extensively examined in Chapter \ref{Hubbardchapter}, along with some newly discovered integrable deformations.

\subsection{Why and how does the YBE define an integrable model?}
\label{definitionsmonodromy}

In this thesis, the models under investigation are constructed on a spin chain: a one-dimensional lattice. Operators can act on each site of the chain. When discussing an observable $A$, the notation $A_j$ specifies the site(s) where the observable acts non-trivially
\begin{align}
A_j=\mathbb{I}\otimes\dots\otimes \underbrace{A}_{j^{th}\,\text{site}} \otimes\dots\otimes\mathbb{I}.
\end{align}
The index $j$ can refer to a single site or multiple sites simultaneously. $L$ is the dimension of the spin chain, also referred as length or volume, the total Hilbert space is $\bigotimes_{L} V$, $V=\mathbb{C}^n$. $\id$ is the identity operator in $V$.
\\
We now introduce operators that play a crucial role in characterizing the integrable properties of the models.
\paragraph{Lax matrix}
The Lax operator\footnote{We use the symbol $L$ to represent both the length of the spin chain and the Lax matrix. However, it will be clear from the context to which of the two we refer.} $L$ can be defined using the RLL relation
\be
R_{0 0'}(\lambda,\lambda')\ L_{0 n}(\lambda,\lambda_n)\ L_{0' n}(\lambda',\lambda_n)
= L_{0' n}(\lambda',\lambda_n)\ L_{0 n}(\lambda,\lambda_n)\ R_{0 0'}(\lambda,\lambda') \,.
\label{Lalgebra}
\ee 

The Lax operator $L$ acts on one auxiliary space ($0$ or $0'$) and one physical space ($n$). We can represent graphically the Lax operator $L_{0i}(\lambda,\lambda_i)$ and the RLL relation in the following way:

\begin{figure}[h!]
\begin{center}
\hspace*{0.em}\raisebox{-4em}{\begin{tikzpicture}
			\draw[thick,blue] (-3,-2) -- (-1,2);
			\draw[gray, thick] (-1,-2) -- (-3,2);
			\filldraw[blue] (-2,0) circle (3pt) node[anchor=west]{};
			\draw (-3,-2) node[anchor=north east] {$0$};
			\draw (-0.6,-2) node[anchor=north east] {$i$};
			\draw (-2.8,-1.3) node[anchor=north east] {$\lambda$};
			\draw (-.6,-1.3) node[anchor=north east] {$\lambda_i$};
			\draw (1,0) node[anchor=north east] {$;$};
			\draw[blue, thick] (2,-2) -- (6,2);
			\draw[gray, thick] (2,2) -- (6,-2);
			\draw[blue, thick] (3,2) -- (3,-2);
			\filldraw[blue] (4,0) circle (3pt) node[anchor=west]{};
			\draw (2.2,-2) node[anchor=north east] {$0$};
			\draw (3.2,-2) node[anchor=north east] {$0'$};
			\draw (6.2,-2) node[anchor=north east] {$n$};
			\draw (2.4,-1.3) node[anchor=north east] {$\lambda$};
			\draw (3.7,-1.3) node[anchor=north east] {$\lambda'$};
			\draw (6.3,-1.3) node[anchor=north east] {$\lambda_n$};
			\draw (8.2,-2) node[anchor=north east] {$0$};
			\draw (11.2,-2) node[anchor=north east] {$0'$};
			\draw (12.2,-2) node[anchor=north east] {$n$};
			\draw (8.4,-1.3) node[anchor=north east] {$\lambda$};
			\draw (11.0,-1.3) node[anchor=north east] {$\lambda'$};
			\draw (12.3,-1.3) node[anchor=north east] {$\lambda_n$};
			\filldraw[blue] (3,1) circle (3pt) node[anchor=west]{};\filldraw[blue] (3,-1) circle (3pt) node[anchor=west]{};
			\draw (7,0) node[anchor=north east] {$=$};
			\draw[blue, thick] (8,-2) -- (12,2);
			\draw[gray, thick] (12,-2) -- (8,2);
			\draw[blue, thick] (11,-2) -- (11,2);
			\filldraw[blue] (11,1) circle (3pt) node[anchor=west]{};\filldraw[blue] (11,-1) circle (3pt) node[anchor=west]{};\filldraw[blue] (10,0) circle (3pt) node[anchor=west]{};
	\end{tikzpicture}}
\end{center}
	\caption{Graphical representation of the Lax matrix (left) and the RLL relation (right). Blue lines identify the auxiliary spaces.}
\end{figure}
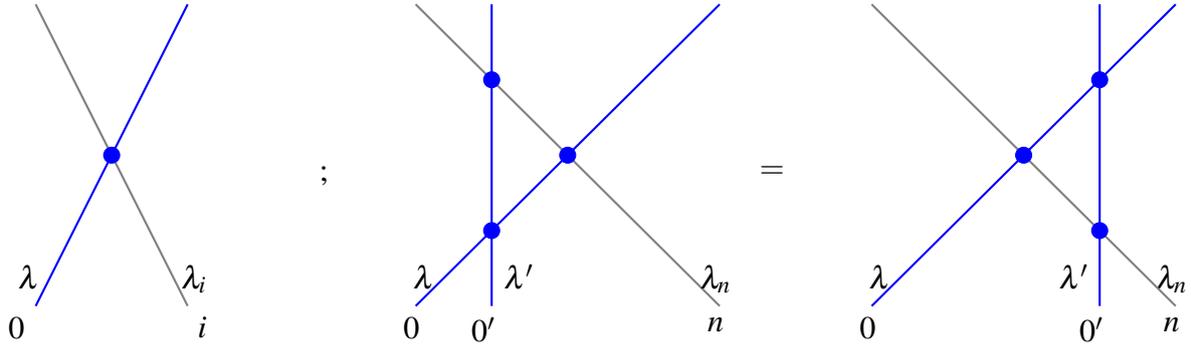

\vspace{3cm}

When the auxiliary space is equivalent to the physical space, it is possible to consider the Lax matrix to be the same as the $R$-matrix. In this situation, we say the spin chain is in the fundamental representation, and the RLL equation is equivalent to the YBE \eqref{eq:YBE}  by taking
\begin{align}
&0 \to 1,
&&0' \to 2,
&&n \to 3.
\end{align}
In this and the following chapters, we  consider the Lax operators and the $R$-matrix as the same objects. The only exception is in Section \ref{sec:medium}, where we  present an example where these two operators are distinct from each other.
\\
If the auxiliary space has dimension 2, the Lax operator $L_{0 n}$ can be represented in matrix form\footnote{The extension to the case of auxiliary space of higher dimension is obvious.}
\be
L_{0 n}(\lambda,\lambda_n) = \left( \begin{array}{cc}
\alpha_{n}(\lambda,\lambda_n)         & \beta_{n} (\lambda,\lambda_n) \\
\gamma_{n}(\lambda,\lambda_n)         & \delta_{n}(\lambda,\lambda_n)
\end{array} \right) 
\,, \qquad n = 1 \,, 2 \,, \ldots \,, L \,,
\ee  
with $\alpha, \beta, \gamma, \delta$ matrices acting on the physical space $n$. The algebra of these operators is encoded into the RLL relation \eqref{Lalgebra}.

\paragraph{Monodromy matrix}
The monodromy matrix $T_{0}(\lambda,\{\lambda_i\})$ is the  product of $L$ operators

\be
T_{0}(\lambda,\{\lambda_i\}) &=& L_{0 L}(\lambda,\lambda_L) \cdots L_{0 1}(\lambda,\lambda_1) \,=\, \left(\begin{array}{cc}
               \alpha_{L} & \beta_{L} \\
				\gamma_{L} & \delta_{L}
			\end{array} \right) \cdots 
 \left(\begin{array}{cc}
                \alpha_{1} & \beta_{1} \\
				\gamma_{1} & \delta_{1}
			\end{array} \right)	
\label{monodromycap1}			
\,.
\ee

In the cases where we can identify the Lax matrix with the $R$-matrix, the monodromy matrix is equivalently defined as\footnote{Sometimes, for convenience, one can also identify $L_{0i}(\lambda,\lambda_i)=R_{0i}(\lambda,\lambda_i')$, where $\lambda_i$ is related to $\lambda_i'$ by a shift.}
\be
T_{0}(\lambda,\{\lambda_i\}) &=& R_{0 L}(\lambda,\lambda_L) \cdots R_{0 1}(\lambda,\lambda_1) \,.
\label{modofromycap1inR}
\ee
The parameters ${\lambda_i}$ are known as inhomogeneities, and each parameter is associated with a particular physical space.  When all the inhomogeneities have the same value $\lambda_i=\theta$ for all $i$,  the spin chain is said to be \textit{homogeneous}.\\
We can represent graphically the monodromy matrix in the following way:

\begin{figure}[h!]{
\begin{center}
\hspace*{-3em}\raisebox{-4em}{\begin{tikzpicture}
			\draw[thick,blue] (0,0) -- (7,0);
			\draw (-0.2,.3) node[anchor=north east] {$0$};
			
			\draw[thick,black] (1,-.5) -- (1,.5);
			\draw (1.2,-0.8) node[anchor=north east] {$L$};

			\draw[thick,black] (2,-.5) -- (2,.5);
			\draw (2.5,-0.8) node[anchor=north east] {$L-1$};			

			\draw[thick,black] (3,-.5) -- (3,.5);
			\draw (3.7,-0.8) node[anchor=north east] {$L-2$};
			
			\draw (4.,0.5) node[anchor=north east] {$...$};

			\draw[thick,black] (5,-.5) -- (5,.5);
			\draw (5.3,-0.8) node[anchor=north east] {$2$};

			\draw[thick,black] (6,-.5) -- (6,.5);
			\draw (6.3,-0.8) node[anchor=north east] {$1$};
	\end{tikzpicture}}
\end{center}
\caption{Graphical representation of the Monodromy matrix. The blue line identifies the auxiliary space.}}
\end{figure}
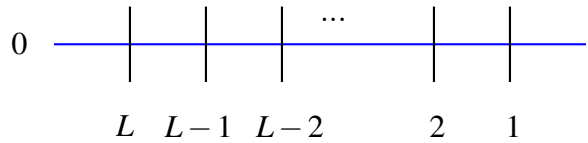

By using \eqref{Lalgebra} repeatedly, we can prove that the monodromy matrix obeys the fundamental relation
\be
R_{0 0'}(\lambda,\lambda')\ T_{0}(\lambda,\{\lambda_i\})\ T_{0'}(\lambda',\{\lambda_i\})
= T_{0'}(\lambda',\{\lambda_i\})\ T_{0}(\lambda,\{\lambda_i\})\ R_{0 0'}(\lambda,\lambda') \,,
\label{fundamental}
\ee 
or in graphical form
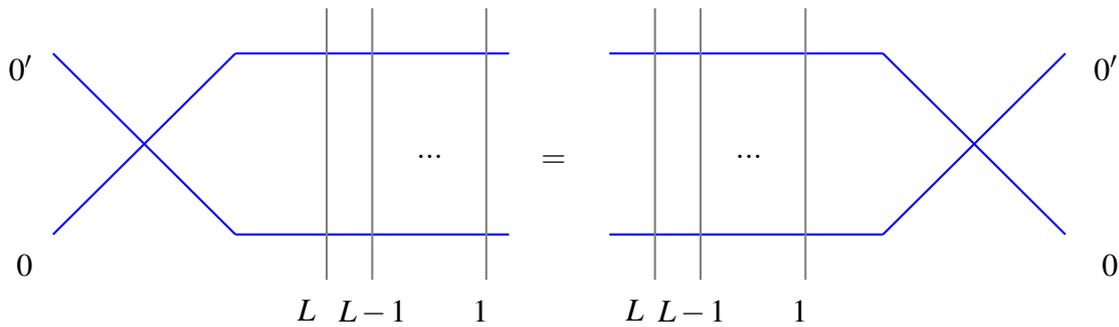
\begin{figure}[h!]{
\begin{center}
	\hspace*{-3em}\raisebox{-4em}{\begin{tikzpicture}[scale = 0.6]
			\draw[thick,blue] (0,0) -- (4,4);
			\draw[thick,blue] (4,4) -- (10,4);
			\draw[thick,blue] (0,4) -- (4,0);
			\draw[thick,blue] (4,0) -- (10,0);

			\draw[thick,gray] (6,-1) -- (6,5);
			\draw (6,-1.2) node[anchor=north east] {$L$};
			\draw[thick,gray] (7,-1) -- (7,5);
			\draw (8,-1.2) node[anchor=north east] {$L-1$};
			\draw[thick,gray] (9.5,-1) -- (9.5,5);
			\draw (9.8,-1.2) node[anchor=north east] {$1$};
			\draw (8.8,2) node[anchor=north east] {$...$};
			
			\draw (-0.2,-0.2) node[anchor=north east] {$0$};
			\draw (-0.2,4.2) node[anchor=north east] {$0'$};			
			\draw (11.5,2) node[anchor=north east] {$=$};

			\draw[thick,blue] (18.2,0) -- (22.2,4);
			\draw[thick,blue] (18.2,4) -- (22.2,0);
			\draw[thick,blue] (12.2,4) -- (18.2,4);
			\draw[thick,blue] (12.2,0) -- (18.2,0);
			
			\draw[thick,gray] (13.2,-1) -- (13.2,5);
			\draw (13.2,-1.2) node[anchor=north east] {$L$};
			\draw[thick,gray] (14.2,-1) -- (14.2,5);
			\draw (15,-1.2) node[anchor=north east] {$L-1$};
			\draw[thick,gray] (16.5,-1) -- (16.5,5);
			\draw (16.8,-1.2) node[anchor=north east] {$1$};
			\draw (15.8,2) node[anchor=north east] {$...$};
			\draw (23.6,-0.2) node[anchor=north east] {$0$};
			\draw (23.6,4.2) node[anchor=north east] {$0'$};
	\end{tikzpicture}}
	\caption{Graphical representation of the RTT relation.}
\end{center}}
\end{figure}
\paragraph{Transfer matrix}
The transfer matrix $t(\lambda,\{\lambda_i\})$ for periodic spin chains is defined as
\be
t(\lambda,\{\lambda_i\}) = \tr_{0} T_{0}(\lambda,\{\lambda_i\}) \,,
\label{transfer}
\ee
where $\tr_0$ is the partial trace over the auxiliary space. $t(\lambda,\{\lambda_i\})$ is an operator acting only on the physical spaces. The graphical representation of the transfer matrix is:

\begin{figure}[h!]{
\begin{center}
\hspace*{-3em}\raisebox{-4em}{\begin{tikzpicture}
			\draw[thick,black] (0,-2.5) -- (0,-1.5);
			\draw (0.2,-2.6) node[anchor=north east] {$j$};
			
			\draw[thick,black] (1,-2.3) -- (1,-1.3);
			\draw (1.5,-2.5) node[anchor=north east] {$j+1$};

			\draw[thick,black] (-1,-2.3) -- (-1,-1.3);
			\draw[thick,black] (-0.6,-2.5) node[anchor=north east] {$j-1$};

			\draw[thick,black] (3.5,-1.5) -- (3.5,-0.5);
			\draw[thick,black] (-3.5,0.5) -- (-3.5,1.5);

			\draw[thick,black] (0,1.5) -- (0,2.5);
\draw (0,2.7) node[anchor=north east] {$1$};

\draw[thick,black] (1,1.3) -- (1,2.3);
\draw (1.5,2.5) node[anchor=north east] {$L$};

\draw[thick,black] (3,0.8) -- (3,1.8);
\draw (3.5,2.2) node[anchor=north east] {$L-1$};

\draw[thick,black] (-1,1.3) -- (-1,2.3);
\draw (-1.2,2.5) node[anchor=north east] {$2$};
			
			\draw [thick,blue](0,0) ellipse (4cm and 2cm);
			
			\end{tikzpicture}}
\end{center}
\caption{Graphical representation of the transfer matrix.}
}\end{figure}
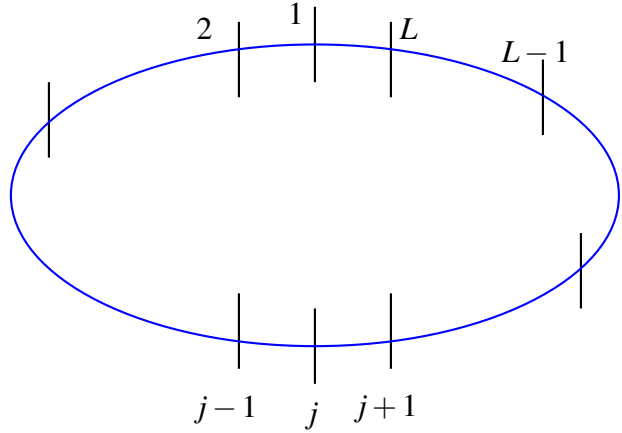

By multiplying \eqref{fundamental} for $R_{00'}(\lambda,\lambda')^{-1}$ from the right, taking the partial trace over both auxiliary spaces $0$ and $0'$ and using the cyclicity of the trace, we find
\be
\left[ t(\lambda,\{\lambda_i\})\,, t(\lambda', \{\lambda_i\}) \right] = 0 \,.
\label{commutativity}
\ee
\paragraph{Conserved charges} Provided that all the inhomogeneities are the same ($\lambda_i=\theta$), we can expand for a spin chain of definite length $L$, the transfer matrix in series
\begin{align}
\log t(\lambda,\theta)=\mathbb{Q}_1\Big(\frac{\theta+\lambda}{2}\Big)+(\lambda-\theta) \,\mathbb{Q}_2\Big(\frac{\theta+\lambda}{2}\Big)+\frac{1}{2} (\lambda-\theta)^2 \, \mathbb{Q}_3\Big(\frac{\theta+\lambda}{2}\Big)+\dots,
\label{expansiontransfer}
\end{align}
and by considering\footnote{To be precise, \eqref{commutativity} implies that also $[\log t(\lambda,\{\lambda_i\}),\log t(\lambda',\{\lambda_i\})]=0$.} \eqref{commutativity} order by order we obtain
\begin{equation}
[\mathbb{Q}_i(\theta),\mathbb{Q}_j(\theta)]=0, \,\,\,\,\,\,\,\,\,\,i,j=1,\dots,\infty,
\label{commutativityallcharges}
\end{equation}
which is the cornestone of integrability. $\mathbb{Q}_i(\theta)$ are the \textbf{conserved charges} characterizing the integrability of the models.\\
If the inhomogeneties are different, the expansion of the transfer matrix can still be performed, but now takes the more involved expression
\begin{align}
\log t(\lambda,\{\lambda_i\})=& \mathbb{Q}_1\Big(\frac{\theta+\{\lambda_i\}}{2}\Big)+\sum_j (\lambda-\lambda_j) \,\mathbb{Q}_2\Big(\frac{\theta+\{\lambda_i\}}{2}\Big)+\nonumber\\&\sum_j \frac{1}{2} (\lambda-\lambda_j)^2 \, \mathbb{Q}_3\Big(\frac{\theta+\{\lambda_i\}}{2}\Big)+\dots.
\end{align}
And the commutation property \eqref{commutativityallcharges} becomes
\begin{equation}
[\mathbb{Q}_i(\theta,\{\lambda_i\}),\mathbb{Q}_j(\theta,\{\lambda_i\})]=0, \,\,\,\,\,\,\,\,\,\,i,j=1,\dots,\infty.
\label{commutativityallcharges2}
\end{equation}
To clarify, in a model characterized by an $R$-matrix that satisfies the YBE, it is always possible to define an infinite set of conserved charges. However, the converse is not always true.  Indeed, some models have an infinite set of conserved charges, but with not known $R$-matrix. An example of such a model is the  Inozemtsev’s spin chain, \cite{klabbers2022coordinate}. In this model, the dynamic is governed by a density Hamiltonian acting on all the sites of the spin chain.  This class of models, including Inozemtsev's spin chain, will not be considered in this thesis. Instead, we consider Hamiltonian density that acts on nearest-neighbour sites or on 3 sites.
\paragraph{Regular $R$-matrix}
We are interested in models where the conserved charges act locally, therefore we should restrict to the case where the $R$-matrix is regular\footnote{We remark that to have a regular $R$-matrix, the spin chain should be homogeneous. Furthermore, the statement that regular $R$-matrix implies that the charges are local can be understood from \eqref{charges} and \eqref{hamiltonian}-\eqref{chargeQ2transfer}.},
\begin{align}
R_{ij}(\lambda,\lambda)=P_{ij},
\label{regularityR}
\end{align}
where $P$ is the permutation operator swapping the site $i$ and $j$ of the chain.  We give the action of $P$ on both vectors $\ket{a}$, $\ket{b}$ and on a  2 site operator $B$:
\begin{align}
&P \ket{a}\otimes \ket{b}=\ket{b}\otimes \ket{a},
&& P_{ij} B_{jk} P_{ij}=B_{ik}.
\label{permutationonvectandoperator}
\end{align}
In general, we can also renormalize the $R$-matrix while still satisfying the Yang Baxter equation (YBE), resulting in $R_{ij}(\lambda,\lambda)\sim P_{ij}$. As mentioned, we are considering the case where the $R$-matrix and the Lax matrix coincide and consequently, the regularity property can be analously given in term of one of these two matrices. In chapter \ref{Hubbardchapter}, we consider one of the case where $L$ and $R$ are two different objects and we define the regularity condition on the Lax matrix, see expression \eqref{regularitylax}.
\\
From now on, we always work with the assumption of regularity, unless explicitly mentioned. For this case, the charges $\bQ_i$ are \textit{local} and have interaction range $i$. This identifies on how many sites of the spin chain the charge $\bQ_i$ is acting non-trivially. As an example, we can consider the case $i=2$
\begin{align}\label{charge2}
\mathbb{Q}_2=\sum_{j} \mathcal{Q}_{j,j+1},
\end{align}
and more generally, each charge $\mathbb{Q}_i$ can be written as a sum of  densities $\mathcal{Q}$ acting non-trivially on $i$ sites of the spin chain. With this in mind, it is important to note that in the expression \eqref{expansiontransfer}, the length $L$ of the spin chain must be bigger than the maximum range of the charge we are interested in. Taking $L$ smaller or equal gives rise to wrapping effect, \cite{Gombor:2022lco}.
\paragraph{Periodic and open boundary conditions}
The boundary conditions of the sum differ if we work with open or closed spin chain.
\\
For example, in \eqref{charge2}
\begin{itemize}
\item closed $\,\,\,\,\mathbb{Q}_2=\sum_{j=1}^L \mathcal{Q}_{j,j+1},\,\,\,\mathcal{Q}_{L,L+1}\equiv \mathcal{Q}_{L,1}$
\item open $\,\,\,\,\mathbb{Q}_2=\sum_{j=1}^{L-1} \mathcal{Q}_{j,j+1}$.
\end{itemize}
In this thesis, we work with closed spin chain (periodic boundary condition).  The scenario changes when dealing with open boundary conditions, \cite{Sklyanin:1988yz}. In such cases, the reflection matrix, often referred to as the $K$-matrix, assumes a prominent role.  The YBE \eqref{eq:YBE} continues to hold, but there are two extra equations called boundary YBE (one for the left and one for the right boundaries).
\paragraph{Charges and transfer matrix: meaning of the first few charges}
By using \eqref{expansiontransfer}, the charges are related to the transfer matrix by
\begin{equation}
\mathbb{Q}_i(\theta)=\partial_{\lambda}^{i-1} \log t(\lambda, \theta)|_{\lambda=\theta}.
\label{charges}
\end{equation}

The charge $\mathbb{Q}_1$ can be identified with the \textit{momentum} of the spin chain. 
We can first consider the case $L=2$ and take the auxiliary space to be the same as the physical one so that the $R$ and the Lax matrix coincide, the transfer matrix \eqref{transfer} is
\be
t(\theta,\theta) = \tr_{0} T_{0}(\theta,\theta) = 
\tr_{0} R_{0 2}(\theta,\theta)\ R_{0 1}(\theta,\theta) 
= \tr_{0}P_{0 2} P_{0 1} = P_{12} 
\label{tat}
\,,
\ee
where we have used the regularity condition \eqref{regularityR}  and the identity 
$\tr_{0} A_{0} P_{0 n} = A_{n}$. 
\\
Given an operator $A$ acting on the first site of the spin chain, we find that
\be
t(\theta,\theta)\  A_{1}\ t(\theta,\theta)^{-1} = 
P_{12}\ A_{1}\ P_{12} = A_{2} \,,
\ee
and this easily generalizes to arbitrary $n$
\be
t(\theta,\theta)\  A_{n}\ t(\theta,\theta)^{-1} = A_{n+1} \,, \qquad 
n = 1 \,, \ldots \,, L \,.
\ee
Remembering that the momentum operator of quantum mechanics follows the property
\be
e^{i a P}\ X\ e^{-i a P} = X + a \,,
\ee
where $P$ and $X$ are the momentum and position operators, and $a$ is a c-number. We can define the momentum operator for a spin 
chain by $e^{i\mathbb{P}} =  t(\theta,\theta)$, which implies
\begin{align}
\mathbb{P}= -i \, \log t(\theta,\theta)=-i \, \mathbb{Q}_1,
\label{momentum}
\end{align}
so the first charge $\mathbb{Q}_1$ is related to the momentum.
\\
For the sake of clarity, we point out that the condition \eqref{commutativity} holds for any $R$-matrix satisfying the YBE. The regularity condition is necessary to ensure the locality of the charges.\\
Traditionally, one takes the charge $\bQ_2$ to be the \textit{Hamiltonian} of the system
\begin{align}
&\bQ_2(\theta)=\mathbb{H}(\theta)=\sum_{j=1}^L \mathcal{H}_{j,j+1}(\theta),
&&\mathcal{H}_{L,L+1} \equiv \mathcal{H}_{L,1}
\end{align}
and given a  regular $R$-matrix, we can construct an integrable spin chain with nearest-neighbour interaction Hamiltonian.
This density is itself related to the $R$-matrix in a very simple way:
\begin{align}
&\mathcal{H}_{12}(\theta)=P_{12}\partial_\lambda R_{12}(\lambda,\theta)|_{\lambda\rightarrow \theta}\,
\label{hamiltonian}
\end{align}
and as given by \eqref{charges}, \cite{Kulish1982}
\begin{align}\label{chargeQ2transfer}
\bQ_2(\theta)=\partial_{\lambda} \log t(\lambda,\theta)|_{\lambda\to \theta}.
\end{align}
For difference form $R$-matrix, the relation is $\mathcal{H}_{12}=P_{12}\partial_\lambda R_{12}(\lambda)|_{\lambda\to 0}$. It becomes clear that we can also investigate the nature of the $R$-matrix, whether it is of difference or of non-difference form, by analysing the charges associated with it. Constant charges correspond to difference form $R$-matrices.\\
Knowing the $R$-matrix gives us information about the Hamiltonian and the dynamics of the system. However, there are alternative methods for constructing higher conserved charges besides using the transfer matrix. For regular $R$-matrices, we can recursively generate the charge $\mathbb{Q}_{i+1}$ from $\mathbb{Q}_i$ using the \textit{boost operator mechanism}, \cite{HbbBoost,Loebbert:2016cdm}. In  chapter \ref{classificationchapter}, we provide a thorough explanation of this mechanism, as we  use it to \textbf{find new integrable models}.

\vspace{1 cm}

In the next section, we follow the approaches of \cite{faddeev1996algebraic,nepomechie1998spin},

\vspace{0.3cm}
$\,\,\,\,\,\,\,\,\,\,$ \textit{"[...] not to begin in full generality but rather to choose a representative example and explain on it all technical features in such a way, that generalization become reasonably evident."}\,\,\,\,\, Faddeev, \cite{faddeev1996algebraic}.

\vspace{0.3cm}

We introduce the well-known Heisenberg spin chain as an example of an integrable model. We utilize this model to initially highlight the significance of integrability in understanding the dynamics of a system.

\section{A typical example: the XXX spin chain}
\label{examplexxxsp}
The XXX model (Heisenberg) is a standard example of integrable spin chain.
\\
Even if this model is really simple, its rich and elegant mathematical structure makes it the prototype of all integrable models. This Hamiltonian was first used for condensed matter applications
where it serves as a model for a ferromagnetic or antiferromagnetic material. Furthermore, it is also very important in high energy physics. In fact, in the AdS/CFT context, it describes the leading order anomalous dimensions for operators in the $SU(2)$ sector of the $\mathcal{N} = 4$ supersymmetric Yang-Mills theory \cite{minahan2003bethe}.

\subsection{Why do we need integrable models?}
Given a physical model, we are interested in solving the spectral problem

\begin{equation}
\mathbb{H}\ket{\psi}=E \ket{\psi}.
\end{equation}
We start to explore the problem considering $\mathbb{H}$ on a spin chain of length $L=2$ and then to generalize to arbitrary length. It will then be clear where the difficulties start to arise and how integrable models can assist in addressing them.

\subsubsection*{Definitions}
We focus on a spin 1/2 chain (the generalization to $s>1/2$ is straightforward and can be found in many reviews or books, for example \cite{faddeev1996algebraic,nepomechie1998spin}).
\\
The Hilbert space is given by $\otimes^L \mathbb{C}^2$ and we use the standard basis
\begin{align}
&\left(
\begin{array}{c}
 1 \\
 0 \\
\end{array}
\right)=\ket{\uparrow},
&&\left(
\begin{array}{c}
 0 \\
 1 \\
\end{array}
\right)=\ket{\downarrow},
\label{standardbasis}
\end{align}
to identify particles of spin up and down, respectively.
\\
The \textbf{density Hamiltonian} of the Heisenberg spin chain is
\begin{align}
\mathcal{H}_{ij}=\frac{J}{4}\Big(X_i X_j+Y_i Y_j+Z_i Z_j-\mathbb{I}\Big),
\label{XXXdensity}
\end{align}
where $X,Y,Z$ are the Pauli Matrices. The inclusion of the identity term is for convenience, as shifting the Hamiltonian globally does not affect the system's dynamics but only leads to a uniform energy level shift. The factor $J$ serves as a normalization constant.
\\
This model is called \textbf{XXX spin chain} because the coefficients in front of each term $XX,\,YY,\,ZZ$ are the same. However, the integrability property remains even when these coefficients are different. For instance, the XXZ spin chain has matching XX and YY coefficients, while the ZZ coefficient differs.  Similarly, the XYZ spin chain features distinct coefficients for all three terms. Other limits can also be taken, like the XX model where the $ZZ$ coefficient is set to $0$.
\\
The density \eqref{XXXdensity} can also be written as
\begin{align}
\mathcal{H}_{ij}=\frac{J}{2}\Big(P_{ij}-\mathbb{I}\Big),
\label{hxxzintermofP}
\end{align}
with $P_{ij}$ the permutation operator acting as defined in \eqref{permutationonvectandoperator} and $\mathbb{I}$ the identity matrix 4x4, or in matrix representation as
\begin{align}
\mathcal{H}_{12}=\frac{J}{2}\left(
\begin{array}{cccc}
 0 & 0 & 0 & 0 \\
 0 & -1 & 1 & 0 \\
 0 & 1 & -1 & 0 \\
 0 & 0 & 0 & 0 \\
\end{array}
\right).
\label{Hamiltonianmatrix}
\end{align}

\subsubsection*{Spectral problem for $L=2$}

We are interested in solving the problem
\begin{align}
\mathbb{H}\ket\psi=(\mathcal{H}_{12}+\mathcal{H}_{21})\ket\psi=E\ket\psi.
\end{align}
Here and in the following we use periodic boundary conditions.\\
It is easy to see by direct inspection that a solution is given by
\begin{align}
&\ket{\psi_{(1,1)}}=\ket{\uparrow}\otimes \ket{\uparrow}=\ket{\uparrow \uparrow},
&&\ket{\psi_{(1,0)}}=\ket{\uparrow \downarrow}+\ket{\downarrow \uparrow},\label{state1XX}\\
&\ket{\psi_{(1,-1)}}=\ket{\downarrow \downarrow},
&&\ket{\psi_{(0,0)}}=\ket{\uparrow \downarrow}-\ket{\downarrow \uparrow}.\label{state2XX}
\end{align}
The states with $\ket{\psi_{(1,i)}}$ have energy $E=0$, while $\ket{\psi_{(0,0)}}$ has energy $E=-2J$.
\\
$J > 0$ is called \textit{antiferromagnetic} regime because the ground state is a spin singlet state; while for $J < 0$ (\textit{ferromagnetic}) there is a degenerate ground state, triplet.
\\
The reason of this result can be also easily understood from the $\alg{su}(2)$ symmetry of the model. In fact, the total spin operator
\begin{align}
\vec{S}=\frac{1}{2}\Big(\vec{\sigma}\otimes\mathbb{I}+\mathbb{I}\otimes \vec{\sigma} \Big),
\end{align}
with $\vec{\sigma}=(X,Y,Z)$, are generators of a reducible 4-dimensional representation of $\alg{su}(2)$.
\\
There exists a unitary matrix $U$ such that
\be
U\ \vec S\ U^{\dagger} = \left( \begin{array}{c|c}
 \vec S_{(S=1)} & \\ 
 \hline
 & \vec S_{(S=0)} 
\end{array} \right) \,,
\ee 
where $\vec S_{(S=1)}$ and $\vec S_{(S=0)}$ generate irreducible 
representations of dimensions 3 and 1, respectively. Since,
\be
\vec S^{2} &=& {1\over 2} \vec\sigma \otimes \vec\sigma 
+ {3\over 2} \id \otimes \id \,,
\ee
the two-site Hamiltonian can be 
expressed in terms of $\vec S^{2}$,
\be
\mathcal{H}_{12} = {J\over 4} \left( \vec\sigma \otimes \vec\sigma  
- \id  \right)
= {J\over 2} \left(  \vec S^{2} - 2 \, \id \right)
\,
\ee
and using the fact that
$\vec S^{2} | S \,, S^{z} \rangle = S ( S + 1) | S \,, S^{z} \rangle $, we 
see that
\be
\mathcal{H}_{12} | S \,, S^{z} \rangle = {J\over 2} \left(  S (S + 1) - 2 \right) 
| S \,, S^{z} \rangle
\,, \quad  S^{z} = -S \,, \ldots \,, S \,; \quad S = 0 \,, 1 \,.
\ee
In this way, we recover the result found earlier: for $S=1$ the energy is $E=0$, and for $S=0$ the 
energy is $E=-J$ (the 2 in the energy eigenvalue given after \eqref{state2XX} comes from the fact that $\mathbb{H}=2\mathcal{H}_{12}$).
\\
We remark that symmetry arguments are of course not available for all physical models and in some cases, only a brute force computation can be performed.

\subsubsection*{Spectral problem for arbitrary $L$ and why integrability is useful}

\begin{figure}[h!]
	\centering
	\includegraphics[height=5cm]{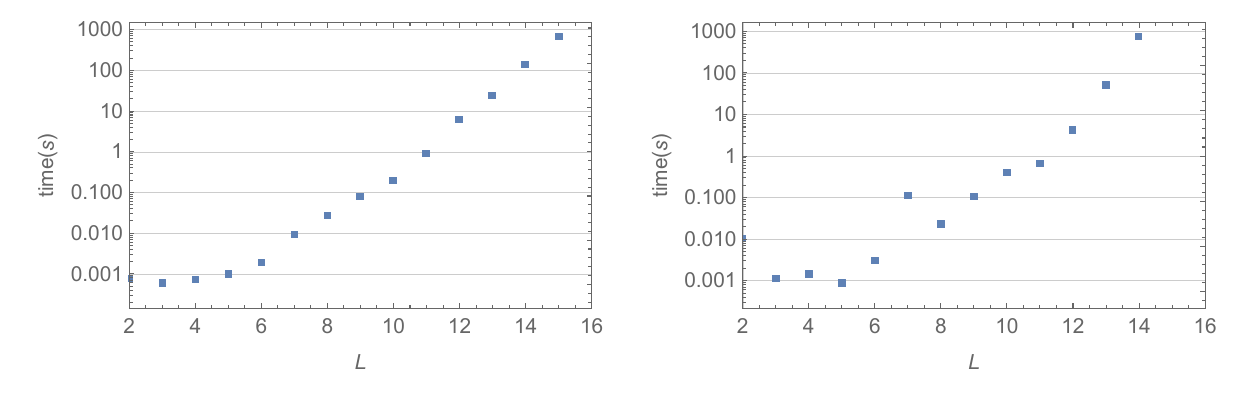}
	\caption{Brute force diagonalization of the Heisenberg Hamiltonian \eqref{Hamiltonianmatrix} for different $L$. The left graph shows the computation time for "Eigenvalues", while the right graph displays the computation time for "Eigenvectors".}
\label{comptime}
	\end{figure}

The problem of diagonalizing the Hamiltonian, now corresponds to a problem of diagonalizing a matrix of dimension $2^L\times 2^L$. While this problem can be solved using a computer, it becomes more difficult as the chain length increases, eventually reaching a point where it becomes impractical. To give a more precise idea, we performed the diagonalization of the Hamiltonian \eqref{Hamiltonianmatrix} by using the software Mathematica 12.3. The eigenvalues and eigenvectors were computed separately using the commands "Eigenvalues" and "Eigenvectors". We set the numerical value $J=-1/2$, and we plot in Figure \ref{comptime} the computation time needed by the program to finish as a function of the length of the spin chain.

From the graph, it is evident that the computation time for the diagonalization was relatively short (the maximum time is approximately 1000 seconds or around 17 minutes). However, when attempting to increase the length of the spin chain by one, we encountered significant difficulties within the time frame we had. The computation of the "Eigenvectors" for $L=15$ was not completed after 7 hours, and a similar issue occurred with "Eigenvalues" for $L=16$. Therefore, explicit diagonalization rapidly becomes computationally challenging.  However, there are optimized algorithms available if one is interested in specific eigenvalues, such as the smallest or the largest.  Or one can use supercomputers.  Nevertheless, it would be highly advantageous to have a method to obtain the complete spectrum efficiently.
\\
If the model is integrable, there exist some techniques that allow to transform the problem of exactly diagonalize a matrix, to the problem of solving a set of polynomial equations. For example, in \cite{marboe2017fast}, the authors developed a  method to generate all possible  solutions (analytically) of this polynomial equations for the rational GL(N|M) spin chain in the case where the quantum numbers involved are not too large, proving the enormous advantage of using integrable models. The analytical solution is in general not available, however using numerical technique, the spectrum can be computed for spin chain with higher length compared to the ones obtained by brute force diagonalization.\\
Furthermore, for many physical purposes, it is interesting to study the behaviour of the eigenvalues in the \textbf{thermodynamic limit} ($L\to \infty$), \cite{van2016introduction}, which is definitely not achievable by direct diagonalization. For all these reasons, \textbf{integrability} becomes handy. For integrable models, the thermodynamic limit can be obtained by involving the so called \textit{string-hypothesis}.
\\
In the following section, we discuss a technique called the \textbf{Algebraic Bethe Ansatz} (or Quantum Inverse Scattering method), \cite{faddeev/takhtajan1}. However, this is not the only available technique to exactly diagonalize the Hamiltonian, and to mention a few: Coordinate Bethe ansatz, \cite{Bethe:1931hc}; Analytic Bethe ansatz, \cite{reshetikhin}; Baxter's $Q$-operator method;
 a method that allows to directly work in the thermodynamic limit, \cite{jimbo1994algebraic};  Quantum Spectral Curve \cite{gromov2017introduction}.
\\
Now, we apply the algebraic Bethe ansatz to the Heisenberg spin chain. Furthermore, in Appendix \ref{BAB3chapter}, we explore a more advanced technique called nested Algebraic Bethe ansatz, which shares similar foundational elements to the approach discussed here, but it applies also for systems with higher dimensional local Hilbert space.
\section{Algebraic Bethe Ansatz}
\label{ABAtheory}
The Algebraic Bethe Ansatz is one of the techniques used to find the eigenvalues and the eigenvectors of the Hamiltonian belonging to an integrable model.  It allows us to obtain not only the eigenvalues of the transfer matrix but also, by taking their logarithmic derivatives, through the use of equation \eqref{charges}, the eigenvalues of all conserved charges.
\\
The $R$-matrix\footnote{Now, we consider the $R$-matrix as a known object, in chapter \ref{classificationchapter} we explain how to obtain it from the Hamiltonian.} of the XXX spin chain is 

\begin{equation}
R_{12}(\lambda)=\lambda\, \mathbb{I}+ i\, P_{12} = \left(
\begin{array}{cccc}
 \lambda +i & 0 & 0 & 0 \\
 0 & \lambda  & i & 0 \\
 0 & i & \lambda  & 0 \\
 0 & 0 & 0 & \lambda +i \\
\end{array}
\right),
\label{RXXX}
\end{equation}
and it is of difference form type.
\\
To verify that this is indeed the $R$-matrix associated with the XXX spin chain, we can substitute it into the expression for the Hamiltonian \eqref{hamiltonian} and we obtain
\begin{align}
&\mathcal{H}_{12}\sim P_{12} \partial_\lambda R_{12}(\lambda)|_{\lambda\to 0}=P_{12},
\label{HXXXunnormalized}
\end{align}
which corresponds to \eqref{hxxzintermofP} up to a normalization factor and a shift by the identity. These transformations are always allowed\footnote{The transformations preserve the integrability of the model, however they may swap between the ferromagnetic and the antiferromagnetic regime. We discuss the allowed transformations in detail in section \ref{identification}.}. If an $R$-matrix satisfies the YBE, then  $g R$ also does. Similarly, shifting the matrix by an operator proportional to the identity is permissible and only results in a global energy shift. We also notice that, the regularity condition holds (any normalization pre-factor is not important) and the conserved charges are local.

\subsection*{The general idea}
The starting point of the Algebraic Bethe ansatz is to find the \textit{reference state} of the model. From this state, by recognising a certain creation operator, we can build the other excited states that belong to the spectrum. This idea is the same as the one used to solve the problem of the harmonic oscillator.
\\
For simplicity, we consider the case of a homogeneous spin chain, which corresponds to setting\footnote{Indeed, a more general condition is to have $\lambda_i = k$ with $k \in \mathbb{C}$. However, we can simplify the analysis by setting $k = 0$ through a reshifting of $\lambda$.} the inhomogeneities $\lambda_i$ in equation \eqref{monodromycap1} to zero.
\subsection*{Commutation relations: From the RTT}

The monodromy matrix $T_{0}(\lambda)$ \eqref{modofromycap1inR} can be written as
\be\label{monodromy22}
T_{0}(\lambda) = R_{0L}(\lambda-\frac{i}{2})\dots  R_{01}(\lambda-\frac{i}{2})=
\left( \begin{array}{cc}
A(\lambda) & B(\lambda)  \\
C(\lambda) & D(\lambda)
\end{array} \right) 
\,,
\ee
where the matrices $A, B, C, D$ are related to the $\alpha_i, \beta_i, \gamma_i, \delta_i$ of \eqref{monodromycap1}. $T_0(\lambda)$ can be seen as a $2 \times 2$ matrix living in the auxiliary space and $A, B, C, D$ are operators on the physical space $\otimes^L\mathbb{C}^2$. We identify the Lax matrix with the $R$-matrix with a shift in  the spectral parameter\footnote{This choice is done in order to have a more symmetric final result for the Bethe equations.}.
\\
Using this expression of the monodromy matrix and the $R$-matrix from \eqref{RXXX} in the fundamental relation RTT (\ref{fundamental}), we can recover the following commutation relations
\begin{align}
&\left[ B(\lambda) \,, B(\lambda') \right] = 0 \,,&&\left[ C(\lambda) \,, C(\lambda') \right] = 0\,,\label{BBop}
\end{align}
\be 
A(\lambda)\ B(\lambda') &=& {a(\lambda' - \lambda)\over b(\lambda' - \lambda)}
B(\lambda')\ A(\lambda) - 
{c(\lambda' - \lambda)\over b(\lambda' - \lambda)} B(\lambda)\ A(\lambda')
\,,\label{algebra2} \\
D(\lambda)\ B(\lambda') &=& {a(\lambda - \lambda')\over b(\lambda - \lambda')}
B(\lambda')\ D(\lambda) - 
{c(\lambda - \lambda')\over b(\lambda - \lambda')} B(\lambda)\ 
D(\lambda') \,,
\label{algebra}
\ee
where 
\begin{align}\label{abcmodelxxx}
&a=\lambda+i,
&&b=\lambda,
&&c=i.
\end{align}
Based on \eqref{BBop}, we interpret $B$ and $C$ respectively as creation and annihilation operators\footnote{The choice between $B$ and $C$ as the creation or annihilation operator is motivated by the expression of the reference state, as clarified in the following paragraph.}.
\subsection*{The reference state and the action of the transfer matrix}

From \eqref{monodromy22}, the transfer matrix $t(\lambda)$ is simply
\begin{equation}
t(\lambda)=A(\lambda)+D(\lambda).
\end{equation}
We consider the \textit{ferromagnetic} regime with $J<0$. The \textbf{reference state} is the one with all spins aligned\footnote{We remind that the reference state has degeneracy $3$. We chose here only one of the possible ground states.},
\be
\omega_{+} =  \underbrace{{1 \choose 0}
\otimes \cdots \otimes {1 \choose 0}}_{L} = \ket{{\uparrow \uparrow \dots \uparrow}}\,,
\label{vacuumxxz}
\ee
which is an eigenstate of $A(\lambda)$ and $D(\lambda)$ and is annihilated by $C(\lambda)$,
\be
A(\lambda)\ \omega_{+} = \left( \lambda + {i\over 2} \right)^{L} 
\omega_{+} \,, \qquad 
D(\lambda)\ \omega_{+} = \left( \lambda - {i\over 2} \right)^{L} 
\omega_{+} \,, \qquad 
C(\lambda)\ \omega_{+} = 0 \,. 
\label{properties}
\ee
Furthermore, this state is not an eigenstate of $B(\lambda)$.\\
The state \eqref{vacuumxxz} is also called \textit{vacuum state} or, since it is degenerate, \textit{pseudo-vacuum} state.
\\
In this particular case, determining the reference state was straightforward due to the exact diagonalization of the Hamiltonian for $L=2$ and the simplicity of the model. However, it should be noted that for more complex models, this may not always be the case.
\\
Let us outline the general steps involved in finding the ground state of a theory:

\begin{itemize}
\item[1.] Write the transfer matrix for a spin chain of $L=2$ and $L=3$;
\item[2.] Perform numerical diagonalization of the matrix to obtain its eigenvectors;
\item[3.] Among those eigenvectors, select the ones that were not eigenvectors of $B(\lambda)$;
\item[4.] Among those, select the ones that are eigenvectors of $A(\lambda)$ and $D(\lambda)$;
\item[5.] Among those, select the ones that factorize, so that we can write them as $\kappa \otimes \kappa \otimes \dots \otimes \kappa$, with $\kappa$ being a vector in $\mathbb{C}^2$.
\end{itemize}

The plan is to carry out these steps on a short spin chain, with $L= 2$ or $3$, making it possible to use direct diagonalization. We then aim to extend what we learn to understand how things work for a spin chain of any length. We explain the reasons for steps 3. and 4. in the footnote \ref{footnotevacuum}. The step 5. is optional. If it is possible to realize it, it is easier to generalize the construction for arbitrary $L$.\\
 Sometimes there is also the requirement that the reference state is annihilated by the $C(\lambda)s$. This condition is met in the case of the Heisenberg spin chain we're examining. Nevertheless, unlike conditions 3. and 4., it does not need to be satisfied for all the models. In fact, it depends if in the RTT relation the $C$ terms appears. This point will be clarified in the Appendix \ref{BAB3chapter}, after the expression \eqref{comrelcond}. 
\\
It is useful to notice that in some integrable models like the anisotropic XYZ chain a simple
reference state is not known in the generic situation, and one has to use other methods to
solve them, \cite{cao2014spin}.

\subsection*{Bethe states}
Due to the property \eqref{BBop}, we can identify the operators $B$s as creation operators and we use them to construct the so-called \textit{Bethe states} or \textit{Bethe vectors},
\be
|\lambda_{1} \,, \ldots \,, \lambda_{M} \rangle =
B(\lambda_{1}) \cdots B(\lambda_{M})\ \omega_{+} \,,
\label{vec}
\ee
where $\lambda_i$ are the \textit{Bethe roots}\footnote{We remark that before we identified with $\lambda_i$ the inhomogeneities. Now $\lambda_i$ are the Bethe Roots. Since we set all the inhomogeneities to $0$, there should be no ambiguity in the notation.}. We can refer to this excited state as a state of $M$ magnons. A key step consists in requiring that \textit{this state is an eigenvector of the transfer matrix}. We will illustrate how this requirement leads to the determination of the Bethe roots' expression.

\subsection*{One magnon state}
To understand  how to find this constraint, we can start with the one magnon state
\begin{align}
\ket{\lambda_1}=B(\lambda_1)\omega_+
\label{oneparticlestatexxz}
\end{align}
and we want that
\begin{align}\label{wewantitforone}
t(\lambda)\ket{\lambda_1}=(A(\lambda)+D(\lambda))B(\lambda_1)\omega_+=\Lambda(\lambda; \lambda_1) \ket{\lambda_1},
\end{align}
with $\Lambda(\lambda; \lambda_1)$ the eigenvalue to be determined.
\\
We use the relations \eqref{algebra2} and \eqref{algebra} and obtain
\begin{align}\label{ton1ptstate}
t(\lambda)\ket{\lambda_1}=&\Big[{a(\lambda_1 - \lambda)\over b(\lambda_1 - \lambda)}
B(\lambda_1)\ A(\lambda)+{a(\lambda - \lambda_1)\over b(\lambda - \lambda_1)}
B(\lambda_1)\ D(\lambda) \Big]\omega_+
\non \\
&-\Big[{c(\lambda_1 - \lambda)\over b(\lambda_1 - \lambda)}
B(\lambda)\ A(\lambda_1)+{c(\lambda - \lambda_1)\over b(\lambda - \lambda_1)}
B(\lambda)\ D(\lambda_1) \Big]\omega_+
\end{align}
and we now use \eqref{properties}
\begin{align}
t(\lambda)\ket{\lambda_1}=&\Big[{a(\lambda_1 - \lambda)\over b(\lambda_1 - \lambda)}
\Big(\lambda+\frac{i}{2}\Big)^L +{a(\lambda - \lambda_1)\over b(\lambda - \lambda_1)}\Big(\lambda-\frac{i}{2}\Big)^L
 \Big]B(\lambda_1)\omega_+
\non \\
&-\Big[{c(\lambda_1 - \lambda)\over b(\lambda_1 - \lambda)}\Big(\lambda_1+\frac{i}{2}\Big)^L
+{c(\lambda - \lambda_1)\over b(\lambda - \lambda_1)}\Big(\lambda_1-\frac{i}{2}\Big)^L
 \Big]B(\lambda)\omega_+.
 \label{expressointxxz}
\end{align}
We recognize that in the  first line, the one magnon state\footnote{\label{footnotevacuum}The expression \eqref{expressointxxz} provides the reasoning behind implementing steps 3. and 4. when selecting the ground state. In particular, because we have identified $B$ as the creation operator for the excited state, $\omega_+$ cannot be its eigenstate. Step 4. is justified by the observation that, as seen in \eqref{algebra2} and \eqref{algebra}, the coefficients in front of $B(\lambda')A(\lambda)$ and $B(\lambda')D(\lambda)$ are generally different. Consequently, we must ensure that $\omega_+$ is an eigenstate of both $A$ and $D$ separately.} \eqref{oneparticlestatexxz} appears and that it corresponds to \eqref{wewantitforone}, if we identify
\begin{align}
\Lambda(\lambda;\lambda_1)={a(\lambda_1 - \lambda)\over b(\lambda_1 - \lambda)}
\Big(\lambda+\frac{i}{2}\Big)^L +{a(\lambda - \lambda_1)\over b(\lambda - \lambda_1)}\Big(\lambda-\frac{i}{2}\Big)^L.
\label{eigenvoneparticlexxz}
\end{align}
The terms in the second line of \eqref{expressointxxz}, often referred to as the \textit{unwanted terms}, are required to vanish. This condition imposes a constraint on the rapidity $\lambda_1$ of the magnon.
In particular, we impose
\begin{align}
{c(\lambda_1 - \lambda)\over b(\lambda_1 - \lambda)}\Big(\lambda_1+\frac{i}{2}\Big)^L
=- {c(\lambda - \lambda_1)\over b(\lambda - \lambda_1)}\Big(\lambda_1-\frac{i}{2}\Big)^L ,
\end{align}
or equivalently
\begin{align}
\Bigg(\frac{\lambda_1+\frac{i}{2}}{\lambda_1-\frac{i}{2}}\Bigg)^L
=- {c(\lambda - \lambda_1)\over c(\lambda_1 - \lambda)}{b(\lambda_1 - \lambda)\over b(\lambda - \lambda_1)}=1.
\label{Betheequationoneparticle}
\end{align}

So, for the one-magnon state, the solution of the eigenvalue problem is given by:
\begin{align}
& \text{eigenvector: Bethe vector} \,\,\,\eqref{oneparticlestatexxz},\non\\
&\text{eigenvalue} \,\,\,\eqref{eigenvoneparticlexxz},\non\\
&\text{constraint for Bethe roots $\lambda_1$: Bethe equations}\,\,\,\eqref{Betheequationoneparticle}.\non
\end{align}

\subsection*{Two magnon state}
To derive the expression for $M$ magnons, let's start by examining the case of $M=2$. By understanding this particular case, we can then easily generalize the results to any number of magnons. The Bethe vector for $M=2$ is
\begin{align}
\ket{\lambda_1,\lambda_2}=B(\lambda_1)B(\lambda_2)\omega_+
\end{align}
and similarly to what we did for $M=1$,
\begin{align}
t(\lambda)\ket{\lambda_1,\lambda_2}=(A(\lambda)+D(\lambda))B(\lambda_1)B(\lambda_2)\omega_+=\Lambda(\lambda; \lambda_1,\lambda_2) \ket{\lambda_1,\lambda_2}.
\end{align}
We now should start to use many times the commutation relations \eqref{algebra2} and \eqref{algebra} and then, when $A(\lambda)$ or $D(\lambda)$ are closed to $\omega_+$ we apply \eqref{properties}.

Explicitly,
\begin{align}
A(\lambda)B(\lambda_1)B(\lambda_2)\omega_+=&\frac{a(\lambda_1-\lambda)}{b(\lambda_1-\lambda)}\frac{a(\lambda_2-\lambda)}{b(\lambda_2-\lambda)}\Big(\lambda+\frac{i}{2}\Big)^L B(\lambda_1)B(\lambda_2)\omega_++\non\\
&-\Big(\frac{a(\lambda_1-\lambda)}{b(\lambda_1-\lambda)}\frac{c(\lambda_2-\lambda)}{b(\lambda_2-\lambda)}-\frac{c(\lambda_1-\lambda)}{b(\lambda_1-\lambda)}\frac{c(\lambda_2-\lambda_1)}{b(\lambda_2-\lambda_1)}\Big)\Big(\lambda_2+\frac{i}{2}\Big)^LB(\lambda)B(\lambda_1)\omega_+\non\\
&-\frac{c(\lambda_1-\lambda)}{b(\lambda_1-\lambda)}\frac{a(\lambda_2-\lambda_1)}{b(\lambda_2-\lambda_1)}\Big(\lambda_1+\frac{i}{2}\Big)^LB(\lambda)B(\lambda_2)\omega_+,
\end{align}
which can be re-written as

\begin{align}
A(\lambda)B(\lambda_1)B(\lambda_2)\omega_+=&\Lambda(\lambda;\lambda_1,\lambda_2)\ket{\lambda_1,\lambda_2}+\non\\
&M_2 B(\lambda)B(\lambda_1)\omega_++M_1 B(\lambda)B(\lambda_2)\omega_+,
\end{align}
where
\begin{align}
&\Lambda(\lambda;\lambda_1,\lambda_2)=\frac{a(\lambda_1-\lambda)}{b(\lambda_1-\lambda)}\frac{a(\lambda_2-\lambda)}{b(\lambda_2-\lambda)}\Big(\lambda+\frac{i}{2}\Big)^L,\\
&M_1=-\frac{c(\lambda_1-\lambda)}{b(\lambda_1-\lambda)}\frac{a(\lambda_2-\lambda_1)}{b(\lambda_2-\lambda_1)}\Big( \lambda_1+\frac{i}{2}\Big)^L,\\
&M_2=-\frac{c(\lambda_2-\lambda)}{b(\lambda_2-\lambda)}\frac{a(\lambda_1-\lambda_2)}{b(\lambda_1-\lambda_2)}\Big( \lambda_2+\frac{i}{2}\Big)^L.
\end{align}
We obtain similarly
\begin{align}
D(\lambda)B(\lambda_1)B(\lambda_2)\omega_+=&\tilde{\Lambda}(\lambda;\lambda_1,\lambda_2)\ket{\lambda_1,\lambda_2}+\non\\
&\tilde M_2 B(\lambda)B(\lambda_1)\omega_++\tilde M_1 B(\lambda)B(\lambda_2)\omega_+,
\end{align}
where
\begin{align}
&\tilde{\Lambda}(\lambda;\lambda_1,\lambda_2)=\frac{a(\lambda_1-\lambda)}{b(\lambda_1-\lambda)}\frac{a(\lambda_2-\lambda)}{b(\lambda_2-\lambda)}\Big(\lambda-\frac{i}{2}\Big)^L,\\
&\tilde M_1=-\frac{c(\lambda-\lambda_1)}{b(\lambda-\lambda_1)}\frac{a(\lambda_1-\lambda_2)}{b(\lambda_1-\lambda_2)}\Big( \lambda_1-\frac{i}{2}\Big)^L,\\
&\tilde M_2=-\frac{c(\lambda-\lambda_2)}{b(\lambda-\lambda_2)}\frac{a(\lambda_2-\lambda_1)}{b(\lambda_2-\lambda_1)}\Big( \lambda_2-\frac{i}{2}\Big)^L.
\end{align}

\subsection*{$M$ magnon state}
This expression generalizes to the case of $M$ magnons, the Bethe vector is
\begin{align}\label{bethestateMmagnons}
\ket{\lambda_1,\lambda_2,\dots,\lambda_M}=B(\lambda_1)B(\lambda_2)\dots B(\lambda_M)\omega_+
\end{align}
and
\begin{align}
A(\lambda)B(\lambda_1)B(\lambda_2)\dots B(\lambda_M)\omega_+=&\prod_{j=1}^M\frac{a(\lambda_j-\lambda)}{b(\lambda_j-\lambda)}\Big(\lambda+\frac{i}{2}\Big)^L \ket{\lambda_1\dots\lambda_M}+\non\\
&\sum_{j=1}^M {M}_j B(\lambda)B(\lambda_1)\dots B(\lambda_{j-1})B(\lambda_{j+1})\dots B(\lambda_M)\omega_+,
\label{ABBB}
\end{align}
with
\begin{align}
M_j=-\frac{c(\lambda_j-\lambda)}{b(\lambda_j-\lambda)}\prod_{k\neq j}\frac{a(\lambda_k-\lambda_j)}{b(\lambda_k-\lambda_j)}\Big(\lambda_j+\frac{i}{2} \Big)^L.
\end{align}
We can repeat similar steps for $D(\lambda)$ and we obtain
\begin{align}
D(\lambda)B(\lambda_1)B(\lambda_2)\dots B(\lambda_M)\omega_+=&\prod_{j=1}^M\frac{a(\lambda-\lambda_j)}{b(\lambda-\lambda_j)}\Big(\lambda-\frac{i}{2}\Big)^L \ket{\lambda_1\dots\lambda_M}+\non\\
&\sum_{j=1}^M \tilde{M}_j B(\lambda)B(\lambda_1)\dots B(\lambda_{j-1})B(\lambda_{j+1})\dots B(\lambda_M)\omega_+,
\label{DBBB}
\end{align}
with
\begin{align}
\tilde{M}_j=-\frac{c(\lambda-\lambda_j)}{b(\lambda-\lambda_j)}\prod_{k\neq j}\frac{a(\lambda_j-\lambda_k)}{b(\lambda_j-\lambda_k)}\Big(\lambda_j-\frac{i}{2} \Big)^L.
\end{align}

Adding the results \eqref{ABBB} and \eqref{DBBB}, we find both the expressions of the eigenvalues and the Bethe Ansatz equation.
\\
The expression of the eigenvalue is
\begin{align}
\Lambda(\lambda;\lambda_1,\dots,\lambda_M)=\prod_{j=1}^M\frac{a(\lambda_j-\lambda)}{b(\lambda_j-\lambda)}\Big(\lambda+\frac{i}{2}\Big)^L+\prod_{j=1}^M\frac{a(\lambda-\lambda_j)}{b(\lambda-\lambda_j)}\Big(\lambda-\frac{i}{2}\Big)^L,
\label{Eigenvaluelambda}
\end{align}
with the Bethe roots $\lambda_i$ constrained to be solution of the Bethe ansatz equation (BAE).
\\
We notice that the the unwanted terms cancel against each other $M_j+\tilde{M}_j=0$. In our specific case, the functions $a(\lambda)$, $b(\lambda)$, and $c(\lambda)$ are given by the expressions \eqref{abcmodelxxx}. In particular, $c$ is a constant and $b(\lambda)=-b(-\lambda)$. This leads us to the Bethe equations
\begin{align}
\Bigg(\frac{\lambda_j+\frac{i}{2}}{\lambda_j-\frac{i}{2}} \Bigg)^L=\prod_{k\neq j}\frac{-a(\lambda_j-\lambda_k)}{a(\lambda_k-\lambda_j)}=\prod_{k\neq j}\frac{\lambda_j-\lambda_k+i}{\lambda_j-\lambda_k-i},\,\,\,\,\,\,\,\,\,\,\,j=1,\dots, M.
\label{Betheequationsxxz}
\end{align}
We remind that the Hamiltonian is related to the transfer matrix by \eqref{chargeQ2transfer}, and considering \eqref{monodromy22}, to obtain the Hamiltonian
\be
\mathbb{H} = {J\over 2}\left( i {d\over d \lambda} \log 
t(\lambda)\Big\vert_{\lambda = \frac{i}{2}} 
- L\, \id \right) 
= \sum_{n=1}^{L-1} \mathcal{H}_{n\,, n+1} + \mathcal{H}_{L \,,1} \,,
\label{Hamiltonian}
\ee 
where we considered the normalization and the shift. Furthermore, the derivative is evaluated at $\lambda=i/2$ because we defined the monodromy matrix as \eqref{monodromy22}. In particular, the 2 site Hamiltonian is related to the $R$-matrix by \eqref{HXXXunnormalized} and we obtain
\be
\mathcal{H}_{ij} = {J\over 2} \left( {P}_{ij}\ \partial_\lambda R_{ij}(\lambda)|_{\lambda=0} - 
\id \right)=  {J\over 2} \left( {P}_{ij} - 
\id \right)\,.
\label{two-site-R}
\ee 

In this way, using the expression of the eigenvalue \eqref{Eigenvaluelambda}, and the fact that the momentum and the energy are related to the transfer matrix by   (\ref{momentum}) and
(\ref{Hamiltonian}), we find 
\be
P = {1\over i} \sum_{\alpha=1}^{M} 
\log \left( {\lambda_{\alpha} + {i\over 2} 
\over \lambda_{\alpha} - {i\over 2}} \right)
\quad (\mbox{mod } 2 \pi)
\,, \qquad 
E = - {J\over 2} \sum_{\alpha=1}^{M} {1\over \lambda_{\alpha}^{2} + {1\over 4}}
\,,
\label{energy/momentum}
\ee 
where $\lambda_\alpha$ are the Bethe roots satisfying \eqref{Betheequationsxxz}.

\subsection*{Trick to obtain the Bethe equations}

We observe that in the case of the XXX spin chain, we were able to explicitly compute the commutation relations between the $A$ and $D$ operators with the $B$ operators and identify the unwanted terms. However, for more complex models, this calculation can be lengthy and cumbersome. Fortunately, there exists an alternative method to derive the Bethe equations.
\\
First, to obtain the expression of the eigenvalues \eqref{Eigenvaluelambda}, we explicitly permute the $A$ and $D$ with all the $B$s and then we require the cancellation of the unwanted terms. By disregarding the unwanted terms from the beginning, obtaining the expression for the eigenvalues becomes straightforward. Specifically, from \eqref{algebra2} each time we commute $A$ with one of the $B$, we get one term of the type $\frac{a}{b}$ and similarly from \eqref{algebra} by commuting $D$ with $B$. Finally, from \eqref{properties}, when the $A$ or $D$ acts on the vacuum, it picks up the eigenvalue. By following this procedure, we can easily derive the expression for the eigenvalue. Schematically,
\begin{equation}
\Lambda(\lambda, \{\lambda_i\})=\underbrace{\Big(\prod_{j=1}^M\Big)}_{\text{from\,} A\, B} \underbrace{\Big( \dots \Big)}_{\text{from}\,A \omega_+}+ \underbrace{\Big(\prod_{j=1}^M\Big)}_{\text{from}\, D\, B} \underbrace{\Big( \dots \Big)}_{\text{from}\, D \omega_+}.
\end{equation}

The eigenvalue $\Lambda(\lambda;\lambda_1,\dots,\lambda_M)$  \eqref{Eigenvaluelambda} has a pole when $b(\lambda-\lambda_j)=0$, that is when $\lambda=\lambda_j$. However, it is important to note that the transfer matrix remains non-singular at these points, suggesting that these poles should not exist. To eliminate these poles, we can ensure that the residue of the eigenvalue at these points is zero. From 
\begin{align}
\text{Res}_{\lambda\to \lambda_j} \eqref{Eigenvaluelambda}=-\mathop{\prod_{k =1}}_{k\neq j}^M\frac{a(\lambda_j-\lambda_k)}{b(\lambda_j-\lambda_k)}\Big(\lambda_j+\frac{i}{2}\Big)^L+\mathop{\prod_{k =1}}_{k\neq j}^M\frac{a(\lambda_k-\lambda_j)}{b(\lambda_k-\lambda_j)}\Big(\lambda_j-\frac{i}{2}\Big)^L=0,
\end{align}
we exactly obtain the  Bethe equations \eqref{Betheequationsxxz}.
\\
This technique proves to be valuable, particularly in the case of more complex models, where the commutation relations between $A$ and $B$, or $D$ and $B$, may involve more than three terms. Keeping track of all of them become tortuous. We use this trick in the Appendix \ref{BAB3chapter}, when we apply the nested Bethe ansatz technique to solve one of the new integrable model that we find. However, we remark that, even though this trick works for the known cases, to the best of our knowledge, it has not been proven that the cancellation of the poles is a sufficient condition for obtaining the Bethe equations.

\subsection*{Roots at infinity}

A valid concern arises when two or more states are degenerate, as we may only obtain one of the corresponding Bethe states via the Algebraic Bethe ansatz. How can we obtain the Bethe states corresponding to the same eigenvalue?
\\
Let us start from the beginning. The reason of the degeneracy is the symmetry. The Hamiltonian \eqref{Hamiltonian} commutes with the generators of the global $\alg{su}(2)$ algebra
\begin{align}
&[\mathbb{H},S_x]=[\mathbb{H},S_y]=[\mathbb{H},S_z]=0,
&&S_A=\sum_j A_j,\,\,\,\,\, A = X, Y, Z.
\end{align}
 For example, we can choose the eigenstate of the Hamiltonian to be the eigenstate of  $S_z$ as well. 
\\
Since at each site of the chain there is a $\alg{su}(2)$ algebra, the whole Hilbert space is a tensor product of $L$ copies of the fundamental representation of $\alg{su}(2)$ and can be  decomposed into a direct sum of irreducible representations $V_\alpha$ of the global $\alg{su}(2)$ symmetry,
\begin{align}
&\mathbb{C}^2 \otimes \dots \otimes \mathbb{C}^2  = \oplus_\alpha V_\alpha,
\end{align}
$V_\alpha$ is an invariant subspace of the Hamiltonian and all the states have the same energy.
\\
For example, for $L=2$, the Hilbert space decomposes into a spin-0 and a spin-1 representation. The eigenstates of the Hamiltonian are \eqref{state1XX} and \eqref{state2XX}. The states $\ket{\uparrow \uparrow}$ and $\ket{\uparrow \downarrow}-\ket{ \downarrow \uparrow}$ are \textit{highest weight states} and corresponds to the Bethe states \eqref{vec}. Having found all the highest weight states, one can use the lowering spin operator and obtain all the other states as well. These are called \textit{descendants}.
\\
An equivalent way to obtain the other states is to consider the Bethe roots at infinity. In fact, in the limit $\lambda\to \infty$, the lowering spin operator $S_-$ is related to the $B$ operator by
\begin{align}
&B(\lambda)=c_1\,\lambda^{L-1}S_-+\dots,\,
&&\lambda\to \infty,
\end{align}
with $c_1$ a constant. If we add roots $\lambda_k=\infty$ to any solutions of the Bethe ansatz, we have the Bethe equations still satisfied. In other words, for finite Bethe roots, the Bethe states \eqref{bethestateMmagnons} are highest weight states, and if some of the roots are at infinity they are descendant. We refer to \cite{staudacher2012review} for a more detailed analysis\footnote{The author breaks the $\alg{su}(2)$ symmetry by considering a twisted model and afterward, sending the twist parameter to $0$, he recovers the un-twisted result.}.
\\
We can also check that since
\begin{align}
&[S_z,B(\lambda)]=-B(\lambda),
\end{align}
the Bethe state of $M$ magnons is an eigenstate of $S_z$, with eigenvalue $L/2-M$, and we can
think of this state as having M spins flipped from $\ket{\uparrow}$ to $\ket{\downarrow}$. This observation clarifies the meaning of an $M$ magnon state, highlighting that it corresponds to a state where a certain number of spins have been flipped. Furthermore, since the eigenvalue of $S_z$ for a Bethe state is $ L/2-M$, to get all highest weight states\footnote{We remark that, restricting to $M\le L/2$, we obtain the highest weight state.  To explore the complete spectrum, which includes $M>L/2$, one must repeatedly apply the lowering spin operator.} it is enough to consider only $M \le L/2$ in the Bethe equations.

\subsection*{Completeness of the Bethe ansatz}

It is also interesting to investigate whether the Bethe equations provide all the eigenvalues of the transfer matrix, a property known as the \textit{completeness} of the Bethe ansatz.  The answer to this question is expected to be positive for the XXX and XXZ model, however a rigorous proof is not available\footnote{We refer to \cite{hao2013completeness} for a summary.}.  It is common practice to verify completeness numerically when applying the Bethe ansatz to a specific model. We first diagonalize numerically the transfer matrix for different lengths of the spin chain $L=3,4,5,\dots$ and then compare the obtained eigenvalues with those derived from the Bethe ansatz.
\\
To illustrate the process, let's consider the case of $L=5$ for the XXX spin chain. By direct diagonalization of the Hamiltonian $\mathbb{H}$ in \eqref{Hamiltonian} with $J=-1$, we find the following eigenvalues $\Lambda$ along with their corresponding degeneracies $d$
\begin{align}
&\{\Lambda, d\}=\{\{0, 6\}, \{0.3455, 8\}, \{0.4410, 
  4\}, \{0.9045, 8\}, \{1.0, 
  2\}, \{1.5590, 4\}\}.
\end{align}
We can verify the completeness of the Bethe ansatz by checking if we can obtain the same eigenvalues by solving the Bethe equations \eqref{Betheequationsxxz} and plugging the solutions into the expression for the energy eigenvalues \eqref{energy/momentum}.
\\
As mentioned, it is enough to focus on the case\footnote{We remark that this restriction is due to the $\alg{su}(2)$ symmetry of the model. If we consider a state with $M>L/2$, its energy will not be different than one of the state with $M\le L/2$. At the level of the Energy \eqref{energy/momentum}, this is understood because we are adding constribution of states characterized by a Bethe root at infinity. We recommend to read the discussion at p. 197, 198 of \cite{staudacher2012review} for more details.} $M\le L/2$. In this case, we obtain\footnote{The Bethe equation are solved numerically by using the command FindRoots in Mathematica and specifying the starting point of the searching.}
\begin{align}
&M=0,\,\,\,\,\,\,\,\,\,\,0,\\
&M=1,\,\,\,\,\,\,\,\,\,\, 0.3455,\, 0.9045,\\
&M=2,\,\,\,\,\,\,\,\,\,\,0.4410,\, 1.5590,\, 1,
\end{align}
that reproduce all the eigenvalues found by numerical diagonalization, confirming the completeness of the Bethe ansatz for the XXX spin chain with $L=5$. The common practice to verify completeness is to extend this computation to different values of $L$ and check if all the eigenvalues will be found.

\subsection*{Is this method really powerful?}

As showed in this chapter, the problem to find the eigenvalues of the Hamiltonian become equivalent to solve the Bethe Ansatz equations \eqref{Betheequationsxxz} for the Bethe roots. We may ask why it is beneficial to use this reformulation since the equations become very difficult to solve as the system size and number of magnons increase.  As we showed in Figure \ref{comptime}, as $L$ increases, direct diagonalization becomes impossible even with the help of a powerful computer due to the exponential growth of the matrix size. In particular, we showed that for the Heisenberg XXX model, the direct diagonalization is possible until $L=15$ for the eigenvectors and $L=16$ for the eigenvalues. If we imagine to apply this problem to a physical system, for example a “metal”, the number of atoms in a unit volume is of the order of the Avogadro number $O(10^{23})$ and the direct diagonalization is clearly impossible.
\\
In that case, however, the equations \eqref{Betheequationsxxz} tend to enormously simplify
and one often is able to derive elegant linear integral equations. The case where $L\to \infty$ is called \textit{thermodynamic limit}, \cite{van2016introduction}. In this context,  various methods have been developed involving some assumptions about the nature of the solutions, such as the \textit{string hypothesis} (see, e.g., \cite{faddeev/takhtajan2}).
\\
From the Bethe equations \eqref{Betheequationsxxz}, qualitatively, it can be understood that computing the Bethe roots for small $M$ is not very complicated numerically, even if $L$ is very big. The problem become more challenging as both $L$ and $M$ grow. However, without any sophisticated algorithm, one can check that it is possible to obtain (numerically) some of the eigenvalues even for $L\sim 100$, $M=40$ and bigger.
\\
Furthermore, if we know the Bethe equations of a theory but lack knowledge of the specific spin chain from which it originates, we can obtain information about the theory's original symmetry by examining the quantity of Bethe roots at infinity.
\\
The Bethe ansatz has proven to be an incredibly powerful technique, yielding a wide range of results in various areas of physics. Some notable achievements include determining the spectrum of low-lying states, finite-size corrections, scattering matrices and thermodynamics (finite temperature and magnetic field).  Progress has been made on the computation of correlation functions.

\subsection*{Other models}
The spin 1/2 isotropic Heisenberg chain which we have discussed is the simplest integrable models.  However, it is possible to introduce additional complexities, such as considering inhomogeneities into the spin chain. Since this chapter aimed to provide an introduction, we focused on the simplest case. In the Appendix \ref{BAB3chapter}, we apply this technique to a more complicated model and we also consider inhomogeneities.\\Moreover, by using the appropriate $R$-matrices, one can apply this technique to the  anisotropic XXZ chains, chains with spins in higher-dimensional representations ($S= 1 \,, 3/2 \,, \ldots$) and chains with spins in 
representations of higher-rank algebras $\alg{su}({N})$.  The Bethe ansatz technique can also be applied for open boundary conditions, \cite{Sklyanin:1988yz}.\\The Bethe Ansatz equations have been obtained for many such models.

\chapter{Classification of Integrable models}

\ifpdf
    \graphicspath{{Chapter2/Figs/Raster/}{Chapter2/Figs/PDF/}{Chapter2/Figs/}}
\else
    \graphicspath{{Chapter2/Figs/Vector/}{Chapter2/Figs/}}
\fi

\label{classificationchapter}
Due to the relevance of quantum integrable models, numerous approaches have been developed to \textit{find solutions} to the Yang-Baxter equations.
\\
In this chapter, we give a brief overview of these methods, discussing their strengths, weaknesses, and the types of solutions found.  Then, we provide a detailed explanation of \textit{the boost automorphism method} and how to apply it with a concrete example. One of the main contributions of this thesis is the application of this method to discover \textit{new solutions} of the Yang-Baxter equation with $R$-matrices of non-difference form in different scenarios.  At the end of the chapter, we enumerate all the contexts in which we applied the method.

\section{A brief introduction}
The most intuitive and obvious method to classify integrable models consists in searching for the solution of the Yang-Baxter equation \eqref{eq:YBE}. The latter is a \textit{cubic polynomial} equation whose solution gives, in principle, all possible quantum integrable models\footnote{We remark that we refer here to quantum integrable models characterized by an $R$-matrix solution of the Yang Baxter equation. There exist, in fact, long range interacting models characterized by the conserved charges but for which the $R$-matrix is not known.}. However, since it is technically enormously difficult to solve cubic coupled functional equations, different groups of researchers developed various methods to overcome this problem.  It's worth mentioning that the literature on this topic is extensive, so our reference list is therefore partial. We have included a selection of papers, and interested readers are encouraged to consult the references therein.
\\
We divide the classifications methods into two classes:
\begin{itemize}
\item \textbf{Top-down}: The outcome of these methods is the expression of the $R$-matrix that satisfies the YBE. This $R$-matrix is the key component that enables us to construct the transfer matrix. By taking the logarithmic derivative of the transfer matrix, we can derive all the conserved charges. 
\item \textbf{Bottom-up}: The outcome is the Hamiltonian belonging to an integrable model. These methods impose constraints on the form of one of the conserved charges, namely the Hamiltonian, ensuring that it belongs to an integrable model. Afterwards, this is used to find the corresponding $R$-matrix.
\end{itemize}
The boost automorphism method used in this thesis belong to the bottom-up class and, in the following, after giving a derivation of the boost operator, we clarify with an example how to perform the various steps of the classification of quantum integrable models.

\section{Top-down methods}
\label{topdownmethods}

\subsection*{Impose symmetries on the $R$-matrix}
In physics, it is common to be interested in  models  that exhibit particular symmetries. The symmetries of the Hamiltonian are carried out also to the $R$-matrix. In particular, requiring that $\mathbb{H}$ commutes with the generator of a bialgebra $\mathcal{A}$, translates to the level of the $R$-matrix as $\Delta^{op}(a)R(u,v)=R(u,v)\Delta(a)$, where $\Delta$ and $\Delta^{\rm op}$ denote the coproduct and opposite coproduct related by conjugation on $\mathcal{A}$, respectively. These constraints restrict the number of independent functions of the $R$-matrix and  usually only leave few functions to be fixed with the help of YBE.
\\
\textbf{Type of solutions found:} This approach has been used since early days \cite{Kulish:1981gi,Kulish1982,jimbo1986quantumr,Bazhanov:1986mu,Kuniba:1991yd} to obtain finite-dimensional irreducible representations of $GL(2, \mathbb{C})$ but also more recently was applied in the context of AdS/CFT \cite{Beisert:2005tm,Borsato:2014hja,Borsato:2014exa,Borsato:2015mma,Lloyd:2014bsa,Hoare:2014kma,Garcia:2020vbz,Garcia:2020lrg}.
\\
\textbf{Pros:} Easy to apply.
\\
\textbf{Cons:} It will only produce $R$-matrices characterized by a given symmetry.
\\
As a curiosity, this approach was one of the first used and in the paper \cite{Kulish1982} of 1982 one of the first classification was given. It is interesting to point out a statement of their paper
\begin{figure}[h!]
  \centering
  \includegraphics[height=3.cm]{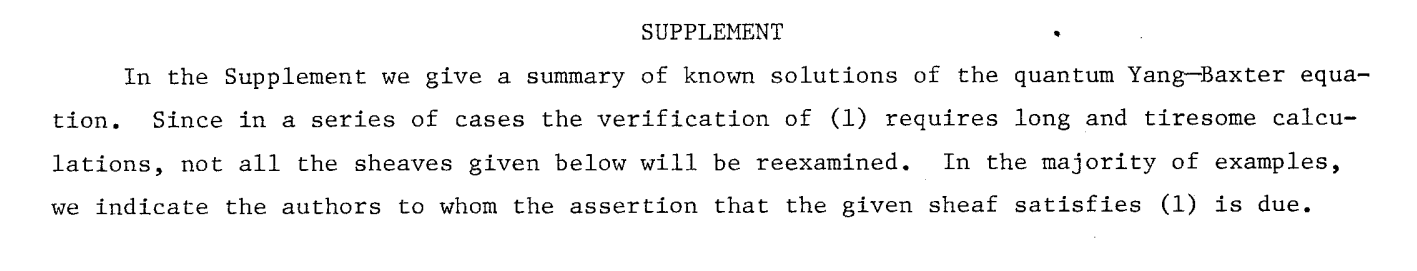}
\end{figure}
\\
Nowadays, with the software Mathematica (and many others) we can check the YBE (equation (1) of \cite{Kulish1982}) in seconds.

\subsection*{Baxterisation procedure}

The Baxterisation procedure uses representations of the braid group, such as the Hecke algebra, the Temperley-Lieb algebra, or the Birman-Murakami-Wenzl algebra, to obtain solutions of the Yang-Baxter equation (YBE) with a spectral parameter.
\\
In fact, it is known that the $S$-matrix of a representation of a braid group is a solution of the Yang-Baxter relation (which is the spectral parameter independent case of the Yang-Baxter equation). The Baxterization was initially proposed in the context of knot theory \cite{Jones:1989ed,Turaev:1988eb,Wu_1993,jones1990baxterization}   and consists in inserting a spectral parameter into a given representation of braid group so that the Yang-Baxter equation is satisfied. We can recover the original representation of the braid group as a limit in the spectral parameter of the Baxterized version.  
\\
\textbf{Type of solutions found:}  Many new $R$-matrices of difference form with dimension 4x4 and 9x9 \cite{Isaev:1995cs,Jimbo:1985vd,cheng1991yang,zhang1991representations,li1993yang,Arnaudon:2002tu,Kulish:2009cx} and recently also of non-difference form of 4x4 type and  higher spin representation of $\alg{sl}(2)$ \cite{Crampe:2016nek,crampe2019back,Crampe:2020slf}.
\\
\textbf{Pros:} Easy to apply
\\
\textbf{Cons:} Only $R$-matrices coming from a representation of an algebra can be found.

\subsection*{Differential approach}
This method  was broadly used in the context of spin chain with open boundary conditions to solve the reflection equation \cite{Sklyanin:1987bi,NepomechieJPA:1991}. It was used in the early days, \cite{Kulish1982} and also, more recently by R. S. Vieira through a series of works \cite{Vieira:2017vnw, Vieira:2019vog}.  In physical models, we can typically assume that the $R$-matrix is differentiable in a neighbourhood of a certain point. This condition is not overly restrictive, in fact for regular spin chain to obtain the conserved charges from the transfer matrix one can take the power series expansion \eqref{expansiontransfer}. 
\\
The method consists in taking the formal derivative of the YBE \eqref{eq:YBE} with respect to one of the spectral parameters and evaluate the derivatives at a fixed point of these variables (for example at zero). Then, since the system of equations is overdetermined, the derivatives and the functions can be treated as independent between each other. In this way, one can select and solve first the easiest equations i.e. those linear in the derivatives and then, plugging back these solutions into the others, the remaining ones become simpler than they were initially. Once all the solutions of the system are found, it is necessary to individually check that the derivatives and functions satisfy the compatibility conditions. These conditions ensure that the solutions are valid and consistent. In some cases, these impose additional constraints, while in other cases they may reveal incorrect or incompatible solutions which should be discarded\footnote{This last case happens if for example we found that one of the entries of the $R$-matrix should be zero, but the corresponding derivative is not zero.}.
\\
\textbf{Type of solutions found:} This method provides a full classification of $R$-matrix of difference form of size $4\times 4$ of $8$-and-lower-vertex models \cite{Vieira:2017vnw}; difference form $R$-matrix of size $9\times 9$ satisfying the so-called \textit{ice rule} condition\footnote{More details on this type of models will be given in section \ref{whatarevertexmodel}.} \cite{Vieira:2019vog}.\\
\textbf{Pros:} The method is complete: given an ansatz for the structure of the $R$-matrix, all the possible solutions of that type will be found. \\
\textbf{Cons:} It quickly becomes unwieldy as the size of the $R$-matrix increases. 

\section{Bottom-up methods}
\label{bottomupmethods}

\subsection*{Models solvable via the coordinate Bethe ansatz}
As discussed in chapter \ref{intro}, the strength of integrable models lies in their exact solvability. The \textit{coordinate Bethe ansatz} is one of the techniques available to solve these models.  Another method to classify integrable models involves the identification of the constraints on the Hamiltonian to ensure that the spectrum can be obtained through the coordinate Bethe ansatz. Having the expression of the Hamiltonian belonging to an integrable model allows the construction of the corresponding $R$-matrix\footnote{The steps on how to obtain the $R$-matrix from the Hamiltonian are given in section \ref{RfromHamiltonian}.}.  
\\
\textbf{Type of solutions found:} In a series of papers \cite{Crampe:2013nha,Fonseca:2014mqa,Crampe:2016she}, the authors classify the integrable Hamiltonians of difference form of dimension $9\times 9$ with 14,17, 19 or 33 vertex.
\\
\textbf{Pros:} Once some integrable models are found, we are sure that they can be solved via the Coordinate Bethe Ansatz.
\\
\textbf{Cons:} Unfeasible for the integrable models that cannot be solved via the coordinate Bethe ansatz\footnote{As already remarked, non all methods to solve integrable models is applicable to all of them. For example, the XYZ spin chain is integrable but it cannot be solved via the standard coordinate Bethe ansatz approach because we cannot define a good reference state.} and it is hard to generalize for $\mathbb{C}^n\otimes \mathbb{C}^n$, $n>3$ local Hilbert space.

\subsection*{Iterative procedure}
This old procedure to find integrable models of difference form type can be considered a mix between the bottom-up and the top-down methods. It consists in fixing the symmetries of the Hamiltonian and reconstruct the $R$-matrix iteratively by imposing the YBE order by order in the spectral parameter. Explicitly, the starting point is the expansion for $R(u)=R^{(0)}+u R^{(1)}+u^2 R^{(2)}+\dots$ . Inserting this into the YBE and considering separately order by order in $u$, gives a set of equations for the $R^{(i)}$. In the models obtained using this method, the solution for $R^{(1)}$ was determined, and it was observed that the matrices $R^{(2)}$ and $R^{(3)}$ are completely determined by the solution\footnote{This is an old conjecture called tangential \textit{star-triangle hypothesis} . It states  that all the higher orders of the $R$-matrix depend on the first one. This is not proved but it is believed to be true, \cite{Idzumi:1994kx}.} of $R^{(1)}$. This allowed to re-sum and find the explicit expression of the $R$-matrix. Since $R^{(1)}$ corresponds to the Hamiltonian \eqref{HXXXunnormalized}, from this method, the Hamiltonian can be computed first and then use the results obtained to re-sum and obtain the $R$-matrix.
\\
\textbf{Type of solutions found:} In \cite{Idzumi:1994kx} and \cite{Martins:2013ipa,Martins:2015ufa} they listed the difference form $R$-matrix of 19-vertex type. 
\\
\textbf{Pros:} The method is complete.
\\
\textbf{Cons:} The re-sum is very complicated for more complex ansatz.

\section{New method: Boost automorphism mechanism}

The main result of this thesis is the use of the \textbf{boost operator} to classify quantum integrable spin chain with $R$-matrix of non-difference form. In this section, we describe in details the method used: we give the derivation of the boost operator \cite{HbbBoost,Loebbert:2016cdm}, explain the main idea behind it and then we clarify the statements with an example.
\\
The boost method was first used to classify $R$-matrix of difference form \cite{deLeeuw:2019zsi,deLeeuw:2019vdb} and then, we generalized it to the case of non-difference form \cite{deLeeuw:2020ahe,classificationybandboost}. In this thesis, we focus on the second case and in particular in application in the context of \textit{Open quantum systems}. We refer to the thesis \cite{ryan2022integrable,pribytok2022automorphic} for applications in AdS/CFT.
\\
As mentioned, there are various historical approaches to classify integrable models and many of them start from the $R$-matrix, solution of the Yang-Baxter equation (YBE). The $R$-matrix is then used to construct the transfer matrix, and by taking the logarithmic derivative, all the conserved charges of the integrable model can be obtained. In particular, the Hamiltonian provides the dynamics of the model.
\\
In our approach, we followed a bottom-up strategy. Instead of starting from the $R$-matrix, we began with the Hamiltonian of the system and used it to derive the corresponding $R$-matrix. This allowed us to directly establish the connection between the Hamiltonian and the integrable structure of the model.
\\
The main \textbf{advantage} of this method lies in its versatility and applicability to a wide range of contexts. In fact, the starting point is to chose an ansatz for the Hamiltonian density $\lH_{12}(\theta)$. The ansatz is dictated by different reasons: we may need to focus on Hamiltonians with a given symmetry or  having some particular properties, for example in chapter \ref{integrableopenquantumsystemchapter} we identify the Hamiltonian to be a \textit{Lindblad superoperator}: the generator of the dynamics of an open quantum systems. While the method can be extended to more general Hamiltonians, it is important to note that the {computational complexity} of the calculations  increase to the point where it may become unfeasible to obtain explicit results. However, with the advancement of computational power, we expect that more general integrable Hamiltonians can be successfully classified using this method.
\\
Obviously, by starting from any type of ansatz for the Hamiltonian, there is no a-priori reason such that it defines an integrable model. This is the point where we apply the method and we restrict the entries of the Hamiltonian such that it belongs to an integrable model. We do this by using the so-called \textit{boost operator}  \cite{Tetelman,HbbBoost,Loebbert:2016cdm}, which is an alternative way to generate the tower of conserved charges for regular integrable models without the need to construct the transfer matrix and expand it. In the following section, we  explicitly write the derivation of the boost operator by following \cite{HbbBoost,Loebbert:2016cdm}. For the moment, we can consider the boost as a machinery that allow us to compute the conserved charge $\mathbb{Q}_3$ by starting from $\mathbb{Q}_2$.
\\
Since our model should be integrable, the two charges $\bQ_2$ (constructed from the density Hamiltonian) and $\bQ_3$ should commute. This will place a number of constraints on the entries of the density $\lH_{12}$ (and consequently on $\mathbb{Q}_2$) in the form of a system of ODEs. We then solve the set of constraints and show that the resulting Hamiltonian defines an integrable system, meaning it can be obtained from a solution of the YBE.

\subsection{Boost method in a nutshell}

\begin{enumerate}
\item Start from an ansatz for the Hamiltonian density $\mathcal{H}_{12}=\mathcal{Q}_2$ (this depends on some functions $h_1, h_2, \dots, h_k$). We remark that we work with the general case of $R$-matrix of non-difference form type. In this case, the Hamiltonian depends on a spectral parameter $\mathcal{H}_{12}=\mathcal{H}_{12}(\theta)$ and $h_i=h_i(\theta)$. For simplicity, we do not write the explicit dependence.
\item Use the boost operator  to construct the density $\mathcal{Q}_3$ (it depends on the same functions $h_1,\, h_2, \,\dots , h_k$).
\item Solve the constraint $[\mathbb{Q}_2,\mathbb{Q}_3]=0$ $\rightarrow$ potentially\footnote{In principle, we should check that all the commutators $[\mathbb{Q}_i,\mathbb{Q}_j]$ vanish. We instead prove integrability by constructing the $R$-matrix.} integrable Hamiltonian $\mathcal{H}_{12}$.
\item Use $\mathcal{H}_{12}$ to find the $R$-matrix via the Sutherland equation.
\item Check that such $R$-matrix  satisfies the YBE.
\item Answer the question: Is the $R$-matrix that we found \textit{new}?
\end{enumerate}

\subsection{Derivation of the boost operator}

In this section, we review the construction of the \textit{boost operator} (also known as ladder operator) for non-difference form models. Our exposition closely follows that of \cite{HbbBoost,Loebbert:2016cdm}. This operator can be used in a recursive relation to generate all the tower of conserved charges without using the transfer matrix.
\\
The term \textit{boost operator} is derived from the fact that, in (1+1)-dimensional continuum field theories, the generators of the Poincaré group are $B, P$ and $H$, which correspond to the generators of Lorentz boosts and spatial and temporal translations, respectively. The last two are the total momentum operator and the Hamiltonian, respectively. They obey the algebra
\begin{align}
&[B,H]=P,
&&[B,P]=H,
&&[H,P]=0.
\end{align}
For some integrable models defined on a lattice, the entire set of conserved quantities satisfy
\begin{align}
&[B,t^{(n)}]=t^{(n+1)},
&&[t^{(n)},t^{(m)}]=0,
\end{align}
with $t^{(0)}$ and $t^{(1)}$ the momentum operator $P$ and the Hamiltonian $H$. The boost acts as a ladder operator for the charges.
\\
In the derivation of the explicit expression of the boost operator, we focus on the case of $R$-matrices of \textbf{non-difference} form $R=R(u,v)\neq R(u-v)$ and, for completeness, we comment on how to apply the method to the difference form case.
\\
Our starting point is the Yang-Baxter equation \eqref{eq:YBE}
\begin{equation}
R_{12}(u_1,u_2)R_{13}(u_1,u_3)R_{23}(u_2,u_3)=R_{23}(u_2,u_3)R_{13}(u_1,u_3)R_{12}(u_1,u_2),
\end{equation}
we take the derivative w.r.t. $u_3$ and we omit the dependence on the spectral parameters. There is no ambiguity, since $R_{ij}=R_{ij}(u_i,u_j)$
\begin{equation}
R_{12}R^\prime_{13}R_{23}+R_{12}R_{13}R^\prime_{23}=R^\prime_{23}R_{13}R_{12}+R_{23}R^\prime_{13}R_{12},
\end{equation}
we identify with $R'(u,v)$ the differentiation w.r.t. $v$. We multiply by $P_{23}$ from the left, we send $u_2\to u_3$ and we use the regularity \eqref{regularityR} and the boundary condition \eqref{hamiltonian}\footnote{Notice that here we are using the boundary condition by taking the derivative w.r.t. the second spectral parameter, so we have an extra minus sign.}

\begin{align}
&R_{ij}(u,u)=P_{ij},
&&\partial_{u_j}R_{ij}(u_i,u_j)|_{u_i\to u_j}=R^\prime_{ij}(u_i,u_j)|_{u_i\to u_j}=-P_{ij}\mathcal{H}_{ij}(u_j).
\label{bcsecondsutherland}
\end{align}
This leds to the well known Sutherland equation 
\begin{equation}
\left[R_{13} R_{12}, \lH_{23}(\theta)\right]  =  R_{13}R^\prime_{12} - R^\prime_{13}R_{12} ,
\label{sutherland22}
\end{equation}
where here we denote\footnote{The Sutherland equation only depends on two spectral parameters since we considered the limit on the YBE where one parameter approaches the other.} $R_{ij}:=R_{ij}(u,\theta)$. We now make the replacement $1\mapsto a$, $2\mapsto k$, $3\mapsto k+1$, obtaining 
\begin{equation}\label{localsuther}
\left[R_{a,k+1} R_{ak}, \lH_{k,k+1}(\theta)\right]  =  R_{a,k+1}R^\prime_{ak} - R^\prime_{a,k+1}R_{ak}.
\end{equation}
We consider an infinite homogeneous spin chain with monodromy matrix\footnote{For clarity, we call the auxiliary space $a$ instead of $0$, since we are using the labels $\infty,\dots,1,0,-1,\dots,-\infty$ for the physical spaces.} $T_a(u,\theta)$ (the infinite version of \eqref{monodromycap1}) given\footnote{Here we identify the auxiliary space with "$a$" instead of "$0$" since "$0$" is one of the site of the physical spin chain.} by 
\begin{equation}
T_a(u,\theta)=\dots R_{a1}R_{a0}R_{a,-1}\dots.
\end{equation}
Now take \eqref{localsuther} and multiply from the left with the product of $R$-matrices $\dots R_{a,k+2}$ and from the right with $R_{a,k-1}\dots$ and then multiply the resulting equation by $k$ and sum over $k$ from $-\infty$ to $\infty$. The two terms on the right hand side of \eqref{localsuther} telescopically cancel\footnote{This happens because we are considering a spin chain of infinite lenght.} and we are left with 
\begin{equation}
\sum_{k=-\infty}^\infty k\, [T_a(u,\theta),\lH_{k,k+1}(\theta)]=\frac{d\, T_a(u,\theta)}{d\theta},
\end{equation}
and by taking the partial trace over the auxiliary space
\begin{equation}
\sum_{k=-\infty}^\infty k\, [t(u,\theta),\lH_{k,k+1}(\theta)]=\frac{d\, t(u,\theta)}{d\theta}.
\end{equation}
Finally, using the expansion \eqref{expansiontransfer} we obtain 
\begin{equation}
\bQ_{r+1}(\theta)=\sum_{k=-\infty}^\infty k\, [\mathcal{H}_{k,k+1}(\theta),\bQ_r(\theta)]+\partial_\theta\bQ_r(\theta),\quad r=1,2,3,\dots .
\label{chargeQrp1}
\end{equation}
We can then identify the \textbf{boost operator} with 

\begin{equation}\label{boostdef}
\mathcal{B}[\mathbb{Q}_2]:=\partial_\theta+\sum_{k=-\infty}^\infty k \lH_{k,k+1}(\theta).
\end{equation}
The infinite sum should be interpreted in a formal sense since what we are interested is not the boost operator itself but rather its commutator with the tower of conserved charges which is perfectly well-defined even for finite chains. In fact, 
\begin{equation}\label{boostconstruct}
\bQ_{r+1}=[\lB[\bQ_2],\bQ_r],\quad r\geq 1.
\end{equation}
To clarify this, we can explicitly compute the boost operator for a chain of finite lenght $L$. This construction will be in fact very useful in the following.
\\
Taking $r=2$ in \eqref{chargeQrp1} and remembering that $\bQ_2=\sum_j \mathcal{H}_{j,j+1}$, we can keep in the commutation relations only the cases where $j=k \pm 1$, all the other terms will trivially vanish and we obtain
\begin{equation}
\bQ_3(\theta)=\sum_{k=-\infty}^\infty k\, [\mathcal H_{k,k+1}(\theta),\mathcal H_{k-1,k}(\theta)+\mathcal H_{k+1,k+2}(\theta)]+\partial_\theta\bQ_2(\theta).
\end{equation}
We shift in the second term of the commutator $k\to k-1$ and, in this way, the linear dependence in $k$ drops out and we get
\begin{equation}
\bQ_3(\theta)=\sum_{k=-\infty}^\infty \, [\mathcal H_{k-1,k}(\theta),\mathcal H_{k,k+1}(\theta)]+\partial_\theta\bQ_2(\theta).
\label{boostQ3}
\end{equation}
The $k$ factor multiplying the sum was problematic because it was breaking the periodicity of the spin chain. Now, we  consider the terms inside the sum as a range three operator, and we write the total charge $\bQ_3$ in a chain of lenght $L$
\begin{equation}
\bQ_3(\theta)=\sum_{k=2}^{L+1} \, [\mathcal H_{k-1,k}(\theta),\mathcal H_{k,k+1}(\theta)]+\partial_\theta\bQ_2(\theta).
\label{boostBQ3finileL}
\end{equation}
In what follows, we use periodic boundary condition
\begin{equation}
\mathcal{H}_{L,L+1}\equiv \mathcal{H}_{L,1}.
\end{equation}

The expression obtained for the boost operator \eqref{boostdef} relies on the assumption that the $R$-matrix is of \textbf{non-difference form} and the Hamiltonian depends on a spectral parameter. In the case of a difference form $R$-matrix, a similar derivation can be carried out, resulting in the absence of the derivative term in \eqref{boostBQ3finileL}.
\section{Detailed explanation with an example}
\label{example}
\subsection*{Step 1. Starting point: ansatz for $\mathcal{H}_{12}$}
Our starting point is a nearest-neighbour Hamiltonian density $\lH_{12}(\theta)$ on $\mathbb{C}^n\otimes \mathbb{C}^n$. We emphasize that the validity of the method does not depend on the chosen ansatz. However, selecting a highly intricate ansatz, may lead to the technical problem of solving a very complicated coupled set of differential equations.
\\
In this example, we  start from an easy ansatz for a  density Hamiltonian in $\mathbb{C}^2\otimes \mathbb{C}^2$:
\begin{align}
\lH_{12}(\theta) = \begin{pmatrix}
0 & 0 & 0 & 0 \\
0 & h_1(\theta) & h_3(\theta) & 0 \\
0 & h_4(\theta) & h_2(\theta) & 0 \\
0 & 0 & 0 & 0 
\end{pmatrix},
\label{Hansatz}
\end{align}
$h_i (\theta)$ are the functions to be determined. To construct the total Hamiltonian $\bQ_2$ from the density, we need to specify the length of the spin chain. We  work\footnote{The reason is the following. A charge $\bQ_i$ is the sum of range $i$ densities. Since we are firstly interested in the commutator $[\bQ_2,\bQ_3]$, this is a sum of densities of range $2+3-1=4$. Starting from a chain of length lower than $4$, the commutator produces some cancellations which do not happen in general, while starting from a chain of higher length is computational more expensive and it does not give any additional information.} with $L=4$. For this reason, the total Hamiltonian is
\begin{equation}
\mathbb{H}(\theta)=\bQ_2(\theta)=\sum_{j=1}^4 \lH_{j,j+1}(\theta),\,\,\,\,\,\,\lH_{4,5}=\lH_{4,1}.
\end{equation}
The last condition defines a periodic chain. 
\subsection*{Step 2. Construction of $\bQ_3$ using the boost operator}

Using the boost operator \eqref{boostBQ3finileL}, the charge $\bQ_3$ is related to $\bQ_2$ as
\begin{equation}
\bQ_3(\theta)=\sum_{j=1}^4[\lH_{j,j+1}(\theta),\lH_{j+1,j+2}(\theta)]+\partial_\theta\bQ_2(\theta)=\sum_{j=1}^4 \lQ_{j,j+1,j+2}(\theta), \label{eq:boostoperator}
\end{equation}
with $\lQ_{j,j+1,j+2}(\theta)$ a range $3$ density.
In matrix form, this is
\begin{align}
\lQ_{123}(\theta) = \begin{pmatrix}
 0 & 0 & 0 & 0 & 0 & 0 & 0 & 0 \\
 0 & 0 & -h_1 h_3 & 0 & -h_3^2 & 0 & 0 & 0 \\
 0 & h_1 h_4 & \dot{h}_1 & 0 & \dot{h}_3-h_2 h_3 & 0 & 0 & 0 \\
 0 & 0 & 0 & \dot{h}_1 & 0 & \dot{h}_3+h_1 h_3 & h_3^2 & 0 \\
 0 & h_4^2 & \dot{h}_4+h_2 h_4 & 0 & \dot{h}_2 & 0 & 0 & 0 \\
 0 & 0 & 0 & \dot{h}_4-h_1 h_4 & 0 & \dot{h}_2 & h_2 h_3 & 0 \\
 0 & 0 & 0 & -h_4^2 & 0 & -h_2 h_4 & 0 & 0 \\
 0 & 0 & 0 & 0 & 0 & 0 & 0 & 0 
\end{pmatrix},
\end{align}
for simplicity we suppress the $\theta$-dependence on the entries and $\partial_\theta h_i= \dot{h}_i$ .

\subsection*{Step 3. Imposing the integrability constraints}

Since we want the model to be integrable, we have to require that the two operators $\bQ_2$ and $\bQ_3$ commute
\begin{align}
[\bQ_2,\bQ_3]=0.
\label{intcondition}
\end{align}
This constraint leads to a set of coupled ordinary differential equations. For the example analysed, those are
\begin{align}
&\dot{h}_3 (h_1+h_2) = (\dot{h}_1+\dot{h}_2) h_3 ,
&&\dot{h}_4 (h_1+h_2) = (\dot{h}_1+\dot{h}_2) h_4 ,
\end{align}
easily solved by
\begin{align}
&h_3 = \frac{c_3}{2} (h_1+h_2),
&& h_4 = \frac{c_4}{2} (h_1+h_2),
\end{align}
for some constants $c_{3,4}$.
\\
In this way, we obtain the Hamiltonian density
\begin{align}\label{HXXZ}
\lH_{12}(\theta) = \begin{pmatrix}
0 & 0 & 0 & 0 \\
0 & h_1 & \frac{c_3}{2} (h_1+h_2) & 0 \\
0 & \frac{c_4}{2} (h_1+h_2) & h_2 & 0 \\
0 & 0 & 0 & 0 
\end{pmatrix}.
\end{align}
This Hamiltonian density "potentially" belongs to an integrable model. At this stage, we have determined that $[\bQ_2,\bQ_3]=0$. However, to establish integrability, it is necessary to show that this commutativity condition holds for all charges, i.e., $[\bQ_i,\bQ_j]=0$ for all $i$ and $j$. To compute the infinite set of commutation relations would be a challenging task, so instead of taking that approach, we can prove the integrability of the model by finding the corresponding $R$-matrix for the density Hamiltonian \eqref{HXXZ}. This approach also provides insight into the logic behind the other bottom-up methods discussed in section \ref{bottomupmethods} for classifying integrable models.

\subsection*{Step 4. Use $\mathcal{H}$ to find the $R$-matrix via the Sutherland equation}
\label{RfromHamiltonian}
To guarantee that the model we found is integrable, we find the $R$-matrix corresponding to the density Hamiltonian. By doing so, we ensure that the model is indeed integrable. 
\\
In order to do this, we start from  the YBE
\begin{align}
R_{12}(u_1,u_2)R_{13}(u_1,u_3)R_{23}(u_2,u_3)=R_{23}(u_2,u_3)R_{13}(u_1,u_3)R_{12}(u_1,u_2)
\end{align} 
and we differentiate with respect to $u_1$ 
\begin{align}
&\dot{R}_{12}(u_1,u_2)R_{13}(u_1,u_3)R_{23}(u_2,u_3)+{R_{12}}(u_1,u_2)\dot{R}_{13}(u_1,u_3)R_{23}(u_2,u_3)=\\
&R_{23}(u_2,u_3)\dot{R}_{13}(u_1,u_3)R_{12}(u_1,u_2)+R_{23}(u_2,u_3)R_{13}(u_1,u_3)\dot{R}_{12}(u_1,u_2)
\end{align} 
where we used the shortcut $\dot{R}_{ij}=\partial_{u_1} R_{ij}(u_1,u_2)$. We now send $u_1\to u_2$ and use the regularity property \eqref{regularityR} the boundary condition \eqref{hamiltonian}
\begin{align}
& R_{ij}(u,u)=P_{ij}, &\dot{R}_{ij}(u,u)=P_{ij}\lH_{ij}(u).
\label{boundaryc}
\end{align}
In this way, one gets another version of the so-called \textit{Sutherland equations} 
\begin{equation}\label{eqn:Sutherland}
\left[R_{13} R_{23}, \lH_{12}(u)\right] = \dot{R}_{13} R_{23} - R_{13} \dot{R}_{23}\, ,
\end{equation}
where  $R_{ij}:=R_{ij}(u,v)$.
Similarly, as explained before, one can get the second Sutherland equation \eqref{sutherland22}
\begin{equation}\label{eqn:Sutherland2}
\left[R_{13} R_{12}, \lH_{23}(v)\right]  =  R_{13}R^\prime_{12} - R^\prime_{13}R_{12} ,
\end{equation}
where now ${R}^\prime_{i,j}=\partial_{u_2} R_{i,j}(u_1,u_2)$. The Sutherland equations \eqref{eqn:Sutherland} and \eqref{eqn:Sutherland2} constitute two sets of ODEs for the entries of the $R$-matrix and the boundary conditions are fixed by \eqref{boundaryc} and \eqref{bcsecondsutherland}.
\\
From Step 3. we found the potentially integrable Hamiltonian, which we use as input in the Sutherland equation. To verify its integrability, we proceed by solving the Sutherland equations for the corresponding $R$-matrix.
\\
In principle, we can start from a general ansatz for the $R$-matrix, but to simplify the problem, we can already impose some constraints on the entries of the $R$-matrix. In particular, we  compute an expansion\footnote{This expansion  provides a transparent explanation for the different signs in the boundary conditions \eqref{bcsecondsutherland} and \eqref{boundaryc}.} of the $R$ up to order $(u-v)^2$
\begin{equation}
R_{12}(u,v)=P_{12}\left(1+ (u-v)\,\mathcal{H}_{12}\left(\frac{u+v}{2}\right)+ \frac{(u-v)^2}{2}\,{\mathcal{H}_{12}}^2\left(\frac{u+v}{2}\right)+\mathcal{O}(u-v)^2\right)
\label{expansion}
\end{equation}
and use it to understand which entries are non-zero and which entries are the same. The expansion of the $R$-matrix at order 2 in $(u-v)$ can be understood from the braiding unitarity property $R(u,v)R(v,u)\propto \id$.
This ansatz on the entries of the $R$-matrix resulted to be correct for all the models analyzed in this thesis.
\\
We should also remember that the YBE continues to hold if one consider a different normalization of the $R$-matrix, $R(u,v)\to f(u,v)R(u,v)$, with $f$ any well defined function in the two parameters $u,v$. For more complicated models, a right choice of the normalization enormously simplifies the solution of the set of differential equations. However, in choosing a normalization, the compatibility with the boundary conditions should still hold.
\\
For this specific example, we used the following ansatz for the $R$-matrix
\begin{align}
R = \begin{pmatrix}
r_1 & 0 & 0 & 0 \\
0 & r_2 & r_3 & 0 \\
0 & r_4 & r_5 & 0 \\
0 & 0 & 0 & r_1
\end{pmatrix} ,
\end{align}
the entries 1,1 and 4,4 are equal as suggested from the expansion \eqref{expansion}. The boundary conditions \eqref{boundaryc} corresponding to this ansatz are
\begin{align}
&r_1(u,u)= 1,
&&r_2(u,u)=0,
&&r_3(u,u)=1,\\
&r_4(u,u)= 1,
&&r_5(u,u)= 0,\\
&\dot{r}_1(u,u)=0,
&&\dot{r}_2(u,u)=\frac{c_4}{2} (h_1(u) + h_2(u)),
&&\dot{r}_3(u,u)=h_2(u),\\
&\dot{r}_4(u,u)= h_1(u),
&&\dot{r}_5(u,u)=\frac{c_3}{2} (h_1(u) + h_2(u)),
\end{align}
where the notation $\dot{r}_{i}(u,u)$ means $\partial_u r_i (u,v)|_{v\to u}$. We are allowed to chose to normalize the entry $r_1(u,v)$ to $1$. Note that not all the choices are admitted, for example $r_5=1$ would have been not compatible with the boundary condition.
\\
From the Sutherland equation \eqref{eqn:Sutherland}, we obtained the set of partial differential equations (PDEs) that are not independent from each other\footnote{Here we have already eliminated the equations that would clearly lead to the same solutions.}
\begin{align}
&c_4 \,r_5=c_3\, r_2,
&&\frac{\dot{r}_4}{r_4}=\frac{\dot{r}_3}{r_3}+h_1-h_2,\\
&\dot{r}_2=\frac{c_4}{2} (h_1+h_2) r_3 r_4,
&&\frac{\dot{r}_3}{r_3}=h_2+\frac{c_3}{2}(h_1+h_2)r_2,\\
&\sqrt{1-c_3 c_4(1-r_3 r_4)}=1+c_3 r_2,
\end{align}
where $\dot{r}_i=\partial_u r_i (u,v)$.\\
In certain cases, solving the set of equations derived from the Sutherland equations proved to be challenging and resulted in non-trivial solutions involving functions such as trigonometric or elliptic Jacobi functions. To simplify the process, a more efficient approach was taken by considering a linear combination of these equations  and using both the Sutherland equations \eqref{eqn:Sutherland}  and \eqref{eqn:Sutherland2}.
\\
For the example analyzed, by solving these equations, imposing the boundary conditions and plugging the result into our ansatz, we obtain
\begin{align}\label{eq:RXXZ}
R =\left(
\begin{array}{cccc}
 1 & 0 & 0 & 0 \\
 0 & \frac{c_4}{\omega  \cot (\omega  H_+)-1} & \frac{\omega  e^{-H_-}}{\omega  \cos (\omega  H_+)-\sin (\omega  H_+)} & 0 \\
 0 & \frac{\omega  e^{H_-}}{\omega  \cos (\omega  H_+)-\sin (\omega  H_+)} & \frac{c_3}{\omega  \cot (\omega  H_+)-1} & 0 \\
 0 & 0 & 0 & 1 \\
\end{array}
\right),
\end{align}
where for simplicity of notation we called 
\begin{align}
&H_\pm = \frac{H_1(u,v) \pm H_2(u,v) }{2}, 
&&H_i(u,v) = \int^u_v h_i(\theta)d\theta,
&&\omega^2=c_3 c_4-1.
\end{align}
\subsection*{Step 5. Check that the $R$-matrix satisties the YBE}
As last step, we check that the $R$-matrix \eqref{eq:RXXZ} is a solution of the Yang-Baxter equation and that the boundary conditions are satisfied. We  conclude that the "potentially" integrable Hamiltonian \eqref{HXXZ} is \textit{actually integrable}.
\\
A comment is now necessary. It is worth noting that the constraint $[\bQ_2,\bQ_3]=0$ alone was sufficient to ensure the integrability of the model, as we were able to find the corresponding $R$-matrix. It will be clear in the following chapters, that for \textit{all} the initial ansatz we chose, this condition was enough to guarantee the integrability of the models. This goes back to an old conjecture of \cite{Grabowski} and to the best of our knowledge it is still unproven. 

\subsection*{Step 6. Is the $R$-matrix that we found new?}
In order to address this question, it is necessary to first examine a related issue.  We return to this point later in section \ref{isthemodelnew}.

\section{How many different models did we discover?}

For the example we have examined, the simplicity of the chosen ansatz results in a single Hamiltonian density that corresponds to an integrable model. However, there is some freedom in choosing the functions $h_1(u)$ and $h_2(u)$ and the constants $c_3,c_4$.
\\
For more complicated ansatz for the density Hamiltonian (and $R$-matrix), this method gives rise to a large redundancy in solutions. However, not all of these models  are independent. In fact, there are different transformations that can be performed on the $R$-matrix (and consequently on the Hamiltonian) that  preserve integrability. We  list them in the next section. As a result, in the following chapters, we present only one representative model from each  class.

\subsection{Identifications}
\label{identification}
Given a regular solution $R(u,v)$ of the Yang-Baxter equation, we can make the following identifications, which preserve integrability and regularity.
\subsubsection*{Local basis transformation (LBT)}
Given $V(u)\in \CC^n$ an invertible matrix,
\begin{align}
R^{(LBT)}(u,v) = \Big[V(u)\otimes V(v)\Big] R(u,v)  \Big[V(u)\otimes V(v)\Big]^{-1},
\end{align}
$R^{(LBT)}(u,v)$ remains a solution of the YBE.
\\
The Hamiltonian associated to this $R$-matrix is
\begin{align}\label{LBTlaw}
\lH^{(LBT)} = \big[\!V\otimes V\big] \lH  \big[\!V\otimes V\big]^{-1}\! - \big[\dot{V} V^{-1}\otimes \id - \id \otimes \dot{V} V^{-1}\big],
\end{align}

where for simplicity we omit the spectral parameter dependence. We notice that terms of the form $A\otimes \id-\id\otimes A$ in the Hamiltonian density can be removed by performing a basis transformation with the matrix $V(u)$ satisfying $\dot{V}=A V$. Adding these telescopic terms to the density Hamiltonian does not affect the total Hamiltonian $\mathbb{H}$ because they cancel each other in periodic spin chain.
\\
The regularity condition of the $R$-matrix is also preserved, in fact
\begin{align}
&R^{(LBT)}(u,u) = \Big[V(u)\otimes V(u)\Big] R(u,u)  \Big[V(u)\otimes V(u)\Big]^{-1}=P.
\end{align}
\subsubsection*{Reparametrization}
$R(g(u),g(v))$ is  still a  solution of the YBE and the regularity condition is  preserved. The Hamiltonian associated to it is
\begin{align}
\lH(u) \mapsto \dot{g} \lH(g(u)).
\end{align}
Notice however that a simple reparametrization of the Hamiltonian without renormalization (and also the other way around) does not preserve the compatibility condition $[\bQ_2,\bQ_3]=0$. We clarify this point in section \ref{whyisthemodelint}.
\subsubsection*{Normalization}
Normalization of the $R$-matrix is also allowed, with the associated Hamiltonian density 
\begin{align}
&R(u,v) \to g(u,v) R(u,v),
&&\lH(\theta) \mapsto \lH(\theta) + \dot{g}(\theta,\theta)\,\id,
\end{align}
where $ \id $ is the identity matrix. To preserve regularity, $g(\theta,\theta)=1$. 

\subsubsection*{Discrete transformations}  
The following discrete transformations are also allowed
\begin{align}
&R(u,v) \to P R(u,v) P, &&\mathcal{H}(\theta)\to P \mathcal{H}(\theta) P\label{discrete1}\\
&R(u,v) \to R^T(u,v), &&\mathcal{H}(\theta)\to P \mathcal{H}^T(\theta) P\label{discrete2}\\
&R(u,v) \to P R^T(u,v)P, &&\mathcal{H}(\theta)\to \mathcal{H}^T(\theta)\label{discrete3}\\
&R(u,v) \to R^* (u,v), &&\mathcal{H}(\theta)\to \mathcal{H}^*(\theta)\label{discrete4}\\
&R(u,v) \to R^\dagger (u,v), &&\mathcal{H}(\theta)\to P \mathcal{H}^\dagger(\theta) P\label{discrete5}
\end{align}
where $^T$ is the transpose of the matrix and $^*$ the complex conjugation element-wise. The regularity condition is also satisfied.
\subsubsection*{Twists}
If $U(u)$ is an invertible $n\times n$ matrix which satisfies $[U(u)\otimes U(v),R_{12}(u,v)]=0$ then it can be shown that 
\begin{equation}\label{twist}
R_{12}(u,v) \to U_2(v)R_{12}(u,v)U_1(u)^{-1}
\end{equation}
is a solution of the YBE. Under this transformation, the Hamiltonian density $\lH_{12}$ is
\begin{equation}\label{twistedH}
\lH_{12}\mapsto U_1 \lH_{12} U_1^{-1}+\dot{U}_1 U_1^{-1}
\end{equation}
and the analogue of the condition $[U(u)\otimes U(v),R_{12}(u,v)]=0$ for the Hamiltonian density can be easily worked out to be 
\begin{equation}\label{twistcond}
[U_1 U_2,\lH_{12}]=\dot{U}_1U_2-U_1\dot{U}_2.
\end{equation}
This relation may also derived by plugging the twisted R-matrix \eqref{twist} and
Hamiltonian \eqref{twistedH} into the Sutherland equations \eqref{eqn:Sutherland} and taking the spectral parameters to be the same. This is not surprising given the similarity between \eqref{twistcond} and the Sutherland equations \eqref{eqn:Sutherland}.
\subsubsection*{Universal and non-universal transformations}
Among the transformations described here, the local basis transformation, reparametrization, normalization and the discrete transformations are \emph{universal} as they do not alter the degeneracies of the eigenvalues and the symmetries of the model. To determine if two models are related by any of these transformations, one can compute the spectrum for spin chains of different lengths and compare the degeneracies.
\\
In contrast, recognizing if two models are related by a twist is more complicated.  This one changes the spectrum and the other physical properties of the integrable model in a non-trivial way. However, on the level of the $R$-matrix, a twist is a simple transformation and the twisted $R$-matrix remains a solution of the YBE.
\\
Some other transformations that preserve integrability are model dependent, for instance Drinfeld twist \cite{drinfeld1983constant}. 

\subsection{Is the model that we found in the example new?}
\label{isthemodelnew}

The Hamiltonian density \eqref{HXXZ} and its corresponding $R$-matrix \eqref{eq:RXXZ}, which we obtained by employing the simple ansatz \eqref{Hansatz} are not new. This is not unexpected considering the simplicity of the ansatz we used.
\\
We now perform the transformations that bring the Hamiltonian \eqref{HXXZ} to a known one. First, we use a local basis transformation to set $h_1=h_2$. This is achieved using the matrix $V(\theta)$
\begin{equation}\label{LBTapply}
V(\theta)={\rm exp}\left(\frac{1}{2}H_-(\theta) Z \right),
\end{equation}
in the transformation law \eqref{LBTlaw}, with $H_\pm(\theta)=\frac{1}{2}\left(H_1(\theta)\pm H_2(\theta)\right)$, $H_i(\theta)=\int^\theta h_i(u)du$ and $Z$ the third Pauli matrix.
\\
Next, we renormalize the Hamiltonian by $\frac{1}{2}\big(h_1+h_2\big)$. In this case, the renomalization does not affect the commutation constraint $[\mathbb{Q}_2,\mathbb{Q}_3]=0$, since the remaining Hamiltonian is constant, \begin{align}\label{Hc3c4}
\lH_{12}(\theta) = \lH_{12}=  \begin{pmatrix}
0 & 0 & 0 & 0 \\
0 & 1 & c_3 & 0 \\
0 & c_4 & 1 & 0 \\
0 & 0 & 0 & 0 
\end{pmatrix}.
\end{align}
Moreover, we use a twist and set $c_3=c_4=c$. Indeed, it is trivial to check that the twist condition \eqref{twistcond} is satisfied for any constant invertible diagonal matrix. In particular we use the twist $U$ 
\begin{equation}
U={\rm diag}\left(\sqrt{c_4},\sqrt{c_3}\right),
\end{equation}
in \eqref{twistedH}, to bring the Hamiltonian density to the form
\begin{align}\label{diffformham}
\lH(\theta) = \begin{pmatrix}
0 & 0 & 0 & 0 \\
0 & 1 & c & 0 \\
0 & c & 1 & 0 \\
0 & 0 & 0 & 0 
\end{pmatrix} .
\end{align}
The Hamiltonian \eqref{HXXZ} presented in the example is equivalent to the XXZ Hamiltonian, which is given by
\begin{align}
&\lH_{XXZ}=\alpha(X_i\,X_{i+1}+Y_i\,Y_{i+1}+\Delta\, Z_i\,Z_{i+1}-\Delta \,\id),
\end{align}
with the identifications $\alpha=-\frac{1}{2\Delta}$ and $\Delta=-\frac{1}{c}$.
\\
At this point, one can apply the same transformations to the $R$-matrix \eqref{eq:RXXZ}. Alternatively, one can obtain the $R$-matrix corresponding to the transformed Hamiltonian by using the Sutherland equation. Both approaches are equivalent and will lead to the same results. We follow the second one.
\\
The Sutherland equations are easily solved since all the coefficients of the Hamiltonian are simply constants. As a consequence, the $R$-matrix corresponding to \eqref{diffformham} is
\begin{align}\label{eq:XXZcons}
R(u) =   e^u\begin{pmatrix}
 \cos \omega u -\frac{\sin \omega u}{\omega}  & 0 & 0 \\
0 & c \frac{\sin \omega u}{\omega} &  1 & 0 \\
0 & 1  & c \frac{\sin \omega u}{\omega} & 0 \\
0 & 0 & 0 &  \cos \omega u -\frac{\sin \omega u}{\omega}
\end{pmatrix},
\end{align}
where now $\omega^2=c^2-1$.
\\
To demonstrate the equivalence between the solution obtained and the solution \eqref{eq:RXXZ}, we can reverse the identifications made to make the Hamiltonian constant.  First, we undo the twist and apply $R_{12}\mapsto U_2^{-1}R_{12}U_1$ to \eqref{eq:XXZcons} and put $c=\sqrt{c_3}\sqrt{c_4}$ so that we arrive at the $R$-matrix for the Hamiltonian \eqref{Hc3c4}. Next, we reparameterize $u\mapsto H_+(u)$, and finally we apply the inverse of the local basis transformation \eqref{LBTapply}, which immediately yields \eqref{eq:RXXZ}. This shows that the two solutions are indeed equivalent and that the model we found with the example is not new.

\paragraph{Difference vs. Non-difference} After using all the identifications, we see that \eqref{eq:RXXZ} is  just an $R$-matrix of \textit{difference form in disguise}. The non-difference nature of the rapidity dependence of the $R$-matrix only resides in the local basis transformations. This can also be considered a way to obtain a non-difference form solution by starting from a difference form one. 
\\
In what follows, we will present our results that will concern $R$-matrix \textit{genuinely of non-difference form} type, so that the non-difference form nature cannot be removed by any of the mentioned identifications. We will present only one matrix for each equivalence class.

\section{Some useful tricks}
Before going into the heart of the classifications, it is worth sharing some tricks we have learned through our exploration of different classes of models. 
\\
While the example discussed here is relatively simple and does not require additional computational details, those tricks were essential to find new models by starting from more complicated ansatz and in  cases where the dimension of the Hilbert space is bigger.
\\
To explain the tricks learned and also the complexity of the problem, it is useful to consider the ansatz used in the chapter \ref{6and8Vmodel}: Hamiltonian of 8-vertex type
\begin{align}
\begin{split}
\mathcal{H}_{12}  =\,& 
h_1  \text{ } \id + h_2 (Z\otimes  \id-  \id \otimes Z) + h_3  \sigma _+\otimes \sigma _-  + 
 h_4 \sigma_-\otimes \sigma _+  \\
& +    h_5 ( Z \otimes  \id +   \id \otimes Z ) + 
h_6 Z \otimes Z  + h_7 \sigma _-\otimes
\sigma _- + h_8 \sigma _+\otimes \sigma _+
\end{split},
\end{align}
where $\sigma_\pm= \frac{1}{2}(X + i\, Y)$ and $h_i=h_i(\theta)$, or in matrix form
\begin{align}
\mathcal{H}_{12}=\left(
\begin{array}{cccc}
 h_1+2 h_5+h_6 & 0 & 0 & h_8 \\
 0 & h_1+2 h_2-h_6 & h_3 & 0 \\
 0 & h_4 & h_1-2 h_2-h_6 & 0 \\
 h_7& 0 & 0 & h_1-2 h_5+h_6 \\
\end{array}
\right).
\end{align}
By plugging this ansatz in the integrability constraint $[\mathbb{Q}_2,\mathbb{Q}_3]=0$ we obtain 21 non-independent equations. To simplify the problem, since we know that solutions related by the Identification given in \ref{identification} are equivalent, we can eliminate some of the redundant degrees of freedom, narrowing down our search to models that are not related by these identifications. For example, without loss of generality we can set $h_1=0$ (shift of the Hamiltonian) and $h_2=0$ (as already discussed, this telescopic terms vanish in a closed spin chain).
\\
Furthermore, we consider a linear combination (with the appropriate coefficients) of 2, 3 or 4 of the differential equations obtained from $[\mathbb{Q}_2,\mathbb{Q}_3]=0$ and start to solve for the simplest one first.
\\
Similar to the discussion in the section \ref{topdownmethods} about the "Differential approach" method, the number of different equations coming from $[\mathbb{Q}_2,\mathbb{Q}_3]=0$ is bigger than the number of variables, so we can consider the derivatives as independent variables. In other word, we can recall $dh_i=\dot{h}_i$. In this way, some of the equations will be linear in $dh_i$. One can select the equations containing those terms and solve for them and then plug back into the remaining equations. We have yet to discover a precise algorithm for determining the optimal order of variables for easier solution finding. In some cases, solving for all the derivatives first and then substituting into the remaining equations works well. Other times, it is more efficient to solve for specific derivatives and functions. After finding the solution, one needs to impose the compatibility condition that the variables and the derivatives are not actually independent.
\\Treating the derivatives as independent variables simplifies the initial stage to solving sets of linear or polynomial equations, deferring the differential equations to a later stage. However, some complications arise also here. For example, for the 8-V type ansatz, we obtained
\begin{equation}
(-h_3 + h_4) h_6 h_7=0.
\end{equation}
There are 3  solutions to this equation
\begin{align}
&h_3=h_4,
&&h_6=0,
&&h_7=0.
\end{align}
We should consider  these 3 branches separately. By taking more general ansatz and increasing the dimension of the Hilbert space, the number of branches increases, rendering the search of a solution a computationally hard problem. For the ansatz of 8-V type, a brute force solution of the equations led to 31 solutions. For more general ansatz,  brute force  is unfeasible and each branch should be analysed separately.
\\
One approach to solve this branch problem is to select a block of $n$ equations to solve and get $m$ solutions. Among the $m$ solutions, there may be some variables that are the same in all the solutions. One can then plug back only this one into the initial system of equations and the complexity of the problem is simplified.
\\
Combining all these tricks allowed us to successfully carry out the classification of integrable models by starting from different ansatz.
\subsection{Ansatz used}
In what follows, we apply the boost automorphism mechanism to classify and discover new integrable models in different contexts.
\begin{itemize}
\item In chapter \ref{integrableopenquantumsystemchapter}, we  assume that the Hamiltonian density has the structure of a Lindblad superoperator. This generates the dynamics of an \textbf{open quantum systems}, a quantum spin chain with Hamiltonian $h$ in contact with a Markovian environment (described by the jump operator $\ell$). We apply the boost method to investigate a specific simple example, revealing the discovery of a novel model.  Additionally, we analyse the connection between Lindbladians and classical Markovian process.
\item In chapter \ref{Lindbladclassificationmodels}, we generalize the ansatz used in chapter \ref{integrableopenquantumsystemchapter}. We consider a general Hamiltonian $h$ and an operator $\ell$ with at most two elements below the diagonal. Then, we consider both $h$ and $\ell$ of 6-Vertex type. Surprisingly, we found two integrable models for which the environment and the system do not require fine-tuning. One of these models, B3, is new and exhibits interesting properties which we analyze. We solve this model via the nested algebraic Bethe ansatz in the Appendix \ref{BAB3chapter}. The other model, B2, is already known and it is related to a deformation of Hubbard model.
\item In chapter \ref{Hubbardchapter}, we show how in some cases we can map Lindblad superoperator to Hermitian matrix. This holds for the Hubbard model. We start from an ansatz with $h$ and $\ell$ of both 8-Vertex type and we discover a new and intriguing model.  This model is related to an elliptic range 3 deformation of the Hubbard model. The $R$-matrix has a very unusual dependence of the spectral parameter,  which, to the best of our knowledge, has not been previously accounted for. The complete classification of models in this class is computationally very complicated and is still a work in progress.
\item In chapter \ref{threeorfour}, we extend the method to higher-dimensional Hilbert spaces. Specifically, we consider the case of $\mathbb{C}^4\otimes\mathbb{C}^4$ with $\mathfrak{su}(2)\oplus \mathfrak{su}(2)$ symmetry. We demonstrate that our method successfully reproduces known models in this class,  such as the $\mathfrak{su}(4)$ Heisenberg XXX spin chain \cite{Kulish1982}, the Hubbard model \cite{essler2005one}, the AdS/CFT $S$-matrix\footnote{This point will be clarified in the footnote \ref{footnotedressingfactor} of section \ref{integrhamsu22}.} \cite{Beisert:2005tm,arutyunov2009foundations,Arutyunov:2006yd} and the related Shastry $R$-matrix \cite{Shastry_1986}. Additionally, we present some new models that emerge from our approach.
\item In chapter \ref{6and8Vmodel}, we consider the case where the local Hilbert space has dimension two and, consequently, the $R$-matrix is of size $4\times4$.  These models satisfy the free fermion condition which we discuss in chapter \ref{ffconditionchapter} and are related to integrable deformation of the AdS/CFT $S$-matrix, discussed in Appendix \ref{adsappendixdef}.
Furthermore,  we prove that all Hermitian 4x4 models with 16 entries, can be brought via the identifications given in section \ref{identification} to models of 8-Vertex type. This implies that obtaining a complete classification of 8-Vertex type models is equivalent to finding all Hermitian integrable models in $\mathbb{C}^2\otimes \mathbb{C}^2$.

\item We have also explored other ansatz that are not discussed in this thesis but can be found in \cite{classificationybandboost}.  These include $9\times 9$ Hamiltonian corresponding to spin 1 chain. We imposed that the Hamiltonian (and also the $R$-matrix) commutes with the Cartan subalgebra of $\mathfrak{su}(3)$. These models are usually referred as 15-Vertex model and satisfy the so called \textit{ice rule condition}. They have interesting connections to the eclectic spin chain,  \cite{garcia2022jordan,garcia2022jordan2,corcoran2022conformal}. 
\end{itemize}
Let us point out that such restrictions are not strictly necessary to implement our approach. In principle, the method allows to classify integrable models starting from the most general Hamiltonian density. Unfortunately, the problem of solving a set of coupled differential equations becomes (with the currect technique and the current computer power) very hard and also due to the fact that our method produces huge number of dependent integrable systems, providing a full classification of all possible  $R$-matrices in a higher dimensional spin chain is difficult and so we limit ourselves to a subset of models which are physically interesting.

\chapter{Integrable open quantum systems}

\ifpdf
    \graphicspath{{Chapter3/Figs/Raster/}{Chapter3/Figs/PDF/}{Chapter3/Figs/}}
\else
    \graphicspath{{Chapter3/Figs/Vector/}{Chapter3/Figs/}}
\fi

\label{integrableopenquantumsystemchapter}

In this chapter, we provide a brief introduction to the properties of open quantum systems,  physical systems that interact with their surrounding \textit{environment}. By introducing specific approximations, which will be detailed later, we can characterize their behavior using the Lindblad master equation. We present a simplified overview of the derivation, referring to the works of Preskill and Cappellaro \cite{preskill1999lecture,cappellaro2011quantum}. For a more detailed and formal treatment, references such as \cite{breuer2002theory,manzano2018harnessing,manzano2020short} are available. 
The dynamics can be equivalently described by using a superoperator $\mathcal{L}$ that acts on a doubled Hilbert space.  
We focus our attention to \textit{integrable open quantum systems}, where the \textit{Lindblad superoperator} $\mathcal{L}$ is one of the conserved charges of an integrable spin chain. While other research groups have explored such systems in the past \cite{ziolkowska2020yang,medvedyeva2016exact}, our work represents the \textbf{first systematic approach} to classify integrable open quantum systems.  We demonstrate with an example how the boost operator can be used to classify integrable open quantum systems and which are the transformations that preserves either integrability or the structure of this superoperator. For our specific application, we consider a spin 1/2 chain, where the Hamiltonian $h$ governs the evolution of the system and the \textit{jump operator} $\ell$ represents the effective action of the environment on the chain. We also describe the connection between the Lindblad equation and the \textit{classical stochastic equation}.
\section{The Lindblad master equation}
\subsection{The Lindblad master equation: derivation} 
\label{derivlindbladeq}
In introductory physics courses, for simplicity, the focus is typically on studying isolated systems, detached from their surrounding environment. However, in  quantum systems, the influence of the environment becomes significant, and it becomes challenging to conduct experiments where the contributions from the environment can be disregarded.\\
In what follow, we use the subscripts $S$, $E$ and $T$ to refer separately to the system, the environment and the total: system + environment. \\
If we consider both the system + environment as a whole, a state in the total Hilbert space is characterized by the density matrix $\rho_T$ and its unitary dynamic is governed by the Von Neumann equation
\begin{align}\label{evolutiontotal}
&\frac{d \rho_T}{d t}=-i [H_T,\rho_T],
&&\rho_T (t)=e^{-i H_T t} \rho_T(0) e^{i H_T t} .
\end{align}
However, in many practical application, the environmental degrees of freedom are incredibly vast and finding a solution of this dynamical equation is an impossible task. Moreover, the focus of interest is often on the evolution of the system itself, while the environment is considered as an uncontrollable entity. Therefore, the main object of study is the dynamics of the density matrix,
\begin{align}
\rho_S(t)=\Tr_E \rho_T(t),
\end{align}
where $\Tr_E$ is the partial trace operation over the environment degree of freedom.
\\
We aim to understand how the environment influences the dynamics of the density matrix $\rho_S$, specifically its time evolution. While it is not obvious that an open system's dynamics can be expressed by a differential equation, we proceed with the assumption that, as defined after equation \eqref{evolutiondiscrete}, the evolution of these dynamics is Markovian, local in time. In such circumstances, the evolution can be described by the Gorini-Kossakowski-Sudarshan-Lindblad master equation. This equation was derived in the 70s by Lindlad \cite{lindbladoriginal} and independently, around the same time, by Gorini-Kossakowski-Sudarshan \cite{gorinioriginal} by using the theory of dynamical semigroups. In the modern literature, there are two main approaches to derive this evolution:
\begin{itemize}
\item \textit{Microscopic derivation}: This involves taking the partial trace $\mathrm{Tr}_E$ in equation \eqref{evolutiontotal} and making a series of approximations to obtain the Markovian Lindblad master equation.
\item \textit{CPTP map}: The focus here is on answering the question: What is the most general way to map a density matrix onto another density matrix? By assuming the evolution to be Markovian, we arrive at the Lindblad equation.
\end{itemize}
We follow here the second approach. The density matrix represents a physical state and must adhere to certain properties. For this reason we need to search for a CPTP map. The "CP" refers to the requirement of \textit{complete positivity}, ensuring the density matrix has non-negative eigenvalues at any time\footnote{Complete positivity is a more general property than positivity. In fact, it should preserve positivity also when acting on a part of a larger system.}.  The "TP" aspect refers to the preservation of the trace\footnote{In particular the trace of the density matrix should be equal to one at any time.}, ensuring that the total probability of finding the system in any state remains unity.
\\
The answer to this question is the Choi-Kraus' theorem. We only state this theorem and refer to \cite{manzano2020short} for the proof.\\
\textbf{Choi-Kraus' theorem.} Any linear map $\mathbb{\mu}: \mathcal{B}(H_S)\to \mathcal{B}(H_S)$ is completely positive and trace preserving iff it can be expressed\footnote{We remark that the Kraus representation is not unique, \cite{preskill1999lecture}.} as
\begin{align}\label{kraussmap}
&\mathbb{\mu}:\,\rho_S(t)=\sum_k M_k(t) \rho_{S} M_k^\dagger (t),
\end{align}
where $\mathcal{B}(H_S)$ is the space of bounded operators acting on the Hilbert space of the system, the operators $M_k \in \mathcal{B}(H_S)$ are called Kraus operator and satisfy
\begin{align}\label{krausscond}
& \sum_k M_k^\dagger M_k=\id.
\end{align}
The sum runs from $0$ to $Q$, with $Q\le N_S^2$, $N_S$ being the dimension of the Hilbert space $H_S$. 
\\
To obtain the dynamic in the form of a master equation, we have to require the \textit{Markovian approximation}: the evolution is local in time. In fact, it is not immediately clear why a first-order differential equation is sufficient to describe the system's dynamics. Let us discuss the reasons behind this.\\
Suppose that we can define the dynamics as 
\begin{align}\label{evolutiondiscrete}
&\dot{\rho}_S(t)=\mathcal{L}[\rho_S(t)],
&\frac{\rho_S(t+dt)-\rho_S(t)}{dt}=\mathcal{L}[\rho_S(t)],\,\,\,\,\,dt \ll 1.
\end{align}
We are assuming that the density matrix at the time $t+dt$ is determined by  its value only at the time $t$. 
This is not guaranteed. In fact,  the total system (system+environment) has unitary evolution, but for part of the total system, the evolution does not necessarily remain local in $t$. During the time evolution, the system and the environment are interacting and the information flows from the system to the environment. The trouble is that, if the environment has memory, it can retain and later releases this information, resulting in non-markovian fluctuation of the system. In other words, if this happens, the state of the system at $t+dt$, is not only affected by the time $t$ but also by older configurations. In this case, we say that the environment is non-markovian. Opposite, if the environment does not have \textit{memory}, it is called markovian and the evolution of the density matrix is local in time.\\
In the real word, perfectly markovian environment does not exist. However, it is a good approximation, to allow the environment to remember the information for an interval of time called $(\Delta t)_{\text{env}}$. After this interval, the information is lost. However, in practice, our measurement instruments have limitations, and we can only resolve the dynamics with a characteristic time $(\Delta t)_{\text{coarse}}$. If $(\Delta t)_{\text{coarse}}\gg (\Delta t)_{\text{env}}$ 
we can neglect the memory of the reservoir since we are unable to resolve its effect. This approximation takes the name of \textbf{Markovian approximation} and it is useful in many practical situation. \\
We refer to the book \cite{carmichael1999statistical} and the work \cite{rivas2010markovian} for a quantitative definition of $(\Delta t)_{env}$. The authors explore various physical scenarios where a numerical solution of the system's dynamics is available and compare it to dynamics obtained through the application of Markovian approximations. For instance, they examine a single harmonic oscillator coupled to an environment of $M$ harmonic oscillators. They define $(\Delta t)_\text{{env}}$ as the full width at half height of the correlation functions involving operators that describe the environment. They demonstrate that, when other parameters are held constant, this quantity decreases as a function of the environmental temperature. Intuitively, at infinite temperature, quantum correlations tend to be eliminated, leading to the emergence of Markovian properties in the environment.
\\
For the range of validity of the Markovian approximation, \eqref{evolutiondiscrete} holds and using  \eqref{kraussmap}, we obtain
\begin{align}\label{mapping}
&\rho_S(t+dt)=\sum_k M_k(dt) \rho_S(t) M_k^\dagger(dt),
&&\rho_S(t+dt)\sim\rho_S(t)+dt\,\mathcal{L}[\rho_S(t)].
\end{align}
If we retains only terms linear in $dt$, we may assume without loss of generality\footnote{In fact, if one starts from $M_j=\id+\sqrt{dt} \alpha_j+ dt \beta_j$, with $\alpha_j$ and $\beta_j$ arbitrary operators and $j=0,1,..., Q<N_S^2$, by using \eqref{krausscond}, the result \eqref{lindbadmultiplef} remains the same. Furthermore, if in the expression for $M_j$, we allow to have terms $dt^w$, $0<w<1$, by using \eqref{mapping} it follows that only $w=1/2$ contributes.} that 
\begin{align}
&M_0=\id+dt (-i \,h + K),
&&M_k=\sqrt{dt}\,\ell_k,\,\,\,k=1,2,\dots,Q< N_S^2,
\end{align}
where $h$ and $K$ are Hermitian operators, $h$, $K$ and $\ell$ are zero-th order in $dt$ and $N_S$ is the dimension of the Hilbert space of the system. Substituting it into \eqref{krausscond}, we find 
\begin{align}
&K=-\frac{1}{2}\sum_{k=1}^{N_S^2} \ell_k^\dagger \ell_k,
\end{align}
and substituting into \eqref{mapping} and considering \eqref{evolutiondiscrete}, we obtain the well-known \textbf{Gorini-Kossakowski-Sudarshan-Lindblad equation} (GKSL) master equation \cite{lindbladoriginal,gorinioriginal}
\begin{equation}\label{lindbadmultiplef}
  \dot{\rho}_S(t)=-i[h,\rho_S]+\sum_k\Big[ \ell_k\rho_S \ell_k^\dagger-\frac{1}{2}\{\ell^\dagger_k \ell_k,\rho_S\}\Big].
\end{equation}
For brevity, in what follows, we refer to this as Lindblad equation, however it should be mentioned that it was independently derived by the other physicists as well.
The first term represents the unitary evolution (thus $h=h_S$ is the usual Hamiltonian of the system). $\ell_k$ are called \textit{quantum jump operators} and the terms in the square bracket represent the potential transitions ("jump") that can occur in the system as a result of its interaction with the reservoir. \\
If the contribution of the environment drops out, $M_0=1-i\, h\, dt$ and all the $M_{k>0}=0$, we recover unitary dynamics. In particular, the relation between the numper of jump and Kraus operators is
\begin{align}
\#\,\,\text{Kraus}\,\,=\,\,\#\text{\,\,jump}+1 ,
\end{align}
with the maximum number or Kraus operator being $N_S^2$, with $N_S$ the dimension of the Hilbert space of the system.
\subsection{The Lindblad master equation: summary}
In what follows, we restrict to the case where there are only two Kraus operators and consequently, only one family of jump operator\footnote{This statement will be clarified in the following section. In fact, in our case, each jump operator acts on nearest-neighbour sites of a spin chain. We refer to them as one family of jump operator since the action of the superoperator in the chain is periodic.}, but the method can be easily generalized to multiple families. For simplicity in the notation, we omit the subscript $S$ in $\rho(t)=\rho_S(t)$.\\
The Lindblad master equation is
\begin{equation}
  \underbrace{\dot{\rho}(t)=-i[h,\rho(t)]}_{\text{von-Neumann equation}}+\underbrace{ \left[
    \ell\rho(t) \ell^\dagger-\frac{1}{2}\{\ell^\dagger \ell,\rho(t)\}\right]}_{\text{dissipator}},
    \label{lindbladeq}
\end{equation}
where $h$ is the Hamiltonian of the system and $\ell$ is the jump operator that describe
the interaction with the environment. The operator $\ell$ is not normalized. In some cases, where the system and environment are not fine tuned, we use $\ell\to \sqrt{U} \ell$, where $U$ measures the strength of the coupling between the system and the environment. 

\subsection{The Lindblad superoperator and the Fock-Liouville space}

\label{sec:ladder}

We can write the Lindblad equation \eqref{lindbladeq} as a linear operator $\mathcal{L}$ acting on a state
\begin{equation}
\dot{\rho}(t)\equiv\mathcal{L}\rho(t), 
\end{equation}
$\mathcal{L}$ is the so-called \textbf{Lindblad super-operator}. This is possible through  an operator-state correspondence.
Fixing an arbitrary basis, the density matrix $\rho=\sum_{a,b}\rho_{ab}\ket{a}\bra{b}$ acting on the space $H$ is interpreted as a vector of the tensor product space (also called Fock-Liouville space)\footnote{We should notice that now the vector density matrix is different from the previous one but, for simplicity of notation, we will keep referring to it as $\rho$. Sometimes, in the literature, the vector $\rho$ is indicated with $\ket{\rho}\rangle$.}
\begin{equation}
  \label{opstate}
  \rho \, \in \,  {H}\otimes {H}^* \equiv {H}^{(1)} {H}^{(2)} ,
\end{equation}

${H}$ is the initial Hilbert space of the system and ${H}^*$ its dual.  For finite dimensional spaces this is not a problem, because there is a natural identification ${H}^*\simeq  {H}$. Since we consider finite system, we do not encounter the problem of infinite dimensional spaces.
\\
We write \eqref{lindbladeq} in components
\begin{equation}
  \dot \rho_{jk}=-ih_{jl}\rho_{lk}+i \rho_{jl}h_{lk}
  +  \left[    \ell_{jl}\rho_{lm} \ell^\dagger_{mk}
    -\frac{1}{2} (\ell^\dagger \ell)_{jl}\rho_{lk}
-\frac{1}{2}\rho_{jl}     (\ell^\dagger \ell)_{lk}\right].
\end{equation}

With the identification \eqref{opstate}, the Lindblad superoperator takes the form
\begin{equation}
  \mL=-i h^{(1)}+i (h^T)^{(2)}+ \left[
\ell^{(1)}  {\ell^{(2)}}^*-\frac{1}{2} \ell^{(1)\dagger} \ell^{(1)}
-\frac{1}{2}  {\ell^{(2)}}^T  {\ell^{(2)}}^*
\right]  .
\label{superop11}
\end{equation}
For any operator $A$ the notation $A^{(1)}$ and $A^{(2)}$ means that it acts only on the first space $A^{(1)}=A\otimes
\id$ or $A^{(2)}=\id\otimes A$. $^T$, $^*$ and $^\dagger$ refer respectively to the transposition, complex conjugation elementwise and transpose conjugate.

\subsection{Lindblad superoperator as a spin chain Hamiltonian}\label{explainnotation}
We want to identify the Lindblad superoperator \eqref{superop11} as a nearest-neighbour operator acting in a spin chain. We consider the case where both the $h$ and $\ell$ operators act as densities in nearest-neighbour site of a spin 1/2 chain of lenght $L$. The total Hilbert space is $H= \bigotimes_L V=\bigotimes_L \mathbb{C}^2$ and the density operators $h_{j,j+1}$ and $\ell_{j,j+1}$,
\begin{align}
&h_{j,j+1}\in {B}(V\otimes V),&&\ell_{j,j+1}\in {B}(V\otimes V),
\end{align}
with ${B}(W)$ the space of bounded operators acting on the Hilbert space $W$, ${B}: W \to W$.\\
The local Fock-Liouville space of the superoperator on a spin chain is $V\otimes V\otimes V^*\otimes V^*$. 
\\
The superoperator defined in \eqref{superop11} takes the form
\begin{align}
  \Li_{j,j+1}=&~-i h^{(1)}_{j,j+1}+i  h^{(2)^T}_{j,j+1}+   \Big(\ell^{(1)}_{j,j+1}  \ell^{(2)*}_{j,j+1}  -\frac{1}{2} \ell^{(1)\dagger}_{j,j+1}  \ell^{(1)}_{j,j+1}-\frac{1}{2}  \ell^{(2)T}_{j,j+1}  \ell^{(2)*}_{j,j+1}\Big).
  \label{superoperatorL2}
\end{align}
It is easy to understand that the  superoperator corresponding to the Lindblad equation \eqref{lindbadmultiplef} with multiple families of jump operator is
\begin{align}
  \Li_{j,j+1}=&~-i h^{(1)}_{j,j+1}+i  h^{(2)^T}_{j,j+1}+  \sum_k \Big(\ell^{(1)}_{k,j,j+1}  \ell^{(2)*}_{k,j,j+1}  -\frac{1}{2} \ell^{(1)\dagger}_{k,j,j+1}  \ell^{(1)}_{k,j,j+1}-\frac{1}{2}  \ell^{(2)T}_{k,j,j+1}  \ell^{(2)*}_{k,j,j+1}\Big) ,
  \label{superopmultiplefamilies}
\end{align}
where $k$ identifies the family.\\
This system is referred to as \textit{bulk Lindbladian} because the jump operators are acting in adjacent sites in the bulk of the spin chain. For spin chain with open boundary condition, it is also common to consider \textit{boundary Lindbladian} where the jump operators act on the boundary of the chain. We do not consider this type of system in this thesis, we refer to \cite{prosen-boundary-lindblad-1} and reference therein, for work in this direction.
\\
We remark an important point necessary for the construction. We want to consider the superoperator $\mathcal{L}$ as acting in a quantum spin chain, so in order to do this, we need to swap two of the middle spaces of the Fock-Liouville space, explicitly
\begin{align}
V\otimes V \otimes V^*\otimes V^*\,\,\,\to\,\,\,V\otimes V^* \otimes V\otimes V^*,
\end{align}
and now the interpretation of the superoperator as acting on a spin chain becomes  clearer 
\begin{align}\label{LspacesVVVV}
&\mathcal{L}_{j,j+1}\,\,\,\,\,\in \,\,\,\,\,\underbrace{V\otimes V^*}_j \otimes \underbrace{V\otimes V^*}_{j+1} . 
\end{align}
To take this into account, we identify with $\sigma$s the matrices acting in $V$ and $\tau$s matrices\footnote{Sometimes in the literature $\tau\to \tilde{\sigma}$ or $\tau\to \bar{\sigma}.$} acting in $V^*$ (similar to \cite{ziolkowska2020yang,medvedyeva2016exact})
\begin{align}
&\,\,\,\,\,\,\,\,\,\,\,\,\,\,\,\,\,\,\,\,\,\,\,\,\,\,\,\,\,\,&& V &&\otimes &&V^*&&\otimes &&V&&\otimes &&V^*&&\,\,\,\,\,\,\,\,\,\,\,\,\,\,\,\,\,\,\,\,\,\,\,\,\,\,\,\,\,\,\non\\
&\,\,\,\,\,\,\,\,\,\,\,\,\,\,\,\,\,\,\,\,\,\,\,\,\,\,\,\,\,\,&& \downarrow &&  &&\downarrow&&  &&\downarrow&&  &&\downarrow&&\,\,\,\,\,\,\,\,\,\,\,\,\,\,\,\,\,\,\,\,\,\,\,\,\,\,\,\,\,\,\non\\
&\,\,\,\,\,\,\,\,\,\,\,\,\,\,\,\,\,\,\,\,\,\,\,\,\,\,\,\,\,\,&& \sigma_j && &&\tau_j && &&\sigma_{j+1}&& &&\tau_{j+1} && &&\,\,\,\,\,\,\,\,\,\,\,\,\,\,\,\,\,\,\,\,\,\,\,\,\,\,\,\,\,\,
\label{notation}
\end{align}
This system is usually referred as two-legs (one for the "bra" and one for the "ket" of the density matrix) spin-ladder system. And the notation $A^{(1)}$ and $A^{(2)}$ should be intepreted as $A^{(1)}$ only containing $\sigma$ and $A^{(2)}$ only $\tau$.
\\
Having in mind this construction, we show how to use the boost operator to find \textbf{new integrable Lindblad superoperators}. The superoperator is associated to a nearest-neighbour (non-Hermitian) Hamiltonian on a spin-chain with local Hilbert space $V\otimes V^*$. This superoperator is integrable if $\mL$ can be written as the derivative of an $R$-matrix which is a solution of the Yang-Baxter equation.
\\
Before showing the method with an example, we briefly mention the convention used.

\subsection{Conventions}
A state of the system is given by the density matrix $\rho$. The explicit matrix representation is basis dependent. In this thesis, in the context of open quantum systems, we work with spin 1/2 chain and we use the standard basis as in \eqref{standardbasis}. We denote a vector in the space as
\begin{equation}
\psi_\uparrow \ket{\uparrow}+\psi_\downarrow \ket{\downarrow} = \left(  \begin{array}{c}
 \psi_{\uparrow} \\
\psi_{\downarrow} \\
\end{array}\right).
\end{equation}
We choose as reference state, a state with all spin down. Particles are spin up. The occupation number is $n_i=\sigma_i^+ \sigma_i^-=\frac{\id + Z_i}{2}.$ 
For nearest-neighbour spins we apply the notation
\begin{align}
  \label{twoseg}
\left(
\begin{array}{c}
 \psi_{\uparrow} \\
\psi_{\downarrow} \\
\end{array}
\right)\otimes \left(
\begin{array}{c}
 \psi_{\uparrow} \\
 \psi_{\downarrow} \\
\end{array}
\right)\equiv \left(
\begin{array}{c}
 \psi_{\uparrow\uparrow} \\
 \psi_{\uparrow\downarrow} \\
 \psi_{\downarrow\uparrow} \\
 \psi_{\downarrow\downarrow} \\
\end{array}
\right).
\end{align}
The density matrix of a two-site segment is given by
\begin{equation}
  \rho_{ab}^{cd}=\psi_{ab} \psi^*_{cd},\qquad a,b,c,d=\uparrow,\downarrow
\end{equation}
and in the superoperator formalism we treat such a density matrix as a
vector with $16$
components ordered as $\rho_1,\rho_2,\dots$. The identification of the components is given by
\begin{align}
  \label{rhoindex}
\rho=\left(
\begin{array}{cccc}
 \rho _{\uparrow  \uparrow}^{\uparrow \uparrow} & \rho _{\uparrow  \uparrow}^{\uparrow    \downarrow}
  & \rho _{\uparrow \uparrow}^{\downarrow \uparrow} & \rho _{\uparrow \uparrow}^{\downarrow \downarrow} \\[1ex]
 \rho _{\uparrow \downarrow}^{\uparrow \uparrow} & \rho _{\uparrow    \downarrow}^{\uparrow \downarrow}
  & \rho_{\uparrow\downarrow}^{\downarrow  \uparrow}    & \rho _{\uparrow \downarrow}^{\downarrow \downarrow} \\[1ex]
 \rho _{\downarrow \uparrow}^{\uparrow \uparrow} & \rho _{\downarrow \uparrow}^{\uparrow \downarrow} & \rho _{\downarrow \uparrow}^{ \downarrow \uparrow} & \rho _{\downarrow \uparrow}^{ \downarrow \downarrow} \\[1ex]
  \rho_{\downarrow \downarrow}^{ \uparrow \uparrow} & \rho_{\downarrow    \downarrow}^{  \uparrow \downarrow}
  & \rho _{\downarrow \downarrow}^{ \downarrow \uparrow} & \rho _{\downarrow \downarrow}^{ \downarrow \downarrow} \\
\end{array}
\right)=
  \begin{pmatrix}
    \rho_1 &  \rho_2 &  \rho_3 &  \rho_4  \\
    \rho_5 &  \rho_6 &  \rho_7 &  \rho_8  \\
    \rho_9 &  \rho_{10} &  \rho_{11} &  \rho_{12}  \\
     \rho_{13} &  \rho_{14} &  \rho_{15} &  \rho_{16}  \\
  \end{pmatrix}.
\end{align}

\section{Finding integrable Lindblad superoperators}
Now, we initiate the \textbf{systematic classification} of integrable open quantum systems using the boost automorphism mechanism. The charge $\mathcal{Q}_{j,j+1}$, traditionally associated with the Hamiltonian of the spin chain, is now the Lindblad superoperator $\mathcal{L}$ \eqref{superoperatorL2}.
We remark that, the operator  $\mathcal{Q}_{j,j+1}=\mathcal{L}_{j,j+1}$ is not Hermitian by construction, the Hermitian operator describing the closed system in the spin 1/2 chain is $h$ in \eqref{superoperatorL2}.
\\For clarity,  we  specifically apply the steps outlined in Section \ref{example} to the case of an open quantum system.

The main steps of the method are:
\begin{itemize}
\item[1.] Start from an ansatz for the densities\footnote{It will become evident in the classification that we did not impose $h$ to be Hermitian. However, we only discover integrable superoperator where $h$ has this property.} $h_{i,i+1}(\theta)$ and $\ell_{i,i+1}(\theta)$. In this thesis, we refer to spin-1/2 chain, both $h$ and $\ell$ are 4x4 matrix. They depend on some un-known functions $h_1(\theta),\dots,l_1(\theta),\dots$.
\item[2.] Plug the ansatz into the superoperator $\mL_{i,i+1}$ defined in \eqref{superoperatorL2} and identify $\mathcal{Q}_2=\mL$.
\item[3.] Use the boost operator \eqref{boostdef} to construct $\mathbb{Q}_3$ (given by \eqref{eq:boostoperator} ).
\item[4.] Impose the integrability constraint $[\mathbb{Q}_2,\mathbb{Q}_3]=0$ and solve the set of differential equations derived by it. This gives a "potentially" integrable superoperator.
\item[5.] Use $\mathcal{H}_{1,2}=\cL_{1,2}$ in the Sutherland equation \eqref{eqn:Sutherland} or \eqref{eqn:Sutherland2} to find the corresponding $R$-matrix.
\item[6.] Check that the $R$ is a solution of the YBE and investigate if the model is new.
\end{itemize}

The steps 1. and 2. impose a very strong constraint: the decomposition of the density $\mathcal{Q}_2$ as a Lindblad superoperator holds for \textit{any} value of $\theta$. In order to have an integrable Lindblad system, it is enough to  find a solution of the Yang-Baxter equation such that $\mathcal{Q}_2$ takes the form of \eqref{superoperatorL2} only for \textit{some} value of $\theta$. 
Unfortunately, solving the general problem is more challenging\footnote{In order to fully solve this problem, one needs to classify all possible integrable models where $\mathcal{Q}_2$ has components on the positions of the non-zero components of $\cL$. Only in a later stage, one can check if the decomposition in term of superoperators hold for any value of $\theta$.} and we leave it for future work. 
\\
We provide further clarification on these steps by considering an example.

\subsection{Example}
\label{exampleforlindblad}
Since we have already provided many details in section \ref{example}, here we only emphasize the steps that differ.

\subsubsection*{1. Ansatz for $h$ and $\ell$}
We start from $h_{i,i+1},\,\ell_{i,i+1}\,\in \complex^2\, \otimes \, \complex^2$. We choose a very simple ansatz where $\ell$ only has two elements and $h$ is of XY-type\footnote{Usually we refer to spin chain of this type if the only non zero coefficients on the Hamiltonian are the $X_iX_{i+1}$ and $Y_i Y_{i+1}$.}
\begin{align}
&\ell_{j,j+1}=
  \begin{pmatrix}
     & & & \\
    &  & & \\
    & 1 & l_1 & \\
       & & & 
  \end{pmatrix} =  \sigma_j^- \sigma^+_{j+1} + \frac{ l_1}{4}\, (\id-Z)_j  (\id+Z)_{j+1} , 
  \label{jumooperatorexample}
\end{align}

\begin{align}
&h_{j,j+1}=
  \begin{pmatrix}
     & & & \\
    &  & h_1 & \\
    & h_2 &  & \\
       & & & 
  \end{pmatrix} = h_1\,  \sigma^+_j \sigma^-_{j+1}  + h_2\,  \sigma^-_j \sigma^+_{j+1} ,
  \label{hsystemexample}
\end{align}
where $h_i = h_i(\theta)$ and $l_1=l_1(\theta)$. As mentioned, we did not impose that $h$ is Hermitian since we did not fix any constraints on $h_1$, $h_2$. For simplicity, we normalize one element of the $\ell$ to one.\\
An important clarification is that, in the more general classifications considered in the following chapter, we did not impose from the beginning that the Hamiltonian $h$ of the closed system is integrable. However, for all the solutions that we find, this was always the case.
\subsubsection*{2. Construction of the superoperator $\cL_{i,i+1}$}
The superoperator $\cL_{i,i+1}$ acts on $V\otimes V^*\otimes V \otimes V^*$ and using the notation \eqref{notation}, the expression \eqref{superoperatorL2} is
\begin{align}\label{Lexample}
\mL_{ij} =&~ -i (h_1\,  \sigma^+_i \sigma^-_j + h_2\,  \sigma^-_i \sigma^+_j)  + i (h_1\,  \tau^-_i \tau^+_j  + h_2\,  \tau^+_i \tau^-_j)  + \nonumber \\
&~ +\big[\sigma^-_i  \sigma^+_j + \frac{ l_1}{4}\, (\id-\sigma^z_i)  (\id+\sigma^z_j)\big]\big[\tau^-_i  \tau^+_j + \frac{ \bar{l_1}}{4}\, (\id-\tau^z_i)  (\id+\tau^z_j)\big] + \nonumber\\
&~ -\frac{1}{2}(l_1\,  \sigma^+_i \sigma^-_j + \bar{l_1}\,  \sigma^-_i \sigma^+_j + \frac{ |l_1|^2}{4}\, (\id-\sigma^z_i)  (\id+\sigma^z_j) + \frac{1}{4}\, (\id+\sigma^z_i)  (\id-\sigma^z_j))\nonumber \\
&~ -\frac{1}{2}(\bar{l_1}\,  \tau^+_i \tau^-_j + l_1\,  \tau^-_i \tau^+_j + \frac{ |l_1|^2 }{4}\, (\id-\tau^z_i)  (\id+\tau^z_j) + \frac{1}{4}\, (\id+\tau^z_i)  (\id-\tau^z_j)), 
\end{align}
where for simplicity we omitted the dependence of the spectral parameter. We identify with $\bar{l}_1$ the complex conjugate of $l_1$ and $|l_1|^2=\bar{l}_1 l_1$. We remark that now $\sigma$ and $\tau$ should be interpreted as \eqref{notation}. For example $\sigma^+_1=\sigma^+\otimes \id \otimes \id \otimes \id $ and $\tau^+_1=\id \otimes\sigma^+\otimes  \id \otimes \id $.
\subsubsection*{3. Construct $\mathbb{Q}_3$ via the boost operator}
With the help of the boost operator, we can compute $\mathbb{Q}_3$ \eqref{eq:boostoperator}, whose density is
\begin{equation}
(\mathcal{Q}_3)_{i,i+1,i+2} =  [\mathcal{L}_{i,i+1}, \mathcal{L}_{i+1,i+2}]-\frac{1}{2}\Big(\dot{\mL}_{i,i+1}+\dot{\mL}_{i+1,i+2}\Big).
\end{equation}
We do not write the explicit expression since it is very long, but the computation is straightforward.
\subsubsection*{4. Impose the integrability constraint}
Plugging $\mathbb{Q}_2$ and $\mathbb{Q}_3$ in the integrability condition \eqref{intcondition} then yields the following simple set of equations
\begin{align}
&2 \,h_1 -i\, l_1 =0,
&&2\, h_2 +i \,\bar{l_1}=0,
&& |l_1|^2 = 1,
\end{align}
\begin{align}
&\dot{l}_1-l_1^2\bar{l_1} = 0 ,\,\,\,\,\,\,\,\,\,\,\dot{l}_1+3 l_1^2\bar{l_1} = 0 ,
&&\to\,\,\,\,\, \dot{l}_1 = 0.
\end{align}

This implies that $h_1$, $h_2$ and $l_1$ are constants and
\begin{align}\label{solutionforsuperoperator}
&l_1 = - i\, e^{i \phi},\,\,\,\,\,
&&h_1 = \frac{i}{2}e^{i\phi} ,
&&h_2 = -\frac{i}{2}e^{-i\phi},
&&\phi \, \in \, \mathbb{R}
\end{align}
or directly in matrix form
\begin{align}
&h=\frac{1}{2}\left(
\begin{array}{cccc}
 0 & 0 & 0 & 0 \\
 0 & 0 & e^{i \phi } & 0 \\
 0 & e^{-i \phi } & 0 & 0 \\
 0 & 0 & 0 & 0 \\
\end{array}
\right),
&&\ell=\left(
\begin{array}{cccc}
 0 & 0 & 0 & 0 \\
 0 & 0 & 0 & 0 \\
 0 & 1 & -i\, e^{i \phi } & 0 \\
 0 & 0 & 0 & 0 \\
\end{array}
\right).
\label{handlmodellindbladexample}
\end{align}

Even if we did not impose it from the beginning, we find that the Hamiltonian $h$ is automatically Hermitian. We plug this solution in the superoperator $\mathcal{L}$ \eqref{superoperatorL2} and we find a "potentially" integrable superoperator.
\subsubsection*{5. Find the $R$-matrix}
We start from an ansatz for the $R$-matrix.
\\
We consider $\mathcal{H}_{12}\to \cL_{12}$ in the expansion \eqref{expansion} and we fix which entries of the $R$-matrix are non-zero and which one are the same. Furthermore, since the superoperator is constant, the $R$-matrix is of difference form. Explicitly, we used the ansatz
\begin{align}\label{rmatA11}
&R(u) = \left(
\begin{array}{cccccccccccccccc}
 r_1 & 0 & 0 & 0 & 0 & 0 & 0 & 0 & 0 & 0 & 0 & 0 & 0 & 0 & 0 & 0 \\
 0 & r_3 & 0 & 0 & r_2 & 0 & 0 & 0 & 0 & 0 & 0 & 0 & 0 & 0 & 0 & 0 \\
 0 & 0 & r_7 & 0 & 0 & 0 & 0 & 0 & r_2 & 0 & 0 & 0 & 0 & 0 & 0 & 0 \\
 0 & 0 & 0 & r_8 & 0 & 0 & 0 & 0 & 0 & 0 & 0 & 0 & r_1 & 0 & 0 & 0 \\
 0 & r_2 & 0 & 0 & 0 & 0 & 0 & 0 & 0 & 0 & 0 & 0 & 0 & 0 & 0 & 0 \\
 0 & 0 & 0 & 0 & 0 & r_1 & 0 & 0 & 0 & 0 & 0 & 0 & 0 & 0 & 0 & 0 \\
 0 & 0 & 0 & r_5 & 0 & 0 & 0 & 0 & 0 & r_4 & 0 & 0 & 0 & 0 & 0 & 0 \\
 0 & 0 & 0 & 0 & 0 & 0 & 0 & r_7 & 0 & 0 & 0 & 0 & 0 & r_2 & 0 & 0 \\
 0 & 0 & r_2 & 0 & 0 & 0 & 0 & 0 & 0 & 0 & 0 & 0 & 0 & 0 & 0 & 0 \\
 0 & 0 & 0 & r_6 & 0 & 0 & r_4 & 0 & 0 & 0 & 0 & 0 & 0 & 0 & 0 & 0 \\
 0 & 0 & 0 & 0 & 0 & 0 & 0 & 0 & 0 & 0 & r_1 & 0 & 0 & 0 & 0 & 0 \\
 0 & 0 & 0 & 0 & 0 & 0 & 0 & 0 & 0 & 0 & 0 & r_3 & 0 & 0 & r_2 & 0 \\
 0 & 0 & 0 & r_4 & 0 & 0 & 0 & 0 & 0 & 0 & 0 & 0 & 0 & 0 & 0 & 0 \\
 0 & 0 & 0 & 0 & 0 & 0 & 0 & r_2 & 0 & 0 & 0 & 0 & 0 & 0 & 0 & 0 \\
 0 & 0 & 0 & 0 & 0 & 0 & 0 & 0 & 0 & 0 & 0 & r_2 & 0 & 0 & 0 & 0 \\
 0 & 0 & 0 & 0 & 0 & 0 & 0 & 0 & 0 & 0 & 0 & 0 & 0 & 0 & 0 & r_1 \\
\end{array}
\right) .
\end{align}
Plugging this ansatz into the Sutherland equation\footnote{It is easy to obtain the difference form version of this equation: $\mathcal{H}(u)\to \cL$, $R_{ij}(u_i,u_j)\to R_{ij}(u)$, $\dot{R}_{ij}(u_i,u_j)\to \partial_u R_{ij}(u)$.} \eqref{eqn:Sutherland}, with the superoperator $\cL_{i,i+1}$ with \eqref{handlmodellindbladexample}, we obtain\footnote{In this case, we  gave the expression of the $R$ in form of a matrix. In the following chapters, we only indicate the non-zero element of the $R$-matrix, with $R_i^j$ the element corresponding to $j$-row and $i$-column.}
\begin{align}
& r_1 = 1,
&&r_4 = e^{-u},
&&r_2 = e^{-\frac{u}{2}},
&&r_8 = 1-e^{-u}
\end{align}
\begin{align}\label{rmatA12}
&&r_3 e^{-i \phi }=i e^{-u} \left(e^u-1\right)=-\, r_7 \, e^{i \phi }=-e^{u/2} e^{i \phi } r_5 = e^{u/2}\, e^{-i \phi }\,r_6.
\end{align}

\subsubsection*{6. Check YBE}
With the software Mathematica, we check that the $R$-matrix satisfies the YBE \eqref{eq:YBE}. 
\\
We tried to map this model to known models by using the identification of the next section, but we were not successful. Despite the simplicity of the ansatz selected, to the best of our knowledge, this model seems to be new. We discuss it in the letter \cite{classificationlind}, where we call it \textbf{model A1}.

\vspace{1 cm}

As in \ref{identification}, there are some transformations that preserve integrability and help to understand if a given model can be mapped to a known one.

\section{Identifications}
\label{identificationlindblad}
The problem of finding integrable superoperators can be reformulated as finding pairs of operators $(h,\ell)$ such that the corresponding superoperator \eqref{superoperatorL2} can be written as the derivative of an $R$-matrix that satisfies the Yang-Baxter equation.
\\
We distinguish between two types of identifications
\begin{itemize}
\item preserving the form of the superoperator;
\item preserving the integrable nature of the model.
\end{itemize}
In our classification, we use both types of transformations to reduce the number of degrees of freedom in our ansatz and make the calculation more feasible.

\subsection{Preserving the form of $\cL$}
\subsubsection*{Shift in $\ell$}
The combination of a shift in $\ell$ with a transformation in $h$ preserves the form of the superoperator. We derive the corresponding transformation on the $h$.\\
Consider a shift in $\ell$ by a term proportional to the identity
\begin{equation}
  \ell\, \to \, \ell+\alpha \id,\qquad \alpha\in\complex.
\end{equation}
This changes in the $\ell$ will be reflected in the superoperator as 
\begin{align}
  \mathcal{L}\rho&=i\,[\rho,h]+ \left(
    \ell\rho \ell^\dagger-\frac{1}{2}\{\ell^\dagger \ell,\rho\}\right)+
 \left(\alpha \rho  \ell^\dagger +\alpha^* \ell\rho -
  \frac{1}{2} \{\alpha^* \ell+\alpha \ell^\dagger,\rho\}\right),\nonumber\\
&=i\,[\rho,h]+ \left(
    \ell\rho \ell^\dagger-\frac{1}{2}\{\ell^\dagger \ell,\rho\}\right)+
  \frac{1}{2} \left( \alpha [\rho,\ell^\dagger] - \alpha^* [\rho,\ell]\right),
 \end{align}
for simplicity we omit the sites $_{i,i+1}$ where the operators act.
\\
This is equivalent to a redefinition of the Hamiltonian
 \begin{equation}
   h\,\to\, h-\frac{i}{2} (\alpha \ell^\dagger -\alpha^* \ell).
 \end{equation}
A simultaneous change
 \begin{align}
 (h,\ell)  \mapsto ( h+\frac{i}{2} (\alpha \ell^\dagger -\alpha^* \ell), \ell+\alpha \, \id)
 \label{transfshiftl}
 \end{align}
 leaves the total superoperator invariant. This can generically be used to set one of the diagonal entries of $\ell$ to zero.

\subsubsection*{Shift in $h$}

Shifting the Hamiltonian of the system by a constant
 \begin{align}
 (h,\ell)  \mapsto ( h+\alpha \, \id, \ell)
 \end{align}
 also leaves the superoperator trivially invariant. It is immediate to see it because the Hamiltonian contribution appears in the commutator with the density matrix. 
 
\subsection{Preserving integrability of $\cL$}
In this section, we show how the transformations of section \ref{identification} act on the superoperator. To see the action on the $R$-matrix, we refer to section \ref{identification}.
\subsubsection*{Local Basis transformations}
If we consider a local basis transformation on the superoperator $\cL$, similar to \eqref{LBTlaw}, we have
\begin{equation}
\mL \to (B \otimes B)\mL (B \otimes B)^{-1}-(\dot B B^{-1}\otimes \id  - \id \otimes \dot B B^{-1})
\end{equation}
This is a well defined transformation that preserves integrability, however it will not always give an interpretation in term of $h$ and $\ell$ separately. If  we restrict to the case $B=A\otimes A$, at the level of $h$ and $\ell$ this implies
\begin{align}
(h,\ell) \mapsto (\,(A\otimes A) \,h \,(A\otimes A)^{-1} , (A\otimes A) \,\ell (A\otimes A)^{-1} \,),
\end{align}

and to keep the structure of the superoperator\footnote{This restriction is due to the terms of the form $\ell^{\dagger}\ell$.}, the matrix $A$ should be unitary, \textit{i.e.} $A^\dag A = \id$. In this way,
\begin{align}
\mathcal{L}\to (A\otimes A\otimes A \otimes A)\, \mathcal{L} \,(A\otimes A\otimes A \otimes A)^{-1},
\end{align}
we omitted the telescopic terms since they cancel on a closed spin chain. So even if the expression of the superoperator changes, integrability is preserved. 

\subsubsection*{Telescopic terms on the $h$}

As a particular case of the telescopic term, we can also consider the identification
\begin{align}
h\to h + \alpha (\sigma_z\otimes \id-\id \otimes \sigma_z),\,\,\alpha \in \complex\,.
\end{align}
This is a particular case of identification at the level of the superoperator
\begin{align}
\cL \to \cL + A\otimes \id-\id \otimes A
\end{align}
with $A=-i\,  (\sigma_z\otimes \id-\id \otimes \sigma_z) $, that preserves the structure of the superoperator.

\subsubsection*{Normalization}
In $\cL$, the dependence of the jump operators is quadratic in $\ell$ and appears as $\ell^{\dagger}\ell$. Any redefinition $\ell\to e^{i\varphi} \ell,\,\,\varphi \in \mathbb{R}$ leaves $\mL$ invariant. The global phases of the jump operators are irrelevant, but relative phases within the components of a given $\ell$ matters.
\\
We can also rescale the superoperator by a positive real\footnote{$\alpha$ should be real, if we suppose to start from $\beta \in \complex$, $h\to \beta h$ and $\ell \to \sqrt {\beta} \ell$, but $\ell^{\dagger}\ell \to |\beta| \ell^\dagger \ell$. } element
\begin{align}
&\mathcal{L}\to \alpha \mathcal{L} ,
&&(h,\ell) \mapsto (\alpha h , \sqrt{\alpha}\, e^{i\phi}\,\ell),\,\,\,\alpha\in \mathbb{R}^+, \,\,\phi\in \mathbb{R}.
\end{align}
The dependence on the spectral parameter in the normalization is discussed in section \ref{identification}.
\subsubsection*{Discrete transformation}
\paragraph{Complex conjugation} 
If $\mL$ is an integrable superoperator, then $\mL^*$ is as well.
\\
This transformation on the $h$ and $\ell$ operator is
\begin{align}
(h,\ell) \mapsto (-h^* , \ell^*).
\end{align}

\paragraph{Parity} If a superoperator $\mathcal{L}$ is integrable, also its parity reverse is, as mentioned in \ref{identification}. This implies
\begin{align}
&\cL \to P \cL P,
&& h \to p \,h \,p\,\,\,\,\,\,\,\,\,\,\text{and}\,\,\,\,\,\,\,\,\,\,\ell \to p\, \ell \,p,
\end{align}
where $P$ and $p$ are the permutation (or parity) operators respectively in $W=V\otimes V^*$ and $V$.\\
In fact, $\mL$ acts on $(V\otimes V^*)_1 \otimes (V\otimes V^*)_2$, where we put the indices 1 and 2 to keep track of the spaces. By permuting spaces, we get $(V\otimes V^*)_2 \otimes (V\otimes V^*)_1$. The action of $P$ in $h^{(1)}$ is

\begin{align}
&h^{(1)}=\sum_{ijkl} h_{ijkl} e_{ij}\otimes \id \otimes e_{kl}\otimes \id
&&\overbrace{\to}^P
&&\sum_{ijkl} h_{ijkl} e_{kl}\otimes \id \otimes e_{ij}\otimes \id = (p h p)^{(1)},
\end{align}
where here we wrote $h^{(1)}$ in components. $e_{ij}$ are unity matrices, with $1$ in position $(i,j)$ and $0$ otherwise.

\paragraph{Transposition}
If $R$ satisfies the Yang-Baxter equation, then the transpose does as well. As mentioned in section \ref{identification}, this translates to the superoperator
\begin{align}
&\cL \to P \cL^T P.
\end{align}
This transformation is useful when we want to check if the model found can be mapped to a known one and it does not have a direct connection at the level of $(h,\ell)$ separately.
\\
In fact, if we want to use the mapping
\begin{align}
&(h,\,\ell)\,
&&\mapsto
&&(h^T,\,\ell^T),
\end{align}
to preserve integrability we have to add the additional constraint
\begin{align}\label{conditionontransposition}
&[\ell^\dagger,\ell]=0.
\end{align}
In this way, it can be shown that
\begin{align}
&\cL^T \,=\, \cL \,\,\Big(\text{with}\,\,\,h^T,\ell^T \Big) .
\end{align}

\paragraph*{Example: add a global phase to $\ell$ + shift in $\ell$} We can apply two of the transformations to the model \eqref{handlmodellindbladexample}. First, we multiply $\ell$ by the phase $i\, e^{-i \phi }$ and then we transform $(h,\ell)$ by using \eqref{transfshiftl} with $\alpha=1$.
\\
An alternative representation for the same superoperator can be obtained with
  \begin{align}
  &   h=0,\qquad
&&  \ell=\left(
\begin{array}{cccc}
 1 & 0 & 0 & 0 \\
 0 & 1 & 0 & 0 \\
 0 & -i \, e^{-i \phi } & 0 & 0 \\
 0 & 0 & 0 & 1 \\
\end{array}
\right).
     \end{align}

\section{Diagonal preserving models: connection with classical stochastic equations}

Before showing the models found in the classification, we explain an important connection between Lindblad superoperators and integrable classical stochastic equations, 
\cite{stochastic-book,Vanicat:2017bas}. 
\\
The Lindblad equation \eqref{lindbladeq} is capable of realizing the \textit{classical flows} on the diagonal of the density matrix. 
The diagonal elements of $\rho$ \eqref{rhoindex} are the \textit{classical probabilities} of finding the system in the given state, and in some cases the operator space spanned by the diagonal elements is kept invariant by the Lindblad superoperator. We refer to these models as \textbf{diagonal preserving}.
\\
We call $\ket{n_1,n_2,\dots,n_L}$, $n_j=\uparrow,\downarrow$ the vectors in the computational basis and $L$ is the length of the spin chain. For diagonal preserving models,  when we project the Lindblad equation \eqref{lindbladeq} to the diagonal elements we obtain
\begin{align}
&\bra{n_1,\dots,n_L}\rho
\ket{n_1,\dots,n_L}\equiv P(n_1,\dots,n_L)
\end{align}
and the flow equation
\begin{align}
&\partial_t \ket{P}=   W \ket{P},
\end{align}
where $  \ket{P}=P(n_1,\dots,n_L) \ket{n_1,\dots,n_L}$ and $ W=\sum_{j=1}^L w_{j,j+1}$ is the \textit{generator of the classical flow},
with matrix elements given by the corresponding projection of $\Li$. 
\\
Even if the model is diagonal preserving, for a generic configuration, the orthogonal complement of the diagonal subspace is not always conserved, so  we can still expect \textit{quantum
effects} in the time evolution. \\
We now show that the model found in the example \eqref{handlmodellindbladexample} is diagonal preserving.

\subsubsection*{Example: the integrable Lindbladiand \eqref{handlmodellindbladexample}}

By direct computation, we show that the model in the example is diagonal preserving. In fact, by projecting the Lindblad superoperator corresponding to \eqref{handlmodellindbladexample}, we obtain the following differential equations
\begin{align}
&\partial_t \ket{\uparrow \uparrow}=0,
&&\partial_t \ket{\uparrow \downarrow}=-\ket{\uparrow \downarrow},
&&\partial_t \ket{\downarrow \uparrow}=\ket{\uparrow \downarrow},
&&\partial_t \ket{\downarrow \downarrow}=0,
\end{align}
and the generator of the classical flow is
\begin{align}
&w_{i,i+1} = \left(
\begin{array}{cccc}
 0 & 0 & 0 & 0 \\
 0 & -1 & 0 & 0 \\
 0 & 1 & 0 & 0 \\
 0 & 0 & 0 & 0 \\
\end{array}
\right) = \sigma^-_i \sigma^+_{i+1} - n_i (1-n_{i+1}) ,
\end{align}
where $n_j=\frac{\id+Z_j}{2}$. This is the generator of the \textit{Totally Asymmetric Simple Exclusion Process} (TASEP) \cite{Vanicat:2017bas}: a particle can only move to the right if the neighbour site is empty. To our best knowledge this is the first realization of the TASEP using an integrable Lindbladian. In this case, the orthogonal complement of the diagonal subspace is not conserved.

\subsubsection*{Diagonal conservation for a non integrable Lindbladian}
The connection between Lindbladians and integrable classical stochastic equations has a rich history by itself and it was studied before the study of integrable Lindbladians (see \cite{stochastic-book} and reference therein).
\\
An example of a diagonal conserving model was discussed in \cite{stoch-coh-mixt} (see also
\cite{eisler-lindblad-xx,essler-piroli-lindblad}).
Here $h_{j,j+1}=0$ and the model has two families of jump operators ($k=1,2$ in \eqref{superopmultiplefamilies}) identified by the letters $R$ or $L$ depending on the direction of motion of the particles\footnote{We recall that in our convenction, the particles are spin up.}
\begin{equation}
  \label{noti}
   \ell^R_{j,j+1}=\sqrt{\varphi_R} \sigma_j^-\sigma_{j+1}^+,\qquad
    \ell^L_{j,j+1}=\sqrt{\varphi_L} \sigma_j^+\sigma_{j+1}^-,
\end{equation}
with $\varphi_{L,R}\ge 0$.
The resulting classical flow was found to be
the Asymmetric Simple Exclusion Process (ASEP), with the generator being
\begin{equation}
  w_{j,j+1}=\varphi_R [\sigma_j^-\sigma_{j+1}^+-  n_j(1-n_{j+1})]+\varphi_L [\sigma_j^+\sigma_{j+1}^--  (1-n_j)n_{j+1}].
\end{equation}
In this model, particles can move to the left or to the right with different probabilities. 
The integrability properties of the
superoperator itself were not investigated in \cite{stoch-coh-mixt,essler-piroli-lindblad}. We checked integrability\footnote{More details on how to check if a model is integrable are given in section \ref{whyisthemodelint} and we refer to \cite{deLeeuw:2022ath} for further details. Here, we briefly say that we checked if it exists a range 3 operator that commutes with it. In the positive case, we argue that the model is integrable.} and we found that the Lindblad system given by \eqref{noti} is not Yang-Baxter
integrable for arbitrary $\varphi_R$ and $\varphi_L$. 
We also studied the two limits $\varphi_R=0, \varphi_L =1 $ and $\varphi_R=1, \varphi_L = 0 $, for which the generators are the TASEP model and we found that in the first case $\cL$ is not integrable while in the second case it is.

\chapter{Classification of integrable open quantum systems}

\ifpdf
    \graphicspath{{Chapter7/Figs/Raster/}{Chapter7/Figs/PDF/}{Chapter7/Figs/}}
\else
    \graphicspath{{Chapter7/Figs/Vector/}{Chapter7/Figs/}}
\fi
\label{Lindbladclassificationmodels}

In this chapter, we present a collection of new models discovered through  the application of the boost automorphism mechanism within the realm of open quantum systems.  This provides the \textbf{first systematic} approach to classify integrable open quantum systems. When a quantum system interacts with a markovian environment, its dynamics can be described by a Lindblad superoperator. We focus on  cases where this superoperator is one of the conserved charges of an integrable model. In most cases, the environmental contribution destroys the integrability property of the system. However, there exist some cases where integrability is preserved and we list some of them in this chapter, along with their properties. We divide the models that we found into two classes:
\begin{itemize}
\item \textbf{Fine tuned models}. This class is physically not very interesting. It is expected that by placing a system in contact with an environment and fine-tuning the latter, the integrability property is preserved.
\item \textbf{Coupled models}. This class is physically more interesting because one can tune the coupling constant to control the impact of the environment according to the specific situation being studied.
\end{itemize}
One of the coupled models\footnote{The notation is explained in the next section. We follow the same names of the letter \cite{classificationlind}, where A and B correspond to two different ansatz that we used.} (model B3) will be deeply analysed, we study the Non Equilibrium Steady States (NESS) and the particle current flowing through  them. We found that this is an integrable example of the pumping effect: there is a finite current even when the coupling constant is very small. This effect was already observed in other systems, for example \cite{pumping}, however this is the first case where it is observed in an integrable model. At the end, we prove that up to identifications,  model B3 is equivalent to the generalized Toda system related to the non-exceptional affine Lie algebra $A_3^{(2)}$. We provide an interpretation for model B3 as two coupled spin-1/2 XXZ chains. We solve this model using the nested Algebraic Bethe ansatz and we report here the expressions of the eigenvalues and the Bethe equations. The details of the computation will be given in Appendix \ref{BAB3chapter}.   Lastly, we discuss model B2 which is an integrable deformation of the Hubbard model. Among all the models found, except from model B2, all of them are of difference form.

\section{Initial ansatz}\label{ansatzopenqs}
To apply our method, we start from an ansatz for both $h$ (the Hamiltonian of the system) and $\ell$ (the jump operator) acting on a spin 1/2 chain $\complex^2\otimes\complex^2$. We allow $h$ and $\ell$ depending on a spectral parameter in order to search for non-difference form models. For completeness we report the expression of the superoperator $\mathcal{L}$, given in \eqref{superoperatorL2},
\begin{align}
  \Li_{j,j+1}=&~-i h^{(1)}_{j,j+1}+i  h^{(2)^T}_{j,j+1}+   \Big(\ell^{(1)}_{j,j+1}  \ell^{(2)*}_{j,j+1}  -\frac{1}{2} \ell^{(1)\dagger}_{j,j+1}  \ell^{(1)}_{j,j+1}-\frac{1}{2}  \ell^{(2)T}_{j,j+1}  \ell^{(2)*}_{j,j+1}\Big).
\end{align}
 We investigate two choices:
\begin{itemize}
\item lower triangular $\ell$ operators with at most two elements below the
diagonal and general $h$ with all the 16 elements,
\item $h$ and $\ell$ operators that both conserve the total $S^z$ quantum number.
\end{itemize}
As previously stated, the classification method is applicable to any generic form of the operators $h$ and $\ell$. However, in order to maintain manageable computational complexity, we had to restrict their expressions. The first ansatz was maily selected for its simplicity. Furthermore, it also contains models with a clear physical significance. For example, if we require that the environment removes particles from the system, the jump operator takes the simple form $\ell_{j,j+1}=\sigma_j^-$ and falls in this category. The second ansatz, as per Chapter \ref{Hubbardchapter}, was chosen because the Hubbard model can be viewed as an open quantum system and it aligns with this ansatz.\\
Following the notation of the letter \cite{classificationlind}, we refer to the model of the first class as "A" and the second one as "B". We show that models A1, A2 and B1 are fine tuned and models B2 and B3 are not. Curiously both ansatz  {\it A} and  {\it B} only allow for diagonal Hamiltonian densities $h$ or it is given by
\begin{equation}
  \label{HXX}
  h_{j,j+1}=\frac{1}{2}\left[e^{i\phi}\sigma_j^+\sigma_{j+1}^-+e^{-i\phi}\sigma_j^-\sigma_{j+1}^+\right],\quad \phi\in\valos.
\end{equation}
This Hamiltonian density describes a free fermionic hopping model. The angle $\phi$ can be understood as a homogeneous twist along the
chain, and the $\phi=0$ point corresponds to the XX spin chain. The model can also be interpreted as the XX chain perturbed by a
Dzyaloshinskii–Moriya interaction term, \cite{Dzyaloshinsky-Moriya-xxz}.
\\
We omit some models with less interesting physical properties, such as those having diagonal $h$ and $\ell$ operators for which the superoperator is trivially integrable.
\section{Partial classification: Fine tuned models}
\subsection{Model A1} This is the model discussed in the example of the previous chapter. It is characterized by $h$ and $\ell$ \eqref{handlmodellindbladexample} and $R$-matrix given in \eqref{rmatA11}-\eqref{rmatA12}. By using the identifications of section \ref{identificationlindblad}, we did not manage to map this model to a known one, so we believe that this is a \textbf{new} model.

\subsection{Model A2}
The $h$ and $\ell$ operators are
\begin{align}
&h=\frac{s }{2}\left(
\begin{array}{cccc}
 0 & 0 & 0 & 0 \\
 0 & 0 & i & 0 \\
 0 & -i  & 0 & 0 \\
 0 & 0 & 0 & 0 \\
\end{array}
\right),
&&\ell=\left(
\begin{array}{cccc}
 1 & 0 & 0 & 0 \\
 0 & 0 & 0 & 0 \\
 0 & s  & 1 & 0 \\
 1 & 0 & 0 & 0 \\
\end{array}
\right) = n_{j+1} +s \sigma_j^-\sigma_{j+1}^++\sigma_j^-\sigma_{j+1}^-,
\end{align}
with $s=\pm1$. $h$ is \eqref{HXX} with $\phi\, =\, s\, \pi/2$. The jump operator describes particle propagation to the right (the term $\sigma_j^-\sigma_{j+1}^+$) and two-body loss ($\sigma_j^-\sigma_{j+1}^-$).
\\
The non-zero entries of the $R$-matrix are
\begin{align}
\nonumber &R^{16}_{16}= R^4_{13}= 1,\\
\nonumber & s R^8_3= s R^{12}_2  =-s R^7_4= -s R^{10}_4 = R^6_1= R^{11}_1= 2 e^{-u} \sinh \left(\frac{u}{2}\right),\\
\nonumber &-s \, R^{12}_{12} = -s\, R^8_8 =R^4_4= R^8_9= R^{12}_5= R^{16}_1= 1-e^{-u},\\
\nonumber &R^2_5= R^3_9= R^8_{14}= R^{14}_8= R^{12}_{15}= R^{15}_{12}= e^{-\frac{u}{2}},\\
\nonumber &R^9_3= R^5_2= e^{-\frac{3 u}{2}},\\
&R_1^1= R^6_6= R^{11}_{11}= R^7_{10}= R^{10}_7= R^{13}_4= e^{-u} ,
\end{align}

and to the best of our knowledge it corresponds to a new integrable model.\\
The model is diagonal preserving with the
generator being
\begin{equation}
w_{j,j+1}=-n_j+\sigma_j^-\sigma_{j+1}^++\sigma_j^-\sigma_{j+1}^-.
\end{equation}
This corresponds to the totally asymmetric limit of the diffusion-annihilation model treated in
\cite{lushnikov-annihilation,lushnikov-annihilation2,asymmetric-two-body-anni-free-fermion}.
To our best knowledge, this is the first time that this generator is
embedded into an integrable Lindbladian.

\subsection{Model B1} \label{modelB1explained}In this case $h=0$ and the jump operator is
\begin{equation}
  \ell=
  \begin{pmatrix}
    s_1 & 0 & 0 & 0  \\
    0 & 0 & 1 & 0 \\
    0 & 1 & 0 & 0 \\
    0 & 0 & 0 & s_2 
  \end{pmatrix},
\end{equation}
where $s_1=\pm 1$, $s_2=\pm 1$. In the case of $s_1=s_2=1$ the superoperator is
equivalent to the Hamiltonian of the $SU(4)$-invariant chain; this
case was listed in \cite{ziolkowska2020yang}. 
\\
The non-zero entries of the $R$-matrix are
\begin{align}
\nonumber &R_1^1= R^{16}_{16}= e^{-u} (u+1),\\
\nonumber&R^6_6= R^{11}_{11}= e^{-u} (s_2  s_1  u+1) ,\\
\nonumber &R^2_5= R^5_2= R^3_9= R^9_3= R^4_{13}= R^{13}_4= R^7_{10}= R^{10}_7= R^8_{14}= R^{14}_8= R^{12}_{15}= R^{15}_{12}= e^{-u},\\
&s_1 R^2_2  = s_1 R^3_3 = s_1 R^5_5  =s_1 R^9_9  = s_2  R^8_8= s_2  R^{12}_{12}= s_2  R^{14}_{14}= s_2  R^{15}_{15}=R^4_4= R^7_7= R^{10}_{10}= R^{13}_{13}= e^{-u} u.
\end{align}
As mentioned earlier, for $s_1=s_2=1$
\begin{align}
R(u)=\frac{u}{e^{u}}\left(1+\frac{1}{u} P \right),
\end{align}
which is the $R$-matrix of the $SU(4)$ invariant chain.
\\
The $R$-matrix has  \textit{difference form} for all four choices, and for $s_1s_2\ne 1$ it seems to be \emph{new}.
\\
The model is \textit{diagonal preserving} in all four cases  and the
generator is
\begin{equation}
  w_{j,j+1}=
  \sigma_j^+\sigma_{j+1}^-+
  \sigma_j^-\sigma_{j+1}^+-
  n_j-n_{j+1}+2n_j n_{j+1}.
\end{equation}
This is the generator of the \textit{Symmetric Simple Exclusion Process}, (SSEP)
\cite{stochastic-book}. To the  best of our knowledge it is the first
time that the SSEP is realized by an integrable Lindbladian.
Even though this model is very simple and its dynamics is generated just by the jump operator, the parameters $s_1$ and $s_2$ have an effect on the off-diagonal sectors of the superoperator, which in turn influences the spectrum. We verified this through direct computation for $L=3,4,5$ since a Bethe ansatz solution is not yet available.

\section{Partial classification: Coupled models}

\subsection{Model B3}
\label{modelB3defined}

Since in this model there is a coupling constant, we can tune the strength of the environment. We analyse this model in more details due to the interesting physical properties. \\
Here, we use the notation of our work \cite{de2022bethe}. This notation allows for simpler expressions compared to the notation we used in the earlier work \cite{classificationlind}. The two notations are connected through the parametrization $\gamma = \tanh\psi$. The coupling constant of the model is proportional to $\gamma$,  so this reparametrization maps strong coupling to strong coupling. Hence strong coupling is $\psi\rightarrow \infty$ and weak coupling corresponds to $\psi \rightarrow 0$.
\subsubsection{$h$ and $\ell$}
Model B3 is again characterized by the Hamiltonian $h$ given by \eqref{HXX}. For simplicity in the notation, we renormalize it
 \begin{align}
&\frac{h_{j,j+1}}{\alpha}=
 \left(
\begin{array}{cccc}
 0 & 0 & 0 & 0 \\
 0 & 0 & e^{i \phi } & 0 \\
 0 & e^{-i \phi } & 0 & 0 \\
 0 & 0 & 0 & 0 \\
\end{array}
\right) = e^{i \phi }\,  \sigma^+_j \sigma^-_{j+1}  + e^{-i \phi }\,  \sigma^-_j \sigma^+_{j+1} ,
\label{hb3}
\end{align}
and the jump operator $\ell$ is 
\begin{align}
\frac{\ell_{j,j+1}}{\beta}=
&\left(
\begin{array}{cccc}
 i \sinh \psi  & 0 & 0 & 0 \\
 0 & i \cosh \psi & e^{-\psi +i \phi } & 0 \\
 0 & e^{\psi -i \phi } & -i \cosh \psi  & 0 \\
 0 & 0 & 0 & i \sinh \psi  \nonumber\\
\end{array}
\right)  \\
& =i \sinh \psi \,(1+2 n_j n_{j+1})+i \,(e^{-\psi } n_j - e^{\psi } n_{j+1})+ e^{-\psi +i \phi } \sigma_j^+\sigma_{j+1}^-+e^{\psi -i \phi } \sigma_j^-\sigma_{j+1}^+,
\label{lb3}
\end{align}
where $\alpha=\cosh ^2\psi \, \text{sech}\,2 \psi $, $\beta=-i \sqrt{\tanh \psi \, \text{sech}\,2 \psi }$ and $n_j=\frac{1}{2}(\id+Z_j)$.
\\
Here $\beta(\psi)$ is the coupling constant.

\subsubsection{R-matrix}

The $R$-matrix is of difference form, with entries
\begin{align}
\nonumber &R_1^1= R^6_6= R^{11}_{11}= R^{16}_{16}= 1,\\
\nonumber &\frac{R^2_2}{e^{i \phi }}= -e^{i \phi }R^3_3 = -e^{i \phi }R^8_8 = \frac{R^{12}_{12}}{e^{i \phi }}= \frac{i (\tanh \psi +1)}{\coth (u-\psi )+\tanh \psi },\\
\nonumber & -e^{i \phi } R^5_5= \frac{R^9_9}{e^{i \phi }}= \frac{R^{14}_{14}}{e^{i \phi }}=-e^{i \phi } R^{15}_{15} = \frac{i (\tanh \psi -1)}{\coth (u-\psi )+\tanh \psi },\\
\nonumber &R^2_5= R^5_2= R^{15}_{12}= R^{12}_{15}= R^8_{14}= R^{14}_8= R^9_3= R^3_9= \text{sech} u \cosh \psi,\\
\nonumber & e^{i \phi } R^4_7= -\frac{R^4_{10}}{e^{i \phi }}= \frac{R^{13}_{10}}{e^{i \phi }}= -e^{i \phi }R^{13}_7 = -i \text{sech }u \cosh \psi  \sinh (u-\psi ) \text{sech}(u+\psi ),\\
\nonumber &-e^{i \phi }R^7_4 = \frac{R^{10}_4}{ e^{i \phi }}= i e^{2 \psi } \text{sech }u \cosh \psi  \sinh (u-\psi ) \text{sech}(u+\psi ),\\
\nonumber &e^{i \phi }R^7_{13} = -\frac{R^{10}_{13}}{e^{i \phi }}= i e^{-2 \psi } \text{sech }u \cosh \,\psi  \sinh (u-\psi ) \text{sech}(u+\psi ),
\\
\nonumber &R^{10}_7= R^7_{10}= \frac{1}{2} \text{sech }u (\cosh \,\psi +\cosh (3 \psi )) \text{sech}(u+\psi ),\\
\nonumber &e^{2 i \phi }R^7_7 = \frac{R^{10}_{10}}{e^{2 i \phi }}= \frac{R^{4}_{4}}{e^{2 i \phi }}= \frac{R^{13}_{13}}{e^{-2 i \phi }}=\tanh \,u \sinh (u-\psi ) \text{sech}(u+\psi ),\\
\nonumber &R^4_{13}= -e^{u-\psi } \text{sech }u \cosh \psi  (\sinh (u-\psi ) \text{sech}(u+\psi )-1),\\
 &R^{13}_4= e^{\psi -u} \text{sech }u \cosh \psi  (\sinh (u-\psi ) \text{sech}(u+\psi )+1).\label{RmatrixmodelB3}
\end{align}
For simplicity, we omitted the dependence on the shifted spectral parameter. Explicitly
\begin{align}
R_i^j=R_i^j(u-\psi).
\end{align}

\subsubsection{Diagonal preserving point}

In the limit $\psi\to \infty$, the model is diagonal preserving and it describes the TASEP model.
For generic values of $\psi$, the model
describes a mixture of quantum and classical transport.

\subsubsection{Particle current}

The particle current $J_k$ can be found from the continuity equation
\begin{align}\label{particlecurrentgeneral}
&\frac{d n_k}{d t}=J_{k-1}-J_{k}
\end{align}
and the Lindblad equation, \cite{stoch-coh-mixt}. It is given by

\begin{equation}\label{particlecurrent}
    J_{k} \,=\, 
 \text{sech}^2\psi \, J_k^0+
 \tanh \psi  (\tanh (2 \psi )+1) n_k\,(\id-n_{k+1})+\tanh \psi  (\tanh (2 \psi )-1)(\id-n_k) n_{k+1},
   \end{equation}
with $J_k^0$ given by 
\begin{equation}
  \label{cohJ}
  J_k^0= \frac{i}{1+\tanh \psi^2} (e^{i\phi}\sigma_k^+\sigma_{k+1}^--e^{-i\phi}\sigma_k^-\sigma_{k+1}^+) .
\end{equation}
$J_k^0$ is the current of the coherent time
evolution dictated by \eqref{HXX}, and the remaining terms describe stochastic transport. The stochastic terms explicitly break the spatial reflection symmetry, representing the current in the ASEP. From here, we observe that sending $\psi \to \infty$, the only remaining term in $J_k$ is $n_k\,(\id-n_{k+1})$, current of the TASEP.

\subsubsection{Non Equilibrium Steady States and Pumping Effect}
An object of particular relevance in the theory of open quantum systems are the Non Equilibrium Steady States (NESS). They correspond to fixed points in the dynamics, 
\begin{align}
&\dot{\rho}_{NESS}=0,
&&\cL \rho_{NESS} = 0 .
\end{align}
From \eqref{lindbladeq} it is easy to see that the identity density matrix is always a NESS, however, searching for other NESS is an interesting and ongoing field of research since their stucture may reveal information about the symmetries of the Lindblad superoperator. We motivate this statement in chapter \ref{NESSCHAPTER}.\\
In the case of model B3, we have observed that for arbitrary values of $\phi$ and $\psi$, the NESS is a \textit{mixed state}, meaning it cannot be described by a single pure state\footnote{The purity allows to determine if a state is pure or mixed. It is defined (for density matrix normalized such that $\Tr(\rho)=1$) as $\Tr(\rho^2)$ and it is equal to $1$ for pure states and $<1$ for mixed states.}. However, when a compatibility condition holds, it degenerates into a pure state. This condition is given by $e^{i (\phi \pm \pi/2)L} = 1$. Specifically, the pure state takes the form of a spin-helix state
\begin{align}
&\rho_h=\ket{\Psi}\bra{\Psi},
&&  \ket{\Psi}=\otimes_{j=1}^L
\frac{1}{\sqrt{2}}  \begin{pmatrix}
1 \\ e^{i j(\phi\pm\pi/2)}
  \end{pmatrix} ,
\end{align}
and $h \ket{\Psi} = 0$, for any value of $\psi$.
\\
The superoperator conserves the particle
number, thus the NESS' in the sectors with fixed spin are given by the appropriate projections of $\rho_h$. Numerical studies on small systems show that in each spin sectors the NESS is unique. For this model, we discovered the existence of  multiple NESS and the multiplicity is attributed to the conservation of particle number. In chapter \ref{NESSCHAPTER}, we analyse a deformation of the Hubbard model and we motivate the multiplicity of the NESS with a hidden symmetry.
\\
The particle current \eqref{particlecurrent} in the projected states can be computed easily, and in the thermodynamic limit we
find
\begin{equation}
\lim_{L\to\infty}  \vev{J_k}=(1+\tanh \psi^2) \vev{n}\big(1-\vev{n}\big).
\end{equation}
We conjecture that this formula holds generally, even if the
compatibility condition is not met. The coupling constant $\beta$ is zero in the limit $\psi\to 0$. In this case, there is still a \textit{finite} particle current. This phenomenon is understood as a \textit{pumping effect}.
The jump operators are coupled to the
coherent current, thus they build up its mean value over time. The current itself is conserved, thus it cannot decay.
Eventually a current carrying state is produced no matter how small the coupling is. This phenomenon was discussed in \cite{pumping,rectification}, and our
model is an integrable example of it.

\subsubsection{Is this model new?}
In this section, we show\footnote{We thank A. Hutsalyuk and A. Liashyk for pointing out this connection.}  that model B3 is \textbf{not new} since it is related to the Generalized Toda System connected to the non-exceptional affine Lie algebras $A_3^{(2)}$ discussed by Jimbo in \cite{jimbo1986quantumr}. More details can be  found in \cite{liashyk2022recurrence}. The $R$-matrix of $A_3^{(2)}$ is
\begin{align}\label{jimboA32}
R^{A_3^{(2)}}(u,v)=\id+\mathbb{P}(u,v)+\mathbb Q(u,v),
\end{align}
with
\begin{align}
&\mathbb{P}(u,v)=\sum_{1\le i,j \le N} p_{ij}(u,v) \,e_{ij}\otimes e_{ji}\,,
&&\mathbb{Q}(u,v)=\sum_{1\le i,j \le N} q_{i'j'}(u,v)\, e_{i'j'}\otimes e_{ji}\,,
\end{align}
\begin{align}
& p_{ij}(u,v) = \left\{\begin{array}{lr}
        f(u,v)-1, & i=j\\
        \tilde{g}(u,v), & i<j\\
        g(u,v), & i>j
        \end{array}\right.,
&q_{ij}(u,v) = q^{\bar{i}-\bar{j}}\left\{\begin{array}{lr}
        f(v \xi,u)-1, & i=j\\
        {g}(v \xi,u), & i<j\\
        \tilde{g}(v \xi,u), & i>j
        \end{array}\right.\,.
\end{align}
In this case $N=4$, $\xi=q^{-4}$, $i' = N+1-i,\,\,\,\, 1 \le i \le N$ and $\bar{i}=3-i$ for $i\ge 3$ and $\bar{i}=2$ for $i<3$ and similar expressions for $j'$ and $\bar{j}$ with $i\leftrightarrow j$,
\begin{align}
&f(u,v) = \frac{q \,u-v\,q^{-1}}{u-v},
&&g(u,v) = \frac{q\, u-u \,q^{-1}}{u-v},
&&\tilde g(u,v) = \frac{q\, v-v\,q^{-1}}{u-v}.
\end{align}
To show that this $R$-matrix is equivalent to the $R$-matrix of model B3, we perform in series some of the transformations given in section \ref{identification}
\begin{itemize}
\item[1.] normalization $R^{A_3^{(2)}}\to \frac{q (u-v)}{q^2 u-v} R^{A_3^{(2)}} = \tilde  R^{A_3^{(2)}}$
\item[2.] twist $(F \otimes F^{-1}) \tilde  R^{A_3^{(2)}}(u^2,v^2)\, (F \otimes F^{-1}) = \tilde{\tilde{R}}^{A_3^{(2)}}(u,v)$ , with $F$ a diagonal matrix with entries $\left\{e^{\psi /2},i e^{\frac{i \phi }{2}},e^{\frac{-i \phi}{2} },i e^{-\frac{\psi }{2}}\right\}$
\item[3.] LBT  $(K(u)\otimes K(v)) \tilde{\tilde{R}}^{A_3^{(2)}}(u,v)(K(u)\otimes K(v))^{-1}$, with \\$K(u)=\frac{1}{4} \left(\frac{1}{u^2}+1\right) (Z_1+Z_2)+\frac{1}{4} \left(1-\frac{1}{u^2}\right) (\id+Z_1\,Z_2)$ .
\end{itemize}
It follows that the $R$-matrix of model B3 is equivalent to the Generalized Toda System $A_3^{(2)}$ if $q=i e^{-\psi}$.
\\
We conclude that this model is {not new}, however the interpretation in term of open quantum system is new. This opens an interesting question whether all the $R$-matrices of the non-exceptional quantum affine Lie algebras can be mapped (after some identifications) to an open quantum system.\\
The difficulty in the discovery of this equivalence is the following: when we begin with the $R$-matrix \eqref{jimboA32} and derive the associated density charge $\mathcal{Q}_2=P \partial_u R^{A_3^{(2)}}(u,v)_{v\to u}$, we attempt to determine the values of $h$ and $\ell$ that would yield the superoperator $\cal{L}$. However, we discover that this correspondence does not occur.  The twist $F$ is the key object that allowed this identification.

\subsubsection{$4-D$ interpretation of the model}
In light of \eqref{notation} and of the discussion in the related section, we can also interpret model B3 as acting on a Hilbert space of local dimension 4. We rewrite the superoperator \eqref{superoperatorL2} by separating the part containing only $\sigma$ ($^{(1)}$) or $\tau$ ($^{(2)}$) operators
\begin{align}
  \Li_{j,j+1}=&~\Big(-i h^{(1)}_{j,j+1} -\frac{1}{2} \ell^{(1)\dagger}_{j,j+1}  \ell^{(1)}_{j,j+1}\Big)+\Big(i  h^{(2)^T}_{j,j+1} -\frac{1}{2}  \ell^{(2)T}_{j,j+1}  \ell^{(2)*}_{j,j+1}\Big)+   \ell^{(1)}_{j,j+1}  \ell^{(2)*}_{j,j+1},
\end{align}
we obtain
\begin{align}
\mathbb{H}_{i,j}={J_{i,j}}^{(1)}+ \tilde J_{i,j}^{(2)}+ { \ell_{i,j}}^{(1)}  {\tilde \ell_{i,j}}^{(2)},
  \label{sup}
\end{align}
where
\begin{align}
J_{i,j}=-i \,h_{i,j}-\frac{1}{2} \ell^{\dagger}_{i,j} \ell_{i,j} ,
\label{Jop}
\end{align}
and $\tilde{J}_{ij}$ is related to $J_{ij}$ by taking the complex conjugate elementwise.\\
The decomposition of the superoperator in term of $h$ and $\ell$ is not unique, as discussed in section  \ref{identificationlindblad}. By using the transformation \eqref{transfshiftl} with $\alpha=-\sinh ^2\psi  \sqrt{\text{csch}(4 \psi )}$, one gets
\begin{align}
&J_{i,j}^{(1)}=i e^{i \phi } (\tanh \psi -1)\sigma_i^+ \sigma_j^--ie^{-i \phi } (\tanh \psi +1)\sigma_i^- \sigma_j^+  - \tanh \psi \,(n_i-n_j)^2,\label{Jij}\\
&\tilde J_{i,j}^{(2)}= -i e^{-i\phi} \, (\tanh \psi -1)\tau_i^+ \tau_j^-+i e^{i \phi }(\tanh \psi +1) \tau_i^- \tau_j^+-\tanh \psi ( \tilde n_i -\tilde n_j)^2,
\end{align}
where we refer to $n_i=\frac{\id+\sigma^z_i}{2}$ and $\tilde{n}_i=\frac{\id+\tau^z_i}{2}$. The  term $\ell_{ij}^{(1)}{\tilde \ell_{ij}}^{(2)}$ in \eqref{sup} is
\begin{align}
\ell_{ij}^{(1)}{\tilde \ell_{ij}}^{(2)}=
\frac{\tanh\psi}{\cosh2\psi}(2 \sinh \psi \, n_i n_j+e^{-\psi }  n_i-i e^{-\psi +i \phi }\sigma_i^+ \sigma_j^--i e^{\psi -i \phi }\sigma_i^- \sigma_j^+-e^{\psi }\, n_j)(c.c)
\label{ells}
\end{align}
where $(c.c)$ has the $\tau$ operators and the coefficients are complex conjugate. 
\\
Interestingly, the expression \eqref{Jij} of $J_{i,j}$  corresponds, up to a twist\footnote{The twist is
$U_i=\left(
\begin{array}{cc}
 1 & 0 \\
 0 & \pm e^{-\psi +i \phi } \\
\end{array}
\right)$.} and a normalization defined in \ref{identification}, to the Hamiltonian of the XXZ chain
\begin{align}
U_i J_{i,j} U_i^{-1} \,=\,(h_{XXZ})_{i,j}\,\sim\, \sigma_i^x \sigma_j^x+\sigma_i^y \sigma_j^y+\Delta \sigma_i^z \sigma_j^z,
\end{align}
where $\Delta=\pm i \sinh \psi $. 
When we send $\psi\rightarrow0$, the Hamiltonian $\mathbb{H}$ simply decomposes into two independent XX spin chains.
\\
To summarize, we see that the model B3 corresponds to two coupled XXZ chains with interaction terms given by the $\ell$s \eqref{lb3}. Applying the same twist $U$ on $\ell$ we get
\begin{align}
U_i \ell_{i,j} U_i^{-1}= \left(
\begin{array}{cccc}
 0 & 0 & 0 & 0 \\
 0 & 1-\tanh \psi  & \mp i\, \text{sech}\psi  & 0 \\
 0 & \mp i\,  \text{sech}\psi  & -\tanh \psi -1 & 0 \\
 0 & 0 & 0 & 0 \\
\end{array}
\right).
\end{align}
In chapter \ref{Hubbardchapter}, we give 4-D interpretation for the Hubbard model, as two XX spin chains coupled. Here, in contrast to the Hubbard model, the coupling constant between the two independent chains is related to the inhomogeneity $\Delta$ in the individual XXZ-spin chains.

\subsubsection{What are the eigenvalues of this model?}

In light of this interpretation, we can consider the Lindblad superoperator as an operator in a four dimensional Hilbert space.  To solve this model, we employed the \textit{nested algebraic Bethe ansatz} technique, which is a more advanced version of the Algebraic Bethe ansatz discussed in chapter \ref{intro}, see \cite{levkovich2016bethe,Slavnov:2019hdn} for recent reviews. After discovering that this model can be mapped to the Lie algebra $A_3^{(2)}$, one could consider to adapt the solution obtained for that particular model (for example in \cite{lima2006algebraic} for periodic boundary condition and \cite{li2007algebraic} for open) for the model B3. However, implementing the twist at the level of the Bethe equations proves to be a highly challenging task. Therefore, we present an independent solution \cite{de2022bethe} that we have derived specifically for this model \eqref{RmatrixmodelB3}. Taking into account the considerable length of the calculation, we have chosen to include it in Appendix \ref{BAB3chapter}, while presenting the main results here. In the appendix, we revisit the definitions provided in section \ref{definitionsmonodromy}, but this time in a higher-dimensional Hilbert space. We define the monodromy, transfer matrix, and reference state, and we establish the commutation relations between the entries of the monodromy matrix based on the RTT relation.  We explain how for model B3 this method closely mirrors the ones for the Hubbard model, \cite{ramos1997algebraic,martins1998quantum} and the $\text{AdS}_5 \times \text{S}_5$  for bound states, \cite{arutyunov2009bound}. However, we observe a distinct feature in our case, as we obtain a \textbf{twisted transfer matrix} for the nested spin chain. Finally, we give the expression of the  eigenvalues of the transfer matrix and the Bethe equations.
\\
Similar to \cite{ramos1997algebraic,martins1998quantum}, we identify three types of creation operators $B_1,\, B_2$ and $B_3$. The first two operators create electrons with spin up and down respectively, while the third operator creates an electron pair. The concept of \textbf{nesting} arises from our two-step procedure: initially considering the entire Hilbert space and subsequently focusing on the smaller Hilbert space $\mathbb{C}^2$.
\\
Unlike in section \ref{ABAtheory}, where we only needed to know the length of the spin chain $L$ and the number of particles $M$ to identify a state, here we also need to specify the types of particles involved. 
A state is characterized by the length $L$ of the spin chain as well as
\begin{align}
\nonumber
&M=\#\, \text{electrons }\\ 
\nonumber
&N=\#\, \text{number of electrons with spin up }
\end{align}
To simplify the expressions, we consider a shift in $u_i$, $v_i$ and $\psi$, in particular
\begin{align}
&\psi \to  {\Psi} +\frac{i \pi }{2},
&&u_i\to  \mathtt{u}_i+\frac{i \pi }{2},
&&v_i\to  \mathtt v_i-\frac{\psi }{2},
\end{align}
and we find the \textbf{eigenvalue} of the transfer matrix,\\

\begin{align}
\nonumber\frac{\Lambda_M(u)}{\mathcal{N}}=&\prod _{i=1}^M \frac{\sinh \left(u-\mathtt v_i+\frac{3 \Psi }{2}\right)}{\sinh\left(u-\mathtt v_i+\frac{\Psi }{2}\right)}+\\
\nonumber&\frac{\lambda_{6V}(u)}{\left(-e^{2 i \phi }\right)^{L-M-N}} \prod _{i=1}^M  \frac{\sinh \left(u-\mathtt v_i-\frac{\Psi }{2}\right)}{\sinh\left(u-\mathtt v_i+\frac{\Psi }{2}\right)} \prod _{j=1}^L \frac{  \sinh \left(u-\mathtt u_j+\psi \right)}{i\,e^{\Psi +i \phi }\sinh\left(u-\mathtt u_j\right)}+\\
&\prod _{i=1}^M \frac{\cosh \left(u-\mathtt v_i-\frac{3 \Psi }{2}\right)}{\cosh \left(u-\mathtt v_i-\frac{\Psi }{2}\right)}\prod _{i=1}^L \frac{ \coth \left(u-\mathtt u_i\right) \sinh \left(u-\mathtt u_i+\Psi \right)}{e^{2 \Psi }\,\cosh \left(u-\mathtt u_i-\Psi \right)},
\end{align}
$\mathcal{N}=\left(-e^{2 i \phi }\right)^{M-N} \left(-e^{\Psi -i \phi }\right)^M$ and
\begin{align}
\frac{\lambda _{6V}(u)}{\left(-e^{2 i \phi }\right)^{L-M}}=&\prod _{i=1}^M-\frac{ \sinh \left(2 \left(u-w_i+\Psi \right)\right)}{e^{2 i \phi }\,\sinh\left(2 \left(u-w_i\right)\right)}+\\
\nonumber&\left(-e^{2 i \phi }\right)^{L-M} \prod _{i=1}^N -\frac{ \sinh \left(2 \left(u-w_i-\Psi \right)\right)}{e^{2 i \phi } \sinh\left(2 \left(u-w_i\right)\right)}\prod _{i=1}^M -\frac{e^{2 i \phi } \sinh \left(2 u-2 \mathtt v_i+\Psi \right)}{\sinh\left(2 u-2 \mathtt v_i-\Psi \right)}.
\end{align}

The "nested" structure is \textit{manifest}, $\Lambda_M$ (the eigenvalue of the main spin chain) depends on  $\lambda_{6V}$ (the eigenvalue of the nested chain). The \textit{rapidities} $\{v\}$ and $\{w\}$ of the particles in the model obey the sets of Bethe equations, for $j=1,\dots,M$
\begin{align}
&\prod _{i=1,i\neq j}^M \frac{\sinh\left(\mathtt v_i-\mathtt v_j-\Psi \right)}{\sinh\left(\mathtt v_i-\mathtt v_j+\Psi \right)}=\prod _{i=1}^L \frac{1}{i\,e^{\Psi +i \phi }}\frac{  \sinh\left(\mathtt u_i-\mathtt v_j-\frac{\Psi }{2}\right)}{\sinh \left(\mathtt u_i-\mathtt v_j+\frac{\Psi }{2}\right)} \prod _{i=1}^N \frac{\sinh \left(2 \mathtt v_j-2 w_i+\Psi \right)}{\sinh\left(2 \mathtt v_j-2 w_i-\Psi \right)}
\end{align}
and the auxiliary Bethe equations, for $j=1,\dots,N$
\begin{equation}
\prod _{i=1,i\neq j}^N \frac{\sinh \left(2 \left(w_i-w_j-\Psi \right)\right)}{\sinh\left(2 \left(w_i-w_j+\Psi \right)\right)}=\left(-e^{2 i \phi }\right)^{L}\prod _{i=1}^M \frac{\sinh\left(2 \mathtt v_i-2 w_j-\Psi \right)}{\sinh\left(2 \mathtt v_i-2 w_j+\Psi \right)}.
\end{equation}
We checked explicitly\footnote{We thank Rafael Nepomechie for providing the code for the numerical computation of the Bethe equations.} that by numerically solving the Bethe Equations for the rapidities, and substituting to the eigenvalues $\Lambda_M(u)$, we find the same eigenvalues as by directly diagonalizing the transfer matrix. We checked for spin chains of small length and up to three particles state. Such validations are typically sufficient to ensure the accuracy and correctness of the results obtained using the nested Bethe ansatz method.\\
We present the result by using the standard Baxter Q-functions 
\begin{align}
\label{Qbaxt1}&Q^{[a]}_{\mathtt u}(x) = \prod_{i=1}^L \sinh[x-{\mathtt u}_i-a\Psi] , &&\tilde{Q}^{[a]}_{\mathtt u}(x) = \prod_{i=1}^L \cosh[x-{\mathtt u}_i-a\Psi], \\
&Q^{[a]}_{\mathtt v}(x) = \prod_{i=1}^M \sinh[x-{\mathtt v}_i-a\Psi], &&\tilde{Q}^{[a]}_{\mathtt v}(x) = \prod_{i=1}^M \cosh[x-{\mathtt v}_i-a\Psi], \\
\label{Qbaxt2}&Q^{[a]}_w(x) = \prod_{i=1}^N \sinh[x-w_i-a\Psi], && \tilde Q^{[a]}_w(x) = \prod_{i=1}^N \cosh[x-w_i-a\Psi].
\end{align}
The eigenvalue is
\begin{align}
\frac{\Lambda_M(u)}{\mathcal{N}}=&\frac{{Q_\mathtt{v}}^{[-3/2]}}{{Q_\mathtt{v}}^{[-1/2]}}+e^{-2\Psi L}\frac{{\tilde Q_\mathtt{v}}^{[3/2]}}{{\tilde Q_\mathtt{v}}^{[1/2]}} \frac{{\tilde Q_\mathtt{u}}^{[0]}}{{ Q_\mathtt{u}}^{[0]}}\frac{{Q_\mathtt{u}}^{[-1]}}{{\tilde Q_\mathtt{u}}^{[1]}}+\frac{ \left(-i e^{-\Psi -i \phi }\right)^L}{\left(-e^{2 i \phi }\right)^{L-M-N}} \frac{{Q_\mathtt{v}}^{[1/2]}}{{Q_\mathtt{v}}^{[-1/2]}}\frac{{Q_\mathtt{u}}^{[-1]}}{{Q_\mathtt{u}}^{[0]}}\lambda_{6V}(u),
\end{align}
\begin{align}
\lambda _{6V}(u)\frac{e^{2 i \phi N}}{\left(-e^{2 i \phi }\right)^{L-M}}=\frac{Q^{[-1]}_w\tilde{Q}^{[-1]}_w}{Q^{[0]}_w\tilde{Q}^{[0]}_w}+\left(-e^{2 i \phi }\right)^{L} \frac{Q^{[1]}_w\tilde{Q}^{[1]}_w}{Q^{[0]}_w\tilde{Q}^{[0]}_w}\frac{Q^{[-1/2]}_\mathtt{v}\tilde{Q}^{[-1/2]}_\mathtt{v}}{Q^{[1/2]}_\mathtt{v}\tilde{Q}^{[1/2]}_\mathtt{v}}
\end{align}\\
where for simplicity we used $Q^{[a]}_t(u)=Q^{[a]}_t$.\\
The Bethe equations for the main chain take the simple form
\begin{align}
\frac{{Q_\mathtt{v}}^{[-1]}}{{Q_\mathtt{v}}^{[1]}}=-\left(\frac{-i}{e^{\Psi+i \phi }}\right)^L \frac{{Q_\mathtt{u}}^{[-1/2]}}{{Q_\mathtt{u}}^{[1/2]}}\frac{{Q_w}^{[-1/2]}}{{Q_w}^{[1/2]}}\frac{{\tilde Q_w}^{[-1/2]}}{{\tilde Q_w}^{[1/2]}},
\end{align}
$j=1,\dots,M$ and for the nested chain
\begin{equation}
\frac{{Q_w}^{[-1]}{\tilde Q_w}^{[-1]}}{{Q_w}^{[1]} {\tilde Q_w}^{[1]}}=-\left(-e^{2 i \phi }\right)^L\frac{{Q_\mathtt{v}}^{[-1/2]} {\tilde Q_\mathtt{v}}^{[-1/2]}}{{Q_\mathtt{v}}^{[1/2]}{\tilde Q_\mathtt{v}}^{[1/2]}}.
\label{beqopnest}
\end{equation}
where  $Q^{[a]}_t(w_j)=Q^{[a]}_t$ and $j=1,\dots,N$.
\\

\vspace{0.5 cm}

The obtained expression for the eigenvalues opens up opportunities to investigate various properties of this model. One interesting avenue is to apply the \textit{thermodynamic Bethe ansatz} and examine the behaviour of the eigenvalues on an infinite spin chain. Of particular interest is also the study of the scaling of the eigenvalue with the second largest real part (with the smallest being 0) with the dimension $L$, which corresponds to the Liouville gap. In a finite system, this quantity is equivalent to the inverse of the relaxation time, as discussed in \cite{vznidarivc2015relaxation}.
\subsection{Model B2}\label{modelB2explained}
Similar to the previous model, Model B2 possesses a coupling constant and exhibits intriguing physical properties.  We first provide  the expressions of $h$ and $\ell$. Next, we analyse  the NESS and we give the expression of the particle current. This model serves as a bridge to the next chapter, as it corresponds to an integrable \textit{deformation} of the Hubbard model (correspondence that will be clarified in the next chapter). Additionally, this is the only model among the ones in the  ansatz presented in section \ref{ansatzopenqs} that does not possess the non-difference form property.
\subsubsection{$h$ and $\ell$}
The Hamiltonian is given by \eqref{HXX} and the jump operator is
 \begin{equation}\label{jumpforB2}
   \frac{\ell(u)}{\beta(u)}=
   \begin{pmatrix}
    \ch(u) & 0 & 0 &0 \\
 0   & 1 & i \,\sh(u) e^{i\, \phi } &0 \\
 0   & -i \,\sh(u) e^{-i \,\phi } & -1 & 0\\
 0   & 0& 0& -\ch(u)
    \end{pmatrix},
  \end{equation}
with $\beta(u)= (\gamma/(2\, \gamma \, \ch (2 u)+2))^{1/4}$
and $\gamma\geq 0$ being a fixed coupling constant.
\subsubsection{R-matrix}

When $u=\phi=0$, this system is equivalent to the XX model with dephasing noise\footnote{Explicitly, the jump operator takes the form $\ell_j=Z_j$. This is called a dephasing noise because the dissipation
destroys the quantum coherence of the states, i.e. off-diagonal density
matrix elements in the diagonal basis of $Z_j$ in the super operator space decay exponentially.}, as discussed in \cite{medvedyeva2016exact}, which, in turn, corresponds to the Hubbard model with an imaginary coupling \cite{essler2005one}. We provide further clarification on this point in chapter \ref{Hubbardchapter} section \ref{explainconnectionB2hub}. When $u\neq 0$, the model can be understood as a twisted version of the inhomogeneous Hubbard model \cite{ziolkowska2020yang,murakami1998new}. The corresponding $R$-matrix can be found in \cite{murakami1998new}, for this reason we do not provide it here.

\subsubsection{Diagonal preserving model}
For generic values of $u$, $\phi$, and $\beta(u)$, the model is not diagonal preserving. However, when $\beta(u)$ is set to $2\sinh(u)$, the model becomes diagonal preserving, and its generator corresponds to the SSEP  model. It is worth noting that model B1, described in section \ref{modelB1explained}, is also diagonal preserving with the SSEP as its generator. This is because model B2 includes model B1 as a special case. Further discussion on this topic can be found in section \ref{sec:large}.
\subsubsection{Particle current and NESS}
The particle current $J_{k}$ can be determined using the continuity relation \eqref{particlecurrentgeneral}, and its expression is given by
\begin{equation}
  J_k=(1-2\beta^2(u) \sh(u)) J_k^{0}
  +\beta^2(u)\sh^2(u)(n_k-n_{k+1}).
\end{equation}
Here
\begin{equation}
  \label{cohJ}
  J_k^0= \frac{i}{2} (e^{i\phi}\sigma_k^+\sigma_{k+1}^--e^{-i\phi}\sigma_k^-\sigma_{k+1}^+) ,
\end{equation}
is the current of the coherent time
evolution dictated by \eqref{HXX}, and the remaining terms describe stochastic transport. \\By analyzing the spectrum of the superoperator $\mathcal{L}$ and in particular the zero eigenvalues, we found that for a given length $L$, there are $L+1$ NESS in the system. One of these NESS is the identity operator, and due to the conservation of particle number in this model, we can obtain the remaining one by projecting into each subsector with a specific particle number. Each subsector possesses a unique NESS.

\chapter{The Hubbard model and its deformation}

\ifpdf
    \graphicspath{{Chapter4/Figs/Raster/}{Chapter4/Figs/PDF/}{Chapter4/Figs/}}
\else
    \graphicspath{{Chapter4/Figs/Vector/}{Chapter4/Figs/}}
\fi

\label{Hubbardchapter}

In this chapter, we introduce the Hubbard model, a toy model used to describe the motion of electrons in the conduction band of a solid. We begin by presenting its definitions in both fermionic and bosonic formulations. Additionally, we explore the existing connection between the Hamiltonian of the Hubbard model and the Lindblad superoperator, which describes the evolution of open quantum systems. Building upon this connection, we present all the cases in which this mapping is possible. Furthermore, we introduce a new nearest-neighbour integrable model characterized by an $R$-matrix that contains an unusual functional dependence on the Jacobi functions (we refer to it as \textit{not quite elliptic}) and we show how to use the \textit{bond-site transformation} to relate it to a range 3 integrable deformation of the Hubbard model. We believe that this is the first range 3 integrable deformation of the Hubbard model. This deformation manifestly breaks the $\alg{u}(1)$ symmetry of the Hubbard model. To validate the significance of our findings, we prove the integrability of the 3 site model and highlight the \textit{unusual} functional dependence of the $R$-matrix.

\section{Introduction}
The Hubbard model \cite{essler2005one} describes the physics of interacting electrons in the lattice. It was introduced by John Hubbard in the sixties in a series of papers "Electron correlations in narrow energy bands", \cite{hubbard1,hubbard2,hubbard3,hubbard4,hubbard5,hubbard6} and has since become an important model in physics because it can be solved exactly. In 1968, Lieb and Wu solved it by using  the coordinate Bethe ansatz \cite{lieb-wu-hubbard}. In one dimension, the Hubbard model is Yang-Baxter integrable, but finding the associated  R-matrix was challenging. It was realized by Kulish and Reshetikhin \cite{Kulish:1981bi} that the R-matrix could not be constructed as a fundamental model of difference form type.  In 1986, B. S. Shastry \cite{Shastry_1986} found the \textit{non difference form} $R$-matrix associated to the spin-model correspondent to the Hubbard model\footnote{The Hubbard model was discovered as a fermionic model; however, it can be mapped to a spin model via a Jordan-Wigner transformation.}.
The proof that this R-matrix actually solves the Yang-Baxter equation was later established by Shiroishi and Wadati \cite{shiroishi1995yang} and it is based on the work of Korepanov \cite{korepanov1989tetrahedral}.
The $R$-matrix was subsequently used to built the algebraic Bethe ansatz \cite{martins1998quantum,ramos1997algebraic}. 
\\
The integrability property makes the Hubbard model a playground for different field of research.
\\In \textbf{condensed matter}, for instance, solving the Hubbard model has provided a theoretical laboratory for studying non-perturbative effects in strongly correlated electron systems. This is important because describing the microscopic behaviour of solids is extremely challenging. The Hubbard model has been instrumental in investigating phenomena such as ferromagnetism, different forms of antiferromagnetism, unconventional superconductivity, charge-density waves, electronic liquid crystalline phases, and topologically ordered phases (such as "spin liquids")  \cite{arovas2022hubbard}. By studying the Hubbard model in specific instances, researchers have been able to explore properties that were previously not accessible. Transport properties of the model have been a subject of interest for many decades (see for example \cite{Fujimoto-drude}). Recently, the
so-called integrable quantum quenches have also been considered in the 1D Hubbard model
\cite{hubbard-integrable-quench}, using information about exact overlaps
\cite{gombor2020boundary1,gombor2021boundary}. 
\\
The Hubbard model finds also application in the context of \textbf{AdS/CFT} \cite{Rej_2006}. In fact, the $S$-matrix relevant for the AdS/CFT correspondence \cite{Beisert_2007,Martins:2007hb} is related to the Shastry's $R$-matrix. This remarkable relation sheds some new light on the symmetry algebra of the Hubbard model. It was known for a long time that the Hubbard model exhibits $\SU(2)\oplus \SU(2)$ symmetry \cite{heilmann1971violation,Essler:1991wg}. By using the map to string theory these could be seen as coming from a centrally extended superalgebra from which the Hubbard model can be obtained in a certain limit \cite{deLeeuw:2015ula}. Moreover, this observation recently leads to the formulation of the \textit{quantum spectral curve} for the Hubbard model \cite{Ekhammar:2021pys}.
\\
Over the years, many \textit{integrable generalizations} of the Hubbard model have appeared, the models of Bariev and Alcaraz
\cite{bariev-alcaraz--hubbard-tj,Beisert:2008tw}, the Essler-Korepin-Schoutens model 
\cite{essler-korepin-schoutens},  and multi-component generalizations
\cite{maassarani-models,maassarani-hubbard,multileg-hubbard}. 
\\
In recent times, the Hubbard model has found applications in the field of \textit{open quantum systems}. This model can be formulated as two interacting XX spin chains connected by a dephasing term. In 2016, taking this into account, Medvedyeva, Essler and Prosen \cite{medvedyeva2016exact} mapped the Hubbard model with complex interaction strength to a Lindbladian: a superoperator generating the dynamics of an open quantum system.

\section{Generalities}
\subsection{Fermionic and bosonic formulations}

The Hamiltonian of the Hubbard model \cite{essler2005one} in the fermionic formulation is
\begin{equation}
  \label{HubbardH}
  \mathbb{H}=\sum_j  \left[  
    (c^{\uparrow}_j)^\dagger c^{\uparrow}_{j+1}+(c^{\uparrow}_{j+1})^\dagger c^{\uparrow}_{j}+
      (c^{\downarrow}_j)^\dagger c^{\downarrow}_{j+1}+(c^{\downarrow}_{j+1})^\dagger c^{\downarrow}_{j}+
    U n^\uparrow_j n^\downarrow_j\right],
\end{equation}
where $U\in\valos$ is the coupling constant of the model. $(c^{\uparrow,\downarrow}_j)^\dagger$, $c^{\uparrow,\downarrow}_j$ are the standard creation and annihilation operators which satisfy the canonical anti-commutation relations
\begin{align}\label{commrelosc}
& \{c^\alpha_j,c^\beta_k\} =   \{(c^\alpha_j)^\dagger,(c^\beta_k)^\dagger\}=0 ,\\
&\{c^\alpha_j,(c^{\beta}_k)^\dagger\} = \delta^{\alpha,\beta}\delta_{j,k} ,
\end{align}

where $j,k$ refer to the local Hilbert spaces , $\alpha,\beta=\uparrow,\downarrow $ and $n$ is the local particle number operator $n_j^\alpha=c_j^{\alpha \dagger}c_j^{\alpha}$.
\\
The local Hilbert space is spanned by the four vectors
\begin{align}
 & \ket{\emptyset}=e_1,
 && \ket{\uparrow}=(c^\uparrow)^\dagger \ket{\emptyset}=e_2,
 && \ket{\downarrow}=(c^\downarrow)^\dagger \ket{\emptyset}=e_3,
&&  \ket{\updownarrow}=(c^\downarrow)^\dagger (c^\uparrow)^\dagger \ket{\emptyset}=e_4,
\label{basis}
\end{align}
with $e_i$ the standard basis.
\\
The Hamiltonian $\mathbb{H}$ commutes with
\begin{align}
  \label{particles}
&    N=\sum_j (n_j^\uparrow+n_j^\downarrow),
   && S_z=\sum_j (n_j^\uparrow-n_j^\downarrow),
\end{align}
where $N$ is the total particle number and $S_z$ the magnetization. We can add two magnetic fields so that the interaction term becomes particle/hole symmetric. This choice preserves the integrability of the model and its explicit form is
\begin{equation}
\mathbb{H}'=\sum_j  \left[  
  (c^{\uparrow}_j)^\dagger c^{\uparrow}_{j+1}+(c^{\uparrow}_{j+1})^\dagger c^{\uparrow}_{j}+
      (c^{\downarrow}_j)^\dagger c^{\downarrow}_{j+1}+(c^{\downarrow}_{j+1})^\dagger c^{\downarrow}_{j}+
    \frac{U}{4} (1-2n^\uparrow_j) (1-2n^\downarrow_j)\right].
\label{HubbardHamiltonianfermion}
\end{equation}
This Hamiltonian enjoys $\SU(2)\oplus \SU(2)$ symmetry.
\\
Since the main result in this chapter is a new integrable deformation of the Hubbard model, it is convenient to give the definition in the \textbf{spin chain context}, working with the ``bosonic'' version of the model. 
\\
The local Hilbert space is  the tensor product
\begin{equation}
  V_j=\complex^2\otimes \complex^2
  \label{hilbertspace}
\end{equation}
with the full Hilbert space being the tensor product 
\begin{equation}
 V=\otimes_{j=1}^L V_j,
\end{equation}
with $L$ the length of the spin chain. With the same notation of \eqref{notation}, the (inverse)-Jordan Wigner transformation is
\begin{align}\label{JWtransformation}
&  \sigma^-_j=\left[\prod_{k=1}^{j-1}(-1)^{n_k^\uparrow}\right]  c^\uparrow_j, &&
  \tau^-_j=\left[\prod_{k=1}^{j-1}(-1)^{n_k^\downarrow}\right]  c^\downarrow_j, \\
&  \sigma^+_j=  (c^{\uparrow}_j)^\dagger \left[\prod_{k=1}^{j-1}(-1)^{n_k^\uparrow}\right],
&&
\tau^+_j=  (c^{\downarrow}_j)^\dagger \left[\prod_{k=1}^{j-1}(-1)^{n_k^\downarrow}\right],\\
&\sigma^z_j=1-2n_j^\uparrow,&& \tau^z_j=1-2n_j^\downarrow.
\end{align}
In this way, the Hubbard model Hamiltonian \eqref{HubbardHamiltonianfermion} takes the expression\footnote{We remark that we consider the (inverse)-Jordan Wigner transformation for the terms acting on the bulk $(j,j+1)$. Only afterward, we consider periodic boundary conditions on the spin chain. Applying the (inverse)-Jordan Wigner transformation to the boundary term will make unwanted string appear.}
\begin{equation}
  \label{Hv}
  \mathbb{H}''=\sum_j \left[ \sigma^+_j\sigma^-_{j+1}+ \sigma^-_j\sigma^+_{j+1}+ \tau^+_j\tau^-_{j+1}+
    \tau^-_j\tau^+_{j+1}+\frac{U}{4}\sigma^z_j\tau^z_j
    \right],
\end{equation}
where $U$ is still the coupling constant of the model. At $U=0$, the model describes two  XX spin chains which do not interact with each other. For $U\neq 0$, the two chains are coupled by a dephasing term\footnote{The interaction is given by the diagonal matrix $Z_i$. "dephasing" is related to the fact that this term changes the relative phase between different spin states.}.
\\
If we perform a twist transformation\footnote{ We use $\mathbb{H}_1= B_1 \mathbb{H}'' B_1^{-1}$ with the diagonal twist $B$ given by
\begin{align}\label{twistB}
&B=\left(
\begin{array}{cccc}
 -1 & 0 & 0 & 0 \\
 0 & i & 0 & 0 \\
 0 & 0 & i & 0 \\
 0 & 0 & 0 & 1 \\
\end{array}
\right).
\end{align}
This is allowed since the condition \eqref{twistcond}, $[B_1 B_2,H'']=0$ is satisfied.},
the Hamiltonian is transformed into
\begin{equation}
  \label{H1}
  \mathbb{H}_1= \sum_j \left[     h^{(1)}_{j,j+1}+   h^{(2)}_{j,j+1}+  \frac{U}{4}\sigma^z_j\tau^z_j
  \right],
  \end{equation}
  where
  \begin{align}
    \label{Hsigma}
& h^{(1)}_{j,j+1}\equiv i \big(\sigma^+_j\sigma^-_{j+1}-\sigma^-_j\sigma^+_{j+1}\big),
&& h^{(2)}_{j,j+1}\equiv i \big(\tau^+_j\tau^-_{j+1}-\tau^-_j\tau^+_{j+1}\big).
\end{align}
We use the notation $^{(1)}$ and $^{(2)}$ to identify operators only containing $\sigma$ or only $\tau$, respectively, as given by \eqref{notation}. The reason of doing the twist will be explained in section \ref{mappinglindbladhermitian}. In fact, without the twist, the condition \eqref{conditiontomap} is not met and we cannot interpret the Hubbard model as an open quantum system.\\
Here and in the following, we use the notation $\mathbb{H}_k$ with $k=1, 2, 3$ to identify the number $k$ of sites of the spin chain spanned by the interaction term. 
The kinetic term above is the same as \eqref{HXX} and it corresponds to a XX chain perturbed by a Dzyaloshinskii-Moriya interaction term \cite{Dzyaloshinsky-Moriya-xxz},
\begin{equation}
\label{HsigmaDM}
h_{j,j+1}=\frac{1}{2}\left[   \sigma^x_{j} \sigma^y_{j+1}-  \sigma^y_{j} \sigma^x_{j+1}\right].
\end{equation}

\subsection{Symmetries}

Now we discuss the symmetries of the Hubbard model in more detail, focusing on the Hamiltonian \eqref{H1}. The Hubbard
model has both discrete and continuous symmetries.
\paragraph{Discrete} A Shiba transformation is defined on a chain of even length $L$ by
\begin{equation}
  \label{Shiba}
  \begin{split}
    \Sc^{(1)}&= \sigma^y_L\sigma^x_{L-1}    \dots  \sigma^y_2\sigma^x_1,\\
      \Sc^{(2)}&= \tau^y_L\tau^x_{L-1}    \dots  \tau^y_2\tau^x_1.\\
  \end{split}
 \end{equation}
This transformation acts on the total number of particles operator $N$ and on the total magnetization $S_z$ as
\begin{align}
&\Sc^{(2)}\Sc^{(1)}N \Sc^{(1)}\Sc^{(2)}=2L-N,
&&\Sc^{(2)}\Sc^{(1)}S_z \Sc^{(1)}\Sc^{(2)}=S_z.
\end{align}

The combination of two Shiba transformations (see \cite{essler2005one}) is a discrete symmetry.
 A similarity transformation with either $\Sc^{(1)}$ or $\Sc^{(2)}$ preserves the kinetic term of $\mathbb{H}_1$, while changing the sign of the interaction term. Explicitly,
\begin{equation}
  \label{shib1}
  S^{(1)} \mathbb{H}_1 S^{(1)}=S^{(2)} \mathbb{H}_1 S^{(2)}= \sum_j \left[     h^{(1)}_{j,j+1}+ h^{(2)}_{j,j+1}-  \frac{U}{4}\sigma^z_j\tau^z_j\right],
\end{equation}
and
\begin{equation}
  \Sc^{(2)}\Sc^{(1)} \mathbb{H}_1 \Sc^{(1)}\Sc^{(2)} = \mathbb{H}_1.
\end{equation}
\paragraph{Continuous}
The Hubbard model Hamiltonian  enjoys invariance under the continuous group $\SU(2)\oplus \SU(2)$, \cite{heilmann1971violation,Essler:1991wg}. The first $\SU(2)$ corresponds to rotations in spin space, which can be interpreted also as a mixing of the $\sigma$ and
$\tau$ operators. The generators are local in space if we express them using the original fermionic variables. However,
when we work with the spin variables, the Jordan-Wigner strings appear. Formally we have
\begin{align}
  \label{ASU2}
A_z=\sum_j  \frac{\sigma^z_j-\tau^z_j}{2}
\end{align}
and
\begin{align}
  &A_+=\sum_j \bigg[\bigg(\prod_{k<j} \sigma^z_k\tau^z_k\bigg) \sigma^+_j\tau^-_j\bigg],
  &&  A_-=\sum_j \bigg[\bigg(\prod_{k<j} \sigma^z_k\tau^z_k\bigg) \sigma^-_j\tau^+_j\bigg],
  \label{ASU2b}
\end{align}
that satisfy the standard $\SU(2)$ algebra
\begin{equation}
[A_+,A_-]= A_z,\qquad [A_z,A_\pm]=\pm 2 A_\pm .
\end{equation}
For both periodic and open boundary conditions, the following condition holds
\begin{equation}
  [A_z,\mathbb{H}_1]=0,
\end{equation}
while for the off-diagonal generators the symmetry relations
\begin{equation}
  [A_\pm,\mathbb{H}_1]=0
\end{equation}
hold only in the case of open boundary conditions, or formally in the infinite chain limit.
\\
The second $\SU(2)$ follows from the Shiba transformation. The idea is to perform a similarity transformation with either
$\Sc^{(1)}$ or $\Sc^{(2)}$, construct the $\SU(2)$ generators of the modified Hamiltonian, and then transform them back to
the original $\mathbb{H}_1$. In this way, we obtain the $\SU(2)$ generators (also called $\eta$-pairing generators)
\begin{align}
  \label{BSU2a}
B_z=\sum_j  \frac{\sigma^z_j+\tau^z_j}{2},
\end{align}
and
\begin{align}
  \label{BSU2b}
 & B_+=\sum_j \bigg[\bigg(\prod_{k<j} \sigma^z_k\tau^z_k\bigg) \sigma^+_j\tau^+_j\bigg],
&&  B_-=\sum_j \bigg[\bigg(\prod_{k<j} \sigma^z_k\tau^z_k\bigg)\sigma^-_j\tau^-_j\bigg].
\end{align}
Similarly to the $A$s, the  $B_z$ operator commutes with $\mathbb{H}_1$ \eqref{H1} for both periodic and open boundary conditions and $B_\pm$ only in the open boundary
case.  All $A$ operators commute with all $B$ operators, therefore the
symmetry algebra is indeed $\SU(2)\oplus \SU(2)$. In chapter \ref{threeorfour}, we explicitly show how to construct the most general Hamiltonian that exhibits this symmetry.
\\
We would like to mention that the fact that $A_\pm$ and $B_\pm$ only commute with $\mathbb{H}_1$ in the open boundary case comes from the non-periodicity of the $A_\pm$ and $B_\pm$ operators defined in \eqref{ASU2b} and \eqref{BSU2b}.

\section{Relation between Lindblad superoperator and Hermitian Hamiltonians}
\label{mappinglindbladhermitian}

In \cite{medvedyeva2016exact}, the authors mapped  the Hubbard model to a Lindblad superoperator in presence of dephasing noise. We present their findings here and further explore the conditions under which this correspondence holds. Considering as starting point the Hubbard Hamiltonian \eqref{H1}, we can then identify
\begin{align}\label{hlhubb}
&  h_{j,j+1}=i \left(\sigma_j^+\sigma_{j+1}^--\sigma_j^-\sigma_{j+1}^+\right),
&&\ell_{j,j+1}=\, Z_j.
\end{align}
The operators in \eqref{hlhubb} satisfy
\begin{align}\label{conditiontomap}
&h = -h^T,
&&\ell=\ell^*=\ell^T,
&&\ell^2 = \ell^\dagger \ell= \id.
\end{align}
These properties allow to rewrite the expression of the Hamiltonian \eqref{H1}

\begin{equation}\label{haftermapping}
 \mathbb{H}\,=\,\,\sum_j      h^{(1)}_{j,j+1}-   {h^{(2)T}_{j,j+1}}  
   + \frac{U}{4} \ell^{(1)}_{j,j+1}  \ell^{(2)}_{j,j+1}.
  \end{equation}
We can add a term proportional to the identity\footnote{This term only shift all the energy levels by a constant and do not alter the integrability property.}, complexify the coupling constant $U\to i\,U$ and then multiply $H$ by $i$, in this way we obtain
\begin{align}
  \Li_{j,j+1}=&~-i h^{(1)}_{j,j+1}+i  h^{(2)T}_{j,j+1} +  \frac{U}{4} \Big(\ell^{(1)}_{j,j+1}  \ell^{(2)}_{j,j+1}  - \id\Big) ,
\end{align}
which is equivalent to the expression of the superoperator \eqref{superoperatorL2} by using \eqref{conditiontomap} and where the dependence on the coupling constant $U$ is explicit. This proves how to map the Hubbard Hermitian Hamiltonian to a Lindblad superoperator.
\\
More generally, we emphasize that if the \textbf{conditions} \eqref{conditiontomap} are met, 
a Lindblad superoperator can be always mapped to an Hermitian operator. It is easy to see that under the conditions \eqref{conditiontomap}, the map also works in the other directions: allowing to write the Hermitian version of the Lindblad superoperator.

\subsection{Model B2}\label{explainconnectionB2hub}
Now we clarify why  model B2 of section \ref{modelB2explained} is related to the Hubbard model. The $h$ of model B2 coincides\footnote{In section \ref{modelB2explained} we stated that the point $\phi=0$ coincides with two XX spin chains and dephasing interaction. We now consider the point $\phi=\pi/2$ to have the open quantum system interpretation, in particular to have the conditions \eqref{conditiontomap} satisfied. It is clear that the two models are related by a twist.} with \eqref{hlhubb} for $\phi=\pi/2$ and the $\ell$ can be written as (up to renormalization)
\begin{align}
&\ell_{12}=Z_1+\eta (X_1 X_2+Y_1 Y_2)+\eta^2 Z_2,
\end{align}
where $\eta=-\tanh \left(\frac{u}{2}\right)$. $\eta$ is the deformation parameter and for $\eta = 0$ (or equivalently $u=0$) it reconducts to the Hubbard model \eqref{hlhubb}.  This deformation was found  in \cite{murakami1998new} and we presented it in the letter \cite{classificationlind} in the context of open quantum systems.

\section{Toward a new integrable deformation of the Hubbard model}
In this section, we outline the procedure that leads us to discover a \textit{novel} integrable deformation of the Hubbard model, characterized by interaction terms spanning three consecutive sites of the spin chain.
\\
We identify this deformation within the framework of open quantum systems, starting from the Lindbladian $\cL$  \eqref{superoperatorL2} as the initial ansatz for the Boost automorphism mechanism. We started from an ansatz for both $h$ and $\ell$ of 8-Vertex type\footnote{We give more details on this type of models in section \ref{whatarevertexmodel}. In these $4\times 4$ models the only non-zero entries may be in positions: 11, 14, 22, 23, 32, 33, 41, 44 of the matrix.}. The complete classification of integrable models using this ansatz is still a work in progress\footnote{This ansatz is more intricate than the ones employing a 6-vertex model for both $h$ and $\ell$ as done in chapter \ref{Lindbladclassificationmodels}. The complexity arises from the two additional entries in the $\ell$ operator (quadratic in the Lindblad superoperator), which results in highly intricate equations for the integrability constraints $[\mathbb{Q}_2,\mathbb{Q}_3]=0$.}. Nonetheless, this model already holds considerable interest on its own. Initially, its connection to the Hubbard model may not be apparent. However, we demonstrate how to perform a \textit{bond-site transformation} that preserves integrability and transforms it into a next-to-nearest neighbor model, representing a new integrable deformation of the Hubbard model. To the best of our knowledge, this is the first range 3 integrable deformation of Hubbard.

\subsection{$h$ and $\ell$ of the two site model}

Although we discovered this model in the context of open quantum systems, we noticed that the $h$ and $\ell$ operators meet the conditions described in equation \eqref{conditiontomap} of section \ref{mappinglindbladhermitian}. As a result, we chose to present its Hermitian version to make the connection with the Hubbard model more transparent. We refer to this model as two site model indicating that the interaction term spans two sites of the chain.
\\
The integrable Hamiltonian that we found is
\begin{equation}
  \label{H2}
  \mathbb{H}_2=\sum_j \big[h^{(1)}_{j,j+1}+h^{(2)}_{j,j+1}+ l^{(1)}_{j,j+1}l^{(2)}_{j,j+1}\big],
\end{equation}
where  
\begin{align}\label{handlnorepar}
&h=\frac{dn}{k}(\sigma^+_j \sigma^-_{j+1}-\sigma^-_j \sigma^+_{j+1})
&&l=\left(
\begin{array}{cccc}
 i \, sn & 0 & 0 & 1 \\
 0 & 0 & -cn & 0 \\
 0 & -cn & 0 & 0 \\
 1 & 0 & 0 & -i\,sn \\
\end{array}
\right),
\end{align}
$sn, \,cn$ and $dn$ are the Jacobi functions $\text{sn}\left(\frac{2 i u}{k}|k^2\right),\,\text{cn}\left(\frac{2 i u}{k}|k^2\right)$ and $\text{dn}\left(\frac{2 i u}{k}|k^2\right)$. The models depend on two parameters $k$ and $u$. $\mathbb{H}_2$ is integrable if $k$ and $u$ are chosen such that the entries of $l$ are real.
\\
To clarify the meaning of $k$ and $u$, we use a different normalization and reparametrization that bring the kinetic term to \eqref{Hsigma} (independent from the parameters) and the $l$ term to $\ell$:
\begin{align}  \label{Lsigma2}
 \ell_{j,j+1} &=\frac{\sin\theta}{2}(Z_j+Z_{j+1})+\cos\theta (\sigma^{-}_{j}  \sigma^-_{j+1}+\sigma^+_{j} \sigma^+_{j+1})+ (\sigma^{-}_{j}\sigma^+_{j+1}+\sigma^+_{j} \sigma^-_{j+1})\nonumber \\
&=\begin{pmatrix}
      \sin\theta & 0& 0 & \cos\theta \\
    0 & 0& 1 & 0 \\
    0 & 1& 0 & 0 \\
      \cos\theta & 0& 0 & -\sin\theta\\
  \end{pmatrix}.
\end{align}

This new reparametrization is related to the old one by
\begin{align}\label{reparametrization22}
&u = -\frac{i k \sec ^2\theta  \text{cn}^{-1}\left(-\sec \theta \left|k^2\right.\right)}{4 U} ,
&&k = \frac{2 i U \cos ^2\theta }{\sqrt{U^2 \sin ^2(2 \theta )+1}}.
\end{align}
The advantage is that now the dependence on the coupling constant $U$ is factorized, and we obtain 
\begin{equation}\label{Hnewrep}
  \mathbb{H}_2=\sum_j \big[h^{(1)}_{j,j+1}+h^{(2)}_{j,j+1}+2 U \ell^{(1)}_{j,j+1}\ell^{(2)}_{j,j+1}\big].
\end{equation}

$\mathbb{H}_2$ depends on both $U$ (coupling constant) and $\theta$, a parameter. We remark that this model does not contain the Hubbard model for any value of $U$ and $\theta$. However, in section \ref{from2to3sitemodelsection} we will show how to use a bond site transformation to bring it to a range 3 model which contains the Hubbard model. The parameter $\theta$ is related to the deformation of the Hubbard model. In order for $\mathbb{H}_2$ to be Hermitian, both parameters need to be real. We restrict the range of $\theta$ to the fundamental domain of $[-\pi, \pi]$.
\\
The $\ell$ operator is of 8-vertex type\footnote{More details on this type of models are given in section \ref{whatarevertexmodel}.} \cite{baxter2016exactly}: particle number is not conserved, but particle creation and
annihilation only happen in pairs (due to the terms $\sigma^{+}_{j}  \sigma^+_{j+1}$ and $\sigma^{-}_{j}  \sigma^-_{j+1}$). $\ell$ obeys the properties
\begin{align}
&  \left(\ell_{j,j+1}\right)^2\propto 1, \label{propertyL2}
\\
&   [\ell_{j,j+1},\ell_{j+1,j+2}]\ne 0  \label{noncomm2}
\end{align}
and it is free fermionic. This property can be understood from the representation \eqref{Lsigma2}:
performing a Jordan-Wigner transformation we  find terms which are only bilinear in the fermionic operators. This point will be clarified in section \ref{fermionicformulationellipticmodel}.\\The spin-1/2 model obtained by considering only the Hamiltonian $\sum_j \ell_{j,j+1}$ are known in the literature as the
  XYh models \cite{XYh-Fan-Wu}. They correspond to a XY spin chain coupled to a magnetic field $Z_i$.  
However, our model involves the coupling  $\ell^{(1)}_{j,j+1}\ell^{(2)}_{j,j+1}$, therefore it is interacting.

\subsection{Symmetries}
\label{symmetriesrange2}
Let us now discuss the symmetries of \eqref{Hnewrep} for a generic value of $\theta$. First of all, we do \textit{not} find any \textit{continuous
symmetries}.
However, there are \textit{discrete symmetries}. In particular, similar to the Hubbard model, the Shiba transformations \eqref{Shiba} preserve the kinetic terms and both of them negate the sign of the coupling constant $U$.
Therefore their combination is a discrete symmetry:
\begin{equation}
  \Sc^{(2)}\Sc^{(1)} \mathbb{H}_2 \Sc^{(1)}\Sc^{(2)}=\mathbb{H}_2.
\end{equation}
Because both interaction matrices ($\ell^{(1)}$ and $\ell^{(2)}$) create/annihilate particles in pairs, the ``fermionic parity'' is conserved for both
sub-chains
\begin{equation}
  \label{Zcomm}
  [Z_\sigma,\mathbb{H}_2]=[Z_\tau,\mathbb{H}_2]=0
\end{equation}
where
\begin{align}\label{Zsigmatau}
&Z_\sigma=\prod_{j=1}^L \sigma^z_j,&&  Z_\tau=\prod_{j=1}^L \tau^z_j.
\end{align}
This property also holds for the three site spin chain given in \eqref{H3}.

\subsection{Why is this model integrable?}
\label{whyisthemodelint}
To establish the integrability of this model, we derived the expression of the $R$-matrix associated with \eqref{handlnorepar}, which is presented in Appendix \ref{Rmatrixsite2Hubbard}. The $R$-matrix satisfies the Yang-Baxter equation\footnote{If the reader is interested in showing this, we would like to remark that by using version 12.3 of Mathematica the check is
straightforward. However, using version 12.0 particular attention should be paid to the choice of the sign of the branch-cut.} and its properties are of particular interest. Firstly, it is of\textbf{ non-difference} form, evident from its dependence on both the sums and differences of Jacobi elliptic functions. Secondly,  it has a very \textbf{non-trivial functional dependence} on the spectral parameters, containing terms that cannot be completely expressed in terms of the usual Jacobi elliptic functions, as they involve expressions such as
\begin{align}
&\sin \frac{1}{2}\Big[\text{am}(u|k^2)-\text{am}(v|k^2)\Big],
&&\sec \frac{1}{2}\Big[\text{am}(u|k^2)-\text{am}(v|k^2)\Big],
\label{noelliptic}
\end{align}
where $\text{am}$ is the Jacobi amplitude. The product of this two functions can be expressed in term of Jacobi functions, in fact
\begin{align}
\tan \frac{1}{2}\Big[\text{am}(u|k^2)-\text{am}(v|k^2)\Big]=&\csc \Big[\text{am}(u|k^2)-\text{am}(v|k^2)\Big]-\cot \Big[\text{am}(u|k^2)-\text{am}(v|k^2)\Big]=\nonumber\\
&\frac{\text{cn}\left(u|k^2\right) \text{cn}\left(v|k^2\right)+\text{sn}\left(u|k^2\right) \text{sn}\left(v|k^2\right)-1}{\text{cn}\left(u|k^2\right) \text{sn}\left(v|k^2\right)-\text{cn}\left(v|k^2\right) \text{sn}\left(u|k^2\right)},
\end{align}
but still one of the two functions in \eqref{noelliptic} escapes the elliptic property. In particular, \textit{it can only be expressed in the Jacobi elliptic functions} $\mathrm{sn,cn,dn}$ \textit{by introducing square roots}.
\\
Explicitly
\begin{align}
\sin \frac{1}{2}\Big[\text{am}(u|k^2)-\text{am}(v|k^2)\Big]=&s_1 \sqrt{\frac{1-\text{cn}\left(u|k^2\right) }{2}}\sqrt{\frac{1+\text{cn}\left(v|k^2\right) }{2}}+\\
&s_2 \sqrt{\frac{1-\text{cn}\left(v|k^2\right) }{2}}\sqrt{\frac{1+\text{cn}\left(u|k^2\right) }{2}},
\end{align}
with $s_1$ and $s_2$ signs that depend on the choice of the branch cut.
\\
To the best of our knowledge, \textbf{this $R$-matrix is new} and we have also not encountered before a model with functional dependence on the square root of Jacobi function.
\\
Let us clarify some points about the integrability of the model \eqref{Hnewrep}. If we take the expression \eqref{Hnewrep} and write $U=U(u)$ and $\theta=\theta(u)$ and construct $\mathbb{Q}_3$ via the boost operator \eqref{boostBQ3finileL}, the condition $[\mathbb{Q}_2,\mathbb{Q}_3]=0 $ gives a set of differential equations for $U$ and $\theta$ that are satisfied by \eqref{reparametrization22}. However, it is important to note that from \eqref{Hnewrep}, with $U$ and $\theta$ kept constants, we cannot obtain the $R$-matrix using the Sutherland equation (as demonstrated in section \ref{example}). This raises a broader question, discussed in \cite{deLeeuw:2022ath,Gombor:2022lco}: how can we determine if a constant Hamiltonian belongs to an integrable model? In fact, suppose we take an Hamiltonian $H(u)$ dependent on the spectral parameter $u$ and we fix $u=\text{constant}$. We construct $\mathbb{Q}_3$ via the boost method. But if we compute $[\mathbb{Q}_2,\mathbb{Q}_3]$ we won't obtain zero since we are missing the derivative term in \eqref{boostBQ3finileL}. In such cases, to verify the integrability of the model, we need to investigate whether it exists a range 2 charge that plays the role of the derivative term in $\mathbb{Q}_3$, resulting in a vanishing commutator $[\mathbb{Q}_2,\mathbb{Q}_3]$. If such a charge exists,  the model \textit{may be} integrable and to prove it, we need to construct the Lax operator.
\\
To prove integrability of $\mathbb{H}_2$ \eqref{Hnewrep} we also need to use the Lax operator, a more fundamental object than the $R$-matrix. The Lax operator is used to construct the conserved charges, as explained in section  \ref{definitionsmonodromy}. To generate the correct Hamiltonian, the Lax operator should be related to the $R$-matrix in the following way\footnote{We would like to remark that the regularity condition for the $R$-matrix in this new parametrization corresponds to consider $u=\mu/\alpha(\mu)$, while for the standard parametrization used in the Appendix \ref{Rmatrixsite2Hubbard} to $u=v$. Same choices of $u$ should be used to reproduce the correct Hamiltonian.} 
\begin{align}
\label{LaxandR}
\La(u,\mu) \equiv R(\alpha\, u,\mu),
\end{align}
where
\begin{align}
\label{relationkUtheta}
&\alpha =\frac{2}{\text{dn}\left(\mu\left|k^2\right.\right)},
&& \mu =\text{cn}^{-1}\left(-\sec \theta \left|k^2\right.\right),
&& k =\frac{2 i U \cos ^2\theta}{\sqrt{1+ U^2 \sin ^2(2 \theta )}}.
\end{align}
This is analogous to the reparametrization \eqref{reparametrization22}.

\section{The bond-site transformation}

Here we use the bond-site transformation to bring the nearest-neighbour model \eqref{Hnewrep} into a form where the Hamiltonian is three site interacting. We then show how this model is a deformation of the Hubbard model.
\\
The bond-site transformation has its origin in the Kramers-Wannier duality \cite{kramers-wannier}, which can be used to determine the critical
point of the Ising model on the square lattice. It can also be applied to the 1D quantum Ising chain, where it acts as a
self-duality \cite{kramers-wannier}.  More recently the same transformation was also used in the ``folded XXZ model''
\cite{folded1,sajat-folded}.

\subsection{Generalities}

\label{generalitiesbst}
To understand the transformation, it is enough to first consider just one copy of the local space $\complex^2$, on which our previous $\sigma^a$
 act (in light of \eqref{notation}). The same argument can be repeated for the $\tau^a$ acting on the second copy of $\mathbb{C}^2$.\\
There are two ways to introduce the bond-site transformation: either formally on the level of the \textit{operators} acting on the Hilbert space or via a \textit{real space} description of the states. We treat both formulations. We first consider models with open boundary conditions.
\\
 On the level of operators, the duality transformation is a particular \textit{Clifford transformation} \cite{clifford1}, a mapping
between operators with the following two requirements:
\begin{itemize}
\item Products of Pauli matrices are mapped to products of Pauli matrices (including possible multiplication with  phases, but without producing linear combinations).
\item The operator algebra is preserved.  
\end{itemize}
The transformation is then defined by the mapping
\begin{align}
  \label{ta1}
&    Z_j \to  X_{j-\frac{1}{2}}  X_{j+\frac{1}{2}}, 
&&    X_j\to \prod_{k=1}^{j}Z_{k-\frac{1}{2}} 
\end{align}
and we can use the operator algebra of the Pauli matrices to extend this mapping to all operators. For example, a product of two
$X$ operators in nearest-neighbour sites is mapped to a single $Z$ matrix, in this way we obtain a symmetric formulation for the
elementary steps\footnote{As stated in \cite{clifford1}, this bond-site transformation can be recognised as the Kramers-Wannier duality of the Ising model.}
\begin{align}
  \label{rules}
&   Z_j \to  X_{j-\frac{1}{2}}  X_{j+\frac{1}{2}},&&   X_j X_{j+1}\to Z_{j+\frac{1}{2}}.
\end{align}

The half shift that we introduced can be understood in the real space interpretation of this transformation: working in the computational basis, we perform a rotation and afterwards we put spin-$1/2$
variables on the bonds\footnote{For example, the bond between the sites $j$ and $j+1$ is $j+1/2$.} between the 
original sites, such that the new variables measure the presence or the absence of a domain wall (kink or
anti-kink). This is why we call these steps a ``bond-site transformation''.
\\
To be more precise, let us assume that the model in question has spin reflection symmetry. Then we can map the Hilbert space
of a chain of length $L$ to that of another chain of length $L-1$, such that for each bond we put an up spin if the two
neighbouring sites have the same orientation, and a down spin if they have different orientation. 
Symbolically
\begin{align}
& \begin{cases}
      \uparrow \uparrow\\
			\downarrow \downarrow
    \end{cases} \to\,\,\, \uparrow,
&&\begin{cases}
      \uparrow \downarrow\\
			\downarrow \uparrow
    \end{cases}   \to \,\,\,\downarrow           .
\end{align}
The original spin pattern can be reconstructed from the bonds up to a global spin reflection step, which preserves all values of the
bonds\footnote{We remark that if the model does not have spin reflection, those statements remain true with the addition that we need to know the state of the first site. Furthermore, in this case the Hamiltonian becomes non-local.}. Denoting the new variables with space positions at half shifts, the mapping on the operatorial level becomes simply
\begin{equation}
  \label{z1}
Z_jZ_{j+1}\to Z_{j+\frac{1}{2}}.
\end{equation}
A single spin flip on the original chain necessarily changes the values on two bonds, thus we obtain the other
elementary transformation rule
\begin{equation}
  \label{z2}
  X_j \to X_{j-\frac{1}{2}} X_{j+\frac{1}{2}}.
\end{equation}
These are not yet identical to the steps \eqref{ta1}-\eqref{rules}. In order to achieve the same formulas, one needs
to perform a global rotation {\it before the bond-site transformation}, which maps
\begin{align}
\label{rotation}
&  Z\to X,&& X\to Z,&& Y\to -Y.
\end{align}
Combining this rotation with \eqref{z1}-\eqref{z2} we obtain the transformation rules \eqref{ta1}-\eqref{rules}. In Appendix \ref{bondsitematrix}, we give an explicit expression of the bond-site transformation in a matrix form.
\\
The advantage of using the formulas \eqref{ta1}-\eqref{rules} is that they describe an \textit{involutive} transformation, so that applying the transformations twice will produce the initial model. However, the transformation is non-local: a subset of local operators remains
local after the mapping, but the 
remaining subset (including $X_i$ by the definition \eqref{ta1}) becomes truly non-local. In \cite{su2022integrable}, the authors give a lists of the terms in the Hamiltonian that remain local after the bond-site transformation. We refer to \cite{clifford1} for the classification of translation invariant Clifford group transformations that preserve locality of all the Hamiltonian and those that lead to non-local terms but preserve locality of certain Hamiltonians. The bond-site transformation belongs to the second class. In the  cases when the local Hamiltonian density  is mapped to local operators, it is possible to define the bond-site transformed model also with periodic boundary conditions. However, in this case the two models are strictly speaking not equivalent. This can be seen on the level of the real space transformation: in the periodic case any state has an even number of domain walls, therefore it is mapped to a state with an even number of down spins. The sectors of the new model with odd down spins do not correspond to the states of the original model. This difference should not affect the thermodynamic properties of the models, but it is crucial for the comparison of finite volume quantities.

\subsection{From two site to three site model}
\label{from2to3sitemodelsection}
Now we apply the bond-site transformation to map the two site model to a three site model. In our case, the transformation \textit{preserves locality}, so we can consider the model with periodic boundary condition.
\\
We apply the transformation  to both $\sigma$ and $\tau$ matrices. For simplicity, we again use the notation of the Pauli Matrices as $X, Y, Z$ with the remark that, when considering them acting on the local Hilbert space $\mathbb{C}^2\otimes \mathbb{C}^2$ they will be either $\sigma$ or $\tau$, respectively, if they act on the first or in the second copy of $\mathbb{C}^2$, as in \eqref{notation}.

\subparagraph{Kinetic term \eqref{HsigmaDM}} To use the transformation \eqref{rules}, first we need to use the rewriting $Y_i \to i \,X_i \,Z_i$ and then the rotation \eqref{rotation}
\begin{equation}
  \begin{split}
&  Y_{j} X_{j+1}-  X_{j} Y_{j+1}=
  i\,(Z_{j+1}- Z_{j})X_j X_{j+1}\qquad\to\qquad\\
  &\hspace{3cm}
  i\,(X_{j+\frac{1}{2}} X_{j+\frac{3}{2}}-X_{j-\frac{1}{2}} X_{j+\frac{1}{2}})
  Z_{j+\frac{1}{2}}=
  Y_{j+\frac{1}{2}} X_{j+\frac{3}{2}}-
  X_{j-\frac{1}{2}} Y_{j+\frac{1}{2}}.
  \end{split}
\end{equation}
The kinetic term is now localized on three sites\footnote{This is evident by considering $j\to j+1/2$. It correspond to consider a new spin chain where the "bonds" now become "sites".}. However, summing over these
contributions for the sites of the periodic chain and redefining $j\to j+\frac{1}{2}$, we see
that {\it the integrated kinetic term is self-dual}. For this particular model, the bond-site
transformation will only change the interaction terms.
\subparagraph{Interaction \eqref{Lsigma2}}
We transform separately the different terms.

First we use the rewriting
\begin{equation}
  \begin{split}
    \sigma^{-}_{j}  \sigma^+_{j+1}+\sigma^+_{j} \sigma^-_{j+1}&=
    \frac{1}{2}\left(    X_jX_{j+1}+ Y_j Y_{j+1}\right) ,\\
    \sigma^{-}_{j}  \sigma^-_{j+1}+\sigma^+_{j} \sigma^+_{j+1}&=
       \frac{1}{2}\left(    X_jX_{j+1}- Y_jY_{j+1}\right) ,\\
  \end{split}
\end{equation}
and then the rotation and the transformation \eqref{rules}
\begin{equation}
  \begin{split}
    Z_{j}+Z_{j+1}\qquad&\to\qquad
      X_{j-\frac{1}{2}}  X_{j+\frac{1}{2}}+   X_{j+\frac{1}{2}}  X_{j+\frac{3}{2}} ,\\
      X_jX_{j+1}\qquad&\to\qquad   Z_{j+\frac{1}{2}} ,\\
      Y_jY_{j+1}=-Z_jZ_{j+1}X_jX_{j+1}\qquad&\to\qquad
      - X_{j-\frac{1}{2}} Z_{j+\frac{1}{2}}X_{j+\frac{3}{2}},
    \end{split} 
\end{equation}
and after $j\to j+1/2$

\begin{equation}
  \begin{split}
    Z_{j}+Z_{j+1}\qquad&\to\qquad
      X_{j}  X_{j+1}+   X_{j+1}  X_{j+2},\\
      X_jX_{j+1}\qquad&\to\qquad   Z_{j+1},\\
      Y_j Y_{j+1}\qquad&\to\qquad
      - X_{j} Z_{j+1}X_{j+2},
    \end{split} 
\end{equation}
we obtain a term that spans three sites of the spin chain.
\\In the next section, we give a summary of the model obtained with the bond-site transformation.

\section{The first range three deformation of the Hubbard model}\label{sec:medium}

\subsection{$h$ and $l$ of the three site model}
We put together the results found in the previous section and we introduce a different parametrization ($\upsilon,\,\kappa$) with the coupling constants $U$ and $\theta$ such that
\begin{align}
\label{equivalenceccu}
&U = \frac{1}{8} \left(\cc^2+1\right)^2 \upsilon ,
&&\theta = -2 i \log \left(\frac{1+i \cc}{\sqrt{\cc^2+1}}\right).
\end{align}
The Hamiltonian \eqref{H2} is now
\begin{equation}
  \label{H3}
  \mathbb{H}_3=\sum_j \big[ h^{(1)}_{j,j+1}+h^{(2)}_{j,j+1}+\frac{\upsilon}{4}\,l^{(1)}_{j,j+1,j+2}l^{(2)}_{j,j+1,j+2}\big] ,
\end{equation}
where
$h$ is \eqref{Hsigma} and
\begin{equation}
  \label{lsigma}
  l_{j,j+1,j+2}= Z_{j+1} + \cc \, (X_{j}+X_{j+2})X_{j+1}- \cc^2 \,X_{j} Z_{j+1}X_{j+2} ,
\end{equation}
 $\mathbb{H}_3$ acts on the Hilbert space $V=\otimes_{i=1}^L V_i=\otimes_{i=1}^L
\big(\mathbb{C}^2 \otimes \mathbb{C}^2\otimes \mathbb{C}^2 \big)$ and the notation $h^{(1)},\,l^{(1)}$ or $h^{(2)},\,l^{(2)}$ identify respectively whether the operators appearing in $h$ and $l$ are respectively $\sigma$ or $\tau$. $\mathbb{H}_3$ signals  that the density of
the  Hamiltonian acts on 3 sites of the spin chain, as it is clear from the subscript $_{j,j+1,j+2}$.  The parameters $\upsilon$ and $\cc$ are the two independent parameters of the model; the model is Hermitian if they are both real. 
\\
This model is a \textbf{deformation of the Hubbard model}. In fact, by sending $\cc=0$ we recover the Hamiltonian \eqref{H1} (with $U=\upsilon$).
\\
Similarly to \eqref{propertyL2}, the operator \eqref{lsigma} satisfies 
\begin{align}
  \label{squarestoone}
  &(l_{j,j+1,j+2})^2=(1+\kappa^2)^2 \id ,
  &&[l_{j,j+1,j+2}(\cc),l_{j+1,j+2,j+3}(\cc')]\neq 0,
\end{align}
the last equality only holds for the case of the Hubbard model $\cc=\cc'=0$.
\\
Furthermore, similarly to \eqref{noncomm2}, $l_{j,j+1,j+2}$ are non-commuting for generic values of $\kappa$,
\begin{align}\label{noncomm}
[l_{j,j+1,j+2}(\cc),l_{j,j+1,j+2}(\cc')]\neq 0.
\end{align}
The equality holds only if $\cc=\cc'$ (trivial) or if $\cc\cc'=-1$.\\
We also checked that  the commutation relation  $[l_{j,j+1,j+2}(\cc),l_{j+2,j+3,j+4}(\cc')]=0$ is valid for any $\cc$ and $\cc'$. The reason is that $l_{j,j+1,j+2}$ is the bond-site transformation of a range two model for which $[\ell_{i,i+1},\ell_{i+2,i+3}]=0$ since the operators act on different sites.

\subsection{Integrability}
\label{range3integrability}
The model given by $\mathbb{H}_3$, found by performing a bond-site transformation of $\mathbb{H}_2$, is integrable: it has an infinite family of commuting local charges.
\\
The integrability property can be proved in two ways:
\begin{itemize}
\item Using the recently developed formalism of \cite{sajat-medium} for
medium range spin chains. It can be checked that the charges can be obtained from the transfer matrix construction. We use this method in the following section and we explicitly find the $R$-matrix. 
\item Showing that the higher charges of the two site Hamiltonian $\mathbb{H}_2$  remain all local\footnote{We remark that with locality we do not mean nearest-neighbour. Higher conserved charges will be characterized by a range of interaction that remains finite even if $L\to\infty$.} if we perform the duality
transformation to the three site family.
\end{itemize}

The second proof is very easy. In fact, it follows from the fact that $\mathbb{H}_2$ commutes with $Z_\sigma$ and $Z_\tau$ given in \eqref{Zsigmatau}. Therefore, it can contain an even number of Pauli matrices which cause a spin flip. The duality transformation
\eqref{rules} produces non-local operators only for an odd number of spin flipping Pauli matrices. Consequently,  all
charges of the two site models remain local after the transformation.
\\
Before discussing the first proof, we give a brief review of the medium range spin chain formalism and comment on how to apply it for model $\mathbb{H}_3$.

\subsubsection{Medium range spin chain -- Short summary of \cite{sajat-medium} - Difference form}
This section is based on the result of the work \cite{sajat-medium}. This is a generalization of the known Quantum Inverse Scattering Method and it is used to describe models with \textit{medium range} interactions. They use the word "medium" to indicate models where the range of interaction is $r\ge 3$, but finite. “medium” is used to distinguish these models both from the nearest-neighbor and the long range cases. The key idea is to enlarge the so-called ``auxiliary space'' (which is typically a
tensor product of copies of the elementary spaces) and to use special Lax operators to allow for the embedding of multi-site (medium range) Hamiltonians into this framework.\\We generalize the definitions of some of the quantities given in chapter \ref{intro} for this particular case.
\\
For an integrable spin chain with three site interactions, the auxiliary space is a tensor product of two copies of the
fundamental vector space. Therefore, the Lax matrix is an operator which acts on three spaces, one physical space and
two auxiliary spaces. It is denoted as ${L}_{a,b,j}(u)$, where $a$ and $b$ are the two auxiliary spaces, and $j$
refers to a physical space. The transfer matrix is defined as
\begin{equation}
  t(u)=\text{Tr}_{a,b} \left[
  L_{a,b,L}(u) \dots L_{a,b,2}(u)    L_{a,b,1}(u)
    \right],
  \end{equation}
where the partial trace is taken over both the auxiliary spaces.
\\
  As for the two site model, in particular see \eqref{charges}, the conserved charges  are defined by taking the logarithmic derivative of the transfer matrix. The Hamiltonian (range 3) is related to the transfer matrix by
  \begin{align}
  \mathbb{H}_3=\mathbb{Q}_3=\partial_u \log t(u)|_{u=0}=t^{-1}(0)\partial_u t(u)|_{u= 0},
  \end{align}
  and similarly the higher conserved charges
    \begin{align}
  \mathbb{Q}_r=\partial_u^{r-2} \log t(u)|_{u=0}.
  \label{highercharges}
  \end{align}

These charges are local if the Lax operator satisfies the initial condition
\begin{equation}\label{regularitylax}
  L_{a,b,j}(0)=P_{a,j}P_{b,j},
\end{equation}
where $P$ stands again for the permutation operator acting respectively (as specified) on one of the auxiliary space and
the physical one. This is the equivalent of the regularity condition for medium range chain.\\
We introduce the "check" operator as
\begin{equation}
{\check{L}_{a,b,j}}(u)=P_{b,j}P_{a,j} {{L}}_{a,b,j}(u).
\label{checkanduncheck}
\end{equation}
and the condition \eqref{regularitylax} translates to
\begin{equation}
  {\check{L}_{a,b,j}}(0)=\id .
\end{equation}

The transfer matrices form a commuting family, which is established from the fundamental intertwining relation:
\begin{equation}
\label{RLL3site0}
\check{R}_{23,45}(u_1,u_2)\check{L}_{123}(u_1)\check{L}_{345}(u_2)=\check{L}_{123}(u_2)\check{L}_{345}(u_1)\check{R}_{12,34}(u_1,u_2).
\end{equation}
Here $\check R (u,v)$ is the checked version of the $R$-matrix which is related to the usual ones by
\begin{equation}
  \check R_{ab,cd}(u,v)=P_{a,c}P_{b,d} R_{ab,cd}(u,v),
\end{equation}
it depends on two spectral parameters, and it acts on a four-fold tensor product
space. Consistency requires the Yang-Baxter equation for the $\check R$-matrix:
\begin{equation}
  \label{RcheckYB}
\check{R}_{34,56}(u_1,u_2)\check{R}_{12,34}(u_1,u_3)\check{R}_{34,56}(u_2,u_3)=\check{R}_{12,34}(u_2,u_3)\check{R}_{34,56}(u_1,u_3)\check{R}_{12,34}(u_1,u_2).
\end{equation}
In order to prove the integrability of a three site Hamiltonian, one needs to provide an explicit solution of
eq. \eqref{RLL3site0}; relation \eqref{RcheckYB} follows automatically.
\subsubsection{Medium range spin chain -- Application to $\mathbb{H}_3$}
We obtained $\mathbb{H}_3$ as a bond-site transformation of $\mathbb{H}_2$, so it is reasonable  to assume that the  Lax matrix is given by the bond-site transformed version of the $2$-site one. 
We follow the steps of Section V.A of
\cite{sajat-medium} summarized in the previous section, but since here the Lax operator will also depend on two spectral parameters, we present the details of this procedure.
\\
The starting point is the $R$-matrix for the 2 site model given in Appendix \ref{Rmatrixsite2Hubbard}, in particular we work with
\begin{equation}
\check{R}_{a,b}(u,v)=P_{a,b} R_{a,b}(u,v).
\end{equation}
We first perform the rotation \eqref{rotation} followed by the transformations \eqref{z1}-\eqref{z2}. We obtain a range-three operator that we will identify as the "checked" Lax matrix
\begin{equation}
  \check{R}_{j,j+1}(u,v)\quad\to\quad   \check{L}_{j,j+1,j+2}(u|v),
  \label{Laxoperator1}
\end{equation}
the Lax will also depend on two parameters.
\\
The intertwining relation takes the form
\begin{equation}
\label{RLL3site}
\check{R}_{23,45}(u_1,u_2)\check{L}_{123}(u_1|u_3)\check{L}_{345}(u_2|u_3)=\check{L}_{123}(u_2|u_3)\check{L}_{345}(u_1|u_3)\check{R}_{12,34}(u_1,u_2).
\end{equation}
In all of the computations of this section, the second spectral parameter of the Lax operator is seen as an outer (spectator) parameter, for
which we do not introduce intertwining relations. This special structure for the three site Lax operator is generic for
models obtained via a bond-site transformation from a two site model.
\\
Note that in \eqref{RLL3site} the Lax operators on
either sides overlap only on three site. The construction of the bond-site transformation implies for example
\begin{equation}
  [\check{L}_{123}(u_1|u_3),\check{L}_{345}(u_2|u_3)]=0.
\end{equation}
This relation comes from the fact that for the two site model, $R$-matrices acting on non-overlapping sites commute, in particular $[\check R_{12}(u_1,u_3),\check R_{34}(u_2,u_3)]=0$. 
\\
In order to prove that a solution exists, we take the checked version of \eqref{eq:YBE} that is
\begin{align}
\check{R}_{12}(u_2,u_3)\check{R}_{23}(u_1,u_3)\check{R}_{12}(u_1,u_2) = \check{R}_{23}(u_1,u_2)\check{R}_{12}(u_1,u_3)\check{R}_{23}(u_2,u_3)
\end{align}
and we take the bond-site transformation, that is
\begin{equation}\label{LLLrange3}
\check{L}_{123}(u_2|u_3)\check{L}_{234}(u_1|u_3)\check{L}_{123}(u_1|u_2)=\check{L}_{234}(u_1|u_2)\check{L}_{123}(u_1|u_3)\check{L}_{234}(u_2|u_3).
\end{equation}
Now multiplying with $\check{L}_{123}(u_2|u_3)^{-1}$ from the left, and with $\check{L}_{234}(u_2|u_3)^{-1}$ from
the right we get 
\begin{equation}
  \check{L}_{234}(u_1|u_3)\check{L}_{123}(u_1|u_2)\check{L}_{234}(u_2|u_3)^{-1}=
  \check{L}_{123}(u_2|u_3)^{-1}\check{L}_{234}(u_1|u_2)\check{L}_{123}(u_1|u_3).
\end{equation}
This makes  possible to define the four-site $\check R$ matrix via two different equations
\begin{equation}
\check{R}_{12,34}(u_1,u_2)=\check{L}_{123}(u_2|u_3)^{-1}\check{L}_{234}(u_1|u_2)\check{L}_{123}(u_1|u_3)
\label{R1234b}
\end{equation}
and
\begin{equation}
\check{R}_{23,45}(u_1,u_2)=\check{L}_{345}(u_1|u_3)\check{L}_{234}(u_1|u_2)\check{L}_{345}(u_2|u_3)^{-1}.
\label{R2345b}
\end{equation}
Note that in the second case, we applied a shift to the indices. Substituting these formulas into \eqref{RLL3site} we see
immediately that the intertwining works as expected.
It can be noticed that formulas \eqref{R1234b}-\eqref{R2345b} depend on $u_3$. Hence, to be precise, one would have to write $\check{R}(u_1,u_2,u_3)$.
However, if we consider the Yang-Baxter equation \eqref{RcheckYB} and relabel the spectral parameters to avoid potential ambiguity, we obtain
\begin{equation}
\check{R}_{34,56}(u,v)\check{R}_{12,34}(u,w)\check{R}_{34,56}(v,w)=\check{R}_{12,34}(v,w)\check{R}_{34,56}(u,w)\check{R}_{12,34}(u,v).
\end{equation}
By direct calculation, this equation is solved for any value of the parameter $u_3$. The Yang-Baxter equation will depend on the usual three spectral parameters $(u,v,w)$ and an additional one $u_3$. In models where the $R$-matrix is of difference form, this freedom actually drops out. However, in our case it does not drop out, therefore the intertwining can be performed by a one-parameter family of $R$-matrices.
\\
We explicitly checked the expressions \eqref{LLLrange3} and \eqref{RcheckYB}, so the integrability of the three site model is also proven.
\subsection{Why is this a new deformation?}
The Hamiltonian $\mathbb{H}_3$ is the \textit{first integrable} range 3 deformation of the Hubbard model. In this normalization, the original Hubbard model is restored for $\cc=0$.
\\
 However, for $\cc\ne 0$, there are
two crucial differences:
\begin{enumerate}
\item The interaction term spans 3 consecutive sites.
\item Particle number conservation is broken.  
\end{enumerate}
The terms including the $\sigma^x$ and $\tau^x$ operators, that are linear or quadratic in the
deformation parameter $\cc$, manifestly break the $\uuu(1)$ symmetries of 
the Hubbard model; they describe correlated particle creation and annihilation processes. In this respect, the model is
a XYZ deformation of Hubbard.
\\
Given the extensive research conducted on the Hubbard model and its various generalization over the years, it is natural to question whether this model is truly new or if it has been previously documented in the literature.  To address this concern, we conducted a comprehensive search of existing literature, but we did not managed to find this model in any of its formulations (see also next Sections). All the previous extensions
and deformations of the Hubbard model had in fact two common properties \cite{bariev-alcaraz--hubbard-tj,essler-korepin-schoutens}:
\begin{enumerate}
\item The fundamental Hamiltonian was always nearest-neighbour interacting.
\item The model had (at least) two local $\uuu(1)$ charges. 
\end{enumerate}
Although our Hamiltonian \eqref{H3} may seem to deviate from these common properties, there is a possibility that it can be seen as a rotated version of a linear combination of two site and three site charges from a known model. In order to exclude this possibility we
performed a search for a generic two site charge $A$ which would commute with our $\mathbb{H}_3$. Explicitly
\begin{equation}
 [\mathbb{H}_3,A]=[\mathbb{H}_3,\sum_j  a_{j,j+1}]=0.
\end{equation}
We found that, for generic coupling constants $\upsilon$ and $\kappa$, the only possibility for the operator density $a_{j,j+1}$ is to be of the
form $a_{j,j+1}=b_j-b_{j+1}+\alpha\,1$, which (after summation over $j$) lead to a trivial global charge. Thus our model
{\it does not have any conserved charges with range less than 
  three}. This calculation allowed us to conclude that our model is not a rotated version of a two site Hamiltonian.
\\
Furthermore, similar to the two site model, also the $R$-matrix of the three site model has the \textit{unusual functional dependence} \eqref{noelliptic}, so it cannot be expressed in terms of Jacobi elliptic functions $\text{sn}$, $\text{cn}$ and $\text{dn}$ without introducing square roots.

\section{Special points and strong coupling}

\subsection{Special points}
\label{specialpointrange23}
\subsubsection{Two site model}
In section \ref{symmetriesrange2} we analysed the symmetry of the Hamiltonian \eqref{Hnewrep} for any values of the parameters $U$ and $\theta$. In this paragraph we analyse some points where there are additional symmetries, which are $\theta=0, \pm \pi/2$.

\paragraph{$\theta=0$ }

In this point, the interaction operator $\ell_{j,j+1}$ \eqref{Lsigma2} is
\begin{equation}
  \label{Lxx}
  \ell_{j,j+1}=X_jX_{j+1}
 \end{equation}
which is represented by an anti-diagonal matrix. This particular model is {\it the bond-site transformation of the
  Hubbard model}.
Accordingly, it possesses two $\uuu(1)$-charges given by $\mathbb{Q}_2^{(1)}$ and $\mathbb{Q}_2^{(2)}$ 
\begin{align}
  \label{Q2X}
&  \mathbb{Q}_2^{(1)}=\sum_j \sigma^x_j \sigma^x_{j+1},&&
    \mathbb{Q}_2^{(2)}=\sum_j \tau^x_j \tau^x_{j+1},
\end{align}
which can  be extended to two $\SU(2)$ algebras.
\\
The known coordinate Bethe Ansatz solution of the Hubbard model \cite{lieb-wu-hubbard} can be used to construct
eigenstates of the model 
\eqref{Hnewrep} with $\ell_{j,j+1}$ given in \eqref{Lxx}. The approach involves performing a bond-site transformation on the eigenstates. However, it should be noted that this computation will only generate states that exhibit an even number of down spins for both the $\sigma$ and the $\tau$ sub-lattices. It follows from the commutation relations \eqref{Zcomm}
that this ``parity'' is indeed consistent with the Hamiltonian. At present it is not known how to treat the odd sub-sectors.  Another motivation to see this is given at the end of section \ref{generalitiesbst}: in the periodic case, every state is characterized by an even number of domain walls, which implies that it is mapped to a state with an even number of down spins. As a result, states with an odd number of down spins do not have a corresponding representation in the original model.

\paragraph{ $\theta=\pm \pi/2$}

For $\theta= \pi/2$, we find the Hamiltonian \eqref{Hnewrep} with the interaction matrix
\begin{equation}
\label{Lsu4sign}
  \ell_{j,j+1}=
  \begin{pmatrix}
    1 & & & \\
    & & 1 & \\
    & 1 & & \\
    & & & -1
  \end{pmatrix}=\sigma_j^+\sigma_{j+1}^-+\sigma_j^-\sigma_{j+1}^++\frac{1}{2}\big(Z_i+Z_{i+1} \big).
\end{equation}
The case with $\theta=-\pi/2$ is not independent from the one just shown: one can apply a unitary off-diagonal
local basis transformation and a re-definition of the coupling constant $U$ to relate the two models. These  cases are special because they enjoy two $\uuu(1)$-symmetries due to the particle conservation: the Hamiltonian $\mathbb{H}_2$ now
commutes with $N$ and $S_z$ given by \eqref{particles}. More generally, it formally commutes with the all the generators
\eqref{ASU2} up to boundary terms. 
\\
Interestingly, this model can be related to model B2 given in section \ref{modelB2explained}, that corresponds to a known deformation of the Hubbard model \cite{murakami1998new}. Model B2 obeys the conditions \eqref{conditiontomap}, consequently we can map to an Hermitian model.\\
By fixing $\phi=\pi/2$ in the Hamiltonian \eqref{HXX} and the jump operator $\ell$ \eqref{jumpforB2}, we obtain the $h$ of \eqref{HsigmaDM} and
\begin{align}
  \label{Lhu2}
  \ell_{j,j+1}=
  \left(
\begin{array}{cccc}
 \ch \,u & 0 & 0 & 0 \\
 0 & 1 & -\sh \,u & 0 \\
 0 & - \sh\, u & -1 & 0 \\
 0 & 0 & 0 & -\ch \,u \\
\end{array}
\right) .
  \end{align}
By rescaling the coupling constant $\beta$ of model B2 to $\beta\to \sqrt{2U }e^{  u}$ and sending $u \to -\, \infty$, $\ell$ reduces to  \eqref{Lsu4sign} and the two models are equivalent.
\\
It is remarkable that two special points of the model B2 are reproduced by \textit{two very
different versions of our models}: the actual Hubbard model ($u=0$) is found as a special point of our three site Hamiltonian $\mathbb{H}_3$, whereas the $u\to -\, \infty$ limit of  can be found in our two site
family $\mathbb{H}_2$. This fact may be a hint that perhaps there is a \textit{larger family} of integrable models which contains all these special points.

\subsubsection{Three  site model}
In this section, we repeat for the three site model the analysis just done for the two site model. Inverting \eqref{equivalenceccu}, we obtain $\cc=\tan \frac{\theta}{2}$. The special points of the three site model are $\cc=0, \pm 1$. We also checked that indeed these points are the only ones that admit a commuting charge which is at most of range\footnote{We remark that in this formalism, a density operator of range 2 can be written in term of $\sigma$ and $\tau$ matrices as $A_{i,i+1}=\sum_{q_1,q_2,q_3,q_4=0}^3 c_{q_1 q_2 q_3 q_4} \sigma_i^{q_1}\sigma_{i+1}^{q_2}\tau_i^{q_3}\tau_{i+1}^{q_4}$, with $\sigma^0$ the identity operator and $\sigma^{1,2,3}$ the set of Pauli matrices.} 2. 

\paragraph{$\cc=0$} In this point the model becomes the Hubbard model, whose properties were already discussed.
%
%
\paragraph{$\cc=\pm1$} In this case, \eqref{lsigma} becomes
\begin{equation}
  \label{lspec1}
  l_{j,j+1,j+2}=\pm (X_{j}+X_{j+2})X_{j+1}+
 Z_{j+1}\left(\id-X_{j}X_{j+2}\right).
\end{equation}
This model possesses exactly two $\uuu(1)$ charges,
\begin{align}
  \label{Q2X}
&  \mathbb{Q}_2^{(1)}=\sum_j \sigma^x_j \sigma^x_{j+1},&&
    \mathbb{Q}_2^{(2)}=\sum_j \tau^x_j \tau^x_{j+1}.
\end{align}
In fact, it can be shown that $[\mathbb{Q}_2^{(1)},\mathbb{H}_3]=[\mathbb{Q}_2^{(2)},\mathbb{H}_3]=0$, if $\mathbb{H}_3$ is computed from \eqref{H3} with the three
site interaction given by \eqref{lspec1}. 
\\
Furthermore, $\mathbb{Q}_2^{(\mu)}$ also  commutes with $\sum_j h_{j,j+1}^{(\mu)}$ and $\sum_j l_{j,j+1,j+2}^{(\mu)}$, for $\mu=1, 2$.  This makes $\mathbb{Q}_2$ a \textit{strong} symmetry of the model. We give more details about strong symmetries and their meaning in chapter \ref{NESSCHAPTER}.

\subsection{Large coupling limits}

\label{sec:large}

Here we investigate the large coupling limits of the models, by considering both the two site and the three site versions. The idea is to take the limit  $U\to \infty$ of the two site Hamiltonian \eqref{Hnewrep} and $\upsilon\to \infty$ of the three site Hamiltonian \eqref{H3}. In these limits the interaction term between the two sub-chains dominate, which is equivalent to setting the kinetic terms equal to zero.
\\
For a generic coupling $\kappa$ and  $\theta$, the non-commutativity in \eqref{noncomm2} and \eqref{noncomm} imply that
the interaction terms generate dynamics in the system. In this way, the strong coupling limit models are also non-trivial
\begin{equation}
\label{H3H2Uinfty}
\mathbb{H}_2^\infty=\sum_j   \ell_{j,j+1}^{(1)} \ell_{j,j+1}^{(2)} ,\qquad  \mathbb{H}_3^\infty= \sum_j l_{j,j+1,j+2}^{(1)}  l_{j,j+1,j+2}^{(2)},
\end{equation}
with $\ell_{j,j+1}$ and $l_{j,j+1,j+2}$ given by \eqref{Lsigma2} and \eqref{lsigma}, respectively.
\\
The two models are the
bond-site transformations of each other. To our best knowledge, these models are also new. \\Their integrability follows
directly from the constructions of the $R$-matrices for the general cases. For the two site model,
the Hamiltonian
  $\mathbb{H}_2^\infty$ is obtained via the substitutions
\begin{align}
&\mathcal{L}(u,\mu)\equiv R(\alpha \,u,\mu),\\
&\alpha =\frac{2 i \cos ^2\theta }{k},
&& \mu =\frac{1}{k}K\left(\frac{1}{k^2}\right),
&& k =i \cot \theta ,
\end{align}
where $K$ is the elliptic integral of the first kind\footnote{In order to get this result, we used the relations of  \cite{dixon1894elementary},
\begin{align}
&\text{dn}\left(v\left|k^2\right.\right)= \text{cn}\left(v\, k\left|\frac{1}{k^2}\right.\right),
&&\text{cn}\left(v\left|k^2\right.\right)= \text{dn}\left(v\, k\left|\frac{1}{k^2}\right.\right),
&&\text{sn}\left(v\left|k^2\right.\right)= \frac{1}{k}\text{sn}\left(v\, k\left|\frac{1}{k^2}\right.\right),
\end{align}
and we chose the branch cut $\sqrt{\sec ^2\theta }\cos \theta=  -1$.
}.
For the three site model, the $R$-matrix can be obtained in the same way as explained above
at the end of Section \ref{range3integrability}.
\\
In the case of the Hubbard model ($\kappa=0$) and its bond-site transformed model ($\theta=0$), the situation is different because the commutators \eqref{noncomm2} and \eqref{noncomm} vanish. 
Both $\mathbb{H}_3^\infty$ and $\mathbb{H}_2^\infty$ become a sum of commuting operators:
\begin{align}
& \mathbb{H}_3^{\infty}\,\,\underbrace{\to}_{\cc\to 0} \,\,\sum_j   \sigma^z_j\tau^z_j
&&\mathbb{H}_2^{\infty}\,\,\underbrace{\to}_{\theta\to 0}\,\, \sum_j   \sigma^x_j\sigma^x_{j+1}\tau^x_j\tau^x_{j+1}
\end{align}
and do not generate non-trivial dynamics.
\\
Furthermore, we noticed that in the case $\theta=\pi/2$ and in the strong coupling limit, the two site Hamiltonian is equivalent to model B1 given in section \ref{modelB1explained}.
\section{Summary of the new models}
\begin{figure}[ht]
  \centering
  \includegraphics[height=9.5cm]{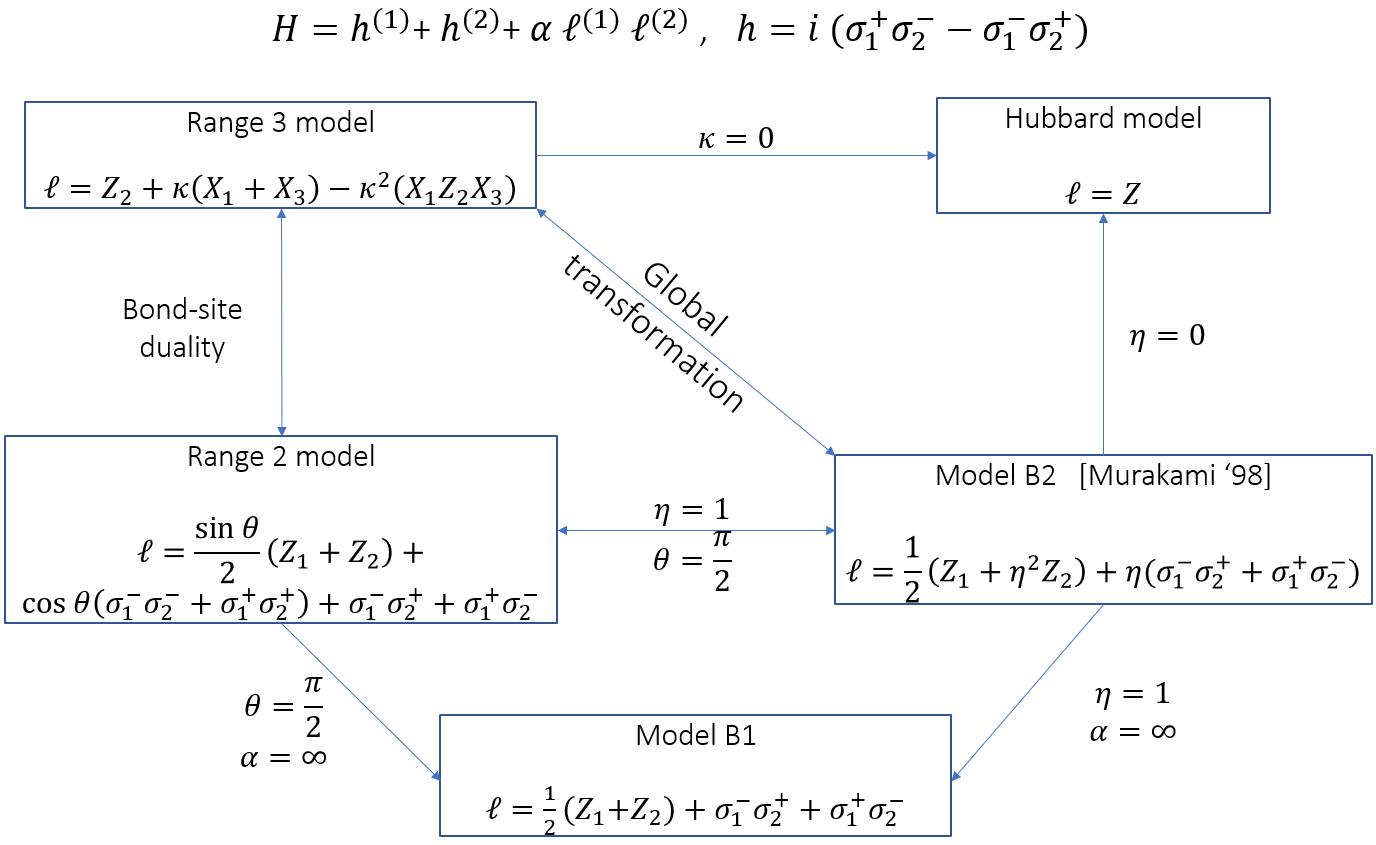}
  \caption{Summary of the connections between the different models we found.}
  \label{fig:example}
\end{figure}

The connections among the different integrable models found is summarized with the graph given in Fig. \ref{fig:example}. Interestingly, we noticed that two special points of the model B2 are reproduced by both the two site and the three site models. Furthermore, the range 3 model and model B2 are also related via a non-local transformation $T(\kappa)$ which we explain in the next chapter, after \eqref{nonlocaltransfltilde}. These observations hint at the possible existence of a \textit{larger family} of integrable models that includes all these special points.

\chapter{Hidden strong symmetries in the Hubbard deformation}
\label{NESSCHAPTER}
\ifpdf
    \graphicspath{{Chapter7/Figs/Raster/}{Chapter7/Figs/PDF/}{Chapter7/Figs/}}
\else
    \graphicspath{{Chapter7/Figs/Vector/}{Chapter7/Figs/}}
\fi

In this chapter, we investigate the symmetry  of the Lindblad superoperator associated with the range 3 deformation of the Hubbard model introduced in chapter \ref{Hubbardchapter}. We begin with a short introduction on the difference between conserved quantities and symmetries in non-Hermitian models. 
We show that in the context of non-Hermitian systems, these two notions become equivalent when a \textit{strong} symmetry is present. The meaning of it will be discussed in the following. Later, we analyze the range 3 model and we discover the presence of multiple NESS. The multiplicity is related to some \textit{hidden strong symmetries} in the form of quasi-local charges.  We compute the NESS exactly in the form of  Matrix Product Operators with fixed bond dimensions and we use this to compute the mean values of some local observables. Furthermore, we prove that the dynamics leads to the emergence of the Gibbs ensemble constructed from the hidden quasi-local charge.

\section{Conserved quantities and symmetries}

Isolated systems exhibit distinct behaviors depending on whether they are integrable or not. In general, an isolated system tends to rapidly \textit{relax} toward an equilibrium state. However, integrable systems behave differently due to the presence of additional conserved quantities. This was demonstrated in the pioneering work of Kinoshita, Wenger, and Weiss, a Quantum Newton's Cradle experiments, \cite{kinoshita2006quantum}. These experiments investigated the non-equilibrium dynamics of Bose gases driven out of equilibrium in one, two, and three dimensions. While the two and three-dimensional systems were observed to quickly relax towards an equilibrium state, the behavior of the one-dimensional system was different. This difference can be attributed to the presence of approximate conservation laws, which makes the system not ergodic, and prevent it from reaching the thermal equilibrium. Driven in part by this motivation, the non-equilibrium dynamics of integrable quantum many-particle systems has been the subject of intense theoretical efforts over the past 20 years, \cite{huse-review,essler-fagotti-quench-review,hfrag-review}. Ergodicity breaking in isolated systems may be due to different mechanisms, all related with \textit{ symmetries} of the system. The symmetries will constraint the dynamics of the system and the ergodicity property is not satisfied anymore.\\
For open quantum systems with Markovian environment, the evolution is dictated by the Lindblad equation. In a generic Lindblad system, there is typically a unique non equilibrium steady state NESS: 
the properties of the initial states are
eventually lost during time evolution. The equivalent of ergodicity breaking  for open quantum systems is the emergence of multiple NESS. Examples of this phenomenon are known \cite{lindblad-unique-1,enej-prosen-spin1degness}. In such cases, the system retains \textit{memory} of the initial state, as different initial density matrices evolve into distinct NESS in the long time limit. This raises the important question:  what are the possible mechanisms that give rise to multiple NESS in a many-body Lindblad system? Similar to the isolated system, non-uniqueness of the NESS is associated with the presence of extra conservation laws. In Lindblad systems, these conserved quantities may arise when the model exhibits what is known as \textit{strong symmetries}.
In fact, as we show in the following, in the context of open quantum systems, conserved charges and symmetries are different objects. However, if a system possesses a strong symmetry, this is also a conserved charge. \\
In \textit{Hermitian quantum mechanics}, an explicit time-independent observable $J = J^\dagger$
 is a conserved quantity (i.e. constant of
motion) if and only if it commutes with the Hamiltonian $H$ of the model. The Noether's theorem holds: to each conserved charge we can associate a continuous symmetry by exponentiating the charge  $U = e^{i\phi J}$, with $\phi$ real. The three statements are then equivalent
\begin{align}
&[J,H]=0,
&&\dot{J}=0,
&&U^\dagger H U = H .
\end{align}
The situation is very different in the case of a Lindbladian, since the Lindblad superoperator (given in \eqref{superoperatorL2}) is \textit{not Hermitian}. In these systems, a symmetry operation \textit{might} or \textit{might not}
lead to a conserved quantity, and not all conserved quantities originate in symmetries, \cite{lindblad-symmetries}. To obtain the expectation value of an operator $J$ ($\Tr ({J \rho})$), one can use the cyclicity propery of the trace and obtain the dynamics in the Heisenberg picture by using the Lindblad equation, 
\begin{align}
\dot J=\mathcal{L}^\dagger(J)= i [h,J]+\ell^\dagger J \ell-\frac{1}{2}\{\ell^\dagger \ell, J\} ,
\end{align}
$J$ is a conserved quantities if $\mathcal{L}^\dagger (J)=0$. To analyze the symmetry, we employ the superoperator formalism.  In this context, we introduce $\mathcal{J}$, which is the superoperator associated with the Hermitian operator $J$ and satisfies $\mathcal{J}=\mathcal{J}^\dagger$. The continuous symmetry generated by $\mathcal{J}$ is $\mathcal{U}=e^{i \phi \mathcal{J}}$. $\mathcal{J}$ is the generator of the symmetry if
\begin{equation}\label{symmetrycondsuper}
[\mathcal{J},\mathcal{L}]=0,
\end{equation}
with $\mathcal{L}$ the Lindblad superoperator and\footnote{The expression of $\mathcal{J}$ can be understood if one consider the expression $[\cdot,A]$ in the superoperator space $\mathcal{A}=A\otimes \id-\id\otimes A^T$. The reason of the transposition can be understood by writing in components the action of $A$ on the right of another operator.} $\mathcal{J}=J\otimes 1-1\otimes J^T$. By substituting the expression of the superoperator $\mathcal{L}$ into the commutation relation \eqref{symmetrycondsuper}, we obtain that one possible way to realize \eqref{symmetrycondsuper} is
\begin{align}
&[J,h]=0,
&&[J,\ell]=\alpha \ell,\,\,\alpha\in \mathbb{R}\,.
\end{align}
In this case, $J$ is a so-called \textit{strong symmetry} and $\La^\dagger (J)=\La (J)=0$. 
\\
These symmetries have been extensively studied in various systems \cite{prosen-strong-symmetries,lindblad-symmetries,enej-prosen-spin1degness,buca-strong-symmetries}.
\\
To summarize, for open quantum systems, the following implications hold
\begin{align}
&\text{if\,\,} [J,h]=0,\, [J,\ell]=0 \text{\,\,\,\,then\,} \,\dot {J}=\mathcal{L}^\dagger(J)=0,  \,\,(\mathcal{J} \text{is conserved}) ,\\
&\text{if\,\,} [J,h]=0,\, [J,\ell]=0 \text{\,\,\,\,then} \,\,\mathcal{U}^\dagger \mathcal{L} \mathcal{U} = \mathcal{L}, \,\,(\mathcal{U}\, \text{is a symmetry}).
\end{align}
Consequently, for Lindbladian systems, the following outcomes can arise:
\begin{itemize}
\item There may exist a conserved quantity which does not commute with everything in $\mathcal{L}$ but is conserved “as a whole”
\item The $\mathcal{U}$  generated by such conserved quantities are not always symmetries of the system
\item A symmetry generator $J$ does not necessarily correspond to a conserved quantity
\item Symmetry generator does not have to commute with everything in $\mathcal{L}$.
\end{itemize}
However, if one can find a non-trivial operator that commutes with everything in $\mathcal{L}$, then the operator is both a symmetry and a conserved quantity of the system and we refer to it as a \textit{strong symmetry}.
It is then natural to ask what conserved quantities and symmetries are useful for. In brief, conserved quantities are associated with the NESS of the system, while symmetries can be utilized to block diagonalize the superoperator $\mathcal{L}$.\\
Of particular interest are strong symmetries that are represented by {\it extensive operators}, i.e.
$Q=\sum_j q(j)$, where $q(j)$ is a quasi-local operator density\footnote{Suppose to have a general translational invariant operator $A=\sum_j a(j)$. $A$ is quasi-local \cite{prosen-enej-quasi-local-review,ilievski2015quasilocal} if the following conditions hold
\begin{itemize}
\item The Hilbert-Schmidt norm satisfies $||A||_{HS}^2\propto L$,
\item For any operator $b$ acting on $k$ site of the spin chain, the overlap $(b,A)$ is independent on $L$, for $L\to\infty$,
\item If we write $a=\sum_{r=1}^L a_{[1,r]}$, in the thermodynamic limit $||a||_{HS}^2=\sum_{r=1}^\infty ||a_{[1,r]}||^2$ is convergent and, in particular $||a_{[1,r]}||_{HS}^2<C e^{-\xi r}, \xi>0$.
\end{itemize}}. 
Previously, these operators were investigated within the framework of the Generalized Gibbs Ensemble,
\cite{prosen-enej-quasi-local-review,JS-CGGE}. Our work is the first one to uncover quasi-local charges in a Lindblad system with local driving in the bulk.

\section{A simple example to warm up: Hubbard Lindbladian} 
In section \ref{mappinglindbladhermitian}, we reviewed the result of \cite{medvedyeva2016exact} where they show how to find the superoperator $\mathcal{L}$ corresponding to the Hubbard model. This model has an extensive strong symmetry given by 
\begin{equation}
  Q_0=\sum_j Z_j,
\end{equation}
which is the global magnetization. In fact, this commutes with both the Hamiltonian and the jump operators given in \eqref{hlhubb}. Accordingly,  the degeneracy of the  NESS is $L+1$, with $L$ being the length of the spin chain.  Each NESS corresponds to a distinct sector of the Hilbert space characterized by a specific magnetization. These sectors can be identified using the $L+1$ projectors $P_N$, which project onto the different magnetization sectors. Alternatively, an over-complete basis for the null-space can be chosen as
\begin{equation}\label{stateshubbard}
\rho(\alpha)\sim  e^{\alpha Q_0}=\prod_j e^{\alpha Z_j}, \qquad \alpha\in\valos .
\end{equation}
These density matrices are linear combinations of $P_N$. By setting $\alpha=0$, the NESS corresponds to the infinite temperature state $\rho\propto \id$, while for $\alpha=\pm \infty$ we obtain the states with particles of all spin up or all down.

\section{Our model: Hubbard deformation} 
Now, we examine the Lindbladian counterpart of the range three integrable Hermitian model discussed in  chapter \ref{Hubbardchapter}.
\\
Since the $h$ and $l$-operator of the range three Hamiltonian \eqref{H3} satisfies the condition \eqref{conditiontomap}, we are allowed to construct the equivalent Lindbladian. Here, we provide the expressions for completeness.
\begin{align}
  \Li_{j,j+1}=&~-i h^{(1)}_{j,j+1}+i  h^{(2)^T}_{j,j+1}+ \nonumber \\ &~ U \Big(l^{(1)}_{j,j+1,j+2}  l^{(2)*}_{j,j+1,j+2}  -\frac{1}{2} l^{(1)\dagger}_{j,j+1,j+2}  l^{(1)}_{j,j+1,j+2}-\frac{1}{2}  l^{(2)T}_{j,j+1,j+2}  l^{(2)*}_{j,j+1,j+2}\Big),
  \label{superoperatorL}
\end{align}
where
$h$ is \eqref{Hsigma} and $l$ is \eqref{lsigma}:
\begin{equation}
  l_{j,j+1,j+2}= Z_{j+1} + \cc \, (X_{j}+X_{j+2})X_{j+1}- \cc^2 \,X_{j} Z_{j+1}X_{j+2} ,
\end{equation}
and satisfy 
\begin{align}\label{relationlop}
&l_{j,j+1,j+2}^\dagger = l_{j,j+1,j+2} ,
&&(l_{j,j+1,j+2})^2 = \id.
\end{align}
In this chapter, we consider the regime $0<\kappa<1$. In fact, other regimes can be treated by special similarity and duality transformations.  Furthermore, for the special  points $\kappa=\pm 1$, as explained in section \ref{specialpointrange23}, the model possesses two extra $\uuu(1)$-charges, which enlarge the null space of the Lindbladian. Those points deserve a separate study and we do not consider them in this thesis. 

\subsection{Fermionic formulation}
\label{fermionicformulationellipticmodel}
The model can alternatively be expressed using fermion operators, with the introduction of Majorana operators\footnote{Those are related to the usual Jordan-Wigner transformation
\begin{align}
&c_j=\Big(\prod_{k<j} Z_k \Big)\sigma_j^-,
&&c_j^\dagger=\sigma_j^+ \prod_{k<j} Z_k ,
\end{align} by $\psi_{2j-1}=c_j^\dagger+c_j$ and $\psi_{2j-1}=i(c_j-c_j^\dagger)$.} $\psi_{2j-1}=X_j \prod_{l<j} Z_l$, $\psi_{2j}=Y_j \prod_{l<j} Z_l$. These operators satisfy $\{ \psi_a,\psi_b \} = 2\delta_{a,b}$. Consequently, the Hamiltonian is given by
\begin{equation} 
h = \sum_{k} \psi_{k-1} \psi_{k+1} \,,  
\label{eq:Hfermions}
\end{equation} 
where the summation is now extended to twice the number of sites in the original spin model.  When considering a spin chain with periodic boundary conditions on $L$ sites, this can be translated into the Majorana language as $\psi_{L+k}=\mathcal{Z} \psi_{k}$,
where $\mathcal{Z} \equiv (-1)^F\equiv \prod_j Z_j$ represents the fermion number parity.
The jump operators \eqref{lsigma} take the form 
\begin{equation}
l = 
\frac{i}{1+\cc^2} \left( \psi_{2j+2} - \cc \,\psi_{2j}  \right) \left( \psi_{2j+1} - \cc \,\psi_{2j+3}  \right) \,.  
\label{eq:lfermions}
\end{equation}

The jump operators break the $U(1)$ symmetry of the original model: they induce particle creation and
annihilation, but due to conservation of $\mathcal{Z}$, creation an annihilation happens \textit{in pairs}. 
\\
While the Hamiltonian \eqref{eq:Hfermions} is bilinear in terms of the Majorana operators and can therefore be diagonalized using free-fermion techniques \cite{XX-original}, the jump operators \eqref{eq:lfermions} introduce quartic terms in the Lindblad equation, and our model is therefore truly interacting. 
 
\subsection{Strong symmetry and NESS}
It is worth to remember  that the range 3 deformed model possesses Yang-Baxter integrability. It is interesting to observe that in the following analysis we do not utilize this property. However, we use the ``superintegrability'' property of the Hamiltonian \eqref{Hsigma}: $h$ has a non-abelian families of conserved charges, which commute with it but not necessarily with one another \cite{superintegrability,ericZn}.

\paragraph{Conserved charges of the Hamiltonian $h$}
We label the set of charges as $[ab]_m$, where $a$ and $b$ can take the label $X$ or $Y$, and for $m \geq 0$, 
\begin{equation} 
[ab]_m \equiv \sum_{j=1}^L a_j \left( \prod_{1\leq k<m} Z_{j+k}   \right) b_{j+m}  \,,
\label{conservedcharges} 
\end{equation} 
where $X_j$, $Y_j$ and $Z_j$ are the Pauli matrices acting on site $j$ of the spin chain and $L$ is the length of the spin chain.
\\
Considering the fermionic formulation of the model \eqref{eq:Hfermions}, the corresponding charges are the set of all possible translationally invariant fermion bilinears.  The presence of an extensive set of local charges often indicates integrability. In this case, the Hamiltonian \eqref{Hsigma} is related to the XX spin chain through a homogeneous twist along the chain. Notably, it exhibits a "superintegrable" nature, where the charges $[ab]_m$ form distinct families that do not commute with each other.
 For instance, the sets of charges $\{[XY]_m\}$ and $\{[YX]_n\}$ commute with one another, but only the combinations $\{[XY]_m - [YX]_m \}$ commute with the charges $\{[XX]_n\}$ or $\{[YY]_n\}$. We also introduce the charge 
 \begin{equation}
\mathcal{Z} =  \prod_{j=1}^L Z_{j}  \,,
 \end{equation}
which commutes with the Hamiltonian as well as with all the charges $[ab]_m$.

\paragraph{Null space of the Lindladian}
By analyzing numerically the spectrum of the Lindblad superoperator for different value of the length $L$, we find that the degeneracy\footnote{We refer here to the values of $\kappa\neq \pm 1$. As mentioned, those two points deserve special attention due to the enlarged symmetries.} of the $0$ eigenvalue is $L+1$.  We motivate this multiple NESS by an unexpected strong symmetry in the system. This symmetry and the associated conserved charge are obtained from the original $Q_0$ of the un-deformed Hubbard
model via a non-local transformation, which is performed by a Matrix Product Operator (MPO).
\\
More specifically, we define the transfer matrix  $T(\cc)$ corresponding to the Hamiltonian \eqref{Hsigma} of our system
\begin{equation}
T(\cc) = \mathrm{Tr}_{0} ( A_{L}(\cc) A_{L-1}(\cc)  \ldots  A_1(\cc) ) \,,
\end{equation}
where "0" is a two-dimensional auxiliary space and $A(\cc)$ takes the form
\begin{equation}
A_j(\cc)
 = \frac{1}{2}
\left(  
\begin{array}{cc}
g^- + g^+Z_j & g^+ X_j -  i g^- Y_j \\
g^- X_j +  i g^+Y_j &   g^+ - g^-Z_j \\ 
\end{array}
 \right)  \,,
\end{equation}
where $g^\pm=\sqrt{1\pm \cc}$. The tranfer matrix $T(\cc)$ forms a mutually commuting family, namely $[T(\cc),T(\cc')]=0$. This property arises from the fact that the Hamiltonian corresponds to the transfer matrix of the XX spin chain\footnote{More specifically, to the transfer matrix based on cyclic representations of the quantum group $U_q(sl_2)$ at $q=i$.}. It can in fact be recast as a series expansion in powers of $\cc$, whose coefficients are the mutually commuting conserved charges of $H$. Specifically, it has the following series expansion around $\cc = 0$:
\begin{equation}
T(\cc) = \mathcal{U} \exp(\mathcal{G}(\cc)) \,,
\label{expansionofT}
\end{equation} 
 where $\mathcal{U}$ is the one-site discrete translation operator, and 
\begin{equation}
\mathcal{G}(\cc) = i \sum_{m \geq 1} \frac{\cc^m}{2m} [YX]_m  \,,
\label{calGdef}
\end{equation}
with $[YX]_m$ given in \eqref{conservedcharges}.\\
The expansion \eqref{calGdef} holds at all orders, even for a system with finite size $L$, as can be checked by computing explicitly the successive logarithmic derivatives of $T(\cc)$ at $\cc=0$. For $L\to \infty$, the series \eqref{calGdef} defines a quasi-local operator for $|\cc|<1$, \cite{ilievski2015quasilocal}. For finite $L$, it can be further rearranged using the properties: $[YX]_{m+L} = - \mathcal{Z} [YX]_m$ for $m\geq 1$, and $[YX]_L = -i L \mathcal{Z}$. A practical expression is 
\begin{equation}
\mathcal{G}(\cc) = \frac{1}{2} \log(\id+\cc^L \mathcal{Z}) +  i \sum_{\substack{m \geq 1\\ m \notin L \mathbb{Z}} } \frac{\cc^m}{2m} [YX]_m,    \,
\end{equation}
which splits between a first term which is Hermitian, and an anti-Hermitian part. From there, we can obtain
\begin{equation}
T(\cc)T(\cc)^{\dagger}  = T(\cc)^{\dagger} T(\cc)  = \id + \cc^L \mathcal{Z} \,
\label{eq:TTdag}
\end{equation}
and equivalently
\begin{equation}
T(\cc)^{-1} = \frac{\id-\cc^L \mathcal{Z}}{1-\cc^{2L}} T(\cc)^{\dagger}  \,.
\label{Tgammam1}
\end{equation}
Hence, in the $L\to \infty$ limit they become inverse of each other. It is useful to also derive this expression using the Matrix Product Operator (MPO) formalism. We can write $T(\cc)T(\cc)^{\dagger}$ as a MPO of bond dimension 4, with auxiliary space $0\otimes 0$, namely 
\begin{equation}
T(\cc)T(\cc)^{\dagger} = \mathrm{Tr}_{0\otimes 0} \Big( \mathcal{M}_{L}(\cc) \mathcal{M}_{L-1}(\cc)  \ldots  \mathcal{M}_1(\cc) \Big) \,,
\label{TTdagMPO}
\end{equation}
where 
\begin{equation}
\mathcal{M}_j(\cc)=\frac{1}{2} \left(
\begin{array}{cccc}
\id+ \sqrt{1-\kappa ^2} Z_j & \sqrt{1-\kappa ^2} X_j+i Y_j & \sqrt{1-\kappa ^2} X_j-i Y_j & \id-\sqrt{1-\kappa ^2} Z_j \\
 -\kappa  X_j & \kappa  Z_j & \kappa  Z_j & \kappa  X_j \\
 -\kappa  X_j & \kappa  Z_j & \kappa  Z_j & \kappa  X_j \\
\id+ \sqrt{1-\kappa ^2} Z_j & \sqrt{1-\kappa ^2} X_j+i Y_j & \sqrt{1-\kappa ^2} X_j-i Y_j & \id-\sqrt{1-\kappa ^2} Z_j \\
\end{array}
\right).
\end{equation}
The MPO is invariant under any change of basis performed in the auxiliary space. We can chose a transformation such that the MPO takes an upper triangular form. In particular, we define $V_0=e^{\frac{i \pi }{4} Y\otimes X}$, where $X$ and $Y$ are now Pauli matrices acting in each copy of the auxiliary space $0$.  \eqref{TTdagMPO} can therefore be recovered by replacing the matrices $\mathcal{M}_j(\cc)$ by $\Big(V_0 \mathcal{M}_j(\cc) V_0^{-1}\Big)$, which takes the form
 \begin{equation}
 \label{Mconjugated}
 V_0 \mathcal{M}_j(\cc) V_0^{-1} = 
 \left(
\begin{array}{cccc}
1 & \sqrt{1-\cc^2} X_j & -i Y_j & - \sqrt{1-\cc^2} Z_j  \\
0 & \cc Z_j  & 0 & \cc X_j \\
0 & 0 & 0 & 0 \\
0 & 0 & 0 & 0 
\end{array} 
  \right) \,.
 \end{equation}
From the block-diagonal form of \eqref{Mconjugated}, it is evident that when taking the trace in \eqref{TTdagMPO}, only the two diagonal terms make a contribution. These terms give rise to the two corresponding terms in \eqref{eq:TTdag}.
\paragraph{Strong symmetry}
Next, we define the deformation of $Q_0$ as
\begin{equation} \label{Qgammadef}
  Q_\cc= T(\cc)^\dagger Q_0 T(\cc),
\end{equation}
which in the $L \to \infty$ limit corresponds to a conjugation relation.
This conjugation can be understood as a quasi-local deformation of $Q_0$,
involving the non-abelian conserved charges of the Hamiltonian \eqref{Hsigma}. $Q_\cc$ remains an extensive
operator, but its operator density $q_\cc(j)=T(\cc)^{\dagger} Z_j T(\cc)$ becomes quasi-local.
\\
By using the MPO formalism, we prove that the operator $Q_\cc$ is a strong symmetry of the Lindbladian: it commutes with the Hamiltonian \eqref{Hsigma}
and also with all the jump operators \eqref{lsigma}. Explicitly, we have to show that
\begin{align}
&[h_{i,i+1},Q_\cc]=0 ,
&&[l_{i,i+1,i+2},Q_\cc]=0,
\end{align}
which is equivalent to show, by using the relation \eqref{eq:TTdag} and the fact that $[Q_0,\mathcal{Z}]=0$,
\begin{align}
&[T(\cc) h_{i,i+1} T(\cc)^\dagger,Q_0]=0 ,
&&[T(\cc) l_{i,i+1,i+2} T(\cc)^\dagger,Q_0]=0.
\end{align}
The first commutation relation trivially holds since $T(\cc) h_{i,i+1} T(\cc)^\dagger = h_{i,i+1}$. For the second one, we use the MPO technique. In particular, we define 

\begin{equation}
T(\cc) B_j T(\cc)^{\dagger} = \mathrm{Tr}_{0\otimes 0} ( \mathcal{M}_{L}(\gamma) \ldots \mathcal{M}^B_{j}(\cc)  \ldots  \mathcal{M}_1(\cc) ) \,,
\label{TBTdagMPO}
\end{equation}
where $B\in\{X,Y,Z\}$. 
We find similarly\footnote{For simplicity, we did not report the expressions of $\mathcal{M}^X_j$, $\mathcal{M}^Y_j$ and $\mathcal{M}^Z_j$, but if needed, the readers can recover them by undoing the rotation.} 
 \begin{equation}
 \label{MXconjugated}
 V_0 \mathcal{M}^X_j(\cc) V_0^{-1} = 
 \left(
\begin{array}{cccc}
0 & \cc Z_j  & 0 & \cc X_j \\
1 & \sqrt{1-\cc^2} X_j & -i Y_j & - \sqrt{1-\cc^2} Z_j  \\
0 & 0 & 0 & 0 \\
0 & 0 & 0 & 0 
\end{array} 
  \right) \,,
 \end{equation}

 \begin{equation}
 \label{MYconjugated}
 V_0 \mathcal{M}^Y_j(\cc) V_0^{-1} = 
 \left(
\begin{array}{cccc}
0 & 0 & 0 & 0 \\
0 & 0 & 0 & 0 \\
 i\sqrt{1-\cc^2}  & i X_j &  \sqrt{1-\cc^2}  Y_j & -i Z_j  \\
 \cc Y_j  & 0  & -i \cc  & 0 \\
\end{array} 
  \right) \,,
 \end{equation}
 
  \begin{equation}
 \label{MZconjugated}
 V_0 \mathcal{M}^Z_j(\cc) V_0^{-1} = 
 \left(
\begin{array}{cccc}
0 & 0 & 0 & 0 \\
0 & 0 & 0 & 0 \\
i \cc Y_j  & 0  & \cc  & 0 \\
 -\sqrt{1-\cc^2}  & - X_j & i \sqrt{1-\cc^2}  Y_j &  Z_j  \\
\end{array} 
  \right) \,.
 \end{equation}

For any three consecutive sites $j,j+1,j+2$, we then have for the jump operator
\begin{equation}
\label{TellT}
T(\cc) l_{j,j+1,j+2} T(\cc)^\dagger = 
\mathrm{Tr}_{0\otimes 0} ( \mathcal{M}_{L}(\cc) \ldots \mathcal{M}^l_{j,j+1,j+2}(\cc)   \ldots  \mathcal{M}_1(\cc) ) \,,
\end{equation} 
where 
\begin{align}
\mathcal{M}^l_{j,j+1,j+2}(\cc) 
\equiv &
\frac{1}{1+\cc^2}
\left(
\mathcal{M}_{j+2} \mathcal{M}^Z_{j+1} \mathcal{M}_{j}
+
\cc (\mathcal{M}^X_{j+2} \mathcal{M}^X_{j+1} \mathcal{M}_{j} 
+
\mathcal{M}_{j+2} \mathcal{M}^X_{j+1} \mathcal{M}^X_{j})
-\right.\nonumber\\
&
\left.\cc^2 \mathcal{M}_{j+2}^X \mathcal{M}^Z_{j+1} \mathcal{M}^X_{j}
\right) \,,
\end{align}
which can be brought to the following form after rotation in the auxiliary space:
\begin{align}
\label{Mellrotated}
V_0 \mathcal{M}^l_{j,j+1,j+2}(\cc)  V_0^{-1}
=\left( 
\begin{array}{cccc}
\tilde{l}_{j,j+1,j+2} & \ldots & \ldots & \ldots \\
0 & \cc^3  Z_j Z_{j+1}Z_{j+2}  \tilde{l}_{j,j+1,j+2}  & 0 & \ldots \\
0 & 0 & 0 & 0 \\
0 & 0 & 0 & 0 
\end{array} \right) \,.
\end{align}
Here we have defined 
\begin{align}\label{ltildedef}
 \tilde{l}_{j,j+1,j+2} \equiv \frac{1}{1+\cc^2}\left(Z_{j+2}+\cc (X_{j+1} X_{j+2} +Y_{j+1} Y_{j+2}) +\cc^2  Z_{j+1}\right)\,,
 \end{align}
 and the $\dots$ denote other combinations of the Pauli matrices which we do not need to consider.
 Indeed, from the triangular structure of \eqref{Mellrotated}, we see again that only the two non-zero diagonal entries give a non-zero contribution to the trace \eqref{TellT}. As a result we find 
\begin{align}
T(\cc) {l}_{j,j+1,j+2} T(\cc)^{\dagger}  =  (\id+\cc^L \mathcal{Z}) \tilde{l}_{j,j+1,j+2}\,,
\end{align} 
or, equivalently,
\begin{align}
T(\cc) {l}_{j,j+1,j+2} T(\cc)^{-1}  = \tilde{l}_{j,j+1,j+2} \,.
\label{nonlocaltransfltilde}
\end{align}
It is important to notice that the jump operator $\tilde{l}_{j,j+1,j+2}$ acts non-trivially on two site of the spin chain and, up to normalization and identification $\kappa=-\coth\big(\frac{u}{2}\big)$, $\phi=\pi/2$, corresponds to the jump operator of model B2 given in \eqref{jumpforB2}. This point clarifies the diagonal connection in the Figure \ref{fig:example}, "Global transformation".
\\
It is easy now to see that all the modified jump operators $\tilde{l}$ commute with the global charge $Q_0=\sum_j Z_j$. \\
Since we proved that $Q_\cc$ commutes with both $h$ and $l$, it is a strong symmetry of the Lindbladian. This implies that it is a \textit{conserved charge} for the Lindbladian time evolution.
\paragraph{NESS}
This result also proves that all powers of $Q_0$, or equivalently, all exponentials of the form $e^{\alpha Q_0}$  are conserved charges for the Lindbladian characterized by $h$ and the generalized jump operator $\tilde{l}$. In particular, since
\begin{align}
&\mathcal{L}(e^{\alpha Q_0})= \mathcal{L}^\dagger(e^{\alpha Q_0})=0
\end{align}
they are (un-normalized) NESS and form a basis for a $L+1$-dimensional space, including the identity.\\Equivalently, undoing the similarity transformation, we find that $
T(\cc)^{-1} e^{\alpha Q_0} T(\cc)$ 
are (un-normalized) NESS for the Lindbladian characterized by $h$ and jump operator ${l}$. Furthermore, considering \eqref{Tgammam1} and the fact that $\mathcal{Z}$ commutes with $h$ and $l$, we can replace $T(\cc)^{-1}$ with $T(\cc)^\dagger$ and obtain

\begin{equation}
  \label{rhodef}
  \rho_{\cc}(\alpha)=T(\cc)^\dagger e^{\alpha Q_0} T(\cc)=
  T(\cc)^\dagger \left[\prod_j e^{\alpha Z_j}\right] T(\cc).
\end{equation}
The states under consideration are density matrices, specifically Hermitian and positive definite. 
They are also \textbf{strong symmetries}.
These states are the NESS of the Lindbladian with fixed deformation parameter $\cc$ and
arbitrary coupling strength $U$. 
The operators $\rho_\cc(\alpha)$, $\alpha\in\valos$   form an overcomplete basis for the null space
of the Lindbladian, which has dimension $L+1$ in a finite volume $L$. This fact can be demonstrated by expanding 
$\rho_\cc(\alpha)$ into a power series in $\alpha$: which yields  the powers of $Q_\cc$ (up to corrections to the
order $\cc^L$). Together with the identity operator, these powers span a space of dimension $L+1$.
\\
Previous studies in the literature have identified steady states represented in MPO form. These instances primarily focused on systems with boundary driving,  \cite{prosen-boundary-lindblad-1, enej-prosen-spin1degness, prosen-exterior-lindblad, prosen-boundary-lindblad-2}.  Our
results are unique because 
we treat a system locally driven in the bulk, and the bond dimension of the MPO is a fixed small number (4 in our case). The operator space
entanglement satisfies an area law.

\paragraph{Frustration free property}
Remarkably, the $\rho_{\cc}(\alpha)$ are related to frustration-free Hamiltonians.  To illustrate this point, we introduce an auxiliary Hermitian superoperator that acts on $\rho$ as follows
\begin{equation}
  M\rho=\sum_j  l_{j,j+1,j+2}  \rho l_{j,j+1,j+2}^\dagger\,.
\end{equation}
In our case, considering $l$ given in \eqref{lsigma}, the strong symmetry and the relations \eqref{relationlop} imply that
$\rho_{\cc}(\alpha)$ are eigenvectors of $M$ with eigenvalue $L$, and that $L$ is the maximal possible
eigenvalue of $M$. By definition, this means that the superoperator $M$ is frustration-free.
\\
A related model with the frustration free property was investigated in \cite{MPS-cluster-model} (see also \cite{frustration-witten}). Their Hamiltonian acts
on the spin-1/2 Hilbert space and it can
be written as
\begin{equation}
  K=\sum_j l_{j,j+1,j+2},
  \label{FFK}
\end{equation}
where $l_{j,j+1,j+2}$ is \eqref{lsigma}. This operator has two extremal states $\ket{\Psi_\pm}$ satisfying the frustration free condition $l_{j,j+1,j+2}\ket{\Psi_\pm}=\pm
\ket{\Psi_\pm}$. It follows that the density matrices $\rho_\pm=\ket{\Psi_\pm}\bra{\Psi_\pm}$ are frustration free
eigenstates of $M$. Furthermore, they are (pure) NESS for our Lindbladian, and they are reproduced by $\rho_{\cc}(\alpha)$ in the
$\alpha\to \pm\infty$ limit. Our procedure to obtain the density matrices $\rho_{\cc}(\alpha)$ can be seen as a generalization of
the methods of \cite{MPS-cluster-model,frustration-witten} to the Lindbladian setting.
\paragraph{Lindbladian scars}
The Hamiltonian $\mathbb{H}_3$ discussed in the previous chapter is related to the Lindbladian we considered here. In particular, we can equivalently write $\mathbb{H}_3$ as
\begin{equation}
\tilde\La\rho\equiv(U^{-1}\La+L)\rho=M\rho+i\, U^{-1}[\rho,H].
\end{equation}

 $\tilde\La$ is Hermitian\footnote{We would like to clarify that we are referring to the Hermiticity property of the action of the superoperator $\tilde{\mathcal{L}}$. This is not the same as Hermiticity or anti-Hermiticity of the commutator $[\rho,H]$.} for $U=iu$, $u\in\valos$. In such a case $\rho_{\cc}(\alpha)$ are still
eigenoperators of $\tilde\La$, they have low spatial entanglement, and they are in the middle of the spectrum for a
generic real $u$. For Hermitian operators, states that exhibit such properties are called quantum many body scars, 
\cite{hfrag-review,fragmentation-scars-review-2}. We propose\footnote{We would like to point out that we are the first to introduce the term "Lindbladian scars". We chose this name because our model arises in the context of open quantum systems, for which the dynamics is governed by the Lindbladian $\mathcal{L}$. As discussed in section \ref{mappinglindbladhermitian}, $\mathcal{L}$ can be mapped to an Hermitian operator ($\tilde{\mathcal{L}}$). The NESS for  $\mathcal{L}$ become Lindbladian scars for $\tilde{\mathcal{L}}$. These states are characterized by low spatial entanglement and they are in the middle of the spectrum.} to call them {\it Lindbladian scars} for our original superoperator
$\La$.  

\paragraph{Mean values} The physical properties of $\rho_{\cc}(\alpha)$ can be demonstrated by computing the mean values of local observables in these states, which can be done using standard MPO techniques. In particular, we are interested in computing the mean value of two operators
\begin{itemize}
\item the magnetization operator $Z_j$
\item the operator $X_jX_{j+1}-Y_jY_{j+1}=2(\sigma^+_j\sigma^+_{j+1}+\sigma^-_j\sigma^-_{j+1})$ (a measure of the breaking of the $U(1)$ symmetry).
\end{itemize}
We start by computing the following objects
\begin{align}
\mathcal{G}(\alpha,\beta) &= \mathrm{Tr} (e^{\alpha Q_0} T(\cc)^{\dagger} e^{\beta Q_0} T(\cc)),  \\
\widetilde{\mathcal{G}}(\alpha,\beta) &= \mathrm{Tr} (e^{\alpha Q_0} T(\cc)^{-1} e^{\beta Q_0} T(\cc)) \,.
\end{align}
All the quantities of interests can be expressed in term of them.
\\
Let us start with $\mathcal{G}(\alpha,\beta)$. Using the MPO formalism above, we can rewrite: 
\begin{align}
\mathcal{G}(\alpha,\beta) = \mathrm{Tr}_{0 \otimes 0} \prod_{j=L}^1  \mathrm{Tr}_j \left(  {\mathcal{M}}_j^{(\alpha,\beta)}(\cc)    \right) \,,
\end{align}
where ${\mathcal{M}}_j^{(\alpha,\beta)}(\cc) \equiv {\mathcal{M}}_j^{(\alpha)}(\cc) e^{\beta Z_j} $ and $\mathcal{M}^{(\alpha)}_j = \cosh\alpha \mathcal{M}_j+  \sinh\alpha \mathcal{M}^Z_j$. After rotation
  \begin{align}
 \label{Malphaconjugated}
& V_0 \mathcal{M}_j^{(\alpha)}(\cc) V_0^{-1} =\nonumber \\
& \left(
\begin{array}{cccc}
\cosh\alpha & \cosh\alpha\, \sqrt{1-\cc^2} X_j & -i \cosh\alpha\, Y_j & - \cosh\alpha\, \sqrt{1-\cc^2} Z_j  \\
0 & \cc \cosh\alpha\, Z_j  & 0 & \cc \cosh\alpha \, X_j \\
i \,\cc \sinh\alpha\,  Y_j  & 0  &  \cc \sinh\alpha  & 0 \\
- \sinh\alpha\, \sqrt{1-\cc^2}  & - \sinh\alpha\, X_j & i\,\sinh\alpha\, \sqrt{1-\cc^2}  Y_j &  \sinh\alpha \,Z_j  
\end{array} 
  \right) \,,
 \end{align}
and
  \begin{align}
 \label{Malphaexpbeta}
&\mathrm{Tr}_j\left( V_0 \mathcal{M}_j^{(\alpha,\beta)}(\cc)  V_0^{-1} \right) = \nonumber\\
&\left(
\begin{array}{cccc}
 2 \cosh \alpha  \cosh \beta  & 0 & 0 & -2 \sqrt{1-\cc ^2} \cosh \alpha  \sinh \beta  \\
 0 & 2 \cc  \cosh \alpha  \sinh \beta  & 0 & 0 \\
 0 & 0 & 2 \cc  \sinh \alpha  \cosh \beta  & 0 \\
 -2 \sqrt{1-\cc ^2} \sinh \alpha  \cosh \beta  & 0 & 0 & 2 \sinh \alpha  \sinh \beta  \\
\end{array}
\right) \,.
 \end{align}
 
The computation of $\mathcal{G}(\alpha,\beta)$ can be performed by diagonalizing \eqref{Malphaexpbeta} in the auxiliary space, leading to: 
 \begin{equation}
 \mathcal{G}(\alpha,\beta) =   (\lambda_1(\alpha,\beta))^L + (\lambda_2(\alpha,\beta))^L+ (\lambda_3(\alpha,\beta))^L+ (\lambda_4(\alpha,\beta))^L \,,  
 \end{equation} 
 where 
\begin{align}
\lambda_1(\alpha,\beta) &= \cosh (\alpha +\beta )+\sqrt{\cosh ^2(\alpha +\beta )-\cc ^2 \sinh (2 \alpha ) \sinh (2 \beta )}\,,
   \cr 
\lambda_2(\alpha,\beta) &= \cosh (\alpha +\beta )-\sqrt{\cosh ^2(\alpha +\beta )-\cc ^2 \sinh (2 \alpha ) \sinh (2 \beta )}\,,
   \cr
\lambda_3(\alpha,\beta) &= 2 \cc  \sinh \alpha  \cosh \beta \,,
   \cr
\lambda_4(\alpha,\beta) &= 2 \cc  \cosh \alpha  \sinh \beta \,,
\label{lambdaeigenvalues}
\end{align} 
are the eigenvalues of \eqref{Malphaexpbeta}.
For later use, we evaluate the function $\mathcal{G}(\alpha,\beta)$ and its derivatives at particular points: 
\begin{align}
\mathcal{G}(0,\beta) &= 2^L(\cosh \beta ^L + (\cc \sinh\beta)^L) ,  \cr 
\frac{1}{ L}\partial_{\alpha}\mathcal{G}(\alpha,\beta)|_{\alpha=0} &= \left(1-\cc ^2\right) 2^L  \tanh \beta  \cosh ^L\beta , \cr
\mathcal{G}(-i \pi/2,\beta) &= (-2 i)^L \left(\cc ^L \cosh ^L\beta +\sinh ^L\beta \right), \cr
\frac{1}{L}\partial_{\alpha}\mathcal{G}(\alpha,\beta)|_{\alpha=-i \pi/2} &= \left(1-\cc ^2\right) (-2 i)^L  \coth \beta  \sinh ^L\beta.
\label{eigenvaluesparticularpoints}
\end{align}

We now turn to $\widetilde{\mathcal{G}}(\alpha,\beta)$. Using \eqref{Tgammam1}, we have 
\begin{align}
\widetilde{\mathcal{G}}(\alpha,\beta) &=\mathrm{Tr} \Big(e^{\alpha Q_0} \frac{\id-\cc^L \mathcal{Z}}{1-\cc^{2L}} T(\cc)^{\dagger}  e^{\beta Q_0} T(\cc)\Big)
=
\frac{\mathcal{G}(\alpha,\beta)-(i\cc)^L \mathcal{G}(\alpha-i \pi/2,\beta)}{1-\cc^{2L}} \,,
\label{expressionG}
\end{align}
where in the last equality we have used the identity $Z_j e^{ \alpha Z_j} = i e^{(\alpha-i \pi/2)Z_j}$.
We can now use these quantities to compute the mean values of $Z_j$ and $X_j X_{j+1} - Y_j Y_{j+1}$.\\
Using translation invariance, we find  that the mean value of the operator $Z_j$ in the state $\rho_\cc(\beta)$ is
\begin{align}
\langle Z_j \rangle_\beta  \equiv \frac{\mathrm{Tr}(Z_j T(\cc)^\dagger e^{\beta Q_0} T(\cc)  )}{\mathrm{Tr}( T(\cc)^\dagger e^{\beta Q_0} T(\cc)  )} 
= \frac{\frac{1}{L}\partial_{\alpha}\mathcal{G}(\alpha,\beta)|_{\alpha=0}}{\mathcal{G}(0,\beta)} =  \frac{(1-\cc^2)\tanh\beta}{1+ (\cc \tanh \beta)^L} \,.
\end{align}

In the large volume limit\footnote{We recall that we restricted to the case $0<\kappa <1$.}, this gives
\begin{equation}
\vev{Z_j}_{L\to\infty}= \left(1-\cc ^2\right)  \tanh \beta .
\end{equation}
In the un-deformed model ($\cc=0$) the mean value is $\tanh\beta$, thus the transformation
\eqref{rhodef} decreases the mean value by a factor that depends only on $\cc$.
\\
The mean value of $X_j X_{j+1} - Y_j Y_{j+1}$ could be similarly obtained from suitably defined generating functions, but we compute it directly here:   
\begin{align}
&\mathrm{Tr}(X_j X_{j+1} T(\cc)^\dagger e^{\beta Q_0} T(\cc)  )
=\non\\
&\,\,\,\,\,\,\,\,\,\,\,\,\,\,\, \mathrm{Tr} \left( \mathrm{Tr}_{0\otimes 0} ( \mathcal{M}^{(0,\beta)}_{L}(\cc) \ldots \mathcal{M}^{(X,\beta)}_{j+1}(\cc)\mathcal{M}^{(X,\beta)}_{j}(\cc)  \ldots  \mathcal{M}^{(0,\beta)}_1(\cc) ) \right)\,,
\end{align} 
where $\mathcal{M}^{(X,\beta)}_{j} = X_j \mathcal{M}^{(0,\beta)}_{j} = \mathcal{M}^{X}_{j} e^{\beta Z_j}$. A similar formula holds with $X\to Y$.

Using
\begin{align}
\mathrm{Tr}_j (V_0 \mathcal{M}^{(0,\beta)}_{j} V_0^{-1}) &=
\left(
\begin{array}{cccc}
 2   \cosh \beta  & 0 & 0 & -2 \sqrt{1-\cc ^2} \sinh \beta  \\
 0 & 2 \cc   \sinh \beta  & 0 & 0 \\
 0 & 0 & 0  & 0 \\
 0  & 0 & 0 & 0 \\
\end{array}
\right)  \,, 
\\
\mathrm{Tr}_j (V_0 \mathcal{M}^{(X,\beta)}_{j} V_0^{-1}) &=
 \left(
\begin{array}{cccc}
0 & 2\cc \sinh\beta  & 0 & 0 \\
2\cosh\beta & 0 & 0 & - 2\sqrt{1-\cc^2} \sinh\beta  \\
0 & 0 & 0 & 0 \\
0 & 0 & 0 & 0 
\end{array} 
  \right) \,, 
\\
\mathrm{Tr}_j (V_0 \mathcal{M}^{(Y,\beta)}_{j} V_0^{-1}) &=
 \left(
\begin{array}{cccc}
0 & 0 & 0 & 0 \\
0 & 0 & 0 & 0 \\
2 i\sqrt{1-\cc^2} \cosh\beta  & 0 &  0 & -2i \sinh\beta  \\
 0  & 0  & -2i \cc  \cosh\beta & 0 \\
\end{array} 
  \right) \,,
\end{align}
we find 
\begin{align}
&\langle X_j X_{j+1}-Y_j Y_{j+1} \rangle_\beta  = \frac{\mathrm{Tr}((X_j X_{j+1}-Y_j Y_{j+1}) T(\cc)^\dagger e^{\beta Q_0} T(\cc)  )}{\mathrm{Tr}( T(\cc)^\dagger e^{\beta Q_0} T(\cc)  )} \\
& = \frac{\cc  \left(2-\cc ^2\right) \tanh\beta +(\cc  \tanh\beta )^{L-1}}{\cc ^L \tanh ^L\beta +1} \,.
\end{align}

The infinite volume limit becomes
\begin{equation}
\vev{X_jX_{j+1}-Y_{j}Y_{j+1}}_{L\to\infty}=\cc(2-\cc^2)\tanh(\beta).
\end{equation}
Having a non-zero mean value for $\cc\neq 0$ is a clear sign of the breaking of the original $U(1)$ symmetry. 

\subsection{Dynamics and Gibbs ensemble}
\begin{figure}[h!]
  \centering
  \includegraphics[scale=0.6]{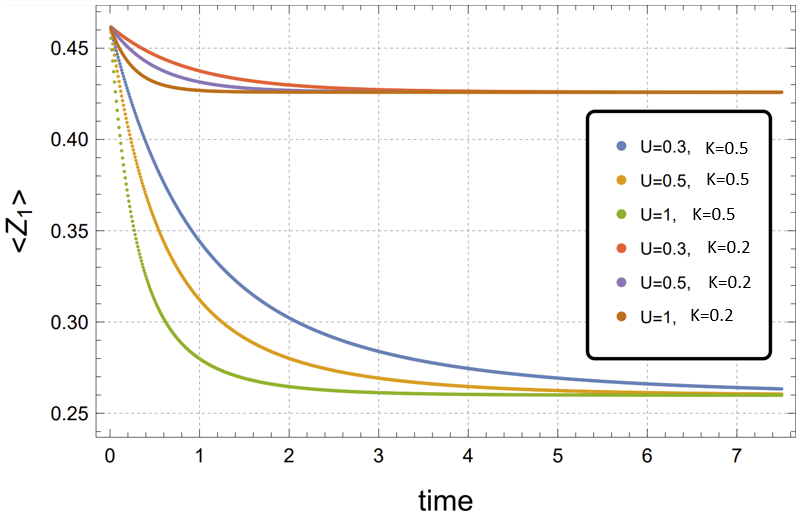}
  \caption{Time evolution of $\vev{Z_1(t)}$ from a selected initial density matrix $\rho_0(\beta)$ \eqref{rho0} with 
    $\beta=0.5$, in a finite volume $L=7$. We choose two 
    different deformation parameters $\cc$ and three coupling strengths $U$. It is seen that the asymptotic values
    depend only on $\cc$ and not on $U$, which influences only the speed of convergence. The asymptotic values agree
    with those predicted by the exact formula \eqref{Q0exact}, therefore they also confirm our postulate about the
    emergence of the Gibbs ensemble.}
  \label{fig:time}
\end{figure}

Now, we consider the real time evolution from selected initial states,
\begin{equation}
  \label{rho0}
\rho(t=0)=  \rho_0(\beta)\equiv \frac{e^{\beta Q_0}}{(2\cosh \beta)^L}.
\end{equation}
These are the (normalized) steady states \eqref{stateshubbard} of the undeformed model ($\cc=0$). They are product operators in real space and in the limit $\beta\to\pm\infty$ they
also include pure states obtained from the reference states with all spins up/down.
\\
We have confirmed that $Q_\cc$ acts as a strong symmetry: both as a symmetry and a conserved charge. As a result, the system retains memory, implying that the long-term dynamics of observables will be affected by the initial state. Additionally, since the resulting non-equilibrium steady states (NESS) are strong symmetries of the model, they are unaffected by the coupling parameter $U$. Thus, we expect that $U$ only influences the rate at which the system converges towards these NESS. This expectation is supported by numerical simulations of real-time dynamics for small system sizes, as shown in Fig. \ref{fig:time}.

 In our Lindblad system, we have a single extensive conserved charge $Q_\cc$. In \textit{analogy} with thermalization in isolated systems, we \textit{postulate} that in large volumes, the resulting steady states can be characterized by a Gibbs ensemble of the following form:
\begin{equation}
  \rho_{\rm G}\sim e^{-\lambda Q_\cc}.
\end{equation}
\\

The conservation of $Q_\cc$ implies that if a Gibbs ensemble $\rho_G\sim e^{-\lambda Q_{\cc}}$ emerges during time evolution, then it has to satisfy
\begin{equation}
  \frac{\text{Tr}\Big(\rho_0Q_\cc\Big)}{\text{Tr}(\rho_0)}=\frac{  \text{Tr}\Big(e^{-\lambda Q_\cc}Q_\cc\Big)}{  \text{Tr}\Big(e^{-\lambda Q_\cc}\Big)}.
\end{equation}
This equation can be used to fix $\lambda$ using knowledge of the initial state.\\
For the initial density matrix \eqref{rho0}, this computation can be performed easily in the infinite volume limit, yielding $\Tr(\rho_0 Q_\cc)=\tanh\lambda=-(1-\cc^2)\tanh\beta$. This result can be used to compute mean values of local observables in the Gibbs Ensemble. We obtain for example the prediction
\begin{equation}
  \label{gibbsres}
\lim_{t\to\infty}  \vev{Z_j(t)}=\text{Tr}\big(\rho_{\rm G} Z_j\big)=-\big(1-\cc^2 \big)\tanh \lambda=(1-\cc^2)^2 \tanh \beta .
\end{equation}

Remarkably, we also performed an exact finite volume computation to find the asymptotic mean values. For the
observable $Z_j$ we find
\begin{align}
\lim_{t\to\infty}\langle Z_j \rangle &= \frac{(\cc ^2-1)^2  \tanh \beta  (1-2 \cc ^L \tanh ^{L-2}\beta +\cc ^{2L})}{(1-\cc ^{2 L})^2} .\,
\label{Q0exact} 
\end{align}
Due to the extensive nature of this calculation, we  report the details in the Appendix \ref{meanvalcompexact}.\\We also confirm these values   by  numerical computation for finite $L$. Furthermore, by considering the large volume limit, and $0<\cc<1$, we recover \eqref{gibbsres}, thus confirming our postulate about the Gibbs Ensemble.

\section{Summary} In this chapter, we have demonstrated that a system with local jump operators can exhibit quasi-local symmetries, which have a profound impact on the real-time dynamics. Our discovery reveal that the NESS of our model can be derived from the NESS of the "Hubbard Lindbladian" through a similarity transformation using a Matrix Product Operator. Notably, this transformation is surprisingly compatible with the local jump operators. It is intriguing to note that the integrability property of the Lindbladians was not utilized in computing the analytical expression of the NESS. However, the superintegrability of the Hamiltonian played a crucial role. One possible place where the role of integrability is hidden is in the derivation of the equation \eqref{Q0exact} to compute the exact finite volume computation of the magnetization.

\chapter{Four dimensional Local Hilbert space and $\alg{su}(2)\oplus \alg{su}(2)$ symmetry}

\ifpdf
    \graphicspath{{Chapter5/Figs/Raster/}{Chapter5/Figs/PDF/}{Chapter5/Figs/}}
\else
    \graphicspath{{Chapter5/Figs/Vector/}{Chapter5/Figs/}}
\fi
\label{threeorfour}

In this chapter, we apply the boost automorphism method to classify integrable models where the local Hilbert space is of dimension 4. A priori,  the Hamiltonian ansatz involves 256 free functions, making it a challenging task to solve the coupled systems of differential equations using current methods. To overcome this, we limit our focus to models with $\alg{su}(2)\oplus \alg{su}(2)$ symmetry and with a non-difference form $R$-matrix.
The reason for this ansatz is that many interesting known physical models such as the Hubbard model (discussed in chapter \ref{Hubbardchapter}), exhibit this symmetry. 
We discover five models which, based on our current knowledge, are new. Furthermore, we demonstrate that the Hubbard model can be classified within our framework.

\section{Different embedding}

To start, we explain how to construct the ansatz for the Hamiltonian. In this section, we closely follow the discussion of the paper \cite{deLeeuw:2019vdb}. In their work, the authors classified models exhibiting $\alg{su}(2)\oplus \alg{su}(2)$ symmetry but characterized by $R$-matrix of  difference form. In the work \cite{classificationybandboost}, we extended it to the non-difference form case.\\
Our interest lies in models where the Hamiltonian exhibits a $\alg{su}(2)\oplus\alg{su}(2)$ symmetry. Given that our analysis is limited to a local Hilbert space with dimension four, we need to systematically classify all four-dimensional representations of this semi-simple Lie algebra. \\
First, we consider the first copy of $\alg{su}(2).$ There are 5 possible four dimensional representation of $\alg{su}(2)$, that are $1\oplus 1\oplus 1\oplus 1, 2\oplus 1 \oplus 1, 2 \oplus 2, 3 \oplus 1, 4$. Here, the numbers indicate the dimensions of each representation within the direct sum.  As we are focusing on non-trivial representations, we exclude $1\oplus 1\oplus 1\oplus 1$. \\After having found the explicit matrix representations of the remaining four cases, we focus on the second copy of $\alg{su}(2)$. For each of these cases, we  consider three general $4\times 4$ matrices and impose two conditions: firstly, that these matrices form an $\alg{su}(2)$ algebra, and secondly, that they commute with the representation of the first $\alg{su}(2)$ factor.  This is enough, up to some trivial similarity transformations, to uniquely fix the representation of the second $\alg{su}(2)$ factor. For this reason, it is enough to label the representation of $\alg{su}(2)\oplus \alg{su}(2)$ by the representation of only one of the factor.\\
If we fix the representation $3\oplus 1$ for the first factor, we find that the representation of the second one is $1\oplus 1\oplus 1\oplus 1$, hence trivial. We do not consider this case.\\
Furthermore, due to the isomorphism $\alg{so}(4) \sim \alg{su}(2)\oplus\alg{su}(2)$, we can discard the representation 4 since it is not independent from $2\oplus 2$. We are only left with $2\oplus 2$ and $2\oplus 1 \oplus 1$, that are independent and non-trivial. The Hubbard model falls into the second class. We remark that in chapter \ref{Hubbardchapter}, we have shown that the Hubbard model has this symmetry. In the current context, we aim to \textit{construct the most general} form of the Hamiltonian dependent on one spectral parameter that exhibits the same symmetry.

We label $t_i^{L/R}$ the first ($L$) and the second ($R$) set of generators in $\alg{su}(2)\oplus \alg{su}(2)$. They satisfy 
\begin{align}
[t_i^{L/R},t_j^{L/R}] = \epsilon_{ijk} t_k^{L/R}.
\end{align}

We denote with $\rho_n(t_i^{L/R})$ the $n$-dimensional irreducible representation of the generators $t_i^{L/R}$. We obtain the following representations.
\paragraph{$2\oplus 1 \oplus 1$}
The two dimensional representation is embedded as a direct sum, in particular
\begin{align}
&\rho_{2\oplus 1 \oplus 1}(t_i^L)=\rho_2(t_i^L)\oplus 0 = \left(
\begin{array}{cc}
 \rho_2 (t_i^L)  & 0 \\
 0 & 0 \\
\end{array}
\right)
\end{align}
and consequently,
\begin{align}
&\rho_{2\oplus 1 \oplus 1}(t_i^R)=0 \oplus \rho_2(t_i^R)= \left(
\begin{array}{cc}
 0  & 0 \\
 0 & \rho_2 (t_i^R) \\
\end{array}
\right) .
\end{align}
This translates to
\begin{align}
&\rho_{2\oplus 1 \oplus 1}(t_i^L \times t_i^R)=\rho_2(t_i^L) \oplus \rho_2(t_i^R).
\end{align}
We fix our ansatz by imposing that for each $\mathcal{A} \in \alg{su(2)}\oplus \alg{su}(2)$, the Hamiltonian density $\mathcal{H}$ satisfies
\begin{align}
[\mathcal{H},\rho_{2\oplus 1 \oplus 1}(\mathcal{A})\otimes 1+1 \otimes \rho_{2\oplus 1 \oplus 1}(\mathcal{A})]=0 .
\end{align}
The Hubbard model belong to this class and the left and right factors correspond to the spin and charge $\alg{su}(2)$ symmetries.
\paragraph{ $2\oplus 2$} In this case, the two dimensional representation is embedded diagonally
\begin{align}
&\rho_{2\oplus 2}(t_i^L)=1\otimes \rho_2(t_i),\\
&\rho_{2\oplus 2}(t_i^R)=\rho_2(t_i)\otimes 1.
\end{align}
This representation is isomorphic to $\alg{so}(4)$ via the basis transformation
\begin{align}
U=\left(
\begin{array}{cccc}
 -i & 0 & 0 & i \\
 1 & 0 & 0 & 1 \\
 0 & i  & i  & 0 \\
 0 & 1 & -1 & 0 \\
\end{array}
\right).
\end{align}
Since the two representations are isomorphic, we only consider $2\oplus 2$.

\vspace{1cm}
We start now to illustrate the integrable models of non-difference form type that we found for the two cases $2\oplus 2$ (we refer to them as $\alg{so}(4)$ type) and $2\oplus 1 \oplus 1$ (2-D representations).

\section{$\alg{so}(4)$ type models}

Fixing
\begin{align}
[\mathcal{H}_{12},\rho_{2\oplus 2}(\mathbf{A})\otimes 1+1 \otimes \rho_{2\oplus 2}(\mathbf{A})]=0 ,
\end{align}
the most general Hamiltonian density underlying this symmetry takes the form
\begin{align}
\lH(\theta) = h_1(\theta) \id+ h_2(\theta) P+ h_3(\theta) K +h_4(\theta) \epsilon_{ijkl}E_{ik} \otimes E_{jl},
\end{align}
where $ \id $ is the $16\times 16$ identity matrix, $P$ is the permutation operator and $K =E_{ij} \otimes E_{ij}$ with $ (E_{ij})_{\alpha\beta}=\delta_{i,\alpha} \delta_{j,\beta}$. Summation over repeated indices is assumed and $ i,j,k,l=1,...,4 $.
\subsubsection{Integrable Hamiltonian}
Using the boost mechanism, we found only one possible integrable Hamiltonian of non-difference form characterized by 
\begin{align}
\lH(\theta)=h_1(\theta) \id+ h_2(\theta) (P - K) +  h_4(\theta) \;\epsilon_{ijkl}\;E_{ik} \otimes E_{jl}.
\label{so4spinchain}
\end{align}
For $h_4(\theta)=0$, this is the non-difference form model corresponding to the usual $\alg{so}(n)$ spin chain with $n=4$, but here the constant coefficients become functions of the spectral parameter. The term proportional to $h_4(\theta)$ is related to the fact that for $n=4$ there is an extra invariant contraction where all indices are contracted with the Levi-Civita symbol. The difference form model corresponding to this one is model 13 of \cite{deLeeuw:2019vdb}. For that model, $h_1,\, h_2,\, h_4$ are constants. However, we remark that if we start from a difference form integrable model and we promote the coefficient to be functions, integrability via the boost method is not guaranteed, see discussion in section \ref{whyisthemodelint}.

\subsection{$R$-matrix}
The $R$-matrix corresponding to \eqref{so4spinchain} can be found by following section \ref{RfromHamiltonian} and it is given by
\begin{align}
&R=e^{H_1(u,v)}\!\bigg[\!\Big(H_2(u,v)-\frac{H_4(u,v)^2}{H_2(u,v)+1}\Big) \id+ P-\frac{H_2(u,v)K-H_4(u,v)\epsilon_{ijkl}E_{ik} \otimes E_{jl} }{H_2(u,v)+1} \bigg],
\label{Rso4}
\end{align}
where $H_i(u,v) = \int^u_v h_i$. Notice that this model is indeed manifestly of non-difference form. Here, we did not use any of the freedom given by the identification (see section \ref{identification}). One function can be absorbed into a normalization ($H_1$), and one can be used in a reparametrization, leaving us with one additional free function.  Since the dependence of the spectral parameters always appear in the form $H_i(u,v)=H_i(u)-H_i(v)$, we can refer to this model as \textit{quasi-difference form}. 

\section{$\su(2)\oplus \su(2)$ symmetry with 2-D representations}
\label{su(2)dpsu(2)}

In this class of model, we consider the four-dimensional representation of $\alg{su}(2)\oplus\alg{su}(2)$ in which both $\alg{su}(2)$'s have two-dimensional representation. This class includes the characteristic spin and charge $\alg{su}(2)$ symmetry of the Hubbard model.

\subsection{Ansatz for the Hamiltonian and $R$-matrix} It is straightforward to show that an $\su(2)\oplus \su(2)$ invariant Hamiltonian  density takes the form
\begin{align}
\lH |\phi_a \phi_b \rangle &= h_1  |\phi_a \phi_b \rangle + h_2  |\phi_b \phi_a \rangle + h_3 \epsilon_{ab}\epsilon_{\alpha\beta} |\psi_\alpha \psi_\beta \rangle, \label{Hphiphi}\\
\lH |\phi_a \psi_\beta \rangle &= h_4   |\phi_a \psi_\beta \rangle + h_5  |\psi_\beta \phi_a \rangle, \label{Hphipsi} \\
\lH |\psi_\alpha \phi_b \rangle &= h_6  |\psi_\alpha \phi_b \rangle + h_{7}  |\phi_b \psi_\alpha \rangle,  \label{Hpsiphi}\\
\lH |\psi_\alpha \psi_\beta \rangle &= h_8  |\psi_\alpha \psi_\beta \rangle + h_9  |\psi_\beta \psi_\alpha \rangle + h_{10} \epsilon_{ab}\epsilon_{\alpha\beta} |\phi_a \phi_b \rangle\label{Hpsipsi},
\end{align}
where $a, b, \alpha, \beta= 1,2$; $\phi_{1,2}$ and $\psi_{1,2}$ span the two independent $\alg{su}(2)$ fundamental representations. 
If we use the canonical basis $e_i, i=1,\dots,4$, with $1$ in position $i$ of the vector, and $0$ otherwise, and we identify
\begin{align}
&\ket{\phi_1}=e_1,
&&\ket{\phi_2}=e_4,
&&\ket{\psi_1}=e_3,
&&\ket{\psi_2}=e_2,
\end{align}
the Hamiltonian is 
\setcounter{MaxMatrixCols}{20}
\begin{align}\label{eq:Hsu2}
\lH = \tiny{
\begin{pmatrix}
h_1+h_2 & 0 & 0 & 0 & 0 & 0 & 0 & 0 & 0 & 0 & 0 & 0 & 0 & 0 & 0 & 0 \\
 0 & h_4 & 0 & 0 & h_7 & 0 & 0 & 0 & 0 & 0 & 0 & 0 & 0 & 0 & 0 & 0 \\
 0 & 0 & h_4 & 0 & 0 & 0 & 0 & 0 & h_7 & 0 & 0 & 0 & 0 & 0 & 0 & 0 \\
 0 & 0 & 0 & h_1 & 0 & 0 & -h_{10} & 0 & 0 & h_{10} & 0 & 0 & h_2 & 0 & 0 & 0 \\
 0 & h_5 & 0 & 0 & h_6 & 0 & 0 & 0 & 0 & 0 & 0 & 0 & 0 & 0 & 0 & 0 \\
 0 & 0 & 0 & 0 & 0 & h_8+h_9 & 0 & 0 & 0 & 0 & 0 & 0 & 0 & 0 & 0 & 0 \\
 0 & 0 & 0 & -h_3 & 0 & 0 & h_8 & 0 & 0 & h_9 & 0 & 0 & h_3 & 0 & 0 & 0 \\
 0 & 0 & 0 & 0 & 0 & 0 & 0 & h_6 & 0 & 0 & 0 & 0 & 0 & h_5 & 0 & 0 \\
 0 & 0 & h_5 & 0 & 0 & 0 & 0 & 0 & h_6 & 0 & 0 & 0 & 0 & 0 & 0 & 0 \\
 0 & 0 & 0 & h_3 & 0 & 0 & h_9 & 0 & 0 & h_8 & 0 & 0 & -h_3 & 0 & 0 & 0 \\
 0 & 0 & 0 & 0 & 0 & 0 & 0 & 0 & 0 & 0 & h_8+h_9 & 0 & 0 & 0 & 0 & 0 \\
 0 & 0 & 0 & 0 & 0 & 0 & 0 & 0 & 0 & 0 & 0 & h_6 & 0 & 0 & h_5 & 0 \\
 0 & 0 & 0 & h_2 & 0 & 0 & h_{10} & 0 & 0 & -h_{10} & 0 & 0 & h_1 & 0 & 0 & 0 \\
 0 & 0 & 0 & 0 & 0 & 0 & 0 & h_7 & 0 & 0 & 0 & 0 & 0 & h_4 & 0 & 0 \\
 0 & 0 & 0 & 0 & 0 & 0 & 0 & 0 & 0 & 0 & 0 & h_7 & 0 & 0 & h_4 & 0 \\
 0 & 0 & 0 & 0 & 0 & 0 & 0 & 0 & 0 & 0 & 0 & 0 & 0 & 0 & 0 & h_1+h_2
\end{pmatrix}},
\end{align}

where the $h_i$'s are dependent on the spectral parameter $\theta$.
\\
Similarly, since the $R$-matrix carries the same symmetry of the Hamiltonian, we write
\begin{align}
R |\phi_a \phi_b \rangle &= r_1  |\phi_a \phi_b \rangle + r_2  |\phi_b \phi_a \rangle + r_3 \epsilon_{ab}\epsilon_{\alpha\beta} |\psi_\alpha \psi_\beta \rangle, \label{Rphiphi}\\
R |\phi_a \psi_\beta \rangle &= r_4   |\phi_a \psi_\beta \rangle + r_5  |\psi_\beta \phi_a \rangle, \label{Rphipsi} \\
R |\psi_\alpha \phi_b \rangle &= r_6  |\psi_\alpha \phi_b \rangle + r_{7}  |\phi_b \psi_\alpha \rangle,  \label{Rpsiphi}\\
R|\psi_\alpha \psi_\beta \rangle &= r_8  |\psi_\alpha \psi_\beta \rangle + r_9  |\psi_\beta \psi_\alpha \rangle + r_{10} \epsilon_{ab}\epsilon_{\alpha\beta} |\phi_a \phi_b \rangle\label{Rpsipsi},
\end{align}
where $r_i=r_i(u,v)$. The matrix representation can be easily obtained by substituting in \eqref{eq:Hsu2} $h_i \to r_i$.
\\
We remark that in chapter \ref{Hubbardchapter}, we wrote the Hubbard model Hamiltonian in terms of $\sigma$ and $\tau$ matrices. The two approaches are in fact equivalent. One can explicitly write the Hamiltonian \eqref{eq:Hsu2} in terms of $\sigma$ and $\tau$ matrices by using \eqref{notation} and perform a Jordan-Wigner transformation to map to the basis of fermionic oscillators ${c}_{j}^\alpha,({c}_{j}^{\alpha})^\dagger$ for each lattice site $j$ with $	\alpha=\uparrow,\downarrow$, satisfying the  anti-commutation relations. The Jordan-Wigner transformation implements the grading on the Hilbert space. In this basis, each 4-dimensional Hilbert space is spanned by
\begin{align}\label{phipsi4dim}
&\ket{\phi_1}=\ket{0},
&&\ket{\phi_2}= ({c}_{j}^\uparrow)^\dagger({c}_{j}^\downarrow)^\dagger \ket{0}=\ket{\updownarrow},
&&\ket{\psi_1}=({c}_{j}^\uparrow)^\dagger \ket{0}=\ket{\uparrow},
&&\ket{\psi_2}=({c}_{j}^\downarrow)^\dagger \ket{0}=\ket{\downarrow},
\end{align}

with $\ket{0}$ the vacuum state satisfying ${c}_{j}^\alpha\ket{0}=0.$ We can then identify $\ket{\phi_1}$ and $\ket{\phi_2}$ as even basis vectors and $\ket{\psi_1}$ and $\ket{\psi_2}$ as odd basis. So for example, the terms $h_5$ and $h_7$ in the Hamiltonian are responsible of the propagation of electrons along the chain. Working with a graded Hilbert space, the Yang-Baxter equation and also the definition of the  tensor product and the trace should be modified accordingly to the graded structure. We do not use the graded formalism in this chapter, but we refer to chapter 12 of \cite{essler2005one} and Appendix A of \cite{deLeeuw:2019vdb} for more details. 
\subsection{Integrable Hamiltonians}\label{integrhamsu22}
Following the step of section \ref{example}, we start from the ansatz for the Hamiltonian \eqref{eq:Hsu2} and we compute the corresponding density $\mathcal{Q}_3$. We then impose the integrability constraint $[\mathbb{Q}_2,\mathbb{Q}_3]=0$ and, after solving the set of equations and eliminating the equivalent models related by the identifications of section \ref{identification}, we find a total of eight integrable models of non-difference form. Six of them have $h_3=0$ and two have $h_3 \neq 0$.

\paragraph{Models with $h_3=0$}
These models are listed in Table \ref{Tab1}.
\setlength{\tabcolsep}{0.4em} 
{\renewcommand{\arraystretch}{1.6}
\begin{table}[h!]
  \begin{center}
  \resizebox{\linewidth}{!}{%
    \begin{tabular}{c||c|c|c|c|c|c|c|c|c|c|} 
      \textbf{$\lH$} & \textbf{$h_1$} & \textbf{$h_2$} & \textbf{$h_3$}& \textbf{$h_4$}& \textbf{$h_5$}& \textbf{$h_6$}& \textbf{$h_7$}& \textbf{$h_8$} & \textbf{$h_9$} &\textbf{$h_{10}$}\\
      \hline
 1 & $\frac{1}{2 \left(\theta ^2-1\right)}$ & $\frac{1}{2}$ & 0& $\frac{\theta }{1-\theta ^2}$ & $\frac{\pm 1}{2} \sqrt{\frac{\theta +1}{\theta -1}}$&$ \frac{\theta }{\theta ^2-1} $&$ \frac{\pm 1}{2} \sqrt{\frac{\theta -1}{\theta +1}}$ & $\frac{1}{2(1- \theta) ^2}$ & $\frac{-1}{2}$ & $c$ \\\hline
 2 & $f $& $h $& 0 &$g$& $\frac{c \, h}{ e^{2 F}}$ & $-g$ &$ \frac{h e^{2 F}}{c } $&$ -f $& $\pm h$ & 0 \\ \hline
 3 & $f$ & $\pm h $& 0&$ g$ & $\frac{c  h }{e^{2 F}}$ & $-g$ & $\frac{h e^{2 F} }{c} $ & $h-f $& 0 & 0 \\ \hline
 4 & $(c_1+2)f $& 0 & 0 & $c_1(f-g) $& $  \frac{c_1(c_1+2)  g}{c_2 e^{2 F}} $& $(c_1+2)\; (f-g)$ &  $ c_2 e^{2 F}g$ &$ c_1 f  $& 0 & 0\\  \hline
 5 & $f$ & 0 & 0 & 0& $g$& 0 & $h$& $-f$ & 0& 0   \\ \hline
 6 &$ f-h$ & 0 & 0 &  $f+h  $&$ \frac{2h }{c \;e^{2F}}$ & $h-f $& $2 c  h e^{2 F}$&$ h-f$ & $\pm 2 h$ & 0 \\\hline
    \end{tabular}}
  \end{center}   
  \caption{All non-difference models with $h_3=0$. $c, c_1,c_2$, are constants, $f,g,h,F$ are $\theta$ dependent functions and $F'=f$.\\}
   \label{Tab1}
\end{table}}

To our best knowledge, except from model 5, \textit{all the models are new} and we discuss some of their physical properties. Model 5 is a quadruple embedding of model 6-vertex B given in the next chapter \ref{8vertexmodels}. This can be seen by applying a constant local basis transformation\footnote{The local basis transformation is $V_j=X_j$, with $X_j$ the first Pauli matrix} to the Hamiltonian of model 6-V B after the redefinition $h_5 \to \frac{1}{2}h_4 h_5$, and by making the identifications at the level of the 6-V B model
\begin{align}
&h_3\to g,
&&h_4\to h,
&&h_4 h_5 \to -f\;.
\end{align}

We notice that, except from model 1, all the others have $h_{10}=0$, so electrons in nearbly sites of the chain cannot merge. 
\\
Furthermore, we analysed for which choices of the free functions and constants,  models 1-6 are \textbf{Hermitian}. We found that model 1 can only be made Hermitian if $c=0$ and the dependence on the spectral parameter drops out, models 2 to 6 are Hermitian if we impose some conditions on the functions, see Table \ref{Tab3}. More generally, we analysed for which choice of the functions and constants, the Hamiltonians and their Hermitian conjugate commutes\footnote{The reason for this choice is the following. If $[\mathcal{H},\mathcal{H}^\dagger]=0$, we can define an Hermitian operator $A=\mathcal{H}+\mathcal{H}^\dagger$. Although it is not guaranteed that $A$ is integrable, since $\mathcal{H}$ is integrable, we can employ one of the integrability techniques to identify a set of eigenvectors. Subsequently, we can form a linear combination of these eigenvectors in such a way that they remain eigenvectors of $A$.  This is only possible since $[A,\mathcal{H}]=[A,\mathcal{H}^\dagger]=0$.}. As expected, the conditions in this second case are more general than the previous case and are given in Table \ref{Tab4}.

\setlength{\tabcolsep}{0.4em} 
{\renewcommand{\arraystretch}{1.4}

\begin{table}[h!]
  \begin{center}
    \begin{tabular}{c||c|} 
\textbf{Model} &  \textbf{Reality conditions}\\\hline
\textbf {1}& $\theta=0,\;\; c=0$\\ \hline
\textbf {2,\;3}& $ e^{4 F^{(r)}}=|c|^2,\;\;f,\,h\, \in\, \mathbb{R}$\\ \hline
\textbf{4}&$\left(e^{4 F^{(r)}}=\frac{{c_1}^{(r)} ({c_1}^{(r)}+2)}{|c_2|^2}\;\;\text{or}\;\;e^{4 F^{(r)}}=\frac{1}{|c_2|^2},{c_1}^{(r)}=-1\right)$,\;\;$c_1,\,f,\,g\, \in\, \mathbb{R}$ \\ \hline
\textbf{5}& $g=h^*$, $f\,\in\,\mathbb{R}$\\ \hline
\textbf{6}& $e^{-4 F^{(r)}}=|c|^2$, $f,\,h\, \in\, \mathbb{R}$\\ \hline
    \end{tabular}
  \end{center}   
  \caption{Conditions required to make models 1-6 Hermitian. We use the superscript $(r)$ to indicate the real part of the functions and constants.}
  \label{Tab3}
\end{table}

\begin{table}[h!]
  \begin{center}
    \begin{tabular}{c||c|} 
\textbf{Model} &  \textbf{Unitarity conditions}\\\hline
\textbf 1  & $\theta^{(r)} = 0,\;\;c=0$\\ \hline
\textbf{2,\;3}& $ e^{4 F^{(r)}}=|c|^2$\\ \hline
\textbf{4}&$e^{4 F^{(r)}}= \frac{{{c_1}^{(i)}}^2+1}{|{c_2}|^2},{c_1}^{(r)}=-1$ or $e^{4 F^{(r)}}= \frac{{c_1}^{(r)} ({c_1}^{(r)}+2)}{|c_2|^2}, {c_1}^{(i)}=0$ or ${c_1}=-1$ \\ \hline
\textbf 5& $\forall\; f,g,h$\\ \hline
\textbf 6& $e^{-4 F^{(r)}}=|c|^2$\\ \hline
    \end{tabular}
  \end{center}   
  \caption{Conditions required to make the full Hamiltonian $\lH$ of models 1-6  satisfy $[\lH,\lH^\dagger]=0$. We use the superscripts $(r)$ and $(i)$ to indicate the real and the complex part of the functions and constants.}
  \label{Tab4}
\end{table}

\paragraph{Models with $h_3 \neq 0$}
\subparagraph{Model 7}
This model is the most general of non-difference form with $h_3\neq 0$. In fact,  model 8 can be obtained from this one by performing a double limit. In order to solve this model, we fix the normalization of the Hamiltonian such that\footnote{This choice is not restrictive. The only case in which $h_{10}$ cannot be normalized to $1$ is when it is equal to zero. Suppose that $h_{10}=0$ from the beginning, we call $\tilde \lH$ the integrable Hamiltonian corresponding to this initial ansatz. $\tilde \lH$ is equivalent to model 1 under the transformation $\left(P\;\tilde \lH \;P \right)^T$. For this reason, $h_{10}=1$ is not restrictive.} $h_{10}=1$.
We set $h_4=h_6=0$ by using the identifications  described in \ref{identification}, after which we get the following set of coupled differential equations
\begin{align}
&h_1 + h_8 = h_2+h_9 = 0,
&& h_8= \frac{(h_5+h_7)^2}{4 h_9}-h_9,
&&h_3= h_5 h_7-h_9^2,
\label{eq:model7}\\
&\dot{h}_5= 2 h_7 h_9-\frac{h_5 (h_5+h_7)^2}{2 h_9},
&&\dot{h}_7= \frac{h_7 (h_5+h_7)^2}{2 h_9}-2 h_5 h_9,
&&\dot{h}_9= h_7^2-h_5^2.
\label{eq:h5h7}
\end{align}
Summing the first two equations of \eqref{eq:h5h7} and taking into account the third one, substituting\footnote{For simplicity, we omit the dependence of $\xi_1, \xi_2, \Xi_1$ and $\Xi_2$ on the spectral parameter.}
\begin{align}
&h_5=\frac{\sqrt{\xi_1} \left({\xi_2}^2-1\right)}{\sqrt{2}\; \xi_2},
&&h_7=\frac{\sqrt{\xi_1} \left({\xi_2}^2+1\right)}{\sqrt{2}\; \xi_2},
\label{eq:subst}
\end{align}
we find that
\begin{align}
&h_9=2\Xi_1,
&&\xi_2= \sigma\frac{\sqrt{\Xi_1} \sqrt{8 \Xi_1+c_1}}{\sqrt{\xi_1}},
\end{align}
where $\Xi_i=\int \xi_i,$ $\sigma=\pm 1$ and $c_1$ is a constant.
To find $\xi_1$, we made the substitution \eqref{eq:subst} in the differential equation for $h_7$ and we get
\begin{align}
&\Xi_1 (8 \Xi_1+c_1) \left(2 \dot{\xi_1} -  c_1 \Xi_1(8\Xi_1+ c_1)\right) = \xi_1^2 (16 \Xi_1+c_1).
\label{eq:a}
\end{align}
For general $c_1$, the equation \eqref{eq:a} can be solved by performing the substitutions $\Xi_1(u)\to w(u)$, $\xi_1(u)\to \dot{w}(u)$ and $\dot{\xi}_1(u)\to \ddot{w}(u)$ and we find that the corresponding
differential equation is solved by elliptic functions
\begin{align}
\xi_1(u)= \frac{i}{8} c_1^2 \text{cs}(z|m) \text{ds}(z|m) \text{ns}(z|m),
\end{align}
where $z=\frac{i}{2}  c_1 (u + c_2)$ and $m=\frac{8 c_3}{c_1^2}$, $c_{2,3}$ are constants.
To summarize, we then get \eqref{eq:model7} together with 
\begin{align}
&h_5-h_7=i\;\frac{\sigma}{2} c_1  \text{ds}(z|m),
&&h_5+h_7=\frac{\sigma}{2} c_1 \text{nc}(z|m) \left(1-\text{ns}(z|m)^2\right),\\
&h_9=-\frac{1}{4} c_1 \text{ns}(z|m)^2.
\label{eq:generalc1}
\end{align}
The Hamiltonian found does not depend on any free functions. Since we know that the Hubbard model exhibits the same symmetry, we should prove that model 7 contains it. In order to prove this, we compared the Hamiltonian that we found with the one derived from requiring\footnote{\label{footnotedressingfactor}To be precise, we started by considering the $S$-matrix given in \cite{Arutyunov:2006yd}. This satisfies the Yang-Baxter equation. We constructed the Hamiltonian corresponding to it by using $\mathcal{H}_{12}=P_{12}\partial_u S_{12}(u,v)|_{u\to v}$, with $S_{12}$ the ungraded version of the $S$-matrix given in \cite{Arutyunov:2006yd}. From the point of view of the YBE, all normalizations are equally valid. In some cases, we can fix the right normalization by requiring some physical properties, for example crossing symmetry. For this reason, when we compared the two Hamiltonians, we find the match between the matrix part. The dressing factor (normalization) cannot be found with the boost method and the analytic expression of it is a very active field of studies of the last decades.} \textit{centrally extended} $\mathfrak{su}(2|2) $ symmetry  \cite{Beisert:2005tm}.  This spin chain have its origin in the planar AdS/CFT correspondence, and it contains the one-dimensional Hubbard model as a special case.

After using an appropriate normalization  and shift, the entries of the Hamiltonian of AdS/CFT are
\begin{align}
&h_1=-h_8,
&&h_2=-h_9,
&&h_3=-\frac{1}{\alpha ^2},
&&h_4=h_6=0,
&&h_{10}=1,
\end{align}
\begin{align}
&h_5=\frac{1-{x^-}^2}{\alpha  (x^--{x^+})}\sqrt{\frac{{x^+}}{x^-}},
&&h_7=\frac{x^+ \dot x ^-}{\dot x^+ x^-} h_5,
&&h_8= \frac{(h_5+h_7)^2}{4 h_9}-h_9,
&&h_9=\frac{1-x^- {x^+}}{\alpha ( x^--{x^+})},
\end{align}
where $\alpha$ is a free constant and $x^+$ and $x^-$ the Zhukovksy variables. $x^\pm$ can be conveniently parametrized using elliptic functions \cite{arutyunov2009foundations} as\footnote{The parameter $\hslash$ here is related to the parameter $g$ of \cite{arutyunov2009foundations} as $\hslash=\frac{2i}{g}$.}
\begin{align}
&x^\pm=-\frac{1}{4}\, i\, \hslash\,  (\text{dn}(\zeta |k)+1) \left(\text{cs}(\zeta |k)\pm i\right), \quad k=\frac{16}{\hslash^2},
\end{align}
 we indeed see that the two Hamiltonian densities are the same under 
\begin{equation}\label{mapAdSCFTtosu2su2}
\hslash\to\alpha\, c_1, \quad \alpha^2\to\frac{2}{c_3} \quad, \quad \zeta\to \frac{i}{2}\, c_1 (c_2+u), \quad \sigma=1.
\end{equation}
The other choice of $\sigma=-1$ is not independent, in fact it can be related to the previous one by a twist\footnote{One can easily see that to make the changes $h_5\to -h_5$ and $h_7\to -h_7$ in \eqref{eq:Hsu2} we can use the following constant twist
\begin{align}
V=\left(
\begin{array}{cccc}
 1 & 0 & 0 & 0 \\
 0 & 1 & 0 & 0 \\
 0 & 0 & -1 & 0 \\
 0 & 0 & 0 & 1 \\
\end{array}
\right),\; W=\left(
\begin{array}{cccc}
 1 & 0 & 0 & 0 \\
 0 & 1 & 0 & 0 \\
 0 & 0 & 1 & 0 \\
 0 & 0 & 0 & -1 \\
\end{array}
\right),\; \; \lH_{\sigma=1}=(V\otimes W) \lH_{\sigma=-1} (V \otimes W)^{-1}.
\end{align}
}. There is no extra freedom in our model,  the constant $c_2$ can be reabsorbed by shifting $u$. Remarkably, our method \textit{naturally leads to the elliptic parameterization} of the AdS/CFT $S$-matrix. 

\subparagraph{Model 8}
Model 8 is not independent from model 7, but it can be obtained as a special limit of it. However, this limit is somewhat singular, so we explicitly present this case.
If we solve \eqref{eq:a} for $c_1=0$, we get $\xi_1(u)= c_2 e^{c_3 u}$ and so
\begin{align}
&h_1=h_4=h_6=h_8=0,
&&h_2=-\frac{2 c_2 e^{c_3 u }}{c_3},
&&h_3=-\frac{c_3{}^2}{16},
&&h_9=\frac{2 c_2 e^{c_3 u }}{c_3},\label{eq:c1equal01}\\
&h_5-h_7=-\sigma\frac{c_3}{2},
&&h_5+h_7=\sigma\frac{4 c_2 e^{c_3 u }}{c_3},
&&h_{10}=1.
\label{eq:c1equal0}
\end{align}
It is interesting to notice that the limit $c_1\to 0$ in the Hamiltonian of model 7 is not well defined because some of the Jacobi functions are divergent in this limit. To find the correct result, we should take the result for general $c_1$ and then follow the steps below
\begin{itemize}
\item[1.] Use the relations that relate the Jacobi functions of modulus $k$ with the ones with modulus $1-\frac{1}{k}$ like: $\text{ns}(i\, x|k)=-i \sqrt{k} \text{cs}\left(x \sqrt{k}|1-\frac{1}{k}\right)$  
\item[2.] Expand for small $c_1$
\item[3.] Rescale $c_3\to c_1$
\item[4.] Perform a second limit for large $u$
\item[5.] Relabel  the constants $c_1$ and $c_2$ to obtain \eqref{eq:c1equal01} and \eqref{eq:c1equal0}.
\end{itemize}

\subsection{Integrable $R$-matrices}
After solving the Sutherland equations, we successfully found a unique regular $R$-matrix corresponding to each integrable Hamiltonian. Most of the equations were straightforward to solve and we do not explicitly write them here. For model 5, the second-order version of the Riccati equation appears, similar to the 6-vertex model B discussed in section \ref{8vertexmodels}.\\
We now present the entries $r_i$ given in \eqref{Rphiphi}-\eqref{Rpsipsi} of the $R$-matrices corresponding to the different integrable Hamiltonians.
\\
In the following $c, c_1, c_2$ are constants, $F_{\pm}=F(u)\pm F(v)$ and similarly for $G$ and $H$,  $r_i=r_i(u,v)$ and $\sigma=\pm 1$. Explicitly, the entries of the $R$-matrices that we got are
\subparagraph{Model 1}
\begin{align}
&r_1=\frac{- r_{10}  \sqrt{1+ v}}{2 c\sqrt{r_{5}}  \sqrt{1+ u}},
&&r_2=\frac{- r_{1}r_9}{r_8},
&&r_3=0,
&&r_4=\pm r_{1} \sqrt{\frac{u+1}{u-1}},
&&r_5=\frac{\sqrt{1-v^2}}{\sqrt{1-u^2}},\\
&r_6=\pm r_1\sqrt{\frac{v-1}{v+1}},
&&r_7=\frac{1}{r_5},
&&r_8=-\frac{r_{4} r_{6}}{r_{1}},
&&r_9=\frac{2  r_{8}}{v-u},
&&r_{10}=c (v-u);
\end{align}
\subparagraph{Model 2}
\begin{align}
&r_1=H_-  e^{F_-} ,
&&r_2=e^{F_-} ,
&&r_3=0,
&&r_4=c H_- e^{-F_+},
&&r_5=e^{ G_-},\\
&r_6=\frac{H_- e^{F_+}}{c},
&&r_7=e^{-G_-},
&&r_8=\pm H_- e^{-F_-},
&&r_9=e^{-F_-},
&&r_{10}=0;
\end{align}
\subparagraph{Model 3}
\begin{align}
&r_1=\pm H_- e^{F_-},
&&r_2=e^{F_-},
&&r_3=0,
&&r_4=\frac{c\; H_-}{ e^{F_+}},
&&r_5=e^{ G_-},\\
&r_6=\frac{H_- e^{F_+}}{c},
&&r_7=e^{-G_-},
&&r_8=0,
&&r_9= \frac{(H_-+1)}{e^{F_-}},
&&r_{10}=0;
\end{align}
\subparagraph{Model 4}
\begin{align}
&r_1=r_3=r_8=r_{10}=0,
&&r_2=\frac{\left((c_1+2) e^{2 G_-}-c_1\right)r_7}{2},
&&r_4= \frac{c_1 (c_1+2) (e^{2 G_-}-1)r_7}{2 c_2 e^{2 F(u)}},\\
&r_5=e^{c_1 (F_--G_-)},\
&&r_6=\frac{{c_2}^2 e^{2 F_+} r_4}{c_1 (c_1+2)},
&&r_7=e^{(2+c_1)(F_--G_-)},\\
&r_9=e^{-2 F_-}r_2;
\end{align}
\subparagraph{Model 5}
\begin{align}
&r_1=0,
&&r_2=\frac{H_- f(v)}{h(v)}+1,
&&r_3= 0,
&&r_4=\frac{1}{H_-}-\frac{r_2 r_9}{H_-},
&&r_5=1,\\
&r_6=H_-,
&&r_7=1,
&&r_8=0,
&&r_9=1-\frac{H_- f(u)}{h(u)},
&&r_{10}=0;
\end{align}
To solve this model we introduced a reparameterization of the spectral parameter, for which
\begin{align}
u\mapsto x(u)=\int ^u\frac{f \dot{h}-h \dot{f}}{h \left(f^2-g h\right)}.
\end{align}
Only taking this into account, the $R$-matrix satisfies the YBE and the boundary conditions. The appearance of a \textit{Riccati type} equation is not a surprise, in fact this model is a quadruple embedding of the 6-vertex B type model given in the next chapter.
\subparagraph{Model 6}
\begin{align}
&r_1=r_3=r_{10}=0,
&&r_2=e^{ F_-+H_-} (1-2 H_-),
&&r_4=\frac{2 H_- e^{H_-}}{c\; e^{F_+}},
&&r_5=e^{F_-+H_-},\\
&r_6=2 c H_- e^{ F_++H_-},
&&r_7=\frac{e^{H_-}}{e^{F_-}},
&&r_8=\pm2 H_- \frac{e^{H_-}}{e^{F_-}},
&&r_9=\frac{e^{H_-}}{e^{F_-}};
\end{align}
\subparagraph{Model 7}
The $R$-matrix for this model is the AdS/CFT $S$-matrix derived in  \cite{Beisert:2005tm,Arutyunov:2006yd} in the string frame. We can recover the $R$-matrix for our model by the mapping \eqref{mapAdSCFTtosu2su2}.

\subparagraph{Model 8}
\begin{align}
&r_1=\frac{e^{-\frac{1}{4} c_3 (u+v)} \left(c_3{}^2 \left(e^{\frac{c_3 u}{2}}-e^{\frac{c_3 v}{2}}\right){}^2-16 c_2 e^{c_3 (u+v)} \sinh \left(\frac{1}{2} c_3 (u-v)\right)\right)}{2 c_3{}^2 \left(e^{\frac{c_3 u}{2}}+e^{\frac{c_3 v}{2}}\right)},
\end{align}
\begin{align}
&r_2=\frac{1}{\cosh \left(\frac{1}{4} c_3 (u-v)\right)},
&&r_3=\frac{1}{4} c_3 \tanh \left(\frac{1}{4} c_3 (u-v)\right),
&&r_5=r_7=\frac{r_2}{r_9}=-\frac{c_3{}^2 r_{10}}{16 r_3}=1,
\end{align}
\begin{align}
&r_4=-\frac{e^{-\frac{1}{4} c_3 (u+v)} \left(e^{\frac{c_3 u}{2}}-e^{\frac{c_3 v}{2}}\right) \left(c_3{}^2-8 c_2 e^{\frac{1}{2} c_3 (u+v)}\right)}{2 c_3{}^2 \sigma },
\end{align}
\begin{align}
&r_6= \frac{8 c_2 e^{\frac{1}{4} c_3 (u+v)} \left(e^{\frac{c_3 u}{2}}-e^{\frac{c_3 v}{2}}\right)}{c_3{}^2 \sigma }-r_4,
&&r_8=\left(r_4+r_6\right) \sigma +r_1.
\end{align}

\section{Comparison between difference and non-difference form models}

\subsection*{$\su(2)\oplus \su(2)$ symmetry}
In order to have a complete classification of the models with $\su(2)\oplus \su(2)$ symmetry with local Hilbert space with dimension 4, we compare the models in Table \ref{Tab1} with the ones in Table 1 of \cite{deLeeuw:2019vdb}, using the allowed identifications, i.e. normalization, shift, rescaling and twists. With this, one can see which non-difference form models constructed here reduce to the difference form given in \cite{deLeeuw:2019vdb}. By doing this comparison, we found the correspondence listed in Table \ref{Tab5}. We should mention that to verify that model 3 of difference form can be obtained from model 4 of non-difference form one needs to perform a limit because the solution is found for $c_2=0$ (pole for $h_5$) and $c_1=-2$. 
Furthermore, we notice that models 2, 4 and 7 of non-difference form generate more than one independent difference form models and that models 1 and 3 do not have a difference form version. 
Finally, models 9, 10 and 11 of \cite{deLeeuw:2019vdb} cannot be obtained from any non-difference form version, so if we want a complete classification of $16\times16$ matrices with $\su(2)\oplus \su(2)$ symmetry we should add those three models.
\begin{table}[h!]
  \begin{center}
    \begin{tabular}{c|c|} 
      \textbf{Difference form}  & \textbf{Non-difference form}\\\hline
1&4 \\  \hline
2&4, 5 \\  \hline
3&4 \\  \hline
4&3 ($h_2=h$) \\  \hline
5&2 ($h_9=h$) \\  \hline
6&6 \\  \hline
7&2 ($h_9=-h$) \\  \hline
8&7  \\  \hline
12&7, 8  \\  \hline
    \end{tabular}
  \end{center}   
  \caption{Correspondence between difference form models in Table 1 of \cite{deLeeuw:2019vdb} and non-difference form of Table \ref{Tab1}.}
  \label{Tab5}
\end{table}
\subsection*{$\alg{so}(4)$ type models}
Similar analysis can be repeated for the models of $\alg{so}(4)$ type. We notice that model 13 of \cite{deLeeuw:2019vdb} is the difference form version of model \eqref{so4spinchain}. However, model 14 ($\mathcal{H}_{12}=K$) is very simple and cannot be recovered as a limit of \eqref{so4spinchain}. To have a complete classification, we should include it.

\section{Summary}

We give here the full classification of integrable models with $\mathfrak{su}(2)\oplus \mathfrak{su}(2)$ symmetry and non-difference form type. To our best knowledge, five of the discovered models are new. The Hamiltonian corresponding to them can be made Hermitian for some choices of the parameters. Furthermore, we showed that the matrix part of the $S$-matrix of ${\rm AdS}_5\times S^5$ integrable system derived in \cite{Beisert:2005tm,Arutyunov:2006yd} requiring centrally extended $\alg{su}(2|2)$ symmetry, can also be found within our classification. However, we remark that with the boost method, it is not possible to reproduce the normalization of the $S$-matrix, known as the dressing factor\footnote{In fact, as mentioned in section \ref{identification}, the normalization is a transformation that preserves the Yang-Baxter equation and consequently, for the purpose of the classifications, we considered models related by different normalizations as the same ones.}.  Our method leads to the elliptic parameterization of it. This model includes the one-dimensional Hubbard model as a special case, as shown in \cite{Beisert_2007}. To achieve a complete classification of both difference and non-difference form integrable models with this specified symmetry, we should add to the models that we found in this chapter, also models 9, 10, 11, and 14 from \cite{deLeeuw:2019vdb}.


\chapter{Two-dimensional local Hilbert space and 8-vertex type model}

\ifpdf
    \graphicspath{{Chapter6/Figs/Raster/}{Chapter6/Figs/PDF/}{Chapter6/Figs/}}
\else
    \graphicspath{{Chapter6/Figs/Vector/}{Chapter6/Figs/}}
\fi
\label{6and8Vmodel}

In this chapter, we apply the boost automorphism method to classify integrable models where the local Hilbert space is $\mathbb{C}^2$. Our specific focus is on the ansatz that employs Hamiltonians of 8-vertex type or lower.  We start with a short introduction on \textit{vertex models} and then we list the models that we found.  Among these models, two can be reduced to a difference form type and are equivalent to the well-known XXZ and XYZ spin chains. The remaining two are of non-difference form: the 6-vertex B can be mapped to the solution A of the paper \cite{6vColored} while the 8-vertex B (to the best of our knowledge) is a \textit{new} model. We do not list integrable models of 7-vertex type, as we have proven that they are always  a particular case of the ones of 8-vertex type. Furthermore, we demonstrate that \textit{any 4x4 Hermitian integrable Hamiltonian}  can be transformed, using the identification outlined in section \ref{identification}, into an 8-vertex model. In the Appendix \ref{adsappendixdef}, we show that the two models of non-difference form type contain the AdS$_2$ and AdS$_3$ integrable models as special case.

\section{Few words on vertex model}
\label{whatarevertexmodel}
There is a correspondence between statistical mechanical lattice models in $D$+1-dimension and $D$-dimensional quantum mechanical models, \cite{eckle2019models}. In a vertex model, each vertex represents an atom or a particle. In 2-D, every particle is connected to four others. These connections are represented by edges, and the direction of an arrow on each bond signifies two possible states for each connected particle.
 We  associate a letter ($a,b,c,d$) to each of the four  edges of the lattice, and these letters can assume values $1$ or $2$. The associated probabilities, denoted as \textit{Boltzmann weights}, are expressed as $R_{ab}^{cd}$. These weights can be represented as entries in a 4x4 matrix, the $R$-matrix. The integrability of the lattice model is determined by whether the $R$-matrix satisfies the Yang-Baxter equation. The problem of finding Yang-Baxter integrable models can be reformulated as finding which Bolzman weight corresponds to a solvable lattice. Solving the corresponding integrable model with one of the integrability techniques, for example the Bethe ansatz, corresponds to calculate the partition function of the vertex model.\\
In higher dimensions, it is also possible to define vertex models. However, in this chapter, our focus is specifically on cases where each edge can have only two possible configurations, distinguished by the orientation of an arrow pointing either inward or outward from the vertex. The number of possibilities is $2^4$, that corresponds to an $R$-matrix with 16 entries. In this chapter,  we restrict to models of 8-vertex type. In this case, there is an even number of arrow pointing toward the vertex and outward it. The  6-vertex models are included into the 8-vertex type and for them, the number of arrow pointing inwards and outwards of each vertex is $2$, see figure \ref{vertexpicture}. 

\begin{figure}[h]
\includegraphics[height=2.2cm]{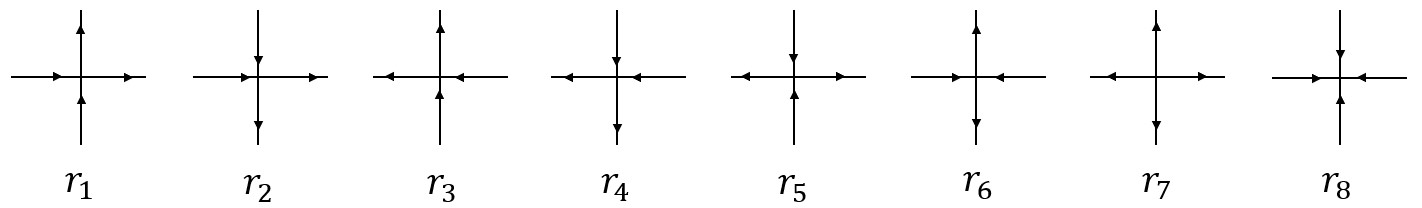}
\caption{Graphical representation of the vertices in a 8-Vertex model}
\label{vertexpicture}
\end{figure}

The condition to have a model of 6-vertex type is also called \textit{ice-rule} condition
\begin{equation}
R^{c d}_{a b}=0 \quad \text{unless} \quad c + d = a + b\,.
\end{equation}

This condition is inspired by the hydrogen-bond configurations observed in a two-dimensional sample of ice, which serves as a prototypical example, \cite{vijay2009studying}. When water freezes, local electrical neutrality requires that each oxygen atom is surrounded by four hydrogen ions such that two hydrogen atoms are closer to the oxygen atom, and two are further away. The arrows in the models identify the proximity or distance of the hydrogen atoms. 
\\
The matrix representation of the vertex given in Figure \ref{vertexpicture} is
\begin{align}
&R  =\left(
\begin{array}{cccc}
 r_1 & 0 & 0 & r_8 \\
 0 & r_2 & r_6 & 0 \\
 0 & r_5 & r_3 & 0 \\
 r_7 & 0 & 0 & r_4 \\
\end{array}
\right),
\label{generalR8vertex}
\end{align}

where $r_i=r_i(u,v)$.
\\We start our search of integrable models having the $R$-matrix of 8-vertex type and consequently also the Hamiltonian, the starting point of the boost automorphism method.

\subsection{$8$-and-lower-vertex models} \label{8vertexmodels}
\subsubsection{Ansatz for $\mathcal{H}$}
The initial ansatz for the Hamiltonian density is
\begin{align}
\begin{split}
\mathcal{H}  =\,& 
h_1  \text{ } \id + h_2 (Z\otimes  \id- \id\otimes Z) + h_3  \sigma _+\otimes \sigma _-  + 
 h_4 \sigma_-\otimes \sigma _+  \\
& +    h_5 ( Z \otimes  \id +   \id \otimes Z ) + 
h_6 Z\otimes Z  + h_7 \sigma _-\otimes
\sigma _- + h_8 \sigma _+\otimes \sigma _+  =\\
&\left(
\begin{array}{cccc}
 h_1+2 h_5+h_6 & 0 & 0 & h_8 \\
 0 & h_1+2 h_2-h_6 & h_3 & 0 \\
 0 & h_4 & h_1-2 h_2-h_6 & 0 \\
 h_7 & 0 & 0 & h_1-2 h_5+h_6 \\
\end{array}
\right) ,
\end{split} 
\label{generalHamiltonian}
\end{align}
where $h_i=h_i(\theta)$.
\\
By using the boost automorphism mechanism, we found four different types of $4\times 4$ Hamiltonians that solve the integrability constraints $[\bQ_2(\theta),\bQ_3(\theta)]=0$. We separate these models in two classes according to their nature: 6 or 8-vertex and we identify with $A$ or $B$ depending if the model can be reduced to difference form (A) or if it is of non-difference form type (B). Before applying the method, we are allowed to remove some of the freedom from the Hamiltonian by using the identification\footnote{The classification remains general since those terms can be added later. However, in this way, the set of equations coming from the integrability constraint $[\mathbb{Q}_2,\mathbb{Q}_3]=0$ is simpler.}. In particular
\begin{itemize}
\item $h_1=0$, corresponds to a shift in the Hamiltonian and it does not affect the commutation relation
\item $h_2=0$, it is the coefficient of a telescopic term that vanishes for closed spin chain.
\end{itemize}
We list the solutions of the integrability constraint $[\mathbb{Q}_2,\mathbb{Q}_3]=0$.
\subsubsection{6-vertex A}
A potentially integrable Hamiltonian is given by
\begin{align}
&h_7=h_8=0,
&&h_3 = c_3 h_6 e^{4 H_5},
&&h_4 = c_4 h_6 e^{-4 H_5},
\label{condition6va}
\end{align}
where $c_{3,4}$ are constants and $\partial_\theta H_5(\theta)=h_5$.
\\
The given Hamiltonian can be transformed into the XXZ spin chain through a series of transformations. As follows:
\begin{align}
\mathcal{H}\,\,\,\to\,\,\,\tilde{\mathcal{H}}=\mathcal{H}+a_1 \id+a_2 (Z\otimes \id-\id\otimes Z)\,\,\,\to\,\,\,\tilde{\tilde{\mathcal{H}}}=B_1 \tilde{\mathcal{H}} B_1^{-1}+\dot{B}_1 B_1^{-1}\,\,\,\to\,\,\,\alpha \tilde{\tilde{\mathcal{H}}},
\end{align}
with
\begin{align}
&a_1=-2 h_5-h_6,
&&a_2=h_5,
&&\alpha=-\frac{1}{2 h_6},
&&B_1=\left(
\begin{array}{cc}
 1 & 0 \\
 0 & \frac{\sqrt{c_3} e^{4 A_5}}{\sqrt{c_4}} \\
\end{array}
\right),
\end{align}
recalling $c=\sqrt{c_3} \sqrt{c_4}$, we obtain
\begin{equation}
\alpha \tilde{\tilde{\mathcal{H}}} = \left(
\begin{array}{cccc}
0 & 0 & 0 & 0 \\
0 & 1 & c & 0 \\
0 & c & 1 & 0 \\
0 & 0 & 0 & 0 
\end{array}
\right) ,
\end{equation}
which is the Hamiltonian density \eqref{diffformham} of the example. The corresponding $R$-matrix is \eqref{eq:XXZcons}.

\subsubsection{6-vertex B}\label{6vertexB}
By considering
\begin{align}
&h_6=h_7=h_8=0,
\end{align}
the Hamiltonian satisfies $[\bQ_2,\bQ_3]=0$ for \textit{any} choice of $h_1, \ldots, h_5$. Three out of these five free functions can be absorbed in the identifications. Specifically, we can set $h_1$ and $h_2$ to zero as mentioned earlier. We can also redefine $h_5$ as $\frac{1}{2} h_4 h_5$ for convenience. After these transformations, we are left with the following form of the Hamiltonian

\begin{align}
&\mathcal{H}=\left(
\begin{array}{cccc}
 h_4\, h_5 & 0 & 0 & 0 \\
 0 & 0 & h_3 & 0 \\
 0 & h_4 & 0 & 0 \\
 0 & 0 & 0 & - h_4 \,h_5 \\
\end{array}
\right)=\frac{h_4 h_5}{2}  (Z_1+Z_2)+h_3 \sigma^+_1 \sigma^-_2+h_4 \sigma^-_1  \sigma^+_2
\end{align}

We normalize the $R$-matrix \eqref{generalR8vertex} such that $r_5=1$ and then it follows from the Sutherland equations \eqref{eqn:Sutherland} that
\begin{align}
&r_7=r_8=0,
&&r_6=1,
&&\dot{r_2} = 
h_4(r_1-h_5 r_2) ,
&&\dot{r}_4 =  -h_4(r_3+h_5 r_4),
&& r_1 r_4 + r_2 r_3 =1,
\end{align}
while $r_4$ satisfies the second order version of the Riccati equation
\begin{align}
\ddot r_4-\frac{\dot{h}_4}{h_4}\dot{r}_4+h_4 r_4\Big[h_3 + \dot{h}_5-h_4h_5^2\Big]=0.
\end{align}
Since we have not yet utilized the freedom associated with reparameterization, we can eliminate the linear term in $r_4$ as follow
\begin{align}
u_i  \mapsto x_i = \int^{u_i}  \frac{\dot h_5}{h_4 h_5^2-h_3}.
\end{align}
This reparametrization removes the explicit dependence on $h_3$. It is then straightforward to solve our system of differential equations to find
\begin{align}
&r_1(x,y) = 1 +  h_5(x)H_4(x,y) ,
&&r_2(x,y) = H_4(x,y) ,\\
&r_3(x,y) =   h_5(x) h_5(y) H_4(x,y) -h_5(x) + h_5(y) ,
&&r_4(x,y) = 1 -h_5(y) H_4(x,y) ,
\end{align}
where again $H_i(x,y) = \int_y^x h_i$.
\\
The $R$-matrix corresponding to this model can be also expressed as
\begin{align}\label{6vBrmatrix}
R = H_4(x,y)
\begin{pmatrix}
h_5(x) & 0 & 0 & 0 \\
0 & 1 & 0 & 0 \\
0 & 0 & h_5(x)h_5(y) & 0 \\
0 & 0 & 0 & -h_5(y) 
\end{pmatrix}
+
\begin{pmatrix}
1 & 0 & 0 & 0 \\
0 & 0 & 1 & 0 \\
0 & 1 & h_5(y) - h_5(x) & 0 \\
0 & 0 & 0 & 1 \\
\end{pmatrix}.
\end{align}
In this way, it is clear that $h_5$ gives rise to the non-difference nature of this solution.  This $R$-matrix satisfies the Yang-Baxter equation with the correct boundary conditions.
In particular, when $h_5$ is constant the $R$-matrix reduces to an $R$-matrix of XXZ type. 
\\
This $R$-matrix is \textit{not new} because it can be mapped by a twist into the solution A of the pure colored Yang-Baxter equation considered in \cite{6vColored}.

\subsubsection{8-vertex A} In the case $h_6\neq0$, the integrability constraint gives that 
\begin{align}
& h_3=h_4 = c_3 h_6,
&&h_5 =0,
&& h_7 = c_7 h_6,
&& h_8 = c_8 h_6,
\end{align}
where $c_i$ are constants. The resulting Hamiltonian, under our identifications, corresponds to the XYZ spin chain \cite{Kulish1982,Vieira:2017vnw}. In particular, we can apply the following transformation

\begin{align}
&\mathcal{H}\,\,\,\to\,\,\,\tilde{\mathcal{H}}=(A\otimes A)\mathcal{H}(A\otimes A)^{-1}\,\,\,\to\,\,\,\frac{1}{h_6} \tilde{\mathcal{H}}=\left(
\begin{array}{cccc}
 1 & 0 & 0 & \sqrt{c_7 c_8} \\
 0 & -1 & c_3 & 0 \\
 0 & c_3 & -1 & 0 \\
 \sqrt{c_7 c_8} & 0 & 0 & 1 \\
\end{array}
\right),
\end{align}
where
\begin{align}
A=\left(
\begin{array}{cc}
 \sqrt[4]{\frac{c_7}{c_8}} & 0 \\
 0 & 1 \\
\end{array}
\right) .
\label{lbt1}
\end{align}
In this way we obtain
\begin{align}
&\frac{1}{h_6}\mathcal{H} = \frac{1}{2} \left(c_3+\sqrt{c_7 c_8} \right)X_i X_{i+1}+\frac{1}{2} \left(c_3-\sqrt{c_7 c_8}\right)Y_i Y_{i+1}+Z_i Z_{i+1},
\end{align}
that is the XYZ spin chain.
\\
The $R$-matrix is more commonly known in the form
\begin{align}\label{xyzrmatrix}
&R(z)=\left(
\begin{array}{cccc}
 q_1(z) & 0 & 0 & q_2(z) \\
 0 & q_3(z) & q_4(z) & 0 \\
 0 & q_4(z) & q_3(z) & 0 \\
 q_2(z) & 0 & 0 & q_1(z) \\
\end{array}
\right) ,
\end{align}
with
\begin{align}
&q_1(z)=\text{sn}\left(\gamma +z\left|k^2\right.\right),
&&q_2(z)=k \text{sn}\left(\gamma \left|k^2\right.\right) \text{sn}\left(z\left|k^2\right.\right) \text{sn}\left(z+\gamma \left|k^2\right.\right),\\
&q_3(z)=\text{sn}\left(z\left|k^2\right.\right),
&&q_4(z)=\text{sn}\left(\gamma \left|k^2\right.\right),
\end{align}
related to our model (up to a shift by the identity)  by the mapping
\begin{align}
&c_3= \frac{4}{\text{cn}\left(\gamma \left|k^2\right.\right)^2 \text{dn}\left(\gamma \left|k^2\right.\right)^2},
&&\sqrt{c_7 c_8}=\frac{4 k \text{sn}\left(\gamma \left|k^2\right.\right)^2}{\text{cn}\left(\gamma \left|k^2\right.\right)^2 \text{dn}\left(\gamma \left|k^2\right.\right)^2},
&&h_6=\frac{1}{2} \text{cn}\left(\gamma \left|k^2\right.\right) \text{dn}\left(\gamma \left|k^2\right.\right) .
\end{align}

\subsubsection{8-vertex B} In the case when $h_6=0$, we find the following differential equations
\begin{align}
&   \frac{\dot{h}_7}{h_7} = \frac{\dot{h}_3+\dot{h}_4}{h_3+h_4} + 4 \frac{h_3-h_4}{h_3+h_4} h_5 ,\label{8VB1}\\
&   \frac{\dot{h}_8}{h_8} = \frac{\dot{h}_3+\dot{h}_4}{h_3+h_4} +  4\frac{h_3-h_4}{h_3+h_4} h_5,\\
& \frac{\dot{h}_5}{h_5} =- \frac{h_3^2-h_4^2}{4 h_5}+ \frac{\dot{h}_3+\dot{h}_4}{h_3+h_4} +4\frac{h_3-h_4}{h_3+h_4} h_5\label{8VB2},
\end{align}
that are solved by
\begin{align}
&h_5= - \frac{1}{4} (h_3+h_4) \tanh (H_3-H_4 + c_5), \label{8vb1}\\
&h_7=c_7 \frac{h_3+h_4}{ \cosh(H_3-H_4+c_5)},\label{8vb2}\\
&h_8=c_8 \frac{h_3+h_4}{ \cosh(H_3-H_4+c_5)},\label{8vb3}
\end{align}
with $c_i$ constant and $\partial_{\theta} H_i=h_i$. Without loss of generality\footnote{For example we can absorb it into the definition of $H_4$}, we can set $c_5=0$.
\\
By using the same diagonal  local basis transformation \eqref{lbt1}, we can set $c_8=c_7$. We can also reparametrize the remaining function with
\begin{align}
& h_3 = \frac{1}{2}\csc(\eta(\theta))(2-\dot{\eta}(\theta)), \\
& h_4 = \frac{1}{2}\csc(\eta(\theta))(2+\dot{\eta}(\theta)),
\end{align}
where $\eta$ is a free function. In this way 
\begin{align}
&H_3-H_4=\log \left(\cot \left(\frac{\eta (\theta)}{2}\right)\right) .
\end{align}
This further results in $h_7=h_8=2 c_7:= k$.
To summarize, we get
\begin{align}\label{8vham}
\mathcal{H}=&\left(
\begin{array}{cccc}
 -\cot (\eta (\theta)) & 0 & 0 & k \\
 0 & 0 & -\frac{1}{2} \left(\dot\eta(\theta)-2\right) \csc (\eta (\theta)) & 0 \\
 0 & \frac{1}{2} \left(\dot\eta(\theta)+2\right) \csc (\eta (\theta)) & 0 & 0 \\
 k & 0 & 0 & \cot (\eta (\theta)) \\
\end{array}
\right)=\nonumber\\
&-\frac{1}{2}  \cot \eta (v)\left(Z_1+Z_2\right)+ k (\sigma^+_1\sigma^+_2+\sigma^-_1\sigma^-_2)+\frac{\eta '(v)+2}{2} \csc (\eta (v))\sigma^-_1\sigma^+_2\nonumber\\
&-\frac{\eta '(v)-2}{2} \csc (\eta (v))\sigma^+_1\sigma^-_2.
\end{align}
As mentioned in the example, we can use the expansion of $R$ \eqref{expansion} in term of the Hamiltonian to restrict the ansatz for the $R$-matrix. In particular, we set $r_5=r_6=1$ and $r_7=r_8$ for the $R$-matrix. The remaining functions are determined from the Sutherland equations and we find
\begin{align}\label{8vBrmatrix1}
r_8(u,v) = k \frac{
\mathrm{sn} (u-v,k^2) \mathrm{cn}(u-v,k^2)}{\mathrm{dn}(u-v,k^2)},
\end{align}
where $\mathrm{sn,cn,dn}$ are the usual Jacobi elliptic functions and
\begin{align}
r_1 &= 
\frac{1}{\sqrt{\sin \eta(u)}\sqrt{\sin\eta(v)}}  \bigg[\sin\eta_+\frac{\mathrm{cn}}{\mathrm{dn}} 
-\cos\eta_+ \mathrm{sn}\bigg], \\
r_2 &= 
\frac{1}{\sqrt{\sin \eta(u)}\sqrt{\sin\eta(v)}}  \bigg[\cos\eta_-\mathrm{sn} +\sin\eta_-\frac{\mathrm{cn}}{\mathrm{dn}} 
\bigg],\\
r_3 &= 
\frac{1}{\sqrt{\sin \eta(u)}\sqrt{\sin\eta(v)}}  \bigg[\cos\eta_-\mathrm{sn}  - \sin\eta_-\frac{\mathrm{cn}}{\mathrm{dn}} 
\bigg],\\
r_4 &= 
\frac{1}{\sqrt{\sin \eta(u)}\sqrt{\sin\eta(v)}}  \bigg[\sin\eta_+\frac{\mathrm{cn}}{\mathrm{dn}} 
+\cos\eta_+ \mathrm{sn}\bigg],\label{8vBrmatrix}
\end{align}
where $\eta_\pm = \frac{\eta(u) \pm \eta(v)}{2}$ and where for simplicity we omitted the dependence on $u-v$ of the Jacobi functions and on the modulus $k^2$, \textit{i.e.} for all of them $\mathrm{sn} = \mathrm{sn}(u-v,k^2)$.\\
It can be checked that this solution satisfies the Yang-Baxter equation and has the correct boundary conditions. To the best of our knowledge, \textit{this model is new} for arbitrary function $\eta$ and constant $k$ and the $R$-matrix is of proper non-difference form type. In the case where $\eta$ is constant, $\eta_-$ vanishes and the model becomes of difference form and reduces to the well-known solution found in \cite{8v,Khachatryan:2012wy,Vieira:2017vnw}.

\subparagraph{Off-diagonal model}
As can be seen from \eqref{8VB1}-\eqref{8VB2}, the cases where $h_5=0$ and $h_3=-h_4$ need special attention due to possible singularities. In this paragraph, we demonstrate that we have not overlooked any solutions, and that model 8-vertex B includes these two models as limiting cases.\\
We can set $h_5=0$ from the beginning and it follows that the Hamiltonian is constant unless $h_3=-h_4$. We can also see this result by considering the solution found in the general case
\begin{align}
&h_5=-\frac{1}{2} \cot (\eta (\theta)).
\end{align}
$h_5=0$ corresponds to take $\eta(\theta)=\frac{\pi}{2}$ and as mentioned in the previous case, if $\eta$ is constant, the Hamiltonian becomes of difference form type.
\\
Also the case where $h_3=-h_4$ can be recovered from 8-v B, however the procedure is less trivial. By applying the method, we found that the entries of the Hamiltonian are
\begin{align}
& h_1=h_2=h_5=h_6=0,
&&h_7 =c_8\; h_8,
&& h_3 =-h_4,
\label{offdiagmodel}
\end{align}
so the Hamiltonian for this model only has off-diagonal entries. It is possible to recover this model from the Hamiltonian of 8-vertex B by following the steps:

\begin{itemize}
\item[1.] We can apply the off-diagonal constant twist to the Hamiltonian density $\lH_{8VB}$ with entries \eqref{8vb1}-\eqref{8vb3},
\begin{align}
U=
\begin{pmatrix}
 0 & a \\
 b & 0 
\end{pmatrix}
\end{align}
to obtain $\tilde \lH_{8VB}\to U_1 \lH_{8VB}U_1^{-1}$. This twist does not satisfy \eqref{twistcond}, so we had to make sure that the integrability condition $[\bQ_2,\bQ_3]=0$ continued to hold also after the twist. This fixes one entry of the twist as $a\to s_1 \sqrt{\frac{c_8}{c7}} b$, with $s_1=\pm 1, \pm i$. We selected $s=i$, but we remark that to get the correct sign we could have chosen also $s=-i$, but not $s=\pm 1$, as will be clarified in the following. Furthermore, we also had to impose the following relations between the entries of the Hamiltonian
\begin{align}
&h_3+h_4=\frac{g(\theta )}{\sqrt{1+ \text{sech}^2(G(\theta ))}}\;,
&&h_3-h_4=g(\theta),
\end{align}
with $g(\theta)$ an arbitrary function.
\item[2.] We apply a diagonal local basis transformation $V(\theta)$. In particular by using \eqref{LBTlaw}, we first fix $\dot{V}V^{-1}$ to eliminate the elements in the (2,2) and (3,3) positions of the Hamiltonian. Then by solving the differential equations, we fixed the matrix $V(\theta)$.
\item[3.] We get an off-diagonal Hamiltonian density and we checked that the sum of the elements at position 2,3 and 3,2 is zero\footnote{This is the point where the choice of $s=\pm i$ was done.}. Moreover the ratio between elements in 1,4 and 4,1 is constant. We can then chose a normalization for example element 3,2 equal to 1 and then the element 1,1 and 4,4 are equal and are function of $g(\theta)$. We then choose a proper reparametrization and bring this to $\lH_{8VB}$.
\end{itemize}
By doing so, we have recovered model \eqref{offdiagmodel} from $\lH_{8VB}$. Since the twist that we used is non-standard, it is unclear how to easily lift it to the level of the $R$-matrix. Nevertheless, it is easy to solve the Sutherland equations for this model directly and we obtain
\begin{align}
R_{\operatorname{off-diag}} = \begin{pmatrix}
 \cosh H_3(u,v) & 0 & 0 & \sin H_7(u,v) \\
 0 & -\sinh H_3(u,v)  & \cos  H_7(u,v)  & 0 \\
 0 & \cos H_7(u,v)  & \sinh  H_3(u,v)  & 0 \\
 \sin H_7(u,v) & 0 & 0 & \cosh H_3(u,v) 
\end{pmatrix}.
\end{align}
We see that it is of quasi-difference form, meaning all of the dependence on the spectral parameters is of the form $H_3(u)-H_3(v)$ and $H_7(u)-H_7(v)$.

\subsection{Hermitian solutions}
In this chapter, we gave the classification of integrable models by starting from an Hamiltonian of 8-vertex type. However, it is interesting to notice that the 8-vertex models analysed include also all the possible integrable Hermitian Hamiltonians of 4x4 type.
\\
To clarify this statement, we start from a Hamiltonian density 4x4,
\begin{align}
\lH = 
h^{(r)} + i\, h^{(i)},
\end{align}
where $h^{(r)}$ and $h^{(i)}$ are the real and imaginary parts of the entries of the Hamiltonian. All the functions are now real-valued and the Hamiltonian depends on 32 independent \textit{ real} functions, which reduces to 16 if we require Hermiticity. 
We can start from this ansatz for the Hamiltonian and require $[\mathbb{Q}_2,\mathbb{Q}_3]=0$. 
By brute force, we can solve these equations for the entries of $h^{(r)}$ and $h^{(i)}$ and  we can discard all solutions that have complex numbers in them. In this way, we find all the potentially integrable Hermitian Hamiltonians of 4x4 type.  At this point, we would have to find the $R$-matrices corresponding to them. However, this was not necessarily. In fact, we prove that all the solutions that we got are related to the 8-vertex form under the identifications of section \ref{identification}.
\\
The classification of models with 8-vertex type is broader than it may initially appear because it includes also all 4x4 Hermitian Hamiltonians. However, it is important to note that this does not imply that the corresponding 8-vertex models themselves are Hermitian.
\\We can show a simple example to understand better this statement. One of the simplest integrable Hermitian Hamiltonian beyond 8-Vertex is
\begin{align}
\mathcal{H}=\left(
\begin{array}{cccc}
 1 & \alpha +i \beta  & \alpha +i \beta  & 0 \\
 \alpha -i \beta  & 0 & 0 & -\alpha -i \beta  \\
 \alpha -i \beta  & 0 & 0 & -\alpha -i \beta  \\
 0 & -\alpha +i \beta  & -\alpha +i \beta  & 1 \\
\end{array}
\right) .
\end{align}
Applying a local basis transformation and then a twist, this Hamiltonian can be brought into a non-Hermitian one of 6-V type.

\section{Summary}
In this chapter, we gave a classification of integrable models starting from an ansatz for the Hamiltonian of 8-vertex type. We discovered four models: two of them were found to be equivalent to the XXZ and XYZ spin chains, while the other two models exhibited a non-difference form. Among them, the 6-vertex B can be mapped to the solution A of the paper \cite{6vColored}, whereas the 8-vertex B (to the best of our knowledge) is a \textit{new} model. In  Appendix \ref{adsappendixdef}, we show that the two models of non-difference form type contain the AdS$_2$ and AdS$_3$ integrable models as special case. Additionally, we demonstrate that any 4x4 Hermitian integrable Hamiltonian can be transformed into an 8-vertex model using the identification outlined in section \ref{identification}. Consequently, the classification we gave, includes all possible 4x4 Hermitian Hamiltonians. 
\chapter{Free fermion conditions}

\ifpdf
    \graphicspath{{Chapter7/Figs/Raster/}{Chapter7/Figs/PDF/}{Chapter7/Figs/}}
\else
    \graphicspath{{Chapter7/Figs/Vector/}{Chapter7/Figs/}}
\fi
\label{ffconditionchapter}
In this chapter, we demonstrate that the integrable models discussed in chapter \ref{6and8Vmodel} exhibit an interesting relation. Specifically, two of these models satisfy a \textit{Baxter relation}, while the other two satisfy a \textit{free fermion condition}. The latter condition is particularly significant as it enables us to express the transfer matrix associated with these models in a diagonal form, simplifying the process of computing eigenvalues and eigenvectors. We explicitly show it for model 6-Vertex B. Moreover, we find the equivalent of the free-fermion condition for some of the  models with $\alg{su}(2)\oplus \alg{su}(2)$ symmetry, discussed in chapter \ref{threeorfour}.

\section{Free fermion conditions}

The $R$-matrix corresponding to the models of chapter \ref{6and8Vmodel} is \eqref{generalR8vertex} and satisfies the condition
\begin{equation}
\frac{(r_1r_4+r_2r_3-r_5r_6-r_7r_8)^2}{r_1r_2r_3r_4}=\text{const}.
\label{general(non)freefermion}
\end{equation}
In particular, models 6-V A and 8-V A are characterized by 
\begin{equation}
\label{classA}
\text{const}\neq 0,\,\,\,\,\,\,\,\,\,\,\text{Baxter  condition}
\end{equation}
while 6-V B and 8-V B by
\begin{equation}
\text{const}=0,\,\,\,\,\,\,\,\,\,\,\text{Free  fermion  condition}.
\end{equation}
This last one can be expressed as
\begin{eqnarray}
\label{free}
r_1 r_4 + r_2 r_3 = r_5 r_6 + r_7 r_8.
\end{eqnarray}
One difference about those two classes of models is that: those of class A are of difference form type while those of class B are not. Another property is that, at the level of the Hamiltonian, class A models contain the term $Z \otimes Z$, while class B models do not.
\\We prove the Baxter and the free fermion conditions both directly and by using the Sutherland equation, which enables potential generalizations of them to other model types.

\subsection{Direct computation}
\subsubsection{Class A: Baxter condition}

Up to basic identifications, models from class A correspond to the XXZ and XYZ integrable spin chain.  The $R$-matrix associated to the 6-V A model is \eqref{eq:XXZcons} and it is easy to check that 
\begin{align}
\frac{(r_1r_4+r_2r_3-r_5r_6-r_7r_8)^2}{r_1r_2r_3r_4} = \frac{4}{c^2}.
\end{align}
For 8-V A the $R$-matrix is \eqref{xyzrmatrix}
and we find
\begin{align}
\frac{(r_1r_4+r_2r_3-r_5r_6-r_7r_8)^2}{r_1r_2r_3r_4} = 4 \text{cn}^2(\gamma ,k^2) \text{dn}^2(\gamma,k^2).
\end{align}
Therefore, we have demonstrated the validity of the Baxter condition.

\subsubsection{Class B: Free Fermion condition} 
\label{explicitv}

For 6-V B, the $R$-matrix is \eqref{6vBrmatrix}, and the entries obey
\begin{equation}
r_1\,r_4+r_2\,r_3=1=r_5\,r_6+r_7\,r_8.
\end{equation}

The entries of the $R$-matrix for the 8-V B model are defined in \eqref{8vBrmatrix1}-\eqref{8vBrmatrix}. As these equations involve Jacobi elliptic functions, we provide a detailed calculation. First, we begin by examining the right-hand side of the equation \eqref{free}
\begin{equation}
r_5r_6+r_7r_8=1+k^2 \frac{\text{sn}^2\,\text{cn}^2}{\text{dn}^2}.
\label{rhsfree8vertexB}
\end{equation}
For the left-hand side, the term $ r_1\,r_4 $ is
\begin{align}
r_1r_4& =\frac{\sin^2\eta_+\frac{\text{cn}^2}{\text{dn}^2}-\cos^2\eta_+\text{sn}^2}{\sin\eta(u)\sin\eta(v)},
\end{align}
and $ r_2\,r_3 $ is
\begin{align}
r_2r_3& =\frac{\cos^2\eta_-\text{sn}^2-\sin^2\eta_-\frac{\text{cn}^2}{\text{dn}^2}}{\sin\eta(u)\sin\eta(v)} .
\end{align}
This leads us to the expression
\begin{align}
r_1r_4+r_2r_3&=\frac{1}{\sin\eta(u)\sin\eta(v)}\left[\left(\sin^2\eta_+-\sin^2\eta_-\right)\frac{\text{cn}^2}{\text{dn}^2}-\left(\cos^2\eta_+-\cos^2\eta_-\right)\text{sn}^2\right]\nonumber\\
&=\frac{\text{cn}^2}{\text{dn}^2}+\text{sn}^2.
\label{lhsfree8vertexB}
\end{align}

By using the identities $ 1-k^2\text{sn}^2=\text{dn}^2 $ and $ 1 = \text{cn}^2+\text{sn}^2 $, we can establish the equality between the two expressions.

\subsection{Prove using the Sutherland equation}
\label{suth}
In this section, we prove that any integrable $R$-matrix of 8-vertex\footnote{By saying 8-vertex, we refer to models which contain up to 8-vertex.} type satisfies the generalised condition \eqref{general(non)freefermion}. We will prove this by using the Sutherland equations \eqref{eqn:Sutherland}-\eqref{eqn:Sutherland2}. We assume that the Hamiltonian associated to the $R$-matrix is also general and it is given by \eqref{generalHamiltonian}. In this section, the integrability property is included in the fact that the $R$-matrix and the Hamiltonian corresponding to a given model obey the Sutherland equations.
\\
We consider the $R$-matrix to be of non-difference form and we use the shortcuts $r_i=r_i(u,v)$ and $h_i=h_i(u)$. 
\\
To prove the relation \eqref{general(non)freefermion}, we follow the steps
\begin{enumerate}
\item Substitute \eqref{generalR8vertex} and \eqref{generalHamiltonian} into the Sutherland equations.
\item Solve for the derivatives $\dot{r_i}$ and $r_i'$ in terms of $ h_i $ and $ r_i $ (without actually solving the differential equations).
\item Then, solve for some of the $ h_i $'s in terms of $r_i $.
\end{enumerate}
Remarkably, by doing this, one obtains the following set of conditions on the coefficients of the $R$-matrix
\begin{align}
& \frac{r_1r_4+r_2r_3-r_5r_6-r_7r_8}{r_2r_4}=f(u),\label{FFBaxterpart1a}\\
& \frac{r_1r_4+r_2r_3-r_5r_6-r_7r_8}{r_1r_3}=g(u)\label{FFBaxterpart2a}\\
& \frac{r_1r_4+r_2r_3-r_5r_6-r_7r_8}{r_3r_4}=g(v),\label{FFBaxterpart1}\\
& \frac{r_1r_4+r_2r_3-r_5r_6-r_7r_8}{r_1r_2}=f(v)\label{FFBaxterpart2},
\end{align}
where $ f(u) $ and $ g(u) $ are functions of $ h_i(u) $.

By multiplying \eqref{FFBaxterpart1a} and \eqref{FFBaxterpart2a} and then also \eqref{FFBaxterpart1} and \eqref{FFBaxterpart2},  we can see that 
\begin{align}
 \frac{(r_1r_4+r_2r_3-r_5r_6-r_7r_8)^2}{r_1r_2r_3r_4} = f(u)g(u) = f(v)g(v),
\end{align}
and hence proving that equation \eqref{general(non)freefermion} holds. Since the Sutherland equations \eqref{eqn:Sutherland}-\eqref{eqn:Sutherland2} are obtained as a derivative of the Yang-Baxter equation, we proved that all integrable $R$-matrices of the form  \eqref{generalR8vertex} belong to one of the two classes A and B.

\section{Why is the free fermion condition important?}
\label{analternativetotheABA}
We have proven that all the 8-V type models found in the previous section satisfy either the Baxter condition or the free fermion condition. Specifically, the non-difference form models satisfy the free fermion condition.
\\
In this section, we explain the importance of this condition and why it is highly useful. It also become evident how the name \textit{free fermion} originated. \\Many examples of models following the free fermion condition can be found in the literature. For instance in \cite{Lieb,Lieb2}, the Hamiltonians and transfer matrices of the XY chain and of Ising-type chain were written in a form that manifestly displays their free fermion nature.  To achieve this, a particular transformation of the canonical spin-chain operators is performed. In this section, we follow these ideas and find the appropriate canonical transformation to make the transfer matrices of our models explicitly of free fermion type.

\subsection{6-Vertex B}

We begin with the 6-vertex B model and examine an $R$-matrix of the form \eqref{generalR8vertex} with $r_7=r_8=0$. We choose a basis for the Hilbert space given by one boson $|\phi\rangle$ and one fermion $|\psi\rangle$,
\begin{eqnarray}
|\phi\rangle \equiv |0\rangle, \qquad |\psi\rangle \equiv c^\dagger |0\rangle, \qquad c|0\rangle = 0,
\end{eqnarray}   
where we introduce canonical fermionic creation and annihilation operators $c^\dagger$ and $c$, respectively, such that
\begin{eqnarray}
\{c,c^\dagger\}=1, \qquad \{c,c\} = \{c^\dagger,c^\dagger\}=0,
\end{eqnarray}
or whenever we work with multiple spaces
 \begin{eqnarray}
\{c_i,c_j^\dagger\}=\delta_{ij}, \qquad \{c_i,c_j\} = \{c_i^\dagger,c_j^\dagger\}=0.
\label{canonicalmanysites}
\end{eqnarray} 
We have to consider the fermionic nature of this operator. There are two equivalent approaches to do this. The first involves employing the Jordan-Wigner transformation\footnote{In this case, as we are dealing with a spin 1/2 chain, we consider the transformation of the $\sigma$ matrices only.} \eqref{JWtransformation}. The second approach, which we adopt in this section, involves considering the graded vector space.
\\
In vector representation, we can associate 
\begin{align}
&\ket{\phi}=\ket{1}=e_1,
&&\ket{\psi}=\ket{2}=e_2,
\end{align}
and for the creation operator
\begin{eqnarray}
c^\dagger = E_{21}, \quad c = E_{12}, \quad m \equiv c \,c^\dagger = E_{11}, \qquad n \equiv c^\dagger c = E_{22}, 
\end{eqnarray}

where  $n$ is the usual number operator and $E_{ij}$ are matrices with only one non-zero entry $1$ in position $(i,j)$.
\\
The grading operates in such a way that the anticommutator relations are satisfied. In particular, we can identify $\ket{0}=\ket{\phi}\otimes\ket{\phi}$
\begin{eqnarray}
c_1^\dagger c_2^\dagger |0\rangle = - c_2^\dagger c_1^\dagger |0\rangle,
\end{eqnarray}

which is the equivalent to
\begin{eqnarray}
[E_{21} \otimes \mathds{1}][\mathds{1} \otimes E_{21}] |\phi\rangle \otimes |\phi\rangle = - [\mathds{1} \otimes E_{21}][E_{21} \otimes \mathds{1}]|\phi\rangle \otimes |\phi\rangle.
\end{eqnarray}

The $R$-matrix can be written in terms of oscillators in the following form\footnote{Notice that this $R$-matrix satisfies the graded YBE. For explicit expressions, we recall Appendix A of \cite{deLeeuw:2019vdb}.}
\begin{align}\label{eq:Rosc}
R^{(osc)}_{ij}(u,v) = r_1 m_i m_j + r_2 n_i m_j + r_3m_in_j - r_4 n_i n_j - r_5 c_i c^\dag_j+r_6 c^\dag_i c_j,
\end{align}
where we suppressed the explicit dependence of $r_i$ on $(u,v)$ and $i$ and $j$ indicate the spaces in which the operators are acting. Suppose the $R$-matrix is regular, \textit{i.e.} $R(u,u) = P$, with $ P $ being the graded permutation operator, then the Hamiltonian density is given by the logarithmic derivative of the
$R$-matrix

\begin{align}
\mathcal{H}_{ij}^{(osc)} = P_{ij} \partial_u R^{(osc)}_{ij}(u,v) \Big|_{v=u}.
\end{align}
It is convenient to associate, for simplicity, $h_i = \partial_u r_i|_{u\to v}$. In the oscillator formalism, the Hamiltonian is\footnote{Notice that here we use for convenience a difference ansatz for the Hamiltonian compared to \eqref{generalHamiltonian}. Now, we label the entries of the Hamiltonian exactly as the $R$-matrix up to identification $r_i\to h_i$.}
\begin{align}\label{eq:Hosc}
\mathcal{H}^{(osc)}_{12} = &\ 
h_1  + (h_6-h_1) n_2 - (h_1+h_5)n_1 - (h_1+h_4-h_5-h_6) n_1n_2  +  h_3  c^\dag_2 c_1 +h_2 c^\dag_1c_2   .
\end{align}
On the level of the Hamiltonian, the free fermion condition \eqref{free} imposes that\footnote{The extra signs come from grading, since grading was not considered in \eqref{free}.} 
\begin{align}
h_1+h_4-h_5-h_6 =0,
\end{align}
 which eliminates the $n_1n_2$ term.

\subsection{Solving homogeneous spin-chains with Free Fermions\label{omogenee}}

We consider the Hamiltonian \eqref{eq:Hosc} for a spin-chain of length $L$. We apply our non-local free fermion transformation to diagonalize this Hamiltonian. We note that it is enough to consider the one-particle sector since this will induce a canonical map between the oscillators ${c}, {c}^\dag$ and a new set of operators. The purpose of this mapping is to recast the Hamiltonian in a manifest free fermion form. We denote these new operators as $\chi,\chi^\dag$ and they will still satisfy canonical anticommutation relations. Restricted to this subsector, our Hamiltonian takes the simple form
\begin{align}
\mathbb{H}^{(1pt)} = \begin{pmatrix}
h_6+h_5 & h_2 & 0 & 0 &\hdots & 0 & h_3 \\
h_3 & h_6+h_5 & h_2 & 0 &\hdots & 0 & 0\\
0 & h_3 & h_6+h_5 & h_2 &\hdots & 0 & 0\\
0 & 0 & h_3 & h_6+h_5 & \hdots & 0 & 0\\ 
\vdots & \vdots & \vdots &\vdots & \ddots & \vdots & \vdots\\
0 & 0 & 0 & 0 & \hdots & h_6+h_5 & h_2\\
h_2 & 0 & 0 & 0 & \hdots & h_3 & h_6+h_5
\end{pmatrix}.
\end{align}
The eigenvalues and eigenvectors of $\mathbb{H}^{(1pt)}$ can now be easily computed. By considering spin chains of different lengths, it is easy to convince ourself that the eigenvectors {{} $\vec{\ell}$} and eigenvalues $\lambda$ are of the form
\begin{align}
&\lambda = h_6+h_5 + h_2 {{} z} + h_3{{} z}^{-1} ,
&& {{} \vec{\ell}} = (1,{{} z},{{} z}^2,\ldots {{} z}^{L-1}),
\end{align}
with ${{} z}^L = 1$.
This means that there are exactly $L$ eigenvectors parameterised by the $L$th roots of unity and we can write the canonical transformation using $z=e^{\frac{2 \pi i k}{L}}$ for $k=1,...,L$
\begin{align}
&c_k  = \frac{1}{\sqrt{L}} {{} \sum_{n=1}^L} e^{2\pi i \frac{k n}{L} } \chi_n,
&&c_k^\dag  = \frac{1}{\sqrt{L}} {{} \sum_{n=1}^L} e^{-2\pi i \frac{k n}{L} } \chi_n^\dag.
\label{canonicaltransf}
\end{align}
By considering these transformations into \eqref{eq:Hosc} and summing over all the sites of the spin chain, we obtain the Hamiltonian\footnote{We recall that we are working with periodic boundary conditions.}
\begin{align}
\mathbb{H} = h_1 L + \sum_{n=1}^L \Big[ (h_2 +h_3) \cos \frac{2\pi n }{L}+ i (h_2 - h_3) \sin \frac{2\pi n }{L}  -h_1+h_4\Big] \chi^\dag_n\chi_n,
\end{align}
which is now manifestly diagonal. Moreover, we can show that this canonical transformation also diagonalizes the full transfer matrix.
\\
For a graded model, the trace should be substituted with the supertrace and the transfer matrix is
\begin{align}
t({{} \theta_0},\vec{\theta}) = {{}\mbox{str}_0 \big[ R_{01} (\theta_0 , \theta_1) \ldots R_{0L} (\theta_0 , \theta_L) \big].}
\end{align}
The parameters $\theta_i$ are the local inhomogeneities and $\theta_0$ is the spectral parameter associated with the auxiliary space. In case all the rapidities coincide $\theta_i = \theta$, then the first logarithmic derivative corresponds to the nearest-neighbour Hamiltonian \eqref{eq:Hosc}. For generic inhomogeneities, all the conserved charges have interaction range $L$. In this paragraph, we consider only the homogeneous case and we refer to the paper \cite{freefermionpaper} for the inhomogeneous one.
\\
We computed the expression of the transfer matrix for different values of the length of the chain and we recognized the general structure for any $L$. We observe that the free fermion condition was needed only from $L=4$. By computing the transfer matrix for different values of $L$, we recognize the following structure
\begin{align}
t_L = -\prod_{k=0}^{L-1}\bigg[ (r_1 - r_3 e^{\frac{2\pi i k}{L}}) M_{k+1} + (r_2 + r_4 e^{\frac{2\pi i k}{L}})  N_{k+1} \bigg],
\end{align}
where
\begin{align}
\label{asdefin}
&N_i=\chi^\dagger_i \chi_i,
&&M_i= \chi_i \chi^\dagger_i.
\end{align}

This transfer matrix is now manifestly diagonal in the new operators $\chi$ and $\chi^\dagger$ and takes a factorised form, which reminds us of separation of variables. Because of this factorised form, we can exponentiate it and read off the conserved charges
\begin{align}
t_L =- \exp \Big[{{} i \,\psi_L} +  \sum_k {{} i\, \omega_k} N_k\Big],
\end{align}
where 
\begin{align}
&{{} i \,\psi_L} = \log (r_1^L - r_3^L),
&& {{} i \,\omega_k}= \log\bigg[\frac{r_2 + e^{\frac{2\pi i k}{L}}r_4}{r_1 - e^{\frac{2\pi i k}{L}}r_3}\bigg].
\end{align}
Since all the number operators commute, the exponent is well-defined. By using the free fermion transformation, we saw that  the expression of the transfer matrix is very compact and elegant. Furthermore, it is very intuitive to solve the eigenvalue problem.
\\
We show this with an example. For simplicity, we consider the case of $L=2$. Even if the AdS/CFT models are not the primary focus of this thesis, in Appendix \ref{adsappendixdef} we show that $6$-V B model is related to AdS$_3$. For the case of pure Ramond-Ramond massless AdS$_3$, 
\begin{align}
&\omega_1 = \pi ,
&&\omega_2 = 0 ,
\end{align}
and we obtain
\begin{align}
{
t_2=-e^{i \psi_2}e^{i \omega_1 N_1} = e^{i \psi_2} \Big[1+ N_1 \sum_{n=1}^\infty \frac{( i \omega_1 )^n}{n!} \Big] = e^{i \psi_2} \Big[1+N_1 (e^{i \omega_1} - 1)\Big] = e^{i \psi_2} \Big[ 1-2 N_1 \Big]},
\end{align}
by using the standard property $N_i^\dagger=N_i$. For AdS$_3$, $ e^{i \psi_2}=\cosh \frac{\theta_{12}}{2}$.
\\
This makes the eigenvalue problem very easy since it is a matter of adding free fermion energy. The additive structure of the transfer matrix is now manifest, while it was previously hidden in the standard formalism. Note that we could have added the piece "$0 \,N_2$" inside the square bracket to show that the transformation we have introduced completely diagonalises the transfer matrix on the two physical spaces.\\
 Furthermore, also the eigenvector problem has an intuitive solution. In fact, from \eqref{canonicaltransf}
 \begin{align}
 &\chi_1=\frac{1}{\sqrt{2}}(-c_1+c_2),
 &&\chi_2=\frac{1}{\sqrt{2}}(c_1+c_2),
 \end{align}
it is clear that both $\eta_1$ and $\eta_2$ still annihilate the state $\ket{0}=\ket{\phi}\otimes\ket{\phi}$. This state is an eigenstate of the transfer matrix $t_2$ with eigenvalue $e^{i\psi_2}$.\\
The other eigenstates are
\begin{align}
&\chi_1^\dagger \ket{0}=\frac{1}{\sqrt{2}}\big(\ket{\phi}\otimes\ket{\psi}+\ket{\psi}\otimes\ket{\phi}\big),&&\text{eigenvalue},\,\,-e^{i\psi_2}\\
&\chi_2^\dagger \ket{0}=\frac{1}{\sqrt{2}}\big(\ket{\phi}\otimes\ket{\psi}-\ket{\psi}\otimes\ket{\phi}\big),&&\text{eigenvalue}\,\,e^{i\psi_2},\\
&\chi_1^\dagger \chi_2^\dagger \ket{0}=\ket{\psi}\otimes \ket{\psi},&&\text{eigenvalue}\,\,-e^{i \psi_2}.
\end{align}
Generalize this results for any $L$ is straightforward.\\
In the paper \cite{freefermionpaper}, we comment on the application of the free fermion technique to the 8-Vertex B model. However,we did not find a closed expression for the transfer matrix that works for any length $L$, so we will not report this result here.
\section{Free fermion condition for $ AdS_5 $ sector}\label{sec:16}

In this section, we apply the  approach discussed in Section \ref{analternativetotheABA} 	to obtain a free fermion condition for the case of a 4-dimensional Hilbert space. We focus on the class of model whose Hamiltonian and $R$-matrix exhibit $\mathfrak{su}(2)\oplus \mathfrak{su}(2)$ symmetry \cite{deLeeuw:2019vdb,classificationybandboost}, in particular where both $\alg{su}(2)$ have a two dimensional representation. We discussed these models in chapter \ref{threeorfour}. This class of models is particularly important because it contains some non-difference form model like the {{} $AdS_5 \times S^5$} superstring sigma model\footnote{This statement is clarified in the footnote \ref{footnotedressingfactor} of section \ref{integrhamsu22}.}, the one-dimensional Hubbard model and some difference form one like the $\alg{su}(4)$ spin chain.
\\
To achieve this, we follow the procedure outlined in Section \ref{suth}. We use the ansatz \eqref{eq:Hsu2} for the Hamiltonian density, and similarly for the $R$-matrix we replace the $h$s with $r$s. We first substitute them in the Sutherland equations \eqref{eqn:Sutherland}-\eqref{eqn:Sutherland2} and we solve for the derivatives $\dot r_i$ and $r_i'$ without explicitly solving the differential equations. By carefully choosing the order of solutions, we ensure that divergences are properly handled. To accomplish this, we used the list of models in \cite{classificationybandboost} as test. In particular, if those models did not exhibit a singular point for the specific variable considered, we consider that variable always different from zero. In this way, we can keep the divergences under control and choose the correct branch of solutions\footnote{Some equations in fact can be factorised in the form $f g =0$, with $f$ and $g$ functions of the $r_i$s. We used the test model of \cite{classificationybandboost} to select whether the solution was $f=0$ or $g=0$. By choosing different classes of models, some choices would probably have been different.}.
Specifically, we obtain the following simple conditions
\begin{align}
&h_3(v) r_{10}r_5=h_{10}(v) r_3 r_7,\\
&h_3(u) r_{10} r_7=h_{10}(u) r_3 r_5.
\end{align}
Now, we can proceed by considering the two cases $r_3\neq 0$ and $r_3=0$. These two cases correspond to the analyses conducted  in chapter \ref{threeorfour}, with $h_3\neq 0$ and $h_3=0$.

\subsection{Case $r_3\neq 0$}
\label{r3no0}

This class of models includes the one-dimensional Hubbard model and the $S$-matrix derived from requiring centrally extended $\alg{su}(2|2)$ symmetry of the $AdS_5 \times S^5$ superstring sigma model \cite{Beisert:2005tm, arutyunov2009foundations, Arutyunov:2006yd}. We find
\begin{align}
& \frac{r_4^2-r_1(r_1+r_2)}{r_3r_7}=\frac{h_2(v)}{h_3(v)},\\
& \frac{r_6^2-r_8(r_8+r_9)}{r_3 r_7}=\frac{h_9(v)}{h_{3}(v)},\\
& \frac{r_6^2-r_1(r_1+r_2)}{r_3r_5}=\frac{h_2(u)}{h_3(u)},\\
& \frac{r_4^2-r_8(r_8+r_9)}{r_3 r_5}=\frac{h_9(u)}{h_{3}(u)}.
\end{align}
We remark that since $r_3 \neq 0$, $h_3$ is also non-zero, which avoids the problem of divergences. Furthermore, in all the models from \cite{classificationybandboost}  that fall into this category, $r_5$ and $r_7$ are non-zero. After solving some of the equation obtained from the first Sutherland equation \eqref{eqn:Sutherland} for the $h$s and plugging the solutions into the others, we are left with four conditions
\begin{align}
&r_4 r_6+r_1 r_8=r_3 r_{10} ,\label{mitev1}\\
&r_5 r_7-r_4 r_6=\left(r_1+r_2\right) \left(r_8+r_9\right) ,\label{mitev2}\\
&r_3 r_{10}+r_5 r_7=r_2 r_9 ,\label{mitev3}\\
&\left(r_1+r_8\right) \left(r_1+r_2+r_8+r_9\right)=\left(r_4-r_6\right)^2 \label{nomitev4}.
\end{align}
It can be shown that these four conditions could have been similarly derived from the second Sutherland equations. We notice that \eqref{mitev1}-\eqref{mitev3} were obtained also in \cite{MitevEtAll}. However, we find the \textit{additional} condition \eqref{nomitev4}.
\\
By using the regularity condition $R(u,u)=P$ and from the definition of the Hamiltonian, we can derive that \eqref{mitev1}-\eqref{nomitev4} impose some very simple constraints on the Hamiltonian
\begin{align}
&h_1+h_8=h_4+h_6,\label{condHam1}\\
&h_2=-h_9,\label{condHam2}\\
&h_5 h_7+h_2 h_9= h_3 h_{10}\label{condHam3}.
\end{align}
Specifically, \eqref{condHam1} and \eqref{condHam2} were derived by differentiating \eqref{mitev3} and \eqref{nomitev4}, \eqref{condHam3} by differentiating  \eqref{mitev1} twice. Furthermore, differentiating two times and combining \eqref{mitev1}-\eqref{nomitev4}, in order to get rid of the second derivatives, we also got
$$\left(h_5+h_7\right)^2=\left(h_2-h_9\right) \left(h_1+h_2-h_8-h_9\right).$$

\subsection{ Case $r_3=0$}

One can check that the case $r_3=0$ and $r_{10}\neq 0$  satisfies \eqref{mitev1}-\eqref{nomitev4} as well. Hence, we can restrict to the case where $r_{10}=0$. Analyzing this case leads to a set of factorised equations
\begin{align}
&\frac{r_1}{r_6}=\frac{h_2(v)}{h_7(v)},
&&\frac{r_8}{r_4}=\frac{h_9(v)}{h_5(v)},
&&\frac{r_1 r_5 r_7}{r_2^2r_4}=\frac{h_2(v)}{h_5(v)},\\
&\frac{r_1}{r_4}=\frac{h_2(u)}{h_5(u)},
&&\frac{r_8}{r_6}=\frac{h_9(u)}{h_7(u)}, 
&&\frac{r_1 r_5 r_7}{r_2^2r_6}=\frac{h_2(u)}{h_7(u)}.
\end{align}
Plugging the solutions for the $h$s into the Sutherland equations, we should consider four possible subcases separately: $r_1\neq 0 \, r_8\neq 0$, $r_1\neq 0\, r_8= 0$, $r_1=0\, r_8\neq 0$ and $r_1= 0\, r_8=0$. We observe that the subcase $r_1=0\, r_8\neq0$ can be recovered from $r_1\neq0 \,r_8=0$ by performing an off-diagonal basis transformation\footnote{The action of this off-diagonal basis transformation is to swap $g_1 \leftrightarrow g_8, g_3 \leftrightarrow g_{10}, g_2 \leftrightarrow g_9, g_4 \leftrightarrow g_6, g_5 \leftrightarrow g_7,$ where $g$ is either $h$ or $r$.} on the $R$-matrix and on the Hamiltonian.
\\
The subcases to be considered are then

\subsubsection*{Subcase $r_1\neq 0, r_8\neq 0$}
We obtain
\begin{align}
&r_5 {r_7}=r_2 r_9,\\
&r_4 {r_6} r_9=r_2 r_8^2,\\
&r_1^2 r_9^2=r_2^2 r_8^2.
\end{align}
These imply \eqref{condHam1} on the entries of the Hamiltonian, along with the additional conditions $h_2^2= h_9^2$, $h_5^2 h_7^2 = h_9^2$.

\subsubsection*{Subcase $r_1\neq0, r_8=0$}
In this case, the entries of the $R$-matrix satisfy
\begin{align}
&r_2^2 r_4 r_6=r_1^2 r_5 r_7,\\
&r_4 r_5 r_6 r_7=\left(r_5 r_7-r_2 r_9\right){}^2.
\end{align}
On the Hamiltonian
\begin{align}
& h_5 h_7= h_2^2,\\
&  h_5 h_7= \left(h_1-h_4-h_6+h_8\right){}^2.
\end{align}

\subsubsection*{Subcase $r_1=0, r_8= 0$}
There are two {{} possibilities. If}
\begin{align}
&h_5(v) r_6 r_9=h_7(v) r_2 r_4 \label{caser10r80} 
&&\text{and}
&& h_7(u) r_4 r_9=h_5(u) r_2 r_6
\end{align}
there are no additional conditions on the entries of the $R$-matrix.
If \eqref{caser10r80} are not verified, we obtain the condition
\begin{align}
&r_4 r_6-r_5 r_7+r_2 r_9=0
\end{align}
that implies on the Hamiltonian \eqref{condHam1}.
\\
In the following section, we show how to rewrite the Hamiltonian to make the free fermion nature explicit.

\subsection{Towards a free fermion Hamiltonian}

Here we mainly follow section \ref{analternativetotheABA} and \cite{Lieb,Lieb2}. As for the chapter \ref{threeorfour}, the Hilbert space is four dimensional and is spanned by two bosons $|\phi_{1,2}\rangle$ and two fermions $|\psi_{1,2}\rangle$. Like in \eqref{phipsi4dim}, we introduce two sets of canonical fermionic creation and annihilation operators $c^\dagger_{\alpha,j},c_{\alpha,j}$ where $\alpha =\, \uparrow,\downarrow$ is the spin and  $j$ is the site of the chain (running from $1$ to {{} the chain length $L$}). If we denote the vacuum by $|0\rangle$ such that $c_{\alpha,j}|0\rangle=0$, then our local Hilbert space is spanned by
\begin{align}
&|\phi_1\rangle= |0\rangle,
&&|\phi_2\rangle= c^\dagger_{\uparrow,j} c^\dagger_{\downarrow,j} |0\rangle,
&&|\psi_1\rangle = c^\dagger_{\uparrow,j} |0\rangle,
&&|\psi_2\rangle = c^\dagger_{\downarrow,j} |0\rangle.
\label{phi12psi12}
\end{align}
These oscillators satisfy the usual anti-commutation relations
\begin{align}\label{eq:oscdef}
&\{c^\dagger_{\alpha,i},c_{\beta,j} \} = \delta_{\alpha\beta}\delta_{ij},
&&\{c_{\alpha,i},c_{\beta,j} \} = 0,
&&\{c^\dagger_{\alpha,i},c^\dagger_{\beta,j} \} = 0,
\end{align}
where $\alpha$ and $\beta$ can be either $\uparrow$ and $\downarrow$. The $R$-matrix \eqref{Rphiphi}-\eqref{Rpsipsi} can be written completely in term of oscillators 
\begin{align}
R_{12}^{(osc)} =
&~\sum_{\substack{\{\alpha,\beta\}=\{\uparrow,\downarrow\}, \\ \{\downarrow,\uparrow\}}} \Big[ ( c^\dagger_{\alpha,1}c_{\alpha,2} + c_{\alpha,1} c^\dagger_{\alpha,2})(C_1 + C_2 (n_{\beta,1} - n_{\beta,2})^2 ) + \nonumber\\
&\qquad ( c^\dagger_{\alpha,1}c_{\alpha,2} - c_{\alpha,1} c^\dagger_{\alpha,2})(C_3(n_{\beta,1}-\frac{1}{2}) + C_4 (n_{\beta,2}-\frac{1}{2}) )\Big]\nonumber \\
&~+( c^\dagger_{\uparrow,1}c^\dagger_{\downarrow,1}c_{\uparrow,2}c_{\downarrow,2} +  c_{\uparrow,1}c_{\downarrow,1}c^\dagger_{\uparrow,2}c^\dagger_{\downarrow,2}) C_5 + \nonumber
( c^\dagger_{\uparrow,1}c_{\downarrow,1}c^\dagger_{\downarrow,2}c_{\uparrow,2} +  c^\dagger_{\downarrow,1}c_{\uparrow,1}c^\dagger_{\uparrow,2}c_{\downarrow,2}) C_6 \nonumber\\
&~ + C_7 (n_{\uparrow,1}-\frac{1}{2})(n_{\downarrow,1}-\frac{1}{2}) + C_8 (n_{\uparrow,2}-\frac{1}{2})(n_{\downarrow,2}-\frac{1}{2})\nonumber  \\
&~ + C_9 (n_{\uparrow,1}-n_{\downarrow,1})^2 (n_{\uparrow,2}-n_{\downarrow,2})^2 +\nonumber\\
&~ + (C_5-C_6) (n_{\uparrow,1}n_{\downarrow,1}+n_{\uparrow,2}n_{\downarrow,2}-1)(n_{\uparrow,1}-n_{\uparrow,2})(n_{\downarrow,1}-n_{\downarrow,2})\nonumber \\ 
&~ +\frac{1}{2} C_5 ((n_{\uparrow,1}-n_{\downarrow,2})^2+(n_{\downarrow,1}-n_{\uparrow,2})^2)+C_0,
\end{align}
\noindent
where $n_{\alpha,k}\equiv c^\dagger_{\alpha,k} c_{\alpha,k}$, $\,\,\,C_0,\dots,C_9$ are functions dependent on the {{} parameters} $(u,v)$ and related to the $r_i$ in the following way
\begin{align}
& C_0=\frac{1}{2} \left(\left(r_4+r_6\right) s_2 +r_2\right),\,C_1=\frac{1}{2} \left(r_7-r_5\right),\,C_2=\frac{1}{2} \left(r_5-r_7-s_1 (r_3+r_{10})\right), \nonumber\\
& C_3=\frac{1}{2} \left( s_1 (r_3-r_{10} ) +r_5+r_7\right),\,C_4=\frac{1}{2} \left(r_5+r_7-s_1 (r_3 -r_{10})\right),\,C_5=-r_2,\,C_6=-r_9,\nonumber\\
& C_7=-2 r_6 s_2 +2 r_1+r_2,\, C_8=-2 r_4 s_2 +2 r_1+r_2,\,C_9=-(r_4+r_6) s_2 +r_1+r_2-r_8-r_9,
\end{align}
$s_1$ and $s_2$ are arbitrary signs\footnote{Interestingly, using oscillators, the arbitrariness of $s_1$ and $s_2$ naturally emerges. $s_1$ and $s_2$ can be understood as a local basis transformation ($s_1$) and a twist ($s_2$).}.
\\
By taking the logarithmic derivative of the $R$-matrix, we get the following expression for the Hamiltonian
\begin{align}
\mathcal{H}_{12}^{(osc)} &= (h_2+h_9) \left[\sum _{\alpha =\{\uparrow ,\downarrow \}} n_{\alpha ,2}  n_{\alpha ,1}-\left(n_{\uparrow ,1}+n_{\downarrow ,1}\right)  n_{\uparrow ,2}  n_{\downarrow ,2}-\left(n_{\uparrow ,2}+n_{\downarrow ,2}\right)  n_{\uparrow ,1}  n_{\downarrow ,1} +  \right.\nonumber\\
&\left.~2n_{\downarrow ,1}  n_{\uparrow ,1}  n_{\downarrow ,2}  n_{\uparrow ,2} \right]+(h_1-h_4-h_6+h_8)\Big[ \left(n_{\uparrow ,2}+n_{\downarrow ,2}\right)  \left(n_{\uparrow ,1}+n_{\downarrow ,1}\right) -\nonumber\\
&2\left(n_{\uparrow ,1}+n_{\downarrow ,1}\right)  n_{\downarrow ,2}  n_{\uparrow ,2} -2 \left(n_{\uparrow ,2}+n_{\downarrow ,2}\right) n_{\downarrow ,1}  n_{\uparrow ,1} + 4n_{\downarrow ,1}  n_{\uparrow ,1}  n_{\downarrow ,2}  n_{\uparrow ,2} \Big] + \mathcal{H}_{12}^{(1pt)}+ \nonumber\\
&\mathcal{H}_{12}^{(2pt)} - \left[ (h_5+h_7)s_2+(h_{10}-h_3)s_1 \right]\mathcal{H}_{12}^{(3pt)},
\label{eq:ham1616}
\end{align}
where we explicitly separate $\mathcal{H}_{12}^{(k\, pt)}$ as subsectors with $k$-particles. This is possible since
\begin{align}
&[n_{TOT},\mathcal{H}_{12}^{(osc)}]=0, && n_{TOT}=\sum _{\alpha =\{\uparrow ,\downarrow \}} n_{\alpha,1}+n_{\alpha,2},
\end{align}
so the total number of particles is conserved. The $ k $-particle subsectors are represented by
\begin{equation}\label{eq:n-particle}
	\begin{aligned}
		\mathcal{H}_{12}^{(1pt)} &= \left(\left( c^\dagger_{\uparrow ,1} c_{\uparrow ,2}+ c^\dagger_{\downarrow ,1} c_{\downarrow ,2}\right)h_5+\left( c^\dagger_{\uparrow ,2} c_{\uparrow ,1}+ c^\dagger_{\downarrow ,2} c_{\downarrow ,1}\right)h_7\right)s_2+h_6\left(n_{\uparrow ,1}+n_{\downarrow ,1}\right)\\
		& +h_4\left(n_{\uparrow ,2}+n_{\downarrow ,2}\right) - (h_1+h_2)\left(-1+n_{TOT}\right),\\
		\mathcal{H}_{12}^{(2pt)} &= f(\uparrow ,\downarrow )+f(\downarrow ,\uparrow )-\left(c^\dagger_{\downarrow ,1} c^\dagger_{\uparrow ,1} c_{\downarrow ,2} c_{\uparrow ,2}+c^\dagger_{\downarrow ,2} c^\dagger_{\uparrow ,2} c_{\downarrow ,1} c_{\uparrow ,1}\right)h_2 \\
		& + n_{\downarrow ,2} n_{\uparrow ,2}(2 h_1+h_2-2 h_4) + n_{\downarrow ,1} n_{\uparrow ,1}(2 h_1+h_2-2 h_6),\\
		\mathcal{H}_{12}^{(3pt)} & = n_{\uparrow ,1}  n_{\uparrow ,2}  \left( c^\dagger_{\downarrow ,1}  c_{\downarrow ,2}- c^\dagger_{\downarrow ,2}  c_{\downarrow ,1}\right)+n_{\downarrow ,1}  n_{\downarrow ,2}  \left( c^\dagger_{\uparrow ,1}  c_{\uparrow ,2}- c^\dagger_{\uparrow ,2}  c_{\uparrow ,1}\right),\\
	\end{aligned}
\end{equation}
with $ f $ defined
\begin{equation}\label{eq:4-osc}
	\begin{aligned}
		f(\alpha,\beta) &= \left(h_{10} s_1 -h_5 s_2 \right) n_{\alpha,1} c^\dagger_{\beta,1} c_{\beta,2}-\left(h_3 s_1 +h_5 s_2 \right) n_{\beta,2} c^\dagger_{\alpha,1} c_{\alpha,2}-\left(h_{10} s_1 +h_7 s_2 \right) n_{\alpha,2} c^\dagger_{\beta,2} c_{\beta,1}\\
		&  + \left(h_3 s_1 -h_7 s_2 \right) n_{\beta,1} c^\dagger_{\alpha,2} c_{\alpha,1}+h_9 c^\dagger_{\alpha,1} c^\dagger_{\beta,2} c_{\alpha,2} c_{\beta,1}+h_2 n_{\beta,2} n_{\alpha,1}.
	\end{aligned}
\end{equation}
We focus on the case where $r_3\neq0$ (section \ref{r3no0}) since it contains the one-dimensional Hubbard model and the $S$-matrix of $AdS_5 \times S_5$. By using \eqref{condHam1} and \eqref{condHam2}, we should  diagonalise
\begin{align}
\mathcal{H}_{12}^{(osc)}=\mathcal{H}_{12}^{(1pt)}+\mathcal{H}_{12}^{(2pt)} - ((h_5+h_7)s_2+(h_{10}-h_3)s_1)\mathcal{H}_{12}^{(3pt)}.
\end{align}
Similarly to \eqref{canonicaltransf}, we can write the canonical transformation

\begin{align}
&c_{\alpha,k}  = \frac{1}{\sqrt{{{}L}}} {{} \sum_{n=1}^N} e^{2\pi i \frac{k_\alpha n}{{{}L}} } \chi_{\alpha,n},
&&c_{\alpha,k}^\dag  = \frac{1}{\sqrt{{L}}} {{} \sum_{n=1}^L} e^{-2\pi i \frac{k_\alpha n}{{L}} } \chi_{\alpha,n}^\dag. \label{canonicaltransf16}
\end{align}
This is again the natural map since for periodic chains the one-particle eigenstates are simple plane waves. Then, considering $\mathbb{H}^{(k\,pt)}=\sum_{i=1}^{{}L} \mathcal{H}_{i,i+1}^{(k\,pt)},$ 

we arrive at
\begin{equation}
	\begin{aligned}
\mathbb{H}^{(1pt)}&=\left(h_1+h_2\right){{}L}+{{}L}\sum _{n=1}^{{}L} \left(h_4+h_6-2 \left(h_1+h_2\right)\right)N_n\\
&+s_2 {L} \sum _{n=1}^{L} \left(\left(h_5+h_7\right) \cos \left(\frac{2 \pi  n}{L}\right)+i \left(h_5-h_7\right) \sin \left(\frac{2 \pi  n}{{L}}\right)\right)N_n,
	\end{aligned}
\end{equation}
where $N_n=\sum _{\alpha =\{\uparrow ,\downarrow \}}N_{\alpha,n}$, $N_{\alpha,n}=\chi^\dagger_{\alpha,n}\chi_{\alpha,n}$
and
\begin{equation}
	\begin{aligned}
\mathbb{H}^{(2pt)}&=\sum _{n,m=1}^{L}\left(2 h_2 \cos \left(\frac{2 \pi  (m+n)}{{L}}\right)+\left(2 \left(2 h_1+h_2\right)-2 \left(h_4+h_6\right)\right)\right) N_{\downarrow,n} N_{\uparrow,m}\\
&+\sum _{n,m=1}^{L}  4\, h_2\, \cos^2\left(\frac{ \pi   (n-m)}{{L}}\right)\left(N_{\downarrow,n}N_{\uparrow,m}+N_{\uparrow,n}N_{\downarrow,m}\right)\\
&-\sum _{n,m=1}^{L} 2  \left(s_2\left(h_5+h_7\right) \cos \left(\frac{2 \pi  m}{{{}N}}\right)+i \sin \left(\frac{2 \pi  m}{{{}N}}\right) \left(\left(h_3-h_{10}\right) s_1 +\left(h_5-h_7\right) s_2\right)\right)\\
&\,\,\,\,\,\,\,\left(N_{\downarrow,n}N_{\uparrow,m}+N_{\uparrow,n}N_{\downarrow,m}\right).
	\end{aligned}
\end{equation}
We are left with finding $\mathbb{H}^{(3pt)}$. After applying the canonical transformation \eqref{canonicaltransf16}, we were able to get a closed expression for ${{}L}=4,5$ and we generalized it for arbitrary $L$,
\begin{equation}
	\begin{aligned}
\mathbb{H}^{(3pt)}&=i \sum_{\substack{\{\alpha,\beta\}=\{\uparrow,\downarrow\}, \\ \{\downarrow,\uparrow\}}}\Big[\sum _{n=1}^{{}L} N_{\alpha,n} \sin \left(\frac{2 \pi  n}{{L}}\right)\Big] \Big[\psi \sum _{i=1}^{L} \left(A_{i,i+1,i+{L}-1,i+2}^{(\beta)}+A_{i+2,i+{L}-1,i+1,i}^{(\beta)}\right)\\
&+\left(S_1 \sum _{i =1}^{L} N_{\beta,i } N_{\beta,i +1}+S_2 \sum _{i =1}^{L} N _{\beta,i } N _{\beta,i+2}\right)\Big],
	\end{aligned}
\end{equation}
where $A^{(\beta)}_{a,b,c,d}= \chi^\dagger_{\beta ,a}\, \chi^\dagger_{\beta ,b}\, \chi _{\beta ,c}\, \chi _{\beta ,d}$, $\psi$, $S_1$ and $S_2$ are constants dependent on ${L}$. We can see that $\mathbb{H}^{(3pt)}$ is not diagonal. 
The coefficient of $\mathcal{H}_{12}^{(3pt)}$ in \eqref{eq:ham1616} is $- ((h_5+h_7)s_2+(h_{10}-h_3)s_1)$. Since $\mathbb{H}^{(3pt)}$ is not diagonal, if this coefficient is zero, our Hamiltonian will be explicitly of free fermion type. We evaluated it for {{} the} various models  of \cite{classificationybandboost}, in particular for model 7 ($AdS_5\times  S^5$) and model 8 (which can be obtained from model 7 by taking a double limit). We found that for model 7, $((h_5+h_7)s_2+(h_{10}-h_3)s_1)$ is not 0. For model 8, it can be easily shown that $((h_5+h_7)s_2+(h_{10}-h_3)s_1)=k$, where $k$ is a constant. With a constant diagonal local basis transformation on the Hamiltonian, we can send $h_{10}\to {{} \zeta} h_{10}$ and $h_3\to \frac{h_3}{{{} \zeta}}$, with ${{} \zeta}$ constant, so the coefficient of the term $\mathbb{H}^{(3pt)}$ can be put to zero. In this case the Hamiltonian is diagonal and we obtained that a double limit of the Hamiltonian\footnote{We refer to the footnote \ref{footnotedressingfactor} of section \ref{integrhamsu22} for the meaning of the Hamiltonian of $AdS_5\times  S^5$.} {{} $AdS_5\times  S^5$} is \textit{manifestly free fermion} type.
\\
Furthermore, we can notice that also the Hamiltonians of model 8 and 12 of the difference form classification \cite{deLeeuw:2019vdb} verify the conditions\footnote{We mention that \eqref{condHam1} and \eqref{condHam2} together with $((h_5+h_7)s_2+(h_{10}-h_3)s_1)=0$ are the only conditions used to make the Hamiltonian \eqref{eq:ham1616}  free fermion type.} \eqref{condHam1} and \eqref{condHam2} and with a diagonal local basis transformation the coefficient of $\mathbb{H}^{(3pt)}$ can be set to zero. In these two cases the Hamiltonians are also of free fermion type. For model 12 this is not surprising since it corresponds to the free {{} Hubbard model ({\it i.e.} with only the kinetic term)} \cite{essler2005one}. However, model 8 is a new model of difference form type in which electrons can only propagate when they are in pair. We believe that the free fermion nature of this model may help to analyze the spectrum and we reserve this analysis for future studies.

\section{Summary}
In this chapter, we have established that the models 6-V A and 8-V A,  discussed in chapter \ref{6and8Vmodel}, satisfy the Baxter relation. On the other hand,  models 6-V B and 8-V B exhibit a free fermion condition. The difference between those two classes of model is that class A are of difference form and their Hamiltonian has the term $Z\otimes Z$, while models of  class B are of non difference form. We have demonstrated that the free fermion condition enables us to express the transfer matrix associated with the 6-V B model in a diagonal form, simplifying the computation of eigenvalues and eigenvectors.  This was previously hidden in the standard formalism.  We have explicitly shown this for the case of AdS$_3$ massless sector of the pure Ramond-Ramond flux.  Furthermore, we have identified the equivalent free fermion conditions for certain models with $\alg{su}(2)\oplus \alg{su}(2)$ symmetry discussed in chapter \ref{threeorfour}. We obtained that the Hamiltonian of model 8 (corresponding to a limit of the ones obtained by requiring centrally extended $\alg{su}(2|2)$ symmetry, \cite{arutyunov2009foundations,Arutyunov:2006yd}) exhibits  a free fermion nature and can be made explicitly diagonal. Additionally, we have discovered that a new integrable model of difference form type presented in \cite{deLeeuw:2019vdb} (model 8), also possesses this property. We believe that these findings may provide an intuitive approach to performing the nesting Bethe ansatz, although we reserve these considerations for future studies.
\chapter*{Conclusions and open questions}
\ifpdf
    \graphicspath{{Chapter7/Figs/Raster/}{Chapter7/Figs/PDF/}{Chapter7/Figs/}}
\else
    \graphicspath{{Chapter7/Figs/Vector/}{Chapter7/Figs/}}
\fi
\addcontentsline{toc}{chapter}{\protect\numberline{}Conclusions and open questions}
\markboth{Conclusions and open questions}{}
In this thesis, we used the boost operator to establish a systematic methodology for classifying quantum integrable spin chains characterized by a regular $R$-matrix of non-difference form.
\\
As mentioned,  traditional approaches to classifying integrable models focus on the search of the $R$-matrix, which is a solution to the Yang-Baxter equation, and utilize it to construct the transfer matrix. The transfer matrix generates all the conserved charges of the integrable model. However, our approach differs in that we start with an ansatz for the Hamiltonian $\mathbb{Q}_2$ and impose constraints to ensure its compatibility with an integrable model. The crucial step, as explained in chapter \ref{classificationchapter}, consists in using the boost operator to construct $\mathbb{Q}_3$, which depends on $\mathbb{Q}_2$. The integrability constraints impose that $[\mathbb{Q}_2,\mathbb{Q}_3]=0$. We impose the results of this constraint on $\mathbb{Q}_2$ and subsequently use it to derive the corresponding $R$-matrix. In this way, the integrability of the models found is guaranteed.\\The primary strength of this method lies in its versatility and wide-ranging applicability across various contexts. The selection of the ansatz depends on several factors, such as the  focus on Hamiltonians with specific symmetries or particular properties.

\vspace{0.3cm}

In chapter \ref{integrableopenquantumsystemchapter}, we present one of the main result of this thesis: the application of the boost automorphism method to \textit{classify integrable open quantum systems}. These correspond to physical systems in contact with Markovian environments. The dynamic of the system is described via the Lindblad Master equation. This evolution may be realized by the action of a Lindblad superoperator on the states of the system. Specifically, our focus is on the finding of integrable Lindblad superoperator. Consequently, our ansatz begins by selecting the Hamiltonian ($h$) of the system and the jump operator ($\ell$) that characterizes the effects that the environment has on the system. We establish the \textbf{first systematic classification of integrable Lindblad} superoperator.

\vspace{0.3cm}

Through the application of our method, we have discovered interesting new integrable models. Specifically, in chapter \ref{Lindbladclassificationmodels}, by imposing the 6-V type conditions on both $h$ and $\ell$, we have identified two integrable models that do not require fine-tuning of the system-environment interaction. We call them model B3 and B2. For model B3, we have examined the physical properties. For a choice of the parameters, the Non Equilibrium Steady States (NESS) are pure spin helix states. Furthermore, our analysis reveals that this is an integrable example of the pumping effect: there is a finite particle current flowing through the system even when the coupling constant with the environment is really small. Additionally, we have established the equivalence of model B3 to the generalized Toda system associated with the non-exceptional affine Lie algebra $A_3^{(2)}$. Moreover, we provide a 4-D interpretation of this model as two coupled spin-1/2 XXZ chains. To solve this model, we employ the nested algebraic Bethe ansatz.

\vspace{0.3cm}

By recalling some known results, in chapter \ref{Hubbardchapter}, we demonstrate that the Hubbard model can be expressed in a superoperator form by complexifying the spectral parameter. Specifically, it can be represented as an XX spin chain with dephasing noise. In this thesis, we have expanded upon this existing understanding by identifying all the potential scenarios where the mapping between Hermitian models and Lindblad superoperators is applicable.\\
In our classification, we encounter the Hubbard model itself, along with an integrable deformation (model B2) of it that was previously discovered.\\
Starting from a more general ansatz, in particular imposing the 8-V conditions for both $h$ and $\ell$,  we have derived a novel elliptic model that can be attributed to a \textit{new integrable deformation of the Hubbard model}. To the best of our knowledge, this deformation is new because the entries of the $R$-matrix depend on the square root of the Jacobi function. We believe that this functional dependence has not been encountered previously. Additionally, we are not aware of any integrable deformation of the Hubbard model that span three site of the spin chain.

\vspace{0.3cm}

In chapter \ref{NESSCHAPTER}, we analyzed this deformation from the point of view of open quantum systems. We calculated the NESS of the model and we discovered that there exists $L+1$ NESS. This implies that the system retains memory of its initial state. Different initial density matrices evolve into distinct NESS in the long-time limit. Investigating the origin of this multiplicity, we discover the presence of hidden strong symmetries in the form of quasi-local charges. We compute the NESS exactly as a Matrix Product Operator with fixed bond dimensions.  Furthermore, we established that the system's dynamics give rise to the emergence of the Gibbs ensemble through the influence of these hidden quasi-local charges.

\vspace{0.3cm}

An intriguing avenue for further investigation involves applying integrability techniques to analyze the range 3 model. However, the fact that the number of particles is not conserved makes the application of  the standard nested Bethe ansatz technique ineffective. We expect that some combination of the nested Bethe Ansatz with methods used to solve the XYZ spin chain needs to be used. A perhaps more promising approach consists in finding the nested Bethe ansatz solution associated to a model connected to it by a global transformation and characterized by $h$ to be the XX spin chain and $\tilde{l}$ given in \eqref{ltildedef}. For this model, the $\alg{u}(1)$ symmetry is preserved and therefore the standard nested Bethe ansatz can be applied.
\\Another important aspect is that, while the integrability property of the superoperator $\mathcal{L}$ did not directly influence the analytical computation of the NESS, the superintegrability property of the Hamiltonian $h$ describing the system played a crucial role. Understanding the potential role of other conserved charges in the dynamical evolution of the model would be significant. Since we were able to precisely compute the magnetization for a finite volume chain, we believe that integrability played a role in this computation, even if not explicitly manifested.
\\
Furthermore,  in table \ref{fig:example} we have summarized the diverse connections between the obtained models, in particular between the range 3 model and three NN-type models. This observation suggests the possible existence of a broader family of integrable models that encompass these distinctive points. Investigating this proposition further and employing the boost operator to identify the models that encompass all these limits holds particular interest.
\\
Another significant aspect to consider is that, in some of the models, the diagonal subsector of the superoperator is preserved and it is the generator of a classical Markovian process. Interestingly, we have demonstrated that multiple models can possess the same diagonal subsector, such as those generated by the ASEP model. It would be interesting to explore whether models sharing the same diagonal subsector exhibit common characteristics, for example in the spectrum of their eigenvalues.
\\
As for the application of the boost operator in the context of open quantum systems, an ongoing work focuses on classifying models characterized by an 8-vertex type Hamiltonian, denoted as $h$ and $\ell$. We hope that this search gives rise to other new interesting integrable models.
\\
An interesting observation emerges from our findings: in all the integrable models we have discovered, the Hamiltonian of the system is always of XY-type. However, a formal proof of this statement is currently unavailable.

\vspace{0.3cm}

Moreover, in chapter \ref{threeorfour}, we applied our method to classify models with a local Hilbert space of dimension 4, focusing specifically on models with $\alg{su}(2) \oplus \alg{su}(2)$ symmetry. Within this class, we recovered the matrix part of the $S$-matrix of AdS$_5 \times$ S$^5$, derived by imposing centrally extended $\mathfrak{su}(2|2)$ symmetry using the elliptic parametrization. This model includes the Hubbard model as a special case. To the best of our knowledge, five of the new models that we obtained are new. The properties of these models are worth exploring in future research.
\\
One possible direction to explore is to generalize the ansatz for the density Hamiltonian by relaxing the requirement of having $\alg{su}(2)\oplus \alg{su}(2)$ symmetry. This approach may lead to new deformations of the Hubbard model.

\vspace{0.3cm}

In addition, in chapter \ref{6and8Vmodel} we investigated the cases where the local Hilbert space is $\mathbb{C}^2$ and the Hamiltonian is of 8-vertex type. In fact, we demonstrated that all Hermitian integrable Hamiltonians can be classified within this class. We discovered four models, two of which are of difference form and are equivalent to the XYZ and XXZ spin chains. The other two models are of non-difference form, namely 6-vertex B and 8-vertex B. While 6-vertex B is already known and corresponds to the solution A of \cite{6vColored}, 8-vertex B is a newly discovered model. Interestingly, we showed that those two models are integrable deformation of the blocks with the same chirality of the matrix part of the $S$-matrix of AdS$_2$ and AdS$_3$ integrable models.

\vspace{0.3cm}

Furthermore, in chapter \ref{ffconditionchapter}, we established that the two non-difference form 8-vertex models satisfy the free-fermion condition. This condition allowed us to express the transfer matrix associated with the 6-vertex B model in a diagonal form, simplifying the calculation of its eigenvalues and eigenvectors. We explicitly demonstrated this for the case of the  pure Ramon-Ramon AdS$_3$ massless sector.\\
Moreover, we derived an equivalent free-fermion condition for certain models with $\alg{su}(2)\oplus\alg{su}(2)$ symmetry. For some of these models, similar to the 6-vertex B, we found that their transfer matrix can also be explicitly diagonalized. An intriguing question arises regarding whether these free-fermion conditions for higher-dimensional Hilbert spaces can provide an intuitive approach to performing the nested Bethe ansatz.

\vspace{0.3cm}

Another interesting feature shared by all the ansatz is the fact that constructing integrable models only required imposing the constraint $[\bQ_2,\bQ_3]=0$. Based on this constraint alone, we were consistently able to find the corresponding $R$-matrix that ensures the integrability of the model. This observation relates to an old conjecture \cite{Grabowski} and to the best of our knowledge, it remains unproven.


\begin{spacing}{0.9}


\bibliographystyle{unsrt} 
\cleardoublepage
\bibliography{References/references} 



\end{spacing}


\begin{appendices} 

\chapter{\textit{Nested} Algebraic Bethe ansatz for model B3}
\label{BAB3chapter}

In this appendix, we apply the Nested Algebraic Bethe ansatz method to model B3 given in section \ref{modelB3defined}. As discussed in chapter \ref{intro}, the aim of the Algebraic Bethe ansatz  is to find the eigenvalues of the transfer matrix \cite{faddeev1996algebraic}. From this, in a systematic way, one can construct the eigenvalues of the tower of the conserved charges characterizing the integrable model.
\\
Similar to section \ref{ABAtheory}, here we define the monodromy, the transfer matrix and the reference state specifically for model B3. By using the RTT relation, we give the commutation relations between the entries of the monodromy matrix and their interpretation. Then, we explicitly compute the eigenvalues of the transfer matrix and the Bethe equations, used to determine the momenta of the particles involved in the theory. We do it explicitly  for a state of one and two magnons and explain how to generalize the result to an arbitrary number of particles. Since the Hilbert space dimension is bigger than the ones used in chapter \ref{intro}, we emphasize the main difference between the Nested Algebraic Bethe ansatz and the standard one.

\section{Diagonalization of the transfer matrix}
\label{nestingBA}

\subsection*{Monodromy and transfer matrix. Definitions}

To define the monodromy matrix $T_a(u)$ for a spin chain of length $L$, we need to introduce an auxiliary Hilbert space $V_a$
\begin{equation}
T_a(u)=\prod_{i=1}^L R_{a i}(u_i-u-b),\,\,\,\,\,T_a(u)\in  V_a \otimes\underbrace{V \otimes \dots \otimes V}_{L-times},
\label{monodromy1}
\end{equation}
$u_i$ is the set of inhomogeneities of the chain and $b$ is a constant. The constant is added in a way that makes the resulting expression simpler\footnote{We mention that this constant was also added in the section \ref{ABAtheory}, where we fixed $b=i/2$}.
The transfer matrix is defined as the partial trace ($\Tr_a$) over the auxiliary space of the monodromy matrix,
\begin{equation}
t(u)=\Tr_a T_a (u),\,\,\,\,\,t(u)\in  \underbrace{V \otimes \dots \otimes V}_{L-times}.
\label{transfer1}
\end{equation} 
This matrix generates all the conserved charges characterizing the integrable models, in particular the charge $\mathbb{Q}_2\equiv\mathcal{L}$ which will be identified as the first logarithmic derivative of the transfer matrix, see \eqref{chargeQ2transfer}.

\subsection*{Monodromy and transfer matrix. Constructions}
The monodromy matrix \eqref{monodromy1} in the auxiliary space takes the form of a $4\times 4$ matrix 
\begin{equation}
T_a(u)=\left(
\begin{array}{cccc}
 T_{00} & B_1 & B_2 & B_3 \\
 C_1 & T_{11} & T_{12} & T_{13} \\
 C_2 & T_{21} & T_{22} & T_{23} \\
 C_3 & T_{31} & T_{32} & T_{33} \\
\end{array}
\right),
\label{monodromy}
\end{equation}
where the entries of this matrix are operators acting on the physical space $\underbrace{V \otimes \dots \otimes V}_{L-times}$. For simplicity, we omitted the $(u)$ dependence from all the entries. We notice that since the auxiliary Hilbert space is now bigger than the ones we considered in section \eqref{ABAtheory}, the corresponding monodromy matrix is also bigger. In fact, in \eqref{monodromy22}, we wrote the monodromy matrix as a $2 \times 2 $ matrix.\\
The transfer matrix \eqref{transfer1} is then
\begin{equation}
t(u)=\sum_{i=0}^3 T_{ii}(u).
\end{equation}
\\
The monodromy matrix and the $R$-matrix satisfy the fundamental commutation relations, also known as the RTT-relations,
\begin{equation}
R_{ab}(v-u)T_a(u)T_b(v)=T_b(v)T_a(u)R_{ab}(v-u).
\label{rtt}
\end{equation}
The space where this matrix acts is  $V_a \otimes V_b \otimes \underbrace{V\otimes \dots \otimes V}_{L \,\text{times}}$, with $V_a$ and $V_b$ auxiliary spaces.
\\
By plugging the expression of the $R$-matrix and the monodromy matrices given respectively  in section \ref{modelB3defined} and \eqref{monodromy}, it follows that
\begin{equation}
[B_i(u),B_i(v)]=0,\,\,\,\,\,i=1,2,3,
\label{commrels}
\end{equation}
which gives the immediate interpretation: $B_1$ and $B_2$ (and also $B_3$) are the creation operators for our theory.

\subsection{The reference states and the action of the transfer matrix}
\label{therefstate}
Since model B3 commutes with the spin operator $S_z$, a good choice for the reference state is
\begin{equation}
\ket{0}=\bigotimes_{i=1}^L\left(
\begin{array}{c}
 1 \\
 0 \\ 0 \\ 0 \\
\end{array}
\right).
\end{equation}

The action of the elements of the transfer matrix  on the reference state $\ket{0}$, by fixing the constant $b=\psi$ is
\begin{equation}
T_{00}(u)\ket{0}=\ket{0},
\label{T00vacuum}
\end{equation}
\begin{equation}
T_{11}(u)\ket{0}=\prod_{i=1}^L \frac{ \sinh \left(u-u_i+\psi \right)}{i\,e^{\psi +i \phi }\,\cosh \left(u-u_i\right)}\ket{0},
\label{T11vacuum}
\end{equation}
\begin{equation}
T_{22}(u) \ket{0}=\prod_{i=1}^L   \frac{i\,\sinh \left(u-u_i+\psi \right)}{e^{\psi - i \phi }\,\cosh \left(u-u_i\right)}\ket{0},
\label{T22vacuum}
\end{equation}
\begin{equation}
T_{33}(u) \ket{0}=\prod_{i=1}^L e^{-2 \psi }\frac{\sinh \left(u-u_i+\psi \right)}{\cosh \left(u-u_i-\psi \right)} \tanh \left(u-u_i\right)  \ket{0},
\label{T33vacuum}
\end{equation}
and the following annihilation identities hold
\begin{equation}
C_1 \ket{0}=C_2 \ket{0}=C_3 \ket{0}=0,\,\,T_{ab}\ket{0}=0\,\, (a\neq b =1,2).
\end{equation}
$\{u_i\}$, already introduced in \eqref{monodromy1}, are the set of inhomogeneities\footnote{We remark that for simplicity in the section \ref{ABAtheory}, we fixed all the inhomogeneities of the chain to be zero.} of the spin chain. From now on, we refer to it as \textit{main} spin chain, for a reason that will be clear in the following.\\
The action of the transfer matrix on the vacuum is
\begin{align}
 t(u) \ket{0}=\prod_{i=1}^L &\bigg[e^{-2 \psi } \sinh \left(u-u_i+\psi \right) \bigg(\frac{\tanh \left(u-u_i\right)}{\cosh \left(u-u_i-\psi \right)}-\frac{2 e^{\psi } \sin \phi }{\cosh \left(u-u_i\right)}\bigg)+1\bigg]\ket{0}.
\end{align}
Due to the commutation relations \eqref{commrels}, an excited state can be constructed by acting with the operators $B_1$, $B_2$ and $B_3$ on the vacuum. As an example, a state of two particles with rapidities\footnote{We clarify the notation used in this appendix with the one used in section \ref{ABAtheory}. Here, $\{u_i\}$ are the inhomogeneities of the main spin chain, $u$ is the spectral parameter of the $R$-matrix and $\{v_i\}$ are the rapidities of the particles. In the section \ref{ABAtheory}, we set all the inhomogeneities to 0, $\lambda$ is the spectral parameter and $\{\lambda_i\}$ are the rapidities of the particles.} $v_1$ and $v_2$ is
\begin{equation}
B_1(v_1)B_2(v_2) \ket{0}.
\end{equation}
In what follows, we explicitly construct states of one and two magnons in order to be eigenstates of the transfer matrix.
\\
To understand if a state is an eigenstate of the transfer matrix, we need to find the commutation relations between $T_{ii}(u)$ and the $B$s operators and then act with them on the vacuum via \eqref{T00vacuum}-\eqref{T33vacuum}. The commutation relations can be found from the RTT \eqref{rtt} and we explicitly give them in what follows. Furthermore, the condition that a state is an eigenstate (Bethe vector) fix a constraint on the rapidities $v_i$ of the particles, the Bethe equations.

\subsection{Commutation relations: here comes the nesting}
Before giving the commutation relations between $T_{ii}(u)$ and the $B$s, we focus on the meaning of the operator $B_3$.
\subsubsection{Commutation relation between $B$s}
We now write the RTT-relations \eqref{rtt} in components, it follows that
\begin{equation}
B_{\alpha }(u) B_{\beta }(v)=B_{\delta}(v) B_{\gamma }(u) r_{\alpha \,\beta  }^{\gamma \,\delta}(v-u)-\epsilon _{\alpha ,\beta } \eta(u-v)\left(B_3(v) T_{00}(u)-B_3(u) T_{00}(v)\right),
\label{BBcomm}
\end{equation}
where $\alpha, \beta = 1, 2$ and
\begin{equation}
\eta(u)=i e^{2 \psi } \cosh \psi \, \text{csch}(u-\psi ),\,\,\,\epsilon=(\epsilon_{11},\epsilon_{12},\epsilon_{21},\epsilon_{22})=(0,e^{-i \phi },-e^{i \phi },0).
\label{eta}
\end{equation}
The elements $r_{\alpha \,\beta  }^{\gamma \,\delta}(u)$ can be written in matrix form, 
\begin{align}
&r(u)=r_{\alpha \,\beta  }^{\gamma \,\delta}(u) e_\gamma^\alpha \otimes e_\delta^\beta,
&&r(u)=\left(
\begin{array}{cccc}
 1 & 0 & 0 & 0 \\
 0 & b(u)e^{-2 i \phi} & a(u) & 0 \\
 0 &  a(u) &b(u)e^{2 i \phi} & 0 \\
 0 & 0 & 0 & 1 \\
\end{array}
\right),
\label{rnested}
\end{align}
where $a(u)=\frac{\sinh (2 \psi )}{\sinh (2 (u+\psi ))}$ and $b(u)=\frac{\sinh (2 u)}{\sinh (2 (u+\psi ))}$.\\
It is easy to show that $r(u)$ satisfies the YBE \eqref{eq:YBE} for a spin-1/2 chain. We can consider it as an $R$-matrix of \textit{twisted 6-vertex type}.
\\
This is the first insight of why the Bethe ansatz is called \textit{nested}: in the commutation relations involving different type of particles, the $r$-matrix of a lower dimensional spin chain appears. The same $r$ also appears from the RTT relations involving $T$ and $B$ operators.\\
The commutation relation \eqref{commrels} between the $B$ operators can also be written in components as \eqref{BBcomm} with $\alpha=\beta$.
In this way, we can give an interpretation for $B_1$, $B_2$ and $B_3$. The commutation relations between two fields of type $B_1$ and $B_2$ generate the operator  $B_3$. One can consider the operators $B_1$ and $B_2$ as creation of a particle of spin up and down respectively, while $B_3$ is responsible for the creation of a pair.\\
The $r$-matrix \eqref{rnested} is a twisted version of the ones that appear in the nesting of the Hubbard model and in $AdS_5 \times S^5$, \cite{ramos1997algebraic,martins1998quantum, arutyunov2009bound}.
\subsubsection{Commutation relations between $T_{ii}$ and $B$s}
As mentioned, we need to solve the eigenvalue problem
\begin{equation}
t(u)\ket {M\{v\}}=
\sum_{i=0}^3 T_{ii} (u)\ket {M\{v\}}=\Lambda_M(u)\ket {M\{v\}},
\end{equation}
where $\ket{M\{v\}}$ is a generic state of $M$ excitations with rapidities $\{v\}$. In what follows, for simplicity, we sometimes refer to it as $\ket{M}$. First, we need to find the commutation relations between $T_{ii}$s and $B$s.\\
From the RTT relation, by a brute force calculation, one gets 256 relations, but not all of them are already in a usable form. In particular, we want the right hand side to be normal ordered and have annihilation and diagonal operators on the right most side. In this way, we can apply the same logic as in section \ref{ABAtheory} and remove the unwanted terms. In other words, we want that the commutation relations are of the form
\begin{equation}
T_{ii}(u)B_\alpha(v)=a_1 B_\alpha(v)T_{ii}(u)+a_2 B_\alpha(u) T_{ii}(v)+\dots,
\label{comrelcond}
\end{equation}
the dots ($"\dots"$) contains terms that either annihilate the reference state (for example in the right there is $C_i$) or acts diagonally on it.\\
We can realize this structure by considering linear combinations of some of the 256 relations that we obtained.
\\
Instead of trying the most general linear combination, we first impose that the structure of our commutator relations is the same as the ones in \cite{ramos1997algebraic,martins1998quantum, arutyunov2009bound}. This drastically simplifies the problem and we find
\begin{align}
&T_{00}(u)B_\alpha(v)=&&\theta_\alpha B_\alpha(v)T_{00}(u)+\rho_\alpha B_\alpha(u)T_{00}(v)\label{T00B}\\
\nonumber&T_{\alpha \alpha^\prime}(u)B_\beta (v) =&&  \alpha_{\alpha}B_{\gamma}(v)T_{\alpha \tau}(u)r_{\alpha^\prime \beta}^{\tau \gamma}(v-u)-\psi_\alpha B_{\alpha^\prime} (u) T_{\alpha \beta}(v)-\\
& && (\upsilon T_{\alpha 3} (u)T_{00}(v)+\beta_\alpha B_3 (u)C_\alpha(v)+\gamma_\alpha B_3 (v) C_\alpha(u))\epsilon_{\alpha^\prime \beta}\label{Taa}\\
&T_{33}(u)B_\alpha(v) =&& \zeta_{1,\alpha} B_\alpha(v)T_{33}(u)+\zeta_{2} B_3(v)T_{3\alpha}(u)+\zeta_3 B_3(u)T_{3 \alpha}(v)+\eta \epsilon_{\gamma \eta}T_{\gamma 3}(u)T_{\eta \alpha}(v)\label{T33B},
\end{align}
for simplicity we omitted the spectral dependence of the coefficients $\theta_\alpha, \rho_\alpha, \alpha_\alpha, \psi_\alpha, \dots$, which is $(u-v)$ for all of them.\\
Remarkably, in \eqref{Taa} we again notice the $r$-matrix of twisted 6-vertex type given in \eqref{rnested}. This is another strong insight of why the Bethe ansatz is called nested and the role played by this matrix will be clear in the next paragraph. In fact, in order to solve the Bethe ansatz for  model B3, we first need to solve the Bethe ansatz for the integrable model characterized by the $r$-matrix of twisted 6-vertex type. 
\\The coefficients in the commutation relations\footnote{We mention that the commutation relations found here are independent on the choice of the constant $b$ in \eqref{monodromy1}. This choice will be manifest when we act with the transfer matrix on the reference state.} are
\begin{align}
 & \theta_1=-\frac{i e^{\psi +i \phi } \cosh (u+\psi )}{\sinh u},
&& \rho_1=\frac{i \cosh \psi  e^{\psi +i \phi }}{\sinh u},
&& \theta_1=-e^{2 i \phi } \theta_2,
&& \rho_1=-e^{2 i \phi } \rho_2,
\label{theta}
\end{align}
\begin{align}
\nonumber  & \alpha_1=i  e^{\psi +i \phi }  \cosh \psi \,(\coth u-\tanh \psi ),
&& \psi_1=i e^{\psi +i \phi }  \cosh \psi  \, \text{csch}u,\\
\nonumber & \upsilon = i e^{2 \psi } \cosh \psi  \, \text{csch}(u-\psi ),
&&\beta_1= -e^{3 \psi +i \phi }\cosh ^2\psi\,  \text{csch}u \, \text{csch}(u-\psi ),\\
&\gamma_1= e^{3 \psi +i \phi } \cosh \psi \, \text{csch}u \, \coth (u-\psi )
\label{alpha}
\end{align}
\begin{align}
&\alpha_1=-e^{2 i \phi } \alpha_2,
&&\psi_1=-e^{2 i \phi } \psi_2,
&&\beta_1=-e^{2 i \phi } \beta_2,
&&\gamma_1=-e^{2 i \phi } \gamma_2
\label{condalpha}
\end{align}
\begin{align}
\nonumber &\zeta_{1,1}=\frac{i e^{\psi +i \phi } \cosh (u-2 \psi )}{\sinh (u-\psi )},
&&\zeta_2=\frac{e^{2 \psi } \cosh \psi  \cosh (u-2 \psi )}{\sinh u \sinh (u-\psi )},\\
&\zeta_3=-  e^{u+2 \psi }\frac{\cosh \psi}{\sinh (u-\psi )} \left(\frac{\cosh (u-2 \psi )}{\sinh u}-1\right),
&&\zeta_{1,1}=-e^{2 i \phi } \zeta_{1,2},
\label{zeta}
\end{align}
where the dependence on the spectral parameter is $\omega=\omega(u)$. We notice that $\eta$ and $\epsilon$ in  \eqref{T33B} are the same as the ones in \eqref{eta}.

\subsection{One particle state}
\label{onepstate}
This section and the next one will help to understand the general derivation for arbitrary number of particles.\\
One magnon can be created either by $B_1$ or $B_2$, so the one particle state is a linear combination of these two with weight $F^a$
\begin{equation}
\ket{1\{v\}}=F^a B_a(v) \ket 0,
\end{equation}
where we sum over the repeated index ($a=1,2$), and $\{v\}=v_1=v$ is the rapidity of the magnon.\\
By using \eqref{T00B}-\eqref{T33B} and \eqref{T00vacuum}-\eqref{T33vacuum}, the action of the transfer matrix on one-particle state is
\begin{equation}
T_{00}(u) \ket {1 \{v\}}= \theta_a(u-v) F^a B_a(v) T_{00}(u) \ket{0}+\dots=\theta_a(u-v)  F^a B_a(v) \ket 0+\dots,
\label{T00one}
\end{equation}
and similarly for $T_{33}(u)$
\begin{equation}
T_{33}(u) \ket {1 \{v\}}=\zeta_{1,a}(u-v)\prod_{i=1}^L e^{-2 \psi }\frac{\sinh \left(u-u_i+\psi \right)}{\cosh \left(u-u_i-\psi \right)} \tanh \left(u-u_i\right) F^a\,B_a(v) \ket 0+\dots.
\label{T33one}
\end{equation}
The terms $T_{11}$ and $T_{22}$ require particular analysis. First, it is convenient to write the relations \eqref{T11vacuum} and \eqref{T22vacuum} in the form
\begin{align}
T_{\alpha \tau}(u)\ket{0}=\delta_{\alpha \tau}\prod_{i=1}^L \,(-e^{2 i \phi})^{\alpha-1}f(u,\{u_i\})\ket{0},
\label{Taaone}
\end{align}
where $f(u,\{u_i\})= \frac{ \sinh \left(u-u_i+\psi \right)}{i\,e^{\psi +i \phi }\cosh \left(u-u_i\right)}$.\\
The action of $T_{11}+T_{22}$ is
\begin{align}\label{commutationrelationTaa}
\nonumber  T_{\alpha \alpha}(u)\ket{1}=& F^a T_{\alpha \alpha}(u) B_a(v) \ket 0= F^a \alpha_\alpha (u-v) B_\gamma(v) r_{\alpha a}^{\tau \gamma} (v-u) T_{k\tau}(u) \ket{0}+\dots=\\
& F^a  \alpha_2 (u-v) (-e^{2 i \phi})^{2-L} B_\gamma(v) \prod_{i=1}^L f(u,\{u_i\})  (-e^{2 i \phi})^{\alpha(L-1)}r_{\alpha a}^{\alpha \gamma} (v-u)\ket{0}+\dots,
\end{align}
where in the last line we used $\alpha_k=(-e^{2\,i\,\phi})^{2-k}\alpha_2$ from \eqref{condalpha}.\\
By neglecting the $\dots$ terms for the moment, we see that $\ket{1}$ is an eigenstate of the transfer matrix if
\begin{equation}
F^a \sum_{\alpha=1}^2 (-e^{2 i \phi})^{\alpha(L-1)}r_{\alpha a}^{\alpha \gamma} (v-u) \sim F^\gamma
\end{equation}
and expanding the sum we get
\begin{equation}
F^a (-e^{2 i \phi})^{L-1} (r_{1 a}^{1 \gamma} +(-e^{2 i \phi})^{L-1}r_{2 a}^{2 \gamma}) \sim F^\gamma,
\label{combr}
\end{equation}
so $F^a$ needs to be an eigenvector of the combination of $r$ given in \eqref{combr}. This will be more clear in the case of $M$ particles, but the contractions of the indices in the $r$ define the transfer matrix of the 6-vertex model for a spin chain of length 1. To summarize, if $F$ is an eigenstate of this transfer matrix, $T_{\alpha \alpha}$ acts diagonally on $\ket 1$. The initial problem of finding the eigenvalues of the transfer matrix built from the $R$-matrix of our model, reduces to the auxiliary problem to diagonalize the transfer matrix of 6-vertex type and \textit{here comes the nesting}. The Bethe ansatz for this spin chain is known (for example see Appendix B of \cite{arutyunov2009bound}), however we repeat the analysis with our notation and twist in section  \ref{betheansatznested}.\\
During all the discussion, we ignored the terms $\dots$ in \eqref{commutationrelationTaa}. We now understood that those terms can be removed by imposing some relations for the rapidity $v$. This condition is called \textit{Bethe equation}. For the case of one and two particles, this calculation is still doable\footnote{In section \ref{ABAtheory}, to take confidence with the methods, we explicitly did for the state of one and two magnons and compare with the results obtained by using the shortcut of imposing that the residue at the pole of the eigenvalue is zero.}, but becomes very tedious for the states of more magnons. We followed here the standard shortcut of the residue that gives the same Bethe equations as the explicit calculation. The eigenvalue of the transfer matrix obtained by summing \eqref{T00one}, \eqref{T33one} and \eqref{Taaone} is
\begin{equation}
t(u)B_a \ket{0}=\Lambda_{1,a}(u)B_a\ket{0},
\end{equation}
where $N=1$ if the particle is created by $B_2$ and $N=0$ otherwise,
\begin{align}\label{eigenv1magnonnested}
\frac{\Lambda_{1,a}(u)}{\sigma e^{\psi +i \sigma \phi }}=&\frac{i \cosh \left(u-v+\psi \right)}{\sinh \left(v-u\right)}+\frac{i \cosh \left(u-v-2 \psi \right)}{\sinh \left(u-v-\psi \right)}\prod _{i=1}^L \frac{ \tanh \left(u-u_i\right) \sinh \left(u-u_i+\psi \right)}{e^{2 \psi }\cosh \left(u-u_i-\psi \right)}+\\
& i \left(-e^{2 i \phi }\right)^{N-L+1} \frac{\cosh \left(u-v-\psi \right)}{\sinh \left(u-v\right)} \lambda_{6V}(u) \prod _{j=1}^L \frac{ \sinh \left(-u_j+u+\psi \right)}{i e^{\psi +i \phi }\cosh \left(u-u_j\right)},
\end{align}
where $\sigma=1$ for $a=1$ ($N=0$) and $\sigma=-1$ for $a=2$ ($N=1$).\\
The expression of $\lambda_{6V}(u)$ is given by \eqref{eigenv6vexplicit} in section \ref{betheansatznested}, for completeness we will also write here the one for one particle
\begin{align}
\frac{\lambda _{6V}(u)}{\left(-e^{2 i \phi }\right)^{L-N-1}}=&\left(-e^{2 i \phi }\right)^L\left(\frac{\sinh \left(2 \left(u-w-\psi \right)\right)}{\sinh \left(2 \left(w-u\right)\right)}\right)^N \frac{\sinh \left(2 \left(v-u\right)\right)}{\sinh \left(2 \left(u-v-\psi \right)\right)}+\\
&\left(\frac{\sinh \left(2 \left(w-u-\psi \right)\right)}{\sinh \left(2 \left(u-w\right)\right)}\right)^N.
\end{align}

As we mentioned, we use the shortcut to remove the unwanted terms. The eigenvalue should be regular, the residue  at the pole should vanish. In this case, the eigenvalue \eqref{eigenv1magnonnested} of the transfer matrix has two poles, for $u\to v$ and $u\to v+\psi$. In what follows we  require the cancellation of the residue at $u \to v$, but it can be proved that analysing the residue around the second pole  give a set of equation that can be mapped to the ones we are giving.\\
This leads to the following results. The rapidities should satisfy the following condition
\begin{align}
\left(\frac{\sinh \left(2 \left(w-v-\psi \right)\right)}{\sinh \left(2 \left(v-w\right)\right)}\right)^N \prod _{j=1}^L \frac{ \sinh \left(v-u_j+\psi \right)}{i\, e^{i \phi +\psi }\cosh \left(v-u_j\right)}=1
\label{BEonep}
\end{align}
and
\begin{align}
\frac{\left(-e^{2 i \phi }\right)^L \sinh \left(2 \left(v-w\right)\right)}{\sinh \left(2 \left(w-v-\psi \right)\right)}=1.
\label{beonepnest}
\end{align}
Let us clarify the meaning of all the parameters appearing in the expressions. The $u_i$ are the set of inhomogeneities of the main chain. $v$ is the rapidity of the one magnon state we are considering and satisfies the Bethe equation \eqref{BEonep}. $v$ is also the inhomogeneity in the nested chain. If the magnon is created by $B_2$, $N=1$,  there is also the parameter $w$. This latter is the rapidity of the particle in the nested chain and can be calculated via the auxiliary Bethe equation \eqref{beonepnest}.\\
The block structures of the eigenvalue and the Bethe equations suggest a way to generalize this result to $M$ magnons. Furthermore, one can get the explicit expression of the eigenstate recursively by following the derivation given in \cite{martins1998quantum}.\\
To understand how this result can be generalized for the $M$ magnons state and where the difficulties emerge, we first explicitly derive the expression for 2 magnons. 

\subsection{Two particle state}
\label{2partstates}
The explicitly computation for the state of two magnons is particularly relevant to understand the generalization for the $M$ magnons case. In fact, the general steps here are slightly different than the one for the one-particle state. The state of two-magnons is composed by two parts
\begin{align}
\ket{2\{v\}}=B_{a_1}(v_1)B_{a_2}(v_2)F^{a_1 a_2}\ket{0}+B_3(v_1)g(v_1,v_2)\epsilon_{a_1 a_2}F^{a_1 a_2}T_{00}(v_2) \ket{0},
\label{2particlesstate}
\end{align}
the first one takes into account that the two particles are created by the operators $B_1$ and $B_2$. In a state of two particles there may also be a pair, created by $B_3$. The fermionic nature of the particle is manifest in the $\epsilon$ which accounts the Pauli exclusion principle. The operator $T_{00}(v_2)$ in the second part is put for  dimensionality, in fact the monodromy matrix is normalized such that  the action of it on the vacuum is $1$, as we see in  \eqref{T00vacuum}.\\
We now derive the expression of $g(u)$ and we interpret $F^{a_1 a_2}$ as the eigenvector on the transfer matrix in the nested chain.
\subsubsection{Action of $T_{00}$}
In order to evaluate the action of $T_{00}$ on $\ket {2\{v\}}$ we need an additional commutation relation. In particular
\begin{align}
T_{00}(u)B_3(v)=q_1 B_3(v) T_{00}(u)+q_2 B_3(u) T_{00}(v)+q_3 B_{1}(u) B_{2}(v)+q_4 B_{2}(u) B_{1}(v),
\end{align}
where we omitted the dependence $\omega=\omega(u-v)$ on the coefficients.
\begin{align}
&q_1(u)=\frac{1}{2} e^{2 \psi } \text{csch}u\, (\cosh (2 u+3 \psi )+\cosh (\psi )) \text{csch}(u+\psi ),\\
&q_2(u)=-e^{u+2 \psi } \cosh \psi \, \text{csch}(u+\psi ) (\text{csch }u \cosh (u+2 \psi )-1),
\end{align}
\begin{align}
&q_3(u)=i \,e^{i \phi } \cosh\, \psi \, \text{csch}(u+\psi ),
&&q_4(u)=-i\, e^{-i \phi } \cosh \psi \, \text{csch}(u+\psi ).
\end{align}
By using this relation and \eqref{T00B}, we imposed  that
\begin{align}
T_{00}(u)\ket{2\{v\}}=\lambda_0(u) \ket{2\{v\}},
\end{align}
and uniquely fix the form of $g(v_1,v_2)$. We got
\begin{align}
g(v_1,v_2)=\frac{\theta_{a_1}(u-v_1)\rho_{a_2}(u-v_2)\eta(u-v_1)}{\lambda_0(u)-q_1(u-v_1)},
\end{align}
where $\eta(u)$ was defined in \eqref{eta}. $g(v_1,v_2)$ does not depend on $u$, in fact plugging all the expressions we got
\begin{align}
g(v_1,v_2)=i e^{2 \psi } \cosh \psi  \text{csch}\left(v_1-v_2-\psi \right)=\eta(v_1-v_2).
\label{expressiong}
\end{align}

This fix the expression of the ansatz for the 2 magnons state. The eigenvalue is
\begin{align}
\lambda_{0}(u)=\theta_{1,a_1}(u-v_1)\theta_{1,a_2}(u-v_2).
\label{eigv2p0}
\end{align}
This eigenvalue factorizes as a product of one particle eigenvalue. This gives a strong hint on how to generalize the calculation to the case of $M$ magnons.
\subsubsection{Action of $T_{33}$}
Similarly, to confirm our result, we can act with $T_{33}$ on $\ket{2\{v\}}$. In this case, the commutation relations that we need are
\begin{align}
T_{33}(u)B_3(v)=\eta_1 B_3(v) T_{33}(u)+\eta_2 B_3(u) T_{33}(v)+\eta_3 T_{13}(u) T_{23}(v)+\eta_4 T_{23}(u) T_{13}(v),
\label{commrel1}
\end{align}
where we omitted the dependence $\omega=\omega(u-v)$ and
\begin{align}
&\eta_1(u)=\frac{1}{2} e^{2 \psi } \text{csch }u\, \text{csch}(u-\psi )(\cosh (2 u-3 \psi )+\cosh \psi ) ,\\
&\eta_2(u)=e^{u+2 \psi } \cosh \psi\,  \text{csch}(u-\psi ) (1-\text{csch }u \cosh (u-2 \psi )),
\end{align}
\begin{align}
&\eta_3(u)=i \cosh \psi\,  e^{2 \psi -i \phi } \text{csch}(u-\psi ),
&&\eta_4(u)=-e^{2 i \phi } \eta_3(u)
\end{align}
and
\begin{align}
T_{3a}(u)B_b(v)=\Gamma_{1ab} B_a(v) T_{3b}(u)+\Gamma_{2ab} B_b(v) T_{3a}(u)+\Gamma_{3} \epsilon_{ab}(B_3(v) C_3(u)-T_{00}(v) T_{33}(u)),
\label{commrel2}
\end{align}
\[ 
 \frac{2\,\Gamma_{1ab}(u) }{e^{2 \psi } \text{csch}u \,\text{csch}(u-\psi )}= 
  \begin{dcases*} 
\cosh (2 u-3 \psi )+\cosh \psi , & if  a=b=1,2 \\ 
\cosh \psi +\cosh (3 \psi ), & if  a$\neq$ b=1,2 
  \end{dcases*} 
\]

\[ 
\Gamma_{2ab}(u)= 
  \begin{dcases*} 
0, & if  a=b=1,2 \\ 
e^{2 \psi +2 i \phi }, & if  a=1, b=2\\ 
e^{2 \psi -2 i \phi }, & if  a=2, b=1
  \end{dcases*} 
\]
\begin{align}
\Gamma_3(u)=i\,e^{2 \psi } \cosh \psi  \, \text{csch}(u-\psi ).
\end{align}
By using \eqref{commrel1} and \eqref{commrel2} and the fact that $C_3(u)\ket{0}=T_{3\alpha}(u)\ket{0}=0$, imposing that
\begin{align}
T_{33}(u)\ket{2}=\lambda_3(u) \ket{2},
\end{align}
uniquely fix the form of $g(v_1,v_2)$. We got
\begin{align}
g(v_1,v_2)=\frac{-\zeta_2(u-v_1)\Gamma_3(u-v_2)}{\zeta_{1,a_1}(u-v_1)\zeta_{1,a_2}(u-v_2)-\eta_1(u-v_1)},
\end{align}
where $\eta(u)$ was defined in \eqref{eta}. Also in this case, $g(v_1,v_2)$ is independent on $u$ and coincides with \eqref{expressiong}.
\\
The eigenvalue of $T_{33}(u)$ is
\begin{align}
\lambda_{3}(u)=\zeta_{1,a_1}(u-v_1)\zeta_{1,a_2}(u-v_2)\prod_{i=1}^L e^{-2 \psi }\frac{\sinh \left(u-u_i+\psi \right)}{\cosh \left(u-u_i-\psi \right)} \tanh \left(u-u_i\right).
\label{eigv2p3}
\end{align}
We can notice that also in this case the eigenvalue factorizes as a product of one particle eigenvalue.
\subsubsection{Action of $T_{11}+T_{22}$}
This calculation is really important because it makes clear the appearance of the twisted transfer matrix.\\
The additional commutation relations that we need are
\begin{align}
C_{\alpha}(u)B_\beta (v)=\kappa_{\alpha \beta} e^{2 \psi}B_\beta(v)C_\alpha(u)+z_{1\alpha}T_{00}(v)T_{\alpha\beta}(u)+z_{2\alpha}T_{00}(u)T_{\alpha\beta}(v),
\end{align}
where the dependence of the $z$s on the spectral parameter is $z=z(u-v)$ and
\begin{align}
&\kappa_{11}=e^{2 i \phi }=\kappa_{22}^*,
&&\kappa_{12}=-1=-\kappa_{21},
\end{align}
\begin{align}
&\frac{i z_{11}(u)}{e^{i \phi }}=-i z_{12}(u) e^{i \phi }=i z_{22} (u) e^{i \phi }=-\frac{i z_{21}(u)}{e^{i \phi }}=\frac{\text{csch }u}{\tanh \psi-1}
\end{align}
and
\begin{align}
T_{\text{aa}}(u) B_3(v)= &s B_3(v) T_{11}(u)+s_2 \left(B_3(u) T_{\text{aa}}(v)-B_3(v) T_{\text{aa}}(u)\right)+\\
&s_{3,a} B_a(u) T_{\text{a3}}(v)+s_{4,a} T_{\text{a3}}(u) B_a(v)+s_5 B_3(v) T_{\text{aa}}(u),
\end{align}
where $s_i=s_i(u-v)$ and
\begin{align}
&s= -(\tanh (\psi )-1)^{-2},
&&s_2(u)= \text{csch}^2\,u,
&&s_5=\tanh ^2\psi 
\end{align}
\begin{align}
\frac{s_{3,1}(u)}{i e^{i \phi }}=\frac{s_{4,1}(u)}{i e^{-i \phi }}=\frac{s_{3,2}(u)}{i e^{-i \phi }}=\frac{s_{4,2}(u)}{i e^{i \phi }}= (\tanh \psi -1) \text{ csch }u.
\end{align}
As in the previous cases, we can identify the nested problem
\begin{align}
\sum_{\alpha=1}^2 T_{\alpha\alpha}(u)\ket{2\{v\}}=\lambda_{12}(u)\ket{2\{v\}}.
\end{align}
After a very long calculation, one can separate the terms with two $B$s operators and the part with $B_3$. From the second part, one can derive the expressions of $g(v_1,v_2)$ already derived in the previous two cases.\\
From the first part, we get
\begin{align}
&\alpha_\alpha(u-v_1)\alpha_\alpha(u-v_2)r_{\alpha a_1}^{\tau \gamma}(v_1-u)r_{\tau a_2}^{\eta k}(v_2-u) B_\gamma(v_1) B_k (v_2) T_{\alpha\eta}(u)F^{a_1 a_2}\ket{0}\\
&\sim \lambda_{12}(u)F^{\gamma k}B_\gamma(v_1)B_k (v_2)\ket{0},
\end{align}
which can be simplified by considering that
$\alpha_k=(-e^{2 i \psi})^{2-k} \alpha_2$ and
\begin{align}
T_{\alpha\eta}(u)\ket 0=\delta_{\alpha \eta}\prod_{i=1}^L (-e^{2i \phi})^{\alpha-1}f(u)\ket{0},
\end{align}
and $f(u)=\frac{ \sinh \left(u-u_i+\psi \right)}{i\, e^{\psi +i \phi }\cosh \left(u-u_i\right)}$,
\begin{align}
&\alpha_2(u-v_1)\alpha_2(u-v_2)(-e^{2i \phi})^{2(2-\alpha)}r_{\alpha a_1}^{\tau \gamma}(v_1-u)r_{\tau a_2}^{\alpha k}(v_2-u) B_\gamma(v_1) B_k (v_2) \\&\prod_{i=1}^L (-e^{2i \phi})^{\alpha-1}f(u) F^{a_1 a_2}\ket{0}\sim \lambda_{12}(u)F^{\gamma k}B_\gamma(v_1)B_k (v_2)\ket{0},
\end{align}
similarly to the case for one particle, we found that the vector $F^{ab}$ should be an eigenvector of
\begin{align}
(-e^{2i \phi})^{2(2-\alpha)}r_{\alpha a_1}^{\tau \gamma}(v_1-u)r_{\tau a_2}^{\alpha k}(v_2-u)
\end{align}
which is the twisted transfer matrix of a chain of length 2. If this happens, the action of $\sum_{\alpha=1}^2 T_{\alpha\alpha}$  on $\ket{2 \{v\}}$ is diagonal. From this result, it is now straightforward to generalize it for the case of $M$ magnons, as done in expression\footnote{For the expression in matrix form, see \eqref{transferm}.} \eqref{Mmagnonstransfer}.\\
We found that the eigenvalue of $T_{11}(u)+T_{22}(u)$ is
\begin{align}
\lambda_{12}(u)=\left(-e^{2 i \phi }\right)^{4-L}\alpha_2(u-v_1)\alpha_2(u-v_2)\lambda_{6V}(u)\prod_{i=1}^L\frac{  \sinh \left(u-u_j+\psi \right)}{i\,e^{\psi +i \phi }\,\cosh \left(u-u_j\right)}.
\label{eigv2p}
\end{align}
\subsubsection{Eigenvalue of the transfer matrix}
By summing the results \eqref{eigv2p0}, \eqref{eigv2p3}, \eqref{eigv2p}, we got the eigenvalue of the transfer matrix for the two magnon state and it is
\begin{align}\label{eigenvaluefor2particlenested}
\nonumber
\Lambda_2(u)=&\theta_{a_1}(u-v_1)\theta_{a_2}(u-v_2)+\\
\nonumber
&\zeta_{1,{a_1}}(u-v_1)\zeta_{1,{a_2}}(u-v_2)\prod_{i=1}^L \frac{\sinh \left(u-u_i+\psi \right)}{\cosh \left(u-u_i-\psi \right)} \frac{\tanh \left(u-u_i\right)}{e^{2 \psi }}+\\
&\left(-e^{2 i \phi }\right)^{4-L}\alpha_2(u-v_1)\alpha_2(u-v_2)\lambda_{6V}(u)\prod_{i=1}^L\frac{ \sinh \left(u-u_j+\psi \right)}{i\,e^{\psi +i \phi } \,\cosh \left(u-u_j\right)}.
\end{align}
The structure of this eigenvalue appears in the form of factorized products of single-excitations terms. In section \ref{Mparticlestates}, we will start from it to find the general expression for the eigenvalue of the $M$ particle state and, by using the shortcut of the residue, we will find the Bethe equations.

\subsection{M-particles state}
\label{Mparticlestates}
As already stressed, the eigenvalue of the transfer matrix for the $M$ particle states can be derived by generalizing the expressions for one and 2 magnons respectively in sections \ref{onepstate} and \ref{2partstates}. The expression of the eigenstate for the $M$ particle will involve a combinatorial expression due to the fact that $B_1$ and $B_2$ generate particles, but $B_3$ generates a pair.\\
However, we now show that to find the expression of the eigenvalues and the Bethe equations we do not need to know the $M$ particle state explicitly.\\
Let us consider more closely the eigenvalue \eqref{eigenvaluefor2particlenested} for the case of two particles.
The meaning of this eigenvalue is clear:
\begin{itemize}
\item the terms with $\theta$, $\zeta$ are the coefficients in the commutation relations $T_{00}$ and $T_{33}$ with each $B_a$,
\item the terms $\alpha$ and $\lambda_{6v}$ are in the commutator $T_{11}+T_{22}$ with the $B_a$,
\item the terms with the product $\prod_{i=1}^L$ comes from the action of $T_{ii}$ on the vacuum.
\end{itemize}
The eigenvalues appears as factorized products of single-excitations terms, so this strongly suggest that, even if the exact eigenstate of two particle state is \eqref{2particlesstate}, the eigenvalues can be obtained very naively just considering
\begin{align}
\ket{2\{v\}}\sim F^{ab}B_a(v_1)B_b(v_2)\ket{0}.
\end{align}
With this in mind, we generalize the result to arbitrary number $M$ of magnons. To do this, we act with the transfer matrix on the state
\begin{align}
\ket{M\{v\}}\sim F^{a_1 a_2 \dots a_M}B_{a_1}(v_1)\dots B_{a_M}(v_M)\ket{0},
\end{align}
where $N$ magnons are generated by $B_2$ and we get the following eigenvalue
\begin{align}
\nonumber\frac{\Lambda_M(u)}{\mathcal{N}}=&\prod _{i=1}^M \frac{ \cosh \left(u-v_i+\psi \right)}{\sinh \left(v_i-u\right)}+\\
\nonumber&\frac{\lambda_{6V}(u)}{\left(-e^{2 i \phi }\right)^{L-M-N}} \prod _{i=1}^M  \frac{\cosh \left(u-v_i-\psi \right)}{\sinh \left(u-v_i\right)}\prod _{j=1}^L \frac{  \sinh \left(u-u_j+\psi \right)}{i\,e^{i \phi +\psi }\,\cosh \left(u-u_j\right)}+\\
&\prod _{i=1}^M \frac{ \cosh \left(u-v_i-2 \psi \right)}{\sinh \left(u-v_i-\psi \right)}\prod _{i=1}^L \frac{ \tanh \left(u-u_i\right) \sinh \left(u-u_i+\psi \right)}{e^{2 \psi }\,\cosh \left(u-u_i-\psi \right)},
\end{align}
where $\mathcal{N}=i^M \left(-e^{2 i \phi }\right)^{M-N} \left(-e^{\psi -i \phi }\right)^M$ and $\lambda_{6V}(u)$ is the eigenvalue of the auxiliary problem given in \eqref{eigenv6vexplicit}. For completeness we will also report it here
\begin{align}
\nonumber\frac{\lambda _{6V}(u)}{\left(-e^{2 i \phi }\right)^{L-M}}=&\prod _{i=1}^N \frac{ \sinh \left(2 \left(u-w_i+\psi \right)\right)}{e^{2 i \phi }\sinh \left(2 \left(u-w_i\right)\right)}+\\
&\left(-e^{2 i \phi }\right)^{L-M} \prod _{j=1}^N \frac{\sinh \left(2 \left(u-w_j-\psi \right)\right)}{e^{2 i \phi } \sinh \left(2 \left(u-w_j\right)\right)}\prod _{j=1}^M \frac{e^{2 i \phi } \sinh \left(2 \left(u-v_j\right)\right)}{\sinh \left(2 \left(u-v_j-\psi \right)\right)}.
\end{align}
To find the Bethe equation, we  use the same shortcut of the one particle case. We  impose that the eigenvalue of the transfer matrix is regular, so that the spurious pole  cancels. We derived the Bethe equation by requiring that the residue at the pole $u=v$ cancels. Another set of Bethe equations can be derived from $u=v+\psi$, but those are not independent to the ones found here, but can be mapped to them.\\
We found that the rapidities $\{v\}$ of the main chain should satisfy the constraint
\begin{align}
&\prod _{i=1,i\neq j}^M -\frac{\cosh \left(v_j-v_i+\psi \right)}{\cosh \left(v_j-v_i-\psi \right)}=\prod _{i=1}^L \frac{  \sinh \left(v_j-u_i+\psi \right)}{i\, e^{\psi +i \phi } \cosh \left(v_j-u_i\right)} \prod _{i=1}^N \frac{\sinh \left(2 \left(w_i-v_j-\psi \right)\right)}{\sinh \left(2 \left(v_j-w_i\right)\right)}
\label{BEmainchiain}
\end{align}
for $j=1,\dots, M$, while the $w$'s satisfy the auxiliary Bethe equations \eqref{BEnestedchain}, 
\begin{equation}
\prod _{i=1,i\neq j}^N \frac{\sinh \left(2 \left(w_j-w_i+\psi \right)\right)}{\sinh \left(2 \left(w_j-w_i-\psi \right)\right)}=\left(-e^{2 i \phi }\right)^{L}\prod _{k=1}^M \frac{\sinh \left(2 \left(w_j-v_k\right)\right)}{\sinh \left(2 \left(v_k-w_j+\psi \right)\right)},
\label{BEnestedchain}
\end{equation}
for $j=1,\dots,N$.

\section{General result}
The Bethe equations that we found before take a bit of a unusual form due to the presence of both $\cosh$ and $\sinh$. However, we can rewrite both the Bethe equations and the eigenvalue by considering a shift in $u_i$, $v_i$ and $\psi$,
\begin{align}
&\psi \to  {\Psi} +\frac{i \pi }{2},
&&u_i\to  \mathtt{u}_i+\frac{i \pi }{2},
&&v_i\to  \mathtt v_i-\frac{\psi }{2}.
\end{align}
We remark, as mentioned in section \ref{modelB3defined}, that the coupling constant of the theory is $\gamma = \tanh \psi$ while under this shift $\gamma\to \coth \Psi $. This means that under this map the strong and weak coupling regimes are interchanged. The eigenvalue of the transfer matrix now becomes
\begin{align}
\nonumber\frac{\Lambda_M(u)}{\mathcal{N}}=&\prod _{i=1}^M \frac{\sinh \left(u-\mathtt v_i+\frac{3 \Psi }{2}\right)}{\sinh\left(u-\mathtt v_i+\frac{\Psi }{2}\right)}+\\
\nonumber&\frac{\lambda_{6V}(u)}{\left(-e^{2 i \phi }\right)^{L-M-N}} \prod _{i=1}^M  \frac{\sinh \left(u-\mathtt v_i-\frac{\Psi }{2}\right)}{\sinh\left(u-\mathtt v_i+\frac{\Psi }{2}\right)} \prod _{j=1}^L \frac{  \sinh \left(u-\mathtt u_j+\psi \right)}{i\,e^{\Psi +i \phi }\sinh\left(u-\mathtt u_j\right)}+\\
&\prod _{i=1}^M \frac{\cosh \left(u-\mathtt v_i-\frac{3 \Psi }{2}\right)}{\cosh \left(u-\mathtt v_i-\frac{\Psi }{2}\right)}\prod _{i=1}^L \frac{ \coth \left(u-\mathtt u_i\right) \sinh \left(u-\mathtt u_i+\Psi \right)}{e^{2 \Psi }\,\cosh \left(u-\mathtt u_i-\Psi \right)},
\end{align}
with $\mathcal{N}=\left(-e^{2 i \phi }\right)^{M-N} \left(-e^{\Psi -i \phi }\right)^M$ and for the nested chain
\begin{align}
\nonumber\frac{\lambda _{6V}(u)}{\left(-e^{2 i \phi }\right)^{L-M}}=&\prod _{i=1}^M-\frac{ \sinh \left(2 \left(u-w_i+\Psi \right)\right)}{e^{2 i \phi }\,\sinh\left(2 \left(u-w_i\right)\right)}+\\
&\left(-e^{2 i \phi }\right)^{L-M} \prod _{i=1}^N -\frac{ \sinh \left(2 \left(u-w_i-\Psi \right)\right)}{e^{2 i \phi } \sinh\left(2 \left(u-w_i\right)\right)}\prod _{i=1}^M -\frac{e^{2 i \phi } \sinh \left(2 u-2 \mathtt v_i+\Psi \right)}{\sinh\left(2 u-2 \mathtt v_i-\Psi \right)}.
\end{align}
Under the same shift, the Bethe equations become
\begin{align}
&\prod _{i=1,i\neq j}^M \frac{\sinh\left(\mathtt v_i-\mathtt v_j-\Psi \right)}{\sinh\left(\mathtt v_i-\mathtt v_j+\Psi \right)}=\prod _{i=1}^L \frac{1}{i\,e^{\Psi +i \phi }}\frac{  \sinh\left(\mathtt u_i-\mathtt v_j-\frac{\Psi }{2}\right)}{\sinh \left(\mathtt u_i-\mathtt v_j+\frac{\Psi }{2}\right)} \prod _{i=1}^N \frac{\sinh \left(2 \mathtt v_j-2 w_i+\Psi \right)}{\sinh\left(2 \mathtt v_j-2 w_i-\Psi \right)}
\end{align}
for $j=1,\dots,M$ and for the nested chain
\begin{equation}
\prod _{i=1,i\neq j}^N \frac{\sinh \left(2 \left(w_i-w_j-\Psi \right)\right)}{\sinh\left(2 \left(w_i-w_j+\Psi \right)\right)}=\left(-e^{2 i \phi }\right)^{L}\prod _{i=1}^M \frac{\sinh\left(2 \mathtt v_i-2 w_j-\Psi \right)}{\sinh\left(2 \mathtt v_i-2 w_j+\Psi \right)},
\end{equation}
for $j=1,\dots,N.$
\\
Let us introduce the standard Baxter Q-functions 
\begin{align}
\label{Qbaxt1}&Q^{[a]}_{\mathtt u}(x) = \prod_{i=1}^L \sinh[x-{\mathtt u}_i-a\Psi] , &&\tilde{Q}^{[a]}_{\mathtt u}(x) = \prod_{i=1}^L \cosh[x-{\mathtt u}_i-a\Psi], \\
&Q^{[a]}_{\mathtt v}(x) = \prod_{i=1}^M \sinh[x-{\mathtt v}_i-a\Psi], &&\tilde{Q}^{[a]}_{\mathtt v}(x) = \prod_{i=1}^M \cosh[x-{\mathtt v}_i-a\Psi], \\
\label{Qbaxt2}&Q^{[a]}_w(x) = \prod_{i=1}^N \sinh[x-w_i-a\Psi], && \tilde Q^{[a]}_w(x) = \prod_{i=1}^N \cosh[x-w_i-a\Psi].
\end{align}
The eigenvalue is
\begin{align}
\frac{\Lambda_M(u)}{\mathcal{N}}=&\frac{{Q_\mathtt{v}}^{[-3/2]}}{{Q_\mathtt{v}}^{[-1/2]}}+e^{-2\Psi L}\frac{{\tilde Q_\mathtt{v}}^{[3/2]}}{{\tilde Q_\mathtt{v}}^{[1/2]}} \frac{{\tilde Q_\mathtt{u}}^{[0]}}{{ Q_\mathtt{u}}^{[0]}}\frac{{Q_\mathtt{u}}^{[-1]}}{{\tilde Q_\mathtt{u}}^{[1]}}+\frac{ \left(-i e^{-\Psi -i \phi }\right)^L}{\left(-e^{2 i \phi }\right)^{L-M-N}} \frac{{Q_\mathtt{v}}^{[1/2]}}{{Q_\mathtt{v}}^{[-1/2]}}\frac{{Q_\mathtt{u}}^{[-1]}}{{Q_\mathtt{u}}^{[0]}}\lambda_{6V}(u),
\end{align}
\begin{align}
\lambda _{6V}(u)\frac{e^{2 i \phi N}}{\left(-e^{2 i \phi }\right)^{L-M}}=\frac{Q^{[-1]}_w\tilde{Q}^{[-1]}_w}{Q^{[0]}_w\tilde{Q}^{[0]}_w}+\left(-e^{2 i \phi }\right)^{L} \frac{Q^{[1]}_w\tilde{Q}^{[1]}_w}{Q^{[0]}_w\tilde{Q}^{[0]}_w}\frac{Q^{[-1/2]}_\mathtt{v}\tilde{Q}^{[-1/2]}_\mathtt{v}}{Q^{[1/2]}_\mathtt{v}\tilde{Q}^{[1/2]}_\mathtt{v}}
\end{align}\\
where for simplicity we used $Q^{[a]}_t(u)=Q^{[a]}_t$.\\
The Bethe equations for the main chain are
\begin{align}
\frac{{Q_\mathtt{v}}^{[-1]}}{{Q_\mathtt{v}}^{[1]}}=-\left(\frac{-i}{e^{\Psi+i \phi }}\right)^L \frac{{Q_\mathtt{u}}^{[-1/2]}}{{Q_\mathtt{u}}^{[1/2]}}\frac{{Q_w}^{[-1/2]}}{{Q_w}^{[1/2]}}\frac{{\tilde Q_w}^{[-1/2]}}{{\tilde Q_w}^{[1/2]}},
\end{align}
$j=1,\dots,M$ and for the nested chain
\begin{equation}
\frac{{Q_w}^{[-1]}{\tilde Q_w}^{[-1]}}{{Q_w}^{[1]} {\tilde Q_w}^{[1]}}=-\left(-e^{2 i \phi }\right)^L\frac{{Q_\mathtt{v}}^{[-1/2]} {\tilde Q_\mathtt{v}}^{[-1/2]}}{{Q_\mathtt{v}}^{[1/2]}{\tilde Q_\mathtt{v}}^{[1/2]}}.
\label{beqopnest}
\end{equation}
where  $Q^{[a]}_t(w_j)=Q^{[a]}_t$ and $j=1,\dots,N$.

\section{Bethe Ansatz for the nested chain}
\label{betheansatznested}
In the previous sections, we showed how to diagonalize the transfer matrix via the nested Algebraic Bethe ansatz approach. In this model, the nesting is manifest from the appearance of the transfer matrix for the twisted 6-vertex model. While the 6-vertex model also appears in the Hubbard model \cite{ramos1997algebraic,martins1998quantum} and in the $\text{AdS}_5 \times \text{S}_5$   $S$-matrix for bound states \cite{arutyunov2009bound}, model B3 is different. In fact, the $r$-matrix \eqref{rnested} is a twisted version of the standard one and the transfer matrix is also twisted as will be explained in the following.
\\
In principle, we can solve the nested problem as an independent one, for a spin chain of a given length and with arbitrary number of excitations. What we actually have to use is that the length of the chain is equal to $M$ (total number of magnons) and that the number of excitations of the nested chain is equal to $N$, number of excitation of type $B_2$. The rapidities of the particles of the main chain are the inhomogeneities of the nested chain.
\\
To summarize
\begin{align}
&M=\#B_1+\#B_2=L_{\text{nested}},
&&N=\#B_2=\#B,
\end{align}
where $B$ will be defined in the following as the creator operator of the nested chain.\\
The $r$-matrix of \eqref{rnested} is
\begin{align}
&r(u)=r_{\alpha \,\beta  }^{\gamma \,\delta}(u) e_\gamma^\alpha \otimes e_\delta^\beta,
&&r(u)=\left(
\begin{array}{cccc}
 1 & 0 & 0 & 0 \\
 0 & b(u)e^{-2 i \phi} & a(u) & 0 \\
 0 &  a(u) &b(u)e^{2 i \phi} & 0 \\
 0 & 0 & 0 & 1 \\
\end{array}
\right),
\end{align}
where $a(u)=\frac{\sinh (2 \psi )}{\sinh (2 (u+\psi ))}$, $b(u)=\frac{\sinh (2 u)}{\sinh (2 (u+\psi ))}$ and $\phi \in \mathbb{R}$.\\
$\phi$ is the twisting and if $e^{2 i \phi}=1$, one finds the standard 6-vertex $r$-matrix\footnote{See for example Appendix B of \cite{arutyunov2009bound}}.\\
To construct the transfer matrix, we recall the results for one, two and 3 magnons
\begin{align}
&\text{one magnon:}\,&&(-e^{2 i \phi})^{\alpha(L-1)}r_{\alpha a}^{\alpha \gamma} (v-u)\\
&\text{two magnons:}\,&&(-e^{2 i \phi})^{\alpha(L-2)}r_{\alpha\, a_1}^{\tau\, \gamma} (v_1-u)r_{\tau\, a_2}^{\alpha\, \gamma_2} (v_2-u)\\
&\text{3 magnons:}\,&&(-e^{2 i \phi})^{\alpha(L-3)}r_{\alpha \,a_1}^{\tau_1 \gamma_1} (v_1-u)r_{\tau_1\, a_2}^{\tau_2 \, \gamma_2} (v_2-u)r_{\tau_2 \, a_3}^{\alpha \, \gamma_3} (v_3-u)
\end{align}
which can be easily generalized to the case of $M$ magnons
\begin{align}
(-e^{2 i \phi})^{\alpha(L-M)}r_{\alpha \,a_1}^{\tau_1 \gamma_1} (v_1-u)r_{\tau_1\, a_2}^{\tau_2 \, \gamma_2} (v_2-u)\dots r_{\tau_{M-1} \, a_M}^{\alpha \, \gamma_M} (v_M-u)
\label{Mmagnonstransfer}
\end{align}
and in matrix form
\begin{align}
&T^{(1)}(u)=\tr_a G_a \prod_{i=1}^M r_{a i}(u_i-u),\label{transferm}\\
&G_a=\left(-e^{2 i \phi }\right)^{\frac{3}{2} (M-L)} \left(
\begin{array}{cc}
 \left(-e^{2 i \phi }\right)^{\frac{M-L}{2}} & 0 \\
 0 & \left(-e^{2 i \phi }\right)^{\frac{L-M}{2}} \\
\end{array}
\right).
\end{align}
$T^{(1)}(u)$ is the twisted transfer matrix and $G_a$ is a $\alg{su}(2)$ element. An example of twisted transfer matrix can be found in \cite{arutyunov2011twisting}. It is easy to check that the RTT is satisfied also with this new definition of transfer matrix.\\
We use the notation $^{(1)}$ to identify objects in the nested chain. Since the model preserves spin, as for the XXX spin chain analyzed in section \ref{ABAtheory}, the  reference state is defined as a state with all spin up
\begin{equation}
\ket{0}^{(1)}=\bigotimes_{i=1}^M \ket{\uparrow}=\bigotimes_{i=1}^M \left(
\begin{array}{c}
 1 \\
 0 \\
\end{array}
\right).
\label{vacuumnested}
\end{equation}
The monodromy matrix is
\begin{equation}
T_a^{(1)}(u)=G_a \prod_{i=1}^M r_{ai}(v_i-u)=\left(
\begin{array}{cc}
 A(u) & B(u) \\
 C(u) & D(u) \\
\end{array}
\right),
\end{equation}
\begin{equation}
A(u)\ket{0}^{(1)}=\left(-e^{2 i \phi }\right)^{L-M} \ket{0}^{(1)},
\end{equation}
\begin{equation}
C(u)\ket{0}^{(1)}=0,
\end{equation}
\begin{equation}
D(u)\ket{0}^{(1)}=\left(-e^{2 i \phi }\right)^{2 (L-M)} \prod _{i=1}^M e^{2 i \phi }\, b\left(v_i-u\right)\ket{0}^{(1)}.
\end{equation}
It is easy to check that the reference state \eqref{vacuumnested} is an eigenstate of the transfer matrix with eigenvalue
\begin{align}
\left(-e^{2 i \phi }\right)^{L-M}\bigg[1+\left(-e^{2 i \phi }\right)^{L-M} \prod _{i=1}^M e^{2 i \phi }\, b\left(v_i-u\right) \bigg].
\end{align}
From the RTT relations\footnote{We can notice that the following commutation relations are very similar to the ones obtained in section \eqref{ABAtheory}. However, here we used a different normalization. The map $a\to 1$, $b\to b e^{2i\phi}$, $c\to a$ will allow to find \eqref{commreln} from \eqref{algebra2}-\eqref{algebra}.}
\begin{equation}
[B(u),B(v)]=0,
\end{equation}
\begin{equation}\label{commreln}
A(v) B(u)= \frac{e^{-2 i \phi }}{b(v-u)}B(u) A(v)-e^{-2 i \phi } \frac{a(v-u) }{b(v-u)}B(v) A(u),
\end{equation}
\begin{equation}
D(v) B(u)= \frac{e^{-2 i \phi } }{b(u-v)}B(u) D(v)-e^{-2 i \phi }\frac{a(u-v)}{b(u-v)}B(v) D(u).
\end{equation}
The eigenstate of $N$ particles is given by
\begin{equation}
\ket{N\{w\}}^{(1)}=B\left(w_1\right)\dots B\left(w_N\right)\ket{0}^{(1)}.
\end{equation}
By applying the transfer matrix to a state of $N$ particles we found the following eigenvalue:
\begin{equation}
\frac{\lambda _{6V}(u)}{\left(-e^{2 i \phi }\right)^{L-M}}=\prod _{i=1}^N \frac{e^{-2 i \phi }}{b\left(u-w_i\right)}+\left(-e^{2 i \phi }\right)^{L-M} \prod _{i=1}^N \frac{e^{-2 i \phi }}{b\left(w_i-u\right)}\prod _{j=1}^M e^{2 i \phi } b\left(v_j-u\right)
\label{eigenv6v}
\end{equation}
and considering the expression for $a(u)$ and $b(u)$
\begin{align}
\nonumber\frac{\lambda _{6V}(u)}{\left(-e^{2 i \phi }\right)^{L-M}}=&\prod _{i=1}^N \frac{ \sinh \left(2 \left(u-w_i+\psi \right)\right)}{e^{2 i \phi }\sinh \left(2 \left(u-w_i\right)\right)}+\\
&\left(-e^{2 i \phi }\right)^{L-M} \prod _{i=1}^N \frac{\sinh \left(2 \left(u-w_i-\psi \right)\right)}{e^{2 i \phi } \sinh \left(2 \left(u-w_i\right)\right)}\prod _{j=1}^M \frac{e^{2 i \phi } \sinh \left(2 \left(u-v_i\right)\right)}{\sinh \left(2 \left(u-v_i-\psi \right)\right)}.
\label{eigenv6vexplicit}
\end{align}
This eigenvalue has a simple pole if $u=w_i$. We can require that the residue at the single pole vanishes and we get the set of Bethe equations for the rapidities $w$
\begin{equation}
\prod _{i=1, i\neq j}^N \frac{b\left(w_i-w_j\right)}{b\left(w_j-w_i\right)}=\left(-e^{2 i \phi }\right)^{L-M}\prod _{k=1}^M e^{2 i \phi } b\left(v_k-w_j\right),\,\,\,\,\, j=1,\dots,N
\end{equation}
or explicitly
\begin{equation}
\prod _{i=1,i\neq j}^N \frac{\sinh \left(2 \left(w_j-w_i+\psi \right)\right)}{\sinh \left(2 \left(w_j-w_i-\psi \right)\right)}=\left(-e^{2 i \phi }\right)^{L}\prod _{k=1}^M \frac{\sinh \left(2 \left(w_j-v_k\right)\right)}{\sinh \left(2 \left(v_k-w_j+\psi \right)\right)}.
\label{BEnestedchain}
\end{equation}

\chapter{The $R$-matrix for the site two model related to Hubbard}
\label{Rmatrixsite2Hubbard}

The $R$-matrix corresponding to the model given in \eqref{H2} is 
\begin{align}
R_{i,i+1}\,=\, &\frac{1}{8} \left(r_8+r_2+r_3+r_{10}\right)\Big(1 +\sigma _i^z  \sigma _{i+1}^z  \tau _i^z  \tau _{i+1}^z \Big)+\frac{1}{8} \left(r_8-r_2-r_3+r_{10}\right)\Big(\sigma _i^z  \sigma _{i+1}^z +\tau _i^z  \tau _{i+1}^z \Big)+\nonumber\\
&\frac{1}{8} \left(r_8-r_2+r_3-r_{10}\right)\Big(\sigma _i^z  \tau _{i+1}^z +\sigma _{i+1}^z  \tau _i^z \Big)+\frac{1}{8} \left(r_8+r_2-r_3-r_{10}\right)\Big(\sigma _i^z  \tau _i^z +\sigma _{i+1}^z  \tau _{i+1}^z \Big)+\nonumber\\
&\frac{r_6}{4}\Big(\sigma _i^z +\tau _i^z -\sigma _i^z  \sigma _{i+1}^z  \tau _{i+1}^z -\sigma _{i+1}^z  \tau _i^z  \tau _{i+1}^z\Big)+\frac{r_5}{4}\Big(\sigma _i^z  \sigma _{i+1}^z  \tau _i^z +\sigma _i^z  \tau _i^z  \tau _{i+1}^z -\sigma _{i+1}^z -\tau _{i+1}^z \Big)+\nonumber\\
&\frac{1}{4} \left(r_1+r_4\right)\Big(\sigma _i^x  \sigma _{i+1}^x  \tau _i^x  \tau _{i+1}^x +\sigma _i^y  \sigma _{i+1}^y  \tau _i^y  \tau _{i+1}^y \Big)+\frac{1}{4} \left(r_4-r_1\right)\Big(\sigma _i^x  \sigma _{i+1}^x  \tau _i^y  \tau _{i+1}^y +\sigma _i^y  \sigma _{i+1}^y  \tau _i^x  \tau _{i+1}^x\Big)+\nonumber\\
&\frac{1}{4} \left(r_9+r_{12}\right)\Big( \sigma _{i+1}^z  \tau _i^y  \tau _{i+1}^y-\sigma _i^x  \sigma _{i+1}^x  \tau _i^z  +\sigma _i^y  \sigma _{i+1}^y    \tau _{i+1}^z-\sigma _i^z    \tau _i^x  \tau _{i+1}^x\Big)+\nonumber\\
&\frac{1}{4} \left(r_9-r_{12}\right)\Big(  \sigma _{i+1}^z  \tau _i^x  \tau _{i+1}^x+\sigma _i^x  \sigma _{i+1}^x   \tau _{i+1}^z-\sigma _i^y  \sigma _{i+1}^y  \tau _i^z  -\sigma _i^z   \tau _i^y  \tau _{i+1}^y\Big)+\nonumber\\
&\frac{1}{4} \left(r_7+r_{11}\right)\Big(  \tau _i^x  \tau _{i+1}^x+\sigma _i^x  \sigma _{i+1}^x  +\sigma _i^y  \sigma _{i+1}^y  \tau _i^z  \tau _{i+1}^z+\sigma _i^z  \sigma _{i+1}^z  \tau _i^y  \tau _{i+1}^y\Big)+\nonumber\\
&\frac{1}{4} \left(r_7-r_{11}\right)\Big(\sigma _i^y  \sigma _{i+1}^y +\tau _i^y  \tau _{i+1}^y +\sigma _i^x  \sigma _{i+1}^x  \tau _i^z  \tau _{i+1}^z +\sigma _i^z  \sigma _{i+1}^z  \tau _i^x  \tau _{i+1}^x\Big)
\end{align}

where

\begin{align}
&r_1= -\frac{2 i k g_{u,v}}{\text{dn}_u+\text{dn}_v},
&&r_4=f_{u,v},
&&r_7= 1,\nonumber\\
&i\, r_9= \frac{ \text{cn}_u-\text{cn}_v}{\text{sn}_u+\text{sn}_v},
&&i\, k\, r_{11}= \frac{\text{dn}_u-\text{dn}_v}{\text{sn}_u+\text{sn}_v},
&&\frac{r_{12}}{k}= \frac{\text{cn}_u-\text{cn}_v}{\text{dn}_u+\text{dn}_v},\nonumber
\end{align}
\vspace{-0.5cm}
\begin{align}
&r_5+r_6=-2 i g_{u,v},
&&k\,(r_5-r_6)={(\text{dn}_v-\text{dn}_u) f_{u,v}},\nonumber\\
&r_3+r_{10}+r_8-r_2=2 f_{u,v},
&&r_3+r_{10}+r_2-r_8=\frac{4\, i\, k (\text{cn}_u-\text{cn}_v)}{(\text{dn}_u+\text{dn}_v) (\text{sn}_u+\text{sn}_v) f_{u,v}},\nonumber
\end{align}

\begin{align}
\nonumber
&\frac{r_{10} (\text{dn}_u+i \,k\, \text{cn}_u \text{sn}_u)+r_{8} (\text{dn}_u-i \,k\, \text{cn}_u \text{sn}_u)}{r_4}=\text{sn}_u^2 (\text{dn}_v-\text{dn}_u)+\frac{2 \text{dn}_u}{f_{u,v}^2}+\frac{2 k^2 \text{sn}_u (\text{cn}_u-\text{cn}_u) g_{u,v}}{(\text{dn}_u+\text{dn}_v) f_{u,v}},
\end{align}

\begin{align}
&\frac{r_3 (\text{dn}_u+i \,k\, \text{cn}_u \text{sn}_u)+r_2 (\text{dn}_u-i\, k\, \text{cn}_u \text{sn}_u)}{\,r_4}=\nonumber\\
&\,\,\,\,\,\,\,\,\,\,\,\,\,\,\,\,\,\,\,\,\,\,\,\,\,\,\,\,\,\,\,\,\,r_{12} \Big( \frac{4 \text{dn}_u}{ (\text{sn}_u+\text{sn}_v) f_{u,v}^2}+\text{sn}_u \left({\text{dn}_v}-{\text{dn}_u}{}\right) \Big)+\frac{ f_{u,v} g_{u,v}}{k}\Big(\frac{2 k^2 \text{sn}_u^2}{f_{u,v}^2}-\text{dn}_u \text{dn}_v+\text{dn}_u^2\Big),\nonumber
\end{align}
we defined the shorthand notations for the Jacobi functions:
\begin{align}
&\text{cn}_u=\text{cn}\left(u|k^2\right),
&&\text{sn}_u=\text{sn}\left(u|k^2\right),
&&\text{Am}_u=\text{am}\left(u|k^2\right),\\
&\text{dn}_u=\text{dn}\left(u|k^2\right),
\end{align}
and we defined for simplicity
\begin{align}
&g_{u,v}=\sin \left(\frac{1}{2} (\text{Am}_u-\text{Am}_v)\right),
&&f_{u,v}=\sec \left(\frac{1}{2} (\text{Am}_u-\text{Am}_v)\right).
\end{align}
These two functions are related by the following expression
\begin{align}
f_{u,v} g_{u,v}=&\csc \Big[\text{Am}_u-\text{Am}_v\Big]-\cot \Big[\text{Am}_u-\text{Am}_v\Big]=\nonumber\\
&\frac{\text{cn}_u \text{cn}_v+\text{sn}_u \text{sn}_v-1}{\text{cn}_u \text{sn}_v-\text{cn}_v \text{sn}_u},
\end{align}
however keeping both the $f$ and $g$, the expressions for the entries of the $R$-matrix become more concise.
\\
The actual matrix form reads
\setcounter{MaxMatrixCols}{20}
\begin{align}\label{RH2}
R = 
\begin{pmatrix}
 r_8 & 0 & 0 & 0 & 0 & -r_{12} & 0 & 0 & 0 & 0 & -r_{12} & 0 & 0 & 0 & 0 & r_1 \\
 0 & r_6 & 0 & 0 & r_7 & 0 & 0 & 0 & 0 & 0 & 0 & r_{11} & 0 & 0 & 0 & 0 \\
 0 & 0 & r_6 & 0 & 0 & 0 & 0 & r_{11} & r_7 & 0 & 0 & 0 & 0 & 0 & 0 & 0 \\
 0 & 0 & 0 & r_2 & 0 & 0 & -r_9 & 0 & 0 & -r_9 & 0 & 0 & r_4 & 0 & 0 & 0 \\
 0 & r_7 & 0 & 0 & -r_5 & 0 & 0 & 0 & 0 & 0 & 0 & 0 & 0 & 0 & r_{11} & 0 \\
 -r_{12} & 0 & 0 & 0 & 0 & r_{10} & 0 & 0 & 0 & 0 & r_1 & 0 & 0 & 0 & 0 & r_{12} \\
 0 & 0 & 0 & -r_9 & 0 & 0 & r_3 & 0 & 0 & r_4 & 0 & 0 & r_9 & 0 & 0 & 0 \\
 0 & 0 & r_{11} & 0 & 0 & 0 & 0 & r_5 & 0 & 0 & 0 & 0 & 0 & r_7 & 0 & 0 \\
 0 & 0 & r_7 & 0 & 0 & 0 & 0 & 0 & -r_5 & 0 & 0 & 0 & 0 & r_{11} & 0 & 0 \\
 0 & 0 & 0 & -r_9 & 0 & 0 & r_4 & 0 & 0 & r_3 & 0 & 0 & r_9 & 0 & 0 & 0 \\
 -r_{12} & 0 & 0 & 0 & 0 & r_1 & 0 & 0 & 0 & 0 & r_{10} & 0 & 0 & 0 & 0 & r_{12} \\
 0 & r_{11} & 0 & 0 & 0 & 0 & 0 & 0 & 0 & 0 & 0 & r_5 & 0 & 0 & r_7 & 0 \\
 0 & 0 & 0 & r_4 & 0 & 0 & r_9 & 0 & 0 & r_9 & 0 & 0 & r_2 & 0 & 0 & 0 \\
 0 & 0 & 0 & 0 & 0 & 0 & 0 & r_7 & r_{11} & 0 & 0 & 0 & 0 & -r_6 & 0 & 0 \\
 0 & 0 & 0 & 0 & r_{11} & 0 & 0 & 0 & 0 & 0 & 0 & r_7 & 0 & 0 & -r_6 & 0 \\
 r_1 & 0 & 0 & 0 & 0 & r_{12} & 0 & 0 & 0 & 0 & r_{12} & 0 & 0 & 0 & 0 & r_8 \\
\end{pmatrix},
\end{align}
where we suppressed the dependence on two spectral parameters, i.e. $r_i=r_i (u,v)$. 
\\
This matrix is written in the standard basis $e_i$ where $\{e_1,e_2,e_3,e_4\}=\{\ket{\emptyset},\ket{\uparrow},\ket{\downarrow},\ket{\updownarrow}\}$, see also \eqref{basis}.
\\
The $R$-matrix given in this appendix satisfies the Yang-Baxter equation. By using version 12.3 of Mathematica the check is
straightforward. However, when using version 12.0, special attention should be given to the choice of the sign of the branch-cut.
\\To the best of our knowledge, this $R$-matrix is new and we have not encountered before a model with this functional dependence. In fact, the entries can only be expressed in terms of the Jacobi functions $\text{sn}, \text{cn}$ and $\text{dn}$ by introducing square roots.

\chapter{Bond-site transformation}
\label{bondsitematrix}

In this appendix, we show that the action of the bond site transformation can be obtained by acting with a unitary transformation on the Pauli matrices. We consider the transformation on the spin 1/2 chain, keeping in mind that the Hilbert space we are considering is $H\otimes H^*$. We apply the transformation separately to $\sigma$ and $\tau$ as in \eqref{notation}.
\\
As before, we use the notation: $X,Y,Z$, to identify the Pauli matrices.
\\
We take the bond-site transformation \eqref{z1}-\eqref{z2} and we shift by half integer the right hand side:
\begin{align}
&X_j\to X_j X_{j+1},
&&Z_jZ_{j+1}\to Z_{j+1}.
\end{align}

We can check explicitly what is the action of this transformation when it acts on a basis of Pauli Matrices. In particular, we start from a spin chain of $L=2$. We obtain:
\subsection*{$L=2$}
The Hilbert space is $\mathbb{C}^2\otimes \mathbb{C}^2$. We seach for the expression of a 4x4 matrix $U_{12}$ such that 
\begin{align}
&U_{12} X_1U_{12}^{-1}= X_1 X_2,\\
&U_{12} Z_1 U_{12}^{-1}= Z_1,\\
&U_{12} Z_2U_{12}^{-1}= Z_1 Z_2.
\end{align}
Additionally, we require that $U_{12} X_2U_{12}^{-1}= X_2$. This last choice is motivated by the fact that we are restricting to the case of $L=2$. By explicit computation we find
\begin{equation}\label{U12}
U_{12}=\left(
\begin{array}{cccc}
 1 & 0 & 0 & 0 \\
 0 & 1 & 0 & 0 \\
 0 & 0 & 0 & 1 \\
 0 & 0 & 1 & 0 \\
\end{array}
\right).
\end{equation}

\section*{$L=3$}
For $L=3$, we can repeat the argument and we have now
\begin{align}
&U_{123}X_1U_{123}^{-1}= X_1 X_2,\\
&U_{123}X_2U_{123}^{-1}= X_2 X_3,\\
&U_{123}Z_1U_{123}^{-1}= Z_1,\\
&U_{123}Z_2U_{123}^{-1}= Z_1Z_2,\\
&U_{123}Z_3U_{123}^{-1}= Z_1 Z_2 Z_3,
\label{transf}
\end{align}
and similarly to before,
\begin{align}
U_{123}X_3 U_{123}^{-1}= X_3.
\end{align}
 The matrix $U_{123}$ is now
\begin{equation}
U_{123}=U_{12} U_{23},
\end{equation}
with $U_{12}=\left(
\begin{array}{cccc}
 1 & 0 & 0 & 0 \\
 0 & 1 & 0 & 0 \\
 0 & 0 & 0 & 1 \\
 0 & 0 & 1 & 0 \\
\end{array}
\right)\otimes\id$ and $U_{23}=\id\otimes\left(
\begin{array}{cccc}
 1 & 0 & 0 & 0 \\
 0 & 1 & 0 & 0 \\
 0 & 0 & 0 & 1 \\
 0 & 0 & 1 & 0 \\
\end{array}
\right)$.
\\
We did a similar calculation for $L=4$ and we obtain
\begin{align}
U_{1234}=U_{12} U_{23}U_{34},
\end{align}
where $U_{ij}$ are constructed from the operator $\left(
\begin{array}{cccc}
 1 & 0 & 0 & 0 \\
 0 & 1 & 0 & 0 \\
 0 & 0 & 0 & 1 \\
 0 & 0 & 1 & 0 \\
\end{array}
\right)$, acting in the sites $i,j$ of the chain.\\
We can generalize this result for arbitrary $L$, in particular
\begin{align}
U_{12\dots L}=U_{12}U_{23}\dots U_{L-1,L}.
\end{align}

\subsection*{Our model}
For the model we analyzed in chapter \ref{Hubbardchapter}, we applied the bond site transformation to the densities operator $h_{12}$ and $\ell_{12}$ of the spin 1/2 chain given in \eqref{Hsigma} and \eqref{Lsigma2}. We applied the bond-site transformation to the densities and then we checked that the new operator $\mathbb{H}_3$ remains integrable. In this way, we did not have to specify the action of the bond-site transformation at the boundary of the spin chain.
\\In particular
\begin{equation}
U_{12}U_{23} (\ell_{12}\otimes \id) U_{23}^{-1}U_{12}^{-1}=l_{123},
\end{equation}
and for the Hamiltonian part
\begin{equation}
U_{12}U_{23} (Y_1X_2-X_1Y_2) U_{23}^{-1}U_{12}^{-1}=(Y_2X_3-X_1Y_2).
\end{equation}
The Hamiltonian part $h$ is self dual after the transformation, so we continue to refer to it is $h_{12}$. The $\ell$ operator becomes of range 3 and we refer to it as $l_{123}$.


\chapter{Exact computation of late time expectation values} 
\label{meanvalcompexact}
In this section, we detail the computation of the late-time expectation values of local observables following from the state
\begin{equation}
  \label{rho01}
\rho(t=0)=  \rho_0(\beta)\equiv \frac{e^{\beta Q_0}}{(2\cosh \beta)^L}.
\end{equation}
In particular we focus on the local observable $\langle Z_j \rangle$. Let us start by recalling that an over-complete basis of the $L+1$-dimensional space of NESS can be generated by the $\tilde{\rho}_\cc(\alpha) = T(\cc)^{-1} e^{\alpha Q_0} T(\cc)$. 
Since $Q_0 = \sum_j Z_j$ has $L+1$ distinct eigenvalues of the form $2n-L$ with $n=0,\ldots, L$, we can alternatively define a basis of the space of NESS in terms of the projectors $\widetilde{P}_n = T(\cc)^{-1} P_n T(\cc)$, where 
\begin{equation}
P_n = \frac{1}{L+1} \sum_{k=0}^L e^{i\frac{2\pi k}{L+1}(\frac{L+Q_0}{2}-n)} ,
\label{eq:Pndef}
\end{equation}
is the projector onto the subspace where $Q_0$ has eigenvalue $2n-L$.
\\
As a consequence, any density matrix in the space of NESS can be decomposed as : 
\begin{equation}
\rho_{\rm NESS} = \sum_{n=0}^L \frac{\mathrm{Tr}(\rho_{\rm NESS}\widetilde{P}_n)}{\mathrm{Tr}(\widetilde{P}_n)}  \widetilde{P}_n.
\end{equation} 
Starting from an arbitrary $\rho(t=0)$, we therefore have at late times 
\begin{equation}
\lim_{t\to\infty} e^{\mathcal{L}t} \rho(t=0) =  \sum_{n=0}^L \frac{\mathrm{Tr}(\rho(t=0)\widetilde{P}_n)}{\mathrm{Tr}(\widetilde{P}_n)}  \widetilde{P}_n
\end{equation}
and therefore, for any observable $\mathcal{O}$, 
\begin{equation}
\lim_{t\to\infty} \langle \mathcal{O} \rangle =  \sum_{n=0}^L \frac{\mathrm{Tr}(\rho(t=0)\widetilde{P}_n)}{\mathrm{Tr}(\widetilde{P}_n)}  \mathrm{Tr}(\widetilde{P}_n \mathcal{O}) \,.
\label{eq:limO}
\end{equation}
All the traces involved in \eqref{eq:limO} can be computed using the matrix product operator techniques. 
\\
Focusing on the observable $Z_j$ we need to compute : 
\begin{equation}
\lim_{t\to\infty} \langle Z_j \rangle =  \sum_{n=0}^L \frac{\mathrm{Tr}(\rho(t=0)\widetilde{P}_n)}{\mathrm{Tr}(\widetilde{P}_n)}  \mathrm{Tr}(\widetilde{P}_n Z_j) \,.
\label{eq:limZ}
\end{equation}
The first trace $\mathrm{Tr}(\widetilde{P}_n) = \mathrm{Tr}(P_n)$ can easily be computed without resorting to MPO techniques, as it corresponds to the dimension of the eigenspace of $P_n$ with eigenvalue $2L-n$, however as a warm-up we present its computation using the previously computed function $\widetilde{\mathcal{G}}(\alpha,\beta)$. Using the decomposition \eqref{eq:Pndef} of the projector $P_n$, we have 
\begin{align}
 \mathrm{Tr}(\widetilde{P}_n)  
 &= 
 \frac{1}{L+1} \sum_{k=0}^L 
e^{i\frac{2\pi k}{L+1}(\frac{L}{2}-n)}  
 \mathrm{Tr} \left( T(\cc)^{-1} e^{i \frac{\pi k}{L+1} Q_0} T(\cc) \right)
 \cr 
&=  \frac{1}{L+1} \sum_{k=0}^L 
e^{i\frac{2\pi k}{L+1}(\frac{L}{2}-n)}  
  \widetilde{\mathcal{G}}(0, \frac{i k\pi }{L+1}) =  \frac{{2^L}}{L+1} \sum_{k=0}^L 
e^{i\frac{2\pi k}{L+1}(\frac{L}{2}-n)} (\cos \frac{k\pi}{L+1})^L 
\cr 
&=  {L \choose n}\,.
\label{trP}
\end{align}

The second trace can be similarly evaluated as :
\begin{align}
 \mathrm{Tr}(\widetilde{P}_n Z_j)  
&=  \frac{1}{L+1} \sum_{k=0}^L 
e^{i\frac{2\pi k}{L+1}(\frac{L}{2}-n)}  
\frac{1}{L}  \partial_{\alpha} \widetilde{\mathcal{G}}(\alpha, \frac{i k\pi }{L{+1}})|_{\alpha=0}
\cr 
&=  i 2^L \frac{1-\cc^2}{1-\cc^{2L}}\frac{1}{L+1} \sum_{k=0}^L 
e^{i\frac{2\pi k}{L}(\frac{L}{2}-n)} (\sin\frac{k\pi}{L+1}(\cos\frac{k\pi}{L+1})^{L-1}+ (i \cc)^L \cos\frac{k\pi}{L+1}(\sin\frac{k\pi}{L+1})^{L-1}) 
\cr 
&= - \frac{1-\cc^2}{1-\cc^{2L}} (1-(-1)^{L-n} \cc^L )  \frac{L-2n}{n}  
{L-1 \choose n-1}\,.
\label{TrPnZj}
\end{align}

We now move to the third trace, $ \mathrm{Tr}(\rho(t=0) \widetilde{P}_n )$. Taking the normalized density matrix $\rho(t=0) = e^{ \alpha Q_0} / (2\cosh \alpha)^L $, 
\begin{align}
 \mathrm{Tr}(\rho(t=0) \widetilde{P}_n )  
&=   \frac{1}{(2\cosh\alpha)^L}\frac{1}{L+1} \sum_{k=0}^L 
e^{i\frac{2\pi k}{L+1}(\frac{L}{2}-n)}  
\widetilde{\mathcal{G}}(\alpha, \frac{i k\pi }{L+1})   
\cr 
&=
   \frac{1}{(2\cosh\alpha)^L}\frac{1}{L+1} \frac{1}{1-\cc^{2L}}
   \left(  \mathcal{F}_n(\alpha) - (i\cc)^L \mathcal{F}_n(\alpha-i \pi/2)  \right)   
   \label{Trrho0Pntilde}
\end{align}
where in the last line we have used the expression \eqref{expressionG} of $\widetilde{\mathcal{G}}$ in terms of $\mathcal{G}$, and introduced the functions
\begin{align}
\mathcal{F}_n(\alpha) &\equiv 
 \sum_{k=0}^L 
e^{i\frac{2\pi k}{L+1}(\frac{L}{2}-n)}  
\widetilde{\mathcal{G}}(\alpha, \frac{i k\pi }{L+1})
\non\\ 
&=  \sum_{k=0}^L 
e^{i\frac{2\pi k}{L+1}(\frac{L}{2}-n)}  
\left( 
\lambda_1(\alpha, \frac{i k\pi }{L+1})^L 
+
\lambda_2(\alpha, \frac{i k\pi }{L+1})^L 
+
\lambda_3(\alpha, \frac{i k\pi }{L+1})^L 
+\lambda_4(\alpha, \frac{i k\pi }{L+1})^L 
\right) \non
\\ 
&\equiv 
\mathcal{F}^{(1)}_n(\alpha) 
+ 
\mathcal{F}^{(2)}_n(\alpha)
+ 
\mathcal{F}^{(3)}_n(\alpha)
+
\mathcal{F}^{(4)}_n(\alpha).
\end{align}

Using the expressions \eqref{lambdaeigenvalues} of the eigenvalues $\lambda_i$, the contributions $\mathcal{F}^{(3)}$ and $\mathcal{F}^{(4)}$ are easily evaluated. We find : 
 \begin{align} 
  \frac{1}{({2}\cosh\alpha)^L}\frac{1}{L+1} \frac{1}{1-\cc^{2L}} \mathcal{F}_n^{(3)}(\alpha) &=\frac{1}{{2^L}}\frac{( \cc  \tanh \alpha)^L }{1-\cc^{2L}} \left(
\begin{array}{c}
 L \\
 n \\
\end{array}
\right) \,,
\label{F3}
\\
  \frac{1}{({2}\cosh\alpha)^L}\frac{1}{L+1} \frac{1}{1-\cc^{2L}} \mathcal{F}_n^{(4)}(\alpha) &=\frac{1}{{2^L}} \frac{\cc^L}{1-\cc^{2L}}(-1)^{L-n} \left(
\begin{array}{c}
 L \\
 n \\
\end{array}
\right) 
\,.
\label{F4}
\end{align}

We now move to the contribution $\mathcal{F}^{(1)}+\mathcal{F}^{(2)}$. Using the expression \eqref{lambdaeigenvalues}, 
\begin{align}
&\lambda_1(\alpha,\beta)^L + \lambda_2(\alpha,\beta)^L
=\non\\
& 2 \sum_{\substack{j=0 \\ j~{\rm even}}}^L 
{L \choose j}  \left( \cosh (\alpha +\beta ) \right)^{L-j} \left(\cosh ^2(\alpha +\beta )-\cc ^2 \sinh (2 \alpha ) \sinh (2 \beta )\right)^{j/2} \,.
\end{align}
Hence, 
\begin{align}
& \mathcal{F}^{(1)}_n(\alpha) 
+ 
\mathcal{F}^{(2)}_n(\alpha) = 
\cr 
&
\frac{2}{2^L} \sum_{k=0}^L 
  \sum_{\substack{j=0 \\ j~{\rm even}}}^L 
{L \choose j}  e^{- \alpha L } e^{\frac{- 2 i n k \pi}{L+1}}  
 \left(1+e^{\frac{2 i k \pi}{L+1}} e^{2\alpha} \right)^{L-j} \left((1+e^{\frac{2 i k \pi}{L+1}} e^{2\alpha})^2 -  \cc^2  (1-e^{\frac{4 i k \pi}{L+1}})(1-e^{4\alpha}) \right)^{j/2}\nonumber
 \\ 
&= 
 \frac{e^{- \alpha L }}{2^{L-1}} \sum_{k=0}^L 
  \sum_{l=0}^{L/2} 
  \sum_{a=0}^{L-2l} 
    \sum_{\substack{b_1, b_2 \geq 0  \\ b_1+b_2 \leq l} }
{L \choose 2l}{L-2l \choose a} {l \choose b_1,b_2,l-b_1-b_2} 
 e^{\frac{2 i (a+b_1 + 2 b_2-n) k \pi}{L+1}}  
e^{2 a \alpha}  (2  e^{2 \alpha}  )^{b_1}\nonumber\\
& \frac{(e^{4\alpha}(1-\cc^2)+\cc^2)^{b_2} }{(1-\cc^2 (1-e^{4\alpha}))^{b_1+b_2-l}}
=
 \frac{L+1}{e^{ \alpha L }}\frac{2}{2^{L}} 
  \sum_{l=0}^{L/2} 
    \sum_{\substack{b_1, b_2 \geq 0  \\ b_1+b_2 \leq l} }
    \frac{L!\,e^{2 n \alpha} 
(1-\cc^2 (1-e^{4\alpha}))^{l-b_1} 2^{b_1} }{(2l)! (n-b_1-2b_2)! {(L-2l-n+b_1+2b_2)!}}\nonumber\\
& {l \choose b_1,b_2,l-b_1-b_2}   
\left(\frac{1-\cc^2 (1-e^{-4\alpha})}{1-\cc^2 (1-e^{4\alpha})}\right)^{b_2} \,.
\end{align}

We further expand this expression to write it as a polynomial in $\cc$ (the deformation parameter) and we obtain (after proper re-arranging the sum)

\begin{align}
&\frac{(L+1) \,L!\, e^{2 \alpha  n}}{2^{L-1} e^{\alpha  L}}\non \\
&\sum\frac{l!\, 2^{l-b_3} \left(e^{-4 \alpha }-1\right)^t \left(e^{4 \alpha }-1\right)^{f-t}}{(2 l)!\, t!\, (b_2-t)! (l-b_3)! (f-t)! (t-b_2+b_3-f)! (n-2 b_2+b_3-l)! (2 b_2-b_3-l+L-n)!}\cc ^{2 f} ,
\end{align}
where we used the shortcut 
\begin{align}
&\sum\to \sum_{f=0}^{L/2} \sum_{l=f}^{L/2} \sum_{b_3=f}^l \sum_{b_2=0}^{b_3} \sum_{t=0}^{b_2}.
\end{align}
We used the software Mathematica 12.3 to further simplify this expression and we obtained
\begin{align}
\mathcal{F}^{(1)}_n(\alpha) 
+ 
\mathcal{F}^{(2)}_n(\alpha) = & \kappa\sum_{f=0}^{L/2} \frac{  \left(e^{4  \alpha }-1\right)^f \cc ^{2 f}    (L-f)!  }{f! n! \,e^{ \alpha  (L-2 n)}\,2^{L-n}}\non\\ &{_3\tilde{F}}_2\left(-f,\frac{1-n}{2},-\frac{n}{2};\frac{ L-n-2 f+1}{2},\frac{L-n-2 f+2}{2};\frac{1}{e^{4  \alpha }}\right),
\label{F12}
\end{align}
where $\kappa=(L+1)\sqrt{\pi } L$ and $_3\tilde{F}_2$ is the hypergeometric function regularized.
\\
Reporting the results \eqref{F3}, \eqref{F4} and \eqref{F12} into \eqref{Trrho0Pntilde} (where the terms $\mathcal{F}_n (\alpha-i \pi/2)$ can just be obtained by shifting the argument), we get :

\begin{align}
\text{Tr} (\rho(t=0)\tilde P_n)=&\frac{1}{ {2^L}}\frac{(-1)^n \cc ^L }{\cc ^{2 L}-1}\binom{L}{n} \left((-\cc  \tanh (\alpha ))^L-(-1)^{n} \tanh ^L(\alpha )+(-1)^n \cc ^L-(-1)^{L}\right)+\nonumber\\
&\frac{1}{( {2}\cosh \alpha)^L} \frac{1}{1-\cc^{2L}}\nonumber\\&\sum_{f=0}^{L/2} \frac{\sqrt{\pi } L  \left(1-e^{4  \alpha }\right)^f \cc ^{2 f} (-1)^{f+n} 2^{n-L}  (L-f)! e^{- \alpha  (L-2 n)} \left((-1)^{L+1} \cc ^L+(-1)^n\right)}{f! n!}\nonumber\\
&\, {_3\tilde{F}}_2\left(-f,\frac{1-n}{2},-\frac{n}{2};\frac{ L-n-2 f+1}{2},\frac{L-n-2 f+2}{2};\frac{1}{e^{4  \alpha }}\right).
\label{Trrho0Pn}
\end{align}

The three factors \eqref{trP}, \eqref{TrPnZj} and \eqref{Trrho0Pn} can now be gathered in the initial expression \eqref{eq:limZ} for $\lim \langle Z_j\rangle$. Performing the sum over $n$, we find that the contributions coming from $\mathcal{F}^{(3)}_n$ and $\mathcal{F}^{(4)}_n$ vanish.  
It remains to compute
\begin{align}
 \lim_{t\to \infty}\langle Z_j \rangle=&\frac{\sqrt{\pi } \left(1-\cc ^2\right) e^{  -\alpha L} }{{2^L}\left(\cc ^{2 L}-1\right)^2(\sinh (2 \alpha ) \text{csch}(\alpha ))^{L}}\non\\
 &\sum_{n=0}^L \sum_{f=0}^{L/2}\frac{ 2^n \left(e^{4 \alpha }-1\right)^f \cc ^{2 f} (2 n-L) e^{2 \alpha  n}  (L-f)! \left((-1)^n-(-\cc)^{L}\right)^2}{ (f+1)!  (n+1)!} \nonumber\\
&\, {_3\tilde{F}}_2\left(-f,\frac{1-n}{2},-\frac{n}{2};\frac{ L-n-2 f+1}{2},\frac{L-n-2 f+2}{2};\frac{1}{e^{4  \alpha }}\right) \,.
\label{almostfinalZ} 
\end{align}
This expression looks complicated at first sight, but we will now see that it is equivalent to the expression \eqref{Q0exact} in the main text. Both expressions contain a prefactor $\frac{1-\cc^2}{(1-\cc^{2L})^2}$ which we therefore omit in the following, and compare the remaining polynomials in $\cc$ order by order. Starting from \eqref{Q0exact}, the remaining polynomial takes the form 
\begin{align}
\left(1-\cc ^2\right) \tanh (\alpha ) \left(1+\cc ^{2 L}-2 \cc ^L \tanh ^{L-2}\alpha \right),
\label{Mariusnodenominator}
\end{align}
in particular the exponents of $\cc$ that gives a contribution different from 0 are $0, 2, L, 2L, L+2, 2L+2$. We shall now demonstrate that all other powers indeed vanish in the polynomial associated with expression \eqref{almostfinalZ}.
In \eqref{almostfinalZ}, the coefficients are $2f, 2f+L, 2f+2L$, so the only non-zero contribution should come from $f=0,1,L/2$, the last one only for $L$ even.
\\
By direct computation, we found that to obtain \eqref{Mariusnodenominator} it is enough to sum the contribution of $f=0$ and $f=1$. Let us show that the other terms vanish.
First, we consider the contribution of $f=L/2$ in \eqref{almostfinalZ}. This is proportional to
\begin{align}
\sum_{n=0}^L \frac{(L-2 n) e^{2 \alpha  n} \left((-1)^n-(-\cc)^{L}\right)^2\sin (\pi  n)}{n } \, \, _2F_1\left(-\frac{L}{2},-\frac{n}{2};1-\frac{n}{2};\frac{1}{e^{4 \alpha }}\right)\label{F21}.
\end{align}
Since the function $ _2F_1\left(-\frac{L}{2},-\frac{n}{2};1-\frac{n}{2};e^{-4 \alpha }\right)$ is finite for $n$ odd, \eqref{F21} vanishes due to $\sin (n \pi)$,  while for $n$ even:
\begin{align}
\, _2F_1\left(-\frac{L}{2},-\frac{n}{2};1-\frac{n}{2};x\right)=\sum_{k=0}^{L/2} \frac{(-1)^k n \binom{L/2}{k} x^k}{n-2 k},
\end{align}
and if we substitute this into \eqref{F21}, we get
\begin{align}
\sum_{k=0}^{L/2}\sum_{n=0}^L \frac{(-1)^k (L-2 n) \sin (\pi  n) \binom{\frac{L}{2}}{k} e^{2 \alpha  (n-2 k)}}{n-2 k}=\sum_{n=0}^L i^n (L-2 n) \binom{\frac{L}{2}}{\frac{n}{2}}=0,
\end{align}
where since $\sin (n \pi)$ is zero, we need to only keep the singular term.
\\
It remains to show that all the terms with $f>1$ do not contribute.  
Removing the irrelevant terms and considering $e^{-\alpha}=z$, we should prove that the following term vanishes
\begin{align}
\sum_{n=0}^{L}\frac{(L-2 n) z^{L-2 n} \, }{ (n+1)!  (1-2 f+L-n)!} {_3F_2}\left(-f,\frac{1-n}{2},-\frac{n}{2};\frac{ L-n-2 f+1}{2},\frac{L-n-2 f+2}{2};z^4\right).
\end{align}
Expanding as a series in $z$ and re-shifting the sum over $n$ we obtain
\begin{align}
\sum_{m=0}^\infty \sum_{n=-2m}^{L-2m} \frac{(-1)^{m} \binom{f}{m} (L-4 m-2 n) z^{L-2 n}}{(n+1)! \, (1-2 f+L-n)!}=\sum_{m=0}^\infty \sum_{n=0}^{L} \frac{(-1)^{m} \binom{f}{m} (L-4 m-2 n) z^{L-2 n}}{(n+1)! \, (1-2 f+L-n)!}.
\end{align}
We can now sum over $m$ and we are left with
\begin{align}
\sum_{n=0}^L (L-2 n) \, _1F_0(-f;;1)+4 f \, _1F_0(1-f;;1).
\label{lasteqhyperg}
\end{align}
Considering that
\begin{align}
_1F_0(a;;x)=(1-x)^{-a},
\end{align}
each terms of \eqref{Q0exact} are zero for $f>1$ as stated at the beginning.
\\
To summarize, we proved that \eqref{eq:limZ} is equivalent to the expression \eqref{Q0exact} given in the main text.

\chapter{Integrable deformations of AdS}
\label{adsappendixdef}
In this appendix, we show that the models 6-vertex B and 8-vertex B obtained in chapter \ref{6and8Vmodel} (by starting from an ansatz of 8-Vertex type) correspond to integrable deformations of the  $\mathrm{AdS}_{2}$ and  $\mathrm{AdS}_{3}$ models.
\\
The $S$-matrices of $\mathrm{AdS}_{2}$ and  $\mathrm{AdS}_{3}$ integrable models contain separate same chirality $4\times 4$ blocks that satisfy the Yang-Baxter equation by themselves. We proved that the matrix part of these blocks fit into our classification. Rather than compare $R$-matrices, we will compare the   Hamiltonians. The dependence on the spectral parameter $u$ (rapidity) of the AdS/CFT Hamiltonians is in the Zhukovski variable $x^\pm$ defined by
\begin{align}
& u = \frac{1}{2}\Big[x^+ + \frac{1}{x^+} +x^- + \frac{1}{x^-}\Big],&& \frac{x^+}{x^-}=e^{i p}.
\end{align}

\section*{AdS$_3$} For  $\mathrm{AdS}_{3}$, the Hamiltonian of particles with the same chirality \cite{Sfondrini:2014via,Borsato:2014hja} can be embedded into both the 6-vertex B and 8-vertex B models. For the spin chain frame \cite{Borsato:2013qpa}, the Hamiltonian can be found substituting in the general ansatz  \eqref{generalHamiltonian} 
\begin{align}
&h_2=h_6=h_7=h_8=0,\\
& h_3=\frac{\dot{x}^-}{x^--x^+}, && h_4=\frac{\dot{x}^+}{x^--x^+}, \\
&h_1 = -\frac{1}{2}(h_3+h_4),
&& h_5=\frac{1}{4}(h_3+h_4),
\end{align}
and we take the positive sign in the square root in $h_3$, $h_4$.
\\We now see that there are two possible types of deformation. First, we can match this model with our 6-vertex B type, which leaves us with a continuous family of deformations since we can add arbitrary functions of the spectral parameter to all of the components. An interesting fact is that we can deform independently the blocks with the same chirality. Second, we can also embed the $AdS_3$ Hamiltonian in our 8-vertex B model \eqref{8vham}. This gives a one-parameter elliptic deformation of the $\mathrm{AdS}_3$ model. The embedding is given, for the spin chain frame, by

\begin{align}
& \cot\eta = \frac{1}{2}\frac{\dot{x}^+ + \dot{x}^-}{x^+-x^-},
&&k=0.
\end{align}
This is a novel elliptic deformation.

\paragraph{AdS$_2$} The massive sector of the $\mathrm{AdS}_2\times S^2 \times T^6$ string sigma model \cite{Hoare:2014kma, Hoare:2015kla} is of  8-vertex B  type. To make this explicitly, we should use a local basis transformation to make $h_8=-h_7=k$, and the non-zero components of the Hamiltonian are parametrized as
\begin{align}
& \cot \eta=\frac{k}{2} e^{\frac{-i p}{2} }\left(x^+-\frac{1}{x^-}\right)\frac{ 1+e^{\frac{i p}{2}} }{ 1-e^{\frac{i p}{2}}}.
\end{align}
We conclude that this integrable model only admits a one-parameter deformation by taking $k$ to be non-zero.

\end{appendices}

\printthesisindex 

\end{document}